\pdfoutput=1

\documentclass[11pt]{article}

\usepackage{lmodern}
\usepackage[utf8]{inputenc}
\usepackage[T1]{fontenc}
\usepackage{amsmath}
\usepackage{amssymb}
\usepackage{mathtools}
\usepackage[english]{babel}
\usepackage{url}
\usepackage{booktabs}
\usepackage{longtable}
\usepackage{cellspace}
\usepackage{makecell}
\usepackage{graphicx}
\usepackage{needspace}
\usepackage{xspace}
\usepackage{setspace}
\usepackage{textcase}
\usepackage{multirow}
\usepackage{subcaption}
\usepackage{myjheppub}
\usepackage{setspace}
\usepackage{bm}
\usepackage{scrextend}

\usepackage[format=plain,labelfont=bf,textfont=it]{caption}

\usepackage[dvipsnames]{xcolor}
\newcommand{\RedTextHTML}{A00000}
\newcommand{\BlueTextHTML}{0B476D}
\definecolor{RedText}{HTML}{\RedTextHTML}
\definecolor{BlueText}{HTML}{\BlueTextHTML}
\definecolor{FittingYellow}{HTML}{ADAB11}
\definecolor{BGColorL}{HTML}{F3F3F3}
\definecolor{BGColorH}{HTML}{D9D9D9}

\usepackage{enumitem}
\setlist[description]{%
  itemsep=0pt,               
  font={\normalfont\scshape}, 
}

\usepackage{mmacells} \mmaSet{ morefv={gobble=2}, 
  moredefined={c, a, intConst, intConstBar, g, gBar, gHat, gBarHat, e, ep, b,
    bp, zeta, s, tau, vz, gz, fz, fBarz, bCoeff, zIntegrate, CSimplify,
    CModWeight, CComplexConj, CSort, DiLimCSimplify, TriLimCSimplify,
    TetLimCSimplify, DiHolMomConsId, DiAHolMomConsId, TriHolMomConsId,
    TriAHolMomConsId, TetHolMomConsId, TetAHolMomConsId, BoxHolMomConsId,
    BoxAHolMomConsId, KiteHolMomConsId, KiteAHolMomConsId, CHolCR, CAHolCR,
    DiCSimplify, TriCSimplify, TetCSimplify, TriFay, CSieveDecomp, CBasis,
    CConvertToNablaE, CConvertFromNablaE, CLaurentPoly, CListHSRs, CCheckConv,
    nablaE, nablaEp, nablaBarE, nablaBarEp, nablaB, nablaBp, nablaBarBBar,
    nablaBarBpBar, nablaA, nablaBarA, basisExpandG, momSimplify, useIds, divHSR,
    diHSR, triHSR, tri2ptHSR, tri3ptHSR, tri3ptFayHSR, usey, verbose, basis,
    addIds, CSimplifyArgs, tetrule, prettify}, leftmargin=3.7em, labelsep=0.2em
}

\ExplSyntaxOn
\makeatletter

\newcounter { mmacellsIn }
\newcounter { mmacellsOut }

\clist_put_right:Nn \l_mmacells_fv_keyval_clist
  {
    formatcom* =
      {
        \cs_set:Npx \@inout{\bool_if:NTF \l_mmacells_intype_bool { In } { Out }}
        \setcounter{mmacells\@inout}{\mmaCellIndex}
        \addtocounter{mmacells\@inout}{-1}
        \refstepcounter { mmacells\@inout }
        \cs_if_eq:NNF \lst@label \@empty
          {
            \cs_if_eq:NNTF \lst@@caption \@empty
              {
                \bool_if:NT \l_mmacells_indexed_bool
                  {
                    \cs_set:Npx \@currentlabelname
                      {
                        \bool_if:NTF \l_mmacells_intype_bool { In } { Out }
                        [ \int_use:N \g_mmacells_index_int ]
                      }
                  }
              }
              { \cs_set:Npx \@currentlabelname { \lst@@caption } }
            \label { \lst@label }
          }
      }
  }

\makeatother
\ExplSyntaxOff

\newcommand{\mmainrarrow}{
  \hspace{-1em}
  \tikz[line width=0.7pt,baseline=-0.65ex]{
    \draw[line cap=round,arrows={-Straight Barb[round,length=2pt,width=4pt]}] (0,0) -- (1.7ex,0);
  }
  \hspace{-1em}
}

\newcommand{\mmaoutrarrow}{
  \hspace{-1em}
  \tikz[line width=0.55pt,baseline=-0.58ex]{
    \draw[arrows={-Straight Barb[round,length=2pt,width=4pt]}] (0,0) -- (1.7ex,0);
  }
  \hspace{-1em}
}

\newcommand{\mmaoverline}{\tikz[line width=0.6pt,baseline=0.1ex]{
    \draw[line cap=round] (0,0) -- (1.7ex,0);
  }}

\newcommand{\mmabigso}{\raisebox{-0.1ex}{\scalebox{1.2}{$\bm{[}$}}}
\newcommand{\mmabigsc}{\raisebox{-0.1ex}{\scalebox{1.2}{$\bm{]}$}}}

\newcommand{\mmabigco}{\raisebox{-0.1ex}{\scalebox{0.8}[1.2]{$\bm{\{}$}}}
\newcommand{\mmabigcc}{\raisebox{-0.1ex}{\scalebox{0.8}[1.2]{$\bm{\}}$}}}

\newcommand{\mmanothing}{\te{\scalebox{1}[1.1]{$\bm{\{\}}$}}}

\mmaDefineMathReplacement{=}{\,\,=\,\,}
\mmaDefineMathReplacement{+}{\,\,+\,\,}
\mmaDefineMathReplacement{==}{\,==\,}
\newcommand{\mmam}{\,\,$-$\,\,}

\newcommand{\mmaoutref}[1]{{\small\color{mmaLabel}\sffamily Out[\ref{#1}]}}
\newcommand{\mmainref}[1]{{\small\color{mmaLabel}\sffamily In[\ref{#1}]}}

\newcommand{\temprefa}{}
\newcommand{\temprefb}{}
\newcommand{\temprefc}{}

\newcommand{\verbcent}[1]{\hspace{0.5ex}#1\hspace{0.5ex}}

\newcommand{\mma}[1]{{\bfseries\ttfamily\small #1}}
\newcommand{\mmaHeading}[1]{{\bfseries\ttfamily\large #1}}
\newcommand{\mmawarning}[2]{\mma{#1::}\textcolor{mmaString}{\mma{#2}}}
\newcommand{\mmastr}[1]{{\bfseries\ttfamily\small\textcolor{mmaString}{#1}}}

\usepackage{listings}
\newcommand{\ac}[1]{\textsc{\MakeLowercase{#1}}}
\newcommand{\acp}[1]{\textsc{\MakeLowercase{#1}}{\footnotesize s}}

\usepackage{tikz}
\usetikzlibrary{intersections}
\usetikzlibrary{decorations.markings}
\usetikzlibrary{arrows.meta}
\usetikzlibrary{calc}
  
\tikzset{
  gluon/.style={decoration={coil,
      pre length=2pt,post length=2pt,segment length=3pt},decorate},
  vertex/.style={circle,inner sep=1.5},
  vertexdot/.style={circle,draw,fill,inner sep=0.7},
  label/.style={fill=white,font=\footnotesize},
  mlabel/.style={outer sep=2,font=\footnotesize},
  biglabel/.style={fill=white,font=\normalsize},
  tlabel/.style={fill=white,font=\footnotesize,inner sep=0pt},
  edge/.style={line width=0.5},
  dedge/.style={line width=0.5,decoration=
    {markings,mark=at position #1 with {\arrow[scale=1.3]{latex}}},
    postaction=decorate},
  dedge/.default=0.8,
  mgf/.style={baseline=-0.6ex}}

\makeatletter
\global\let\tikz@ensure@dollar@catcode=\relax
\makeatother

\usepackage[
  style=phys,
  eprint=true,
  biblabel=brackets,
  chaptertitle=false,
  pageranges=false,
  defernumbers=true,
  doi=false,
  backend=biber,
  bibencoding=utf8,
  ]{biblatex}
\bibliography{literature.bib}

\DeclareFieldFormat{doi}{}

\newcommand{\mmapackage}{\NoCaseChange{\texttt{ModularGraphForms}}\xspace}

\newcommand{\mycomment}[1]{}

\let\originalleft\left
\let\originalright\right
\renewcommand{\left}{\mathopen{}\mathclose\bgroup\originalleft}
\renewcommand{\right}{\aftergroup\egroup\originalright}

\makeatletter
\providecommand*{\shuffle}{%
  \mathbin{\mathpalette\myshuffle@{}}%
}
\newcommand*{\myshuffle@}[2]{%
  \sbox0{$#1\vcenter{}$}%
  \kern .15\ht0 
  \rlap{\vrule height .25\ht0 depth 0pt width 2.5\ht0}%
  \raise.1\ht0\hbox to 2.5\ht0{%
    \vrule height 1.75\ht0 depth -.1\ht0 width .17\ht0 %
    \hfill
    \vrule height 1.75\ht0 depth -.1\ht0 width .17\ht0 %
    \hfill
    \vrule height 1.75\ht0 depth -.1\ht0 width .17\ht0 %
  }%
  \kern .15\ht0 
}
\makeatother

\DeclareFontFamily{U}{mathx}{\hyphenchar\font45}
\DeclareFontShape{U}{mathx}{m}{n}{
      <5> <6> <7> <8> <9> <10>
      <10.95> <12> <14.4> <17.28> <20.74> <24.88>
      mathx10
      }{}
\DeclareSymbolFont{mathx}{U}{mathx}{m}{n}
\DeclareFontSubstitution{U}{mathx}{m}{n}
\DeclareMathAccent{\widecheck}{0}{mathx}{"71}

\newcommand{\mfp}{\mathfrak{p}}

\renewcommand{\Re}{\operatorname{Re}}
\renewcommand{\Im}{\operatorname{Im}}

\newcommand{\te}{\text}

\DeclareMathOperator*{\res}{Res}

\newcommand{\eqspace}{\hphantom{{}={}}}

\newcommand{\dd}{\!\mathrm{d}}
\newcommand{\ddd}{\mathrm{d}}

\newcommand{\RR}{\mathbb{R}}
\newcommand{\NN}{\mathbb{N}}
\newcommand{\CC}{\mathbb{C}}

\newcommand{\ZZ}{\mathbb{Z}}

\newcommand{\SL}{\mathrm{SL}}

\newcommand{\EE}{\mathrm{E}}
\newcommand{\GG}{\mathrm{G}}
\newcommand{\GGhat}{\hspace{0.1ex}\widehat{\hspace{-0.1ex}
    \raisebox{0pt}[0.65em]{$\GG$}}}
\newcommand{\GGhatb}{\hspace{0.1ex}\widehat{\hspace{-0.1ex}
    \raisebox{0pt}[0.75em]{$\overline{\raisebox{0pt}[0.7em]{$\GG$}}$}}}
\newcommand{\mmaGGhat}{$\hat{\text{G}}$}
\newcommand{\mmaGGhatb}{$\hat{\raisebox{0pt}[2ex]{
      $\overset{\raisebox{0pt}[-0.2em]{\text{\_}}}{\text{G}}$\hspace{1.5ex}}}$\hspace{-1.4ex}}

\newcommand{\realCusp}{\mathrm{S}}

\newcommand{\BB}{\mathrm{B}}

\newcommand{\smatrix}[1]{\begin{smallmatrix}#1\end{smallmatrix}}
\newcommand{\sbmatrix}[1]{\left[\begin{smallmatrix}#1\end{smallmatrix}\right]}
\newcommand{\spmatrix}[1]{\left(\begin{smallmatrix}#1\end{smallmatrix}\right)}

\DeclareMathOperator{\KN}{KN}

\newcommand{\II}{\mathcal{I}}

\newcommand{\ap}{\alpha'}

\newcommand{\esvtau}[1]{{\cal E}^\mathrm{sv}
  \!\big[\begin{smallmatrix}#1\end{smallmatrix};\tau\big]}

\newcommand{\esvStau}[1]{{\cal E}^\mathrm{sv}
  \!\big[\begin{smallmatrix}#1\end{smallmatrix};-\tfrac{1}{\tau}\big]}

\newcommand{\betasvtau}[1]{
  \beta^\mathrm{sv}\! \big[\begin{smallmatrix}#1\end{smallmatrix}; \tau\big]}

\newcommand{\betasvStau}[1]{\beta^\mathrm{sv}
  \! \big[\begin{smallmatrix}#1\end{smallmatrix};-\tfrac{1}{\tau}\big]}

\newcommand{\vmsmash}[1]{\raisebox{0pt}[0pt]{$#1$}}

\newcommand{\nablab}{\overline{\nabla}{}}

\newcommand{\nabladg}{\nabla_{\hspace{-0.5ex}0}}
\newcommand{\nabladgb}{\overline{\nabla}_{\hspace{-0.5ex}0}}

\newcommand{\piimtau}{\bigg(\!\frac{\pi}{\tau_{2}}\!\parbox{4pt}{$\bigg)$}}
\newcommand{\imtaupi}{\bigg(\!\frac{\tau_{2}}{\pi}\!\parbox{4pt}{$\bigg)$}}
\newcommand{\piimtauil}{\big(\!\frac{\pi}{\tau_{2}}\!\parbox{4pt}{$\big)$}}

\DeclareMathOperator*{\esumsym}{\hspace{0.5em}\sum
  \raisebox{-0.5em}{\makebox[0.2em]{$\scriptstyle \mathrm{E}$}}\hspace{0.3em}}
\newcommand{\esum}{\hspace{-0.5em}\esumsym}

\newcommand{\noblock}{\varnothing}

\newcommand{\cform}[1]{\operatorname{\mathcal{C}\hspace{-3pt}}\big[
  \protect\begin{smallmatrix}#1\protect\end{smallmatrix}\parbox{4pt}{$\big]$}}

\newcommand{\aform}[1]{\operatorname{\mathcal{A}\hspace{-3pt}}\big[
  \protect\begin{smallmatrix}#1\protect\end{smallmatrix}\parbox{4pt}{$\big]$}}

\newcommand{\mmaincform}[1]{c\ensuremath{\mmabigso\protect
    \begin{smallmatrix}#1\protect\end{smallmatrix}\parbox{4pt}{$\mmabigsc$}}}
\newcommand{\mmaoutcform}[1]{C\ensuremath{\big[
  \protect\begin{smallmatrix}#1\protect\end{smallmatrix}\parbox{4pt}{$\big]$}}}
\newcommand{\mmainaform}[1]{a\ensuremath{\mmabigso\protect
    \begin{smallmatrix}#1\protect\end{smallmatrix}\parbox{4pt}{$\mmabigsc$}}}
\newcommand{\mmaoutaform}[1]{A\ensuremath{\big[
  \protect\begin{smallmatrix}#1\protect\end{smallmatrix}\parbox{4pt}{$\big]$}}}

\newcommand{\mmainintconst}[1]{intConst\ensuremath{\mmabigso\protect
    \begin{smallmatrix}#1\protect\end{smallmatrix}\parbox{4pt}{$\mmabigsc$}}}
\newcommand{\mmaoutintconst}[1]{intConst\ensuremath{\big[\protect
    \begin{smallmatrix}#1\protect\end{smallmatrix}\parbox{4pt}{$\big]$}}}

\newcommand{\cformtri}[3]{\operatorname{\mathcal{C}\hspace{-3pt}}\big[\!
    {\setlength{\tabcolsep}{1pt}\setlength\arrayrulewidth{0.8pt}
      \begin{tabular}{c|c|c}
        $\smatrix{#1}$ & $\smatrix{#2}$ & $\smatrix{#3}$
      \end{tabular}
    }
    \!\parbox{4pt}{$\big]$}}

\newcommand{\mmaincformtri}[3]{c\ensuremath{\mmabigso
    \protect\begin{smallmatrix}#1\protect\end{smallmatrix}\text{,}
    \protect\begin{smallmatrix}#2\protect\end{smallmatrix}\text{,}
    \protect\begin{smallmatrix}#3\protect\end{smallmatrix}\parbox{4pt}{$\mmabigsc$}}}

\newcommand{\mmaoutcformtri}[3]{C\ensuremath{\big[\!
    {\setlength{\tabcolsep}{1pt}\setlength\arrayrulewidth{0.8pt}
      \begin{tabular}{c|c|c}
        $\smatrix{#1}$ & $\smatrix{#2}$ & $\smatrix{#3}$
      \end{tabular}
    }
    \!\parbox{4pt}{$\big]$}}}

\newcommand{\mmaoutintconsttri}[3]{intConst\ensuremath{\mmabigso\!
    {\setlength{\tabcolsep}{1pt}\setlength\arrayrulewidth{0.8pt}
      \begin{tabular}{c|c|c}
        $\smatrix{#1}$ & $\smatrix{#2}$ & $\smatrix{#3}$
      \end{tabular}
    }
    \!\parbox{4pt}{$\mmabigsc$}}}

\newcommand{\cformbox}[4]{\operatorname{\mathcal{C}\hspace{-3pt}}\big[\!
    {\setlength{\tabcolsep}{1pt}\setlength\arrayrulewidth{0.8pt}
      \begin{tabular}{c|c|c|c}
        $\smatrix{#1}$ & $\smatrix{#2}$ & $\smatrix{#3}$ & $\smatrix{#4}$
      \end{tabular}
    }
    \!\parbox{4pt}{$\big]$}}

\newcommand{\mmaincformbox}[4]{c\ensuremath{\mmabigso
    \protect\begin{smallmatrix}#1\protect\end{smallmatrix}\text{,}
    \protect\begin{smallmatrix}#2\protect\end{smallmatrix}\text{,}
    \protect\begin{smallmatrix}#3\protect\end{smallmatrix}\text{,}
    \protect\begin{smallmatrix}#4\protect\end{smallmatrix}\parbox{4pt}{$\mmabigsc$}}}

\newcommand{\mmaoutcformbox}[4]{C\ensuremath{\big[\!
    {\setlength{\tabcolsep}{1pt}\setlength\arrayrulewidth{0.8pt}
      \begin{tabular}{c|c|c|c}
        $\smatrix{#1}$ & $\smatrix{#2}$ & $\smatrix{#3}$ & $\smatrix{#4}$
      \end{tabular}
    }
    \!\parbox{4pt}{$\big]$}}}
  
\newcommand{\cformkite}[5]{\operatorname{\mathcal{C}\hspace{-3pt}}\mmabigso\!
    {\setlength{\tabcolsep}{1pt}\setlength\arrayrulewidth{0.8pt}
      \begin{tabular}{c|c||c|c||c}
        $\smatrix{#1}$ & $\smatrix{#2}$ & $\smatrix{#3}$ & $\smatrix{#4}$
        & $\smatrix{#5}$
      \end{tabular}
    }
    \!\parbox{4pt}{$\mmabigsc$}}

\newcommand{\mmaincformkite}[5]{c\ensuremath{\mmabigso
    \protect\begin{smallmatrix}#1\protect\end{smallmatrix}\text{,}
    \protect\begin{smallmatrix}#2\protect\end{smallmatrix}\text{,}
    \protect\begin{smallmatrix}#3\protect\end{smallmatrix}\text{,}
    \protect\begin{smallmatrix}#4\protect\end{smallmatrix}\text{,}
    \protect\begin{smallmatrix}#5\protect\end{smallmatrix}\parbox{4pt}{$\mmabigsc$}}}

\newcommand{\mmaoutcformkite}[5]{C\ensuremath{\big[\!
    {\setlength{\tabcolsep}{1pt}\setlength\arrayrulewidth{0.8pt}
      \begin{tabular}{c|c||c|c||c}
        $\smatrix{#1}$ & $\smatrix{#2}$ & $\smatrix{#3}$ & $\smatrix{#4}$
        & $\smatrix{#5}$
      \end{tabular}
    }
    \!\parbox{4pt}{$\big]$}}}

\newdimen\aboverulesepbuffer
\newdimen\belowrulesepbuffer

\newcommand{\cformtet}[6]{
  \operatorname{\mathcal{C}\hspace{-3pt}}\Bigg[\!
  {\setlength{\aboverulesep}{-1pt}
    \setlength{\belowrulesep}{0pt}
    \setlength{\tabcolsep}{2pt}
    \setlength\arrayrulewidth{0.8pt}
    \begin{tabular}{c||c||c}
      $\begin{smallmatrix}#1\end{smallmatrix}$ &
      $\begin{smallmatrix}#2\end{smallmatrix}$ &
      $\begin{smallmatrix}#3\end{smallmatrix}$ \\[0.7ex]
      \cmidrule(l{2pt}r{5pt}){1-1}\cmidrule(l{2pt}r{5pt}){2-2}
      \cmidrule(l{2pt}r{2pt}){3-3}
      \rule[1.5ex]{0pt}{1ex}$\begin{smallmatrix}#4\end{smallmatrix}$ &
      $\begin{smallmatrix}#5\end{smallmatrix}$ &
      $\begin{smallmatrix}#6\end{smallmatrix}$
    \end{tabular}}
  \!\parbox{4pt}{$\Bigg]$}}

\newcommand{\mmaincformtet}[6]{c\ensuremath{\mmabigso
    \protect\begin{smallmatrix}#1\protect\end{smallmatrix}\text{,}
    \protect\begin{smallmatrix}#2\protect\end{smallmatrix}\text{,}
    \protect\begin{smallmatrix}#3\protect\end{smallmatrix}\text{,}
    \protect\begin{smallmatrix}#4\protect\end{smallmatrix}\text{,}
    \protect\begin{smallmatrix}#5\protect\end{smallmatrix}\text{,}
    \protect\begin{smallmatrix}#6\protect\end{smallmatrix}\parbox{4pt}{$\mmabigsc$}}}

\newcommand{\mmaoutcformtet}[6]{C\ensuremath{\Bigg[
    {\setlength{\aboverulesep}{-1pt}
    \setlength{\belowrulesep}{0pt}
    \setlength{\tabcolsep}{2pt}
    \setlength\arrayrulewidth{0.8pt}
    \begin{tabular}{c|c|c}
      $\begin{smallmatrix}#1\end{smallmatrix}$ &
      $\begin{smallmatrix}#2\end{smallmatrix}$ &
      $\begin{smallmatrix}#3\end{smallmatrix}$ \\[0.7ex]
      \midrule
      \rule[1.5ex]{0pt}{1ex}$\begin{smallmatrix}#4\end{smallmatrix}$ &
      $\begin{smallmatrix}#5\end{smallmatrix}$ &
      $\begin{smallmatrix}#6\end{smallmatrix}$
    \end{tabular}}
    \parbox{4pt}{$\Bigg]$}}}

\usepackage[
  pdfusetitle,
  hidelinks,
  bookmarksnumbered,
  bookmarksdepth=3,
  colorlinks=true,
  pdfstartview=FitV,
  pdfpagemode=UseNone,
  bookmarksopen=true,
  bookmarksopenlevel=1,
  hypertexnames=true,
  pdfhighlight=/O,
  linktocpage=true,
  allcolors=BlueText,
  breaklinks=true,
  pdfencoding=unicode
  ]{hyperref}

\begin{document}

\frenchspacing
\raggedbottom

\numberwithin{equation}{section}


\title{Basis Decompositions and a Mathematica Package\\for Modular Graph Forms}

\author{Jan E.\ Gerken}

\affiliation{Max-Planck-Institut f\"ur Gravitationsphysik,
Albert-Einstein-Institut,
DE-14476 Potsdam, Germany}

\emailAdd{jan.gerken@aei.mpg.de}

\date{\today}

\abstract{Modular graph forms (\acp{MGF}) are a class of non-holomorphic modular
  forms which naturally appear in the low-energy expansion of closed-string
  genus-one amplitudes and have generated considerable interest from pure
  mathematicians. \acp{MGF} satisfy numerous non-trivial algebraic- and
  differential relations which have been studied extensively in the literature
  and lead to significant simplifications. In this paper, we systematically
  combine these relations to obtain basis decompositions of all two- and
  three-point \acp{MGF} of total modular weight $w+\bar{w}\leq12$, starting from
  just two well-known identities for banana graphs. Furthermore, we study
  previously known relations in the integral representation of \acp{MGF},
  leading to a new understanding of holomorphic subgraph reduction as Fay
  identities of Kronecker--Eisenstein series and opening the door towards
  decomposing divergent graphs. We provide a computer implementation for the
  manipulation of \acp{MGF} in the form of the \texttt{Mathematica} package
  \mmapackage which includes the basis decompositions obtained.}

\maketitle{}

\section{Introduction}
\label{sec:introduction}
Scattering amplitudes in string theory have in recent years experienced a rise
in interest due to the rich mathematical structures appearing in their
calculation and their close relations to field-theory amplitudes. Tree-level
string amplitudes at genus zero are by now well under control and many powerful
results have been obtained for genus-one amplitudes as well.

Closed-string genus-one amplitudes are given by integrals of correlators in the
worldsheet conformal field theory (\ac{CFT}) over the moduli space of punctured
tori. In this paper, we systematically study relations between modular graph
forms (\acp{MGF}), a class of functions of the modular parameter
$\tau=\tau_{1}+i\tau_{2}$, $\tau_{1},\tau_{2}\in\RR$ with $\tau_{2}>0$ which
make the computation of the low-energy expansion of the integrals over the
punctures algorithmic and have been studied widely in the
literature~\cite{green2000, green2008, dhoker2015, dhoker2019,
  basu2016,dhoker2015a, zerbini2016, dhoker2016a, basu2016b, basu2016a,
  dhoker2016, kleinschmidt2017, brown2017, brown2017b, dhoker2017, basu2017a,
  brown2017a, basu2017b, gerken2019f, gerken2019e, dhoker2019a, dorigoni2019a,
  dhoker2019b, dhoker2019c, basu2019, zagier2019, berg2019a, hohenegger2019,
  gerken2019d, gerken2020c, basu2020}. Genus-two generalizations of these
techniques were studied in \cite{dhoker2014, dhoker2015b, dhoker2017a,
  dhoker2018, basu2019b}. The low-energy expansion is an expansion in powers of
the inverse string tension $\ap$ or, equivalently, in the Mandelstam variables
\begin{align}
  s_{ij}&=-\frac{\alpha'}{2}k_{i}\cdot k_{j}\,,
  \label{eq:15}
\end{align}
where the $k_{i}$ are the momenta of the asymptotic string states. The resulting
\acp{MGF} $\mathcal{C}_{\Gamma}(\tau)$, introduced in \cite{dhoker2016a}, are
non-holomorphic modular forms, i.e. they transform as
\begin{align}
  \mathcal{C}_{\Gamma}\left(\frac{\alpha\tau+\beta}{\gamma\tau+\delta}\right)
  =(\gamma\tau+\delta)^{a}(\gamma\bar{\tau}+\delta)^{b}
  \mathcal{C}_{\Gamma}(\tau)\,,\qquad
  \begin{pmatrix}
    \alpha & \beta\\
    \gamma & \delta
  \end{pmatrix}
  \in\SL(2,\ZZ)\,,
  \label{eq:47}
\end{align}
where $a$ is the holomorphic modular weight and $b$ is the antiholomorphic
modular weight and the \acp{MGF} are labeled by Feynman-like decorated graphs
$\Gamma$. Similarly to how Feynman diagrams can be translated into nested
integrals over loop momenta, we can read off the representation of \acp{MGF} in
terms of nested lattice sums from the graph $\Gamma$ (the momenta are discrete
since the torus is compact). A second representation of \acp{MGF} is in terms of
torus integrals of Jacobi forms, corresponding to the position-space
representation of $\Gamma$.

\acp{MGF} satisfy many non-trivial relations which are hard to see in the
lattice sum- (or integral-) representation, e.g. the \ac{MGF} that was denoted
by $C_{1,1,1}$ in \cite{zagier} satisfies the relation
\begin{align}
  C_{1,1,1}&=\imtaupi^{3}\sum_{\substack{(m_{1},n_{1})\in\ZZ^{2}\\(m_{2},n_{2})\in\ZZ^{2}}}'
  \frac{1}{|m_{1}\tau+n_{1}|^{2}|m_{2}\tau+n_{2}|^{2}|(m_{1}+m_{2})\tau+n_{1}+n_{2}|^{2}}\nonumber\\
  &=\imtaupi^{3}\sum_{(m,n)\in\ZZ^{2}}' \frac{1}{|m\tau+n|^{6}}
  +\sum_{r=1}^{\infty}\frac{1}{r^{3}}\label{eq:176}\\
  &=\EE_{3}+\zeta_{3}\,,\nonumber
\end{align}
where the prime on the sums indicates that we omit the origin from the summation
domains $\ZZ^{2}$ and the non-holomorphic Eisenstein series $\EE_{s}$ are
defined in \eqref{eq:51}. We use the notation
\begin{align}
  \zeta_{k}=
  \sum_{n=1}^{\infty}
  \frac{1}{n^k},
  \qquad k\in\NN\,,\qquad k\geq2\,,
  \label{eq:28}
\end{align}
for zeta values. Identities of this type were studied
extensively in the literature \cite{dhoker2015, dhoker2019, basu2016b,
  dhoker2019a, dhoker2016a, dhoker2016, basu2016a, basu2017b,
  gerken2019f}. Although \acp{MGF} also satisfy many non-trivial differential
equations, we will focus here mainly on algebraic relations.

\subsection{Summary of results}
In this paper, we derive relations between \acp{MGF} of the form \eqref{eq:176}
by systematically applying known and new manipulation techniques to a large
class of \acp{MGF}. In particular, we extend known techniques for two- and
three-point graphs to a complete treatment of four-point graphs and show that in
the integral representation, the well-known technique of holomorphic subgraph
reduction (\ac{HSR}) \cite{dhoker2016a, gerken2019f} is equivalent to Fay
identities of the Kronecker--Eisenstein series. This yields a more compact and
iterative procedure for performing \ac{HSR} on higher-point graphs than was
previously available in the literature. Furthermore, we give a first systematic
discussion of divergent \acp{MGF} and show how these can be interpreted as
arising form kinematic poles in torus integrals.

By applying these manipulations extensively to all two- and three-point graphs
of total modular weight $a+b\leq12$, we find basis decompositions for all these
graphs starting from just the two well-known decompositions of the banana graphs
$D_{3}=C_{1,1,1}$ and $D_{5}$, cf.~\eqref{eq:396} and \eqref{eq:397}. The
structure of the basis (and in particular its dimension) agrees with the
predictions made previously in the literature \cite{dhoker2019a,
  gerken2020c}. The arguments in \cite{gerken2020c} based on iterated Eisenstein
integrals show in particular that the basis elements obtained are linearly
independent and span the space of \acp{MGF} of arbitrary topology (and the
corresponding modular weight). Furthermore, since the Laurent polynomials of the
basis elements are known~\cite{dhoker2015, dhoker2016}, this allows us to easily obtain the
Laurent polynomials of all the decomposed \acp{MGF}. With the help of the
Laurent polynomials, we construct the five real cusp forms in the space of
\acp{MGF} at weight $(6,6)$, cf.~\eqref{eq:8} and show that no real cusp forms
exist at lower weights.

Using the basis decompositions of \acp{MGF}, we expand the generating series
$Y^{\tau}_{\vec{\eta}}$ of Koba--Nielsen integrals defined in \cite{gerken2020c}
in the basis-\acp{MGF} for two and three points up to order 12, corresponding to
\acp{MGF} of total modular weight at most 12. These expansions were crucial in
determining the dictionary between \acp{MGF} and iterated Eisenstein integrals
in \cite{gerken2020c} and are made available in an ancillary file to the arXiv
submission of this paper.

Finally, we provide the \texttt{Mathematica} package \mmapackage in the
ancillary files of the arXiv submission which automatizes the manipulations
discussed in this paper and contains all basis decompositions for two and
three-point \acp{MGF} of weight $a+b\leq12$. Furthermore, the \mmapackage
package can be used to automatically expand Koba--Nielsen integrals in terms of
\acp{MGF}. This package was used to obtain the expansions of the
$Y^{\tau}_{\vec{\eta}}$ mentioned above.

\subsection{Outline}

This paper is structured as follows: In Section~\ref{sec:mgfs}, we review the
definition of modular graph forms and their different representations as well as
some other important objects. In Section~\ref{sec:math-pack}, we begin the main
body of the paper with a brief overview of the \mmapackage package, followed in
Section~\ref{sec:graph-topol-not} by an introduction of the notation for
\acp{MGF} which we will be using throughout for two-, three- and four-point
graphs. For these graphs, we collect a number of simple manipulations in
Section~\ref{sec:simple-relations} which are largely known in the literature. In
Section~\ref{sec:HSR} we discuss holomorphic subgraph reduction and how it is
related to the Fay identity of the Kronecker--Eisenstein series. In
Section~\ref{sec:sieve-algorithm}, we review the sieve algorithm
\cite{dhoker2016a} and discuss its implementation in the \mmapackage
package. Since using the relations discussed in
Section~\ref{sec:simple-relations} can lead to divergent \acp{MGF}, even if we
start out with only convergent graphs, we discuss divergent \acp{MGF} in
Section~\ref{sec:divMGF}. All the manipulations discussed up to this point are
combined in Section~\ref{sec:conjectured-basis} to obtain basis decompositions
for a large class of \acp{MGF}. Section~\ref{sec:conclusion} contains a
conclusion and outlook. In Appendix~\ref{cha:mma-reference} we give a complete
reference to the \mmapackage package and in Appendix~\ref{sec:3pt-IBPs}, we
discuss further details about kinematic poles of three-point integrals.


\section{Modular graph forms}
\label{sec:mgfs}

In this section, we give a brief review of the structures appearing in the
evaluation of genus-one closed-string integrals and introduce modular graph
forms.

\subsection{Koba--Nielsen integrals and Kronecker--Eisenstein series}
\label{sec:KN-integrals}

After evaluating the \ac{CFT} correlator of the vertex operators, closed string
genus-one amplitudes can be written in terms of integrals of the form
\begin{align}
  \int_{\Sigma_{\tau}^{n-1}}\dd\mu_{n-1}\,\phi^{(a,b)}(\vec{z},\vec{\bar{z}},\tau)
  \KN_{n}(\vec{z},\tau)\,,
  \label{eq:1}
\end{align}
where $\Sigma_{\tau}\subset\CC$ is the torus with modular parameter
$\tau=\tau_{1}+i\tau_{2}$, $\tau_{1},\tau_{2}\in\RR$, parametrized by the
parallelogram spanned by the paths $(0,1)$ and $(0,\tau)$ with opposite edges
identified. We integrate over the puncture positions $z_{i}$ (collectively
denoted by $\vec{z}$\,) using the modular invariant integration measure
\begin{align}
  \dd\mu_{n-1}
  =\prod_{k=2}^{n}\frac{\ddd \Re(z_{k})\wedge \ddd \Im(z_{k})}{\tau_{2}}
  =\prod_{k=2}^{n}\dd v_{k}\wedge\ddd u_{k}\,,
  \label{eq:72}
\end{align}
where we have fixed the origin of the coordinate system to $z_{1}=0$. In
\eqref{eq:72}, we also gave the integration measure in terms of the coordinates
$u$ and $v$, which are aligned with the axes of the parallelogram,
\begin{align}
  u=\frac{\Im(z)}{\tau_{2}}\,,\quad
  v=\Re(z)-\frac{\tau_{1}}{\tau_{2}}\Im(z)\quad\Rightarrow\quad z=u\tau+v\,.
  \label{eq:232}
\end{align}
The function $\phi^{(a,b)}(\vec{z},\vec{\bar{z}},\tau)$ in the integrand of
\eqref{eq:1} depends on the positions $z_{i}$, their complex conjugates
$\bar{z}_{i}$ and the modular parameter $\tau$ and transforms as a
non-holomorphic Jacobi form of weight $(a,b)$ (and vanishing index), i.e.
\begin{align}
  \phi^{(a,b)}\Big(\frac{\vec{z}}{\gamma\tau+\delta},
  \frac{\vec{\bar{z}}}{\gamma\bar{\tau}+\delta},
  \frac{\alpha\tau+\beta}{\gamma\tau+\delta}\Big)
  =(\gamma\tau+\delta)^{a}(\gamma\bar{\tau}+\delta)^{b}
  \phi^{(a,b)}(\vec{z},\vec{\bar{z}},\tau)\,,
  \label{eq:2}
\end{align}
where $\spmatrix{\alpha & \beta \\ \gamma & \delta}\in\SL(2,\ZZ)$. We will give
more details on the form of $\phi$ shortly. The \emph{Koba--Nielsen factor}
$\KN_{n}(\tau)$ in \eqref{eq:1} is defined by (we will from now on drop the
explicit dependence on $\vec{z}$ and $\tau$)~\cite{green1988a}
\begin{align}
  \KN_{n}=\exp\left( \sum_{1\leq i<j}^{n}s_{ij}G_{ij} \right)
  \label{eq:91}
\end{align}
in terms of the Mandelstam invariants \eqref{eq:15} and the Green function
$G_{ij}=G(z_{ij},\tau)=G(z_{i}{-}z_{j},\tau)$ on the torus which satisfies
\begin{align}
  \partial_{z}\partial_{\bar{z}}G(z,\tau)=-\pi\delta^{(2)}(z,\bar{z})
  +\frac{\pi}{\tau_{2}}\,.
  \label{eq:81}
\end{align}
The Green function is doubly periodic in $z$ and can hence be written as a
double Fourier-series in this variable. In this representation, it is given by
\cite{green2000}
\begin{align}
  G(z,\tau)=\frac{\tau_{2}}{\pi}\sum_{(m,n)\in\ZZ^{2}}'
  \frac{e^{2\pi i(mv-nu)}}{|m\tau+n|^{2}}
  =\frac{\tau_{2}}{\pi}\sum_{p}'
  \frac{e^{2\pi i\langle p,z \rangle}}{|p|^{2}}\,,
  \label{eq:85}
\end{align}
where the prime again indicates that the origin is omitted from the sum and we
used the notation
\begin{align}
  p &= m\tau+n &\langle p,z\rangle &=
  mv-nu=\frac{\left(p\bar{z}-\bar{p}z\right)}{2i\tau_{2}}\,.
  \label{eq:86}
\end{align}
In the representation \eqref{eq:85} it is manifest that the Green function is
modular invariant and hence the integral in \eqref{eq:1} transforms as a
non-holomorphic modular form of weight $(a,b)$.

In order to describe the structure of $\phi^{(a,b)}$ in more detail, consider
the \emph{Kronecker--Eisenstein series} \cite{kronecker1881, brown2011}
\begin{align}
  \Omega(z,\eta,\tau)=
  \exp\left( 2\pi i \eta \frac{\Im z}{\tau_{2}}\right)
  \frac{\theta_{1}'(0,\tau) \theta_{1}(z+\eta,\tau)}
  {\theta_{1}(z,\tau)\theta_{1}(\eta,\tau)}\,,
  \label{eq:98}
\end{align}
where $\theta_{1}(z,\tau)$ is the first Jacobi theta function. $\Omega$ is
doubly periodic in $z$ and can therefore be written as a Fourier series,
\begin{align}
  \Omega(z,\eta,\tau)
  = \sum_{p}\frac{e^{2\pi i\langle p,z \rangle}}{p+\eta}\,,
  \label{eq:100}
\end{align}
where we used the notation \eqref{eq:86}. By expanding $\Omega$ in $\eta$, 
\begin{align}
  \Omega(z,\eta,\tau) = \sum_{a\geq 0} \eta^{a-1} f^{(a)}(z,\tau)\,,
  \label{eq:102}
\end{align}
we define the functions $f^{(a)}(z,\tau)$ which have Fourier expansion
\begin{subequations}
  \label{eq:3}
  \begin{align}
    f^{(0)}(z,\tau)&=1 \label{eq:344}\\
    f^{(a)}(z,\tau)
    &=(-1)^{a-1}\sum_{p}' \frac{2\pi i \langle p,z \rangle}{p^{a}}\,,\quad a>0
    \label{eq:103}\\
    \overline{f^{(b)}(z,\tau)}
    &=-\sum_{p}' \frac{2\pi i \langle p,z \rangle}{\bar{p}^{b}}\,,\quad b>0\,.
    \label{eq:104}
  \end{align}
\end{subequations}
Note that for the Fourier series of $f^{(a)}$ with $a\leq2$ are not absolutely
convergent.  From this representation, it is easy to check that the $f^{(a)}$
satisfy the differential equations
\begin{align}
  \partial_{\bar{z}} f^{(a)}(z,\tau) = -\frac{\pi}{\tau_{2}} f^{(a-1)}(z,\tau)
  +\pi\delta_{a,1} \delta^{(2)}(z,\bar z)\, ,\quad a\geq 1 \,.
  \label{eq:364}
\end{align}
This implies
\begin{align}
  \partial_{z}G(z,\tau)=-f^{(1)}(z,\tau)\,,
  \label{eq:345}
\end{align}
upon comparing \eqref{eq:364} to \eqref{eq:81}. The Fourier representation
\eqref{eq:3} of the $f^{(a)}(z,\tau)$ also manifests that they transform as
Jacobi forms of weight $(a,0)$ (and vanishing index),
\begin{align}
  f^{(a)}\Big(\frac{z}{\gamma\tau+\delta},
  \frac{\alpha\tau+\beta}{\gamma\tau+\delta}\Big)
  =(\gamma\tau+\delta)^{a}f^{(a)}(z,\tau)\,,\qquad
  \begin{pmatrix}
    \alpha & \beta\\
    \gamma & \delta
  \end{pmatrix}
  \in\SL(2,\ZZ)\,.
\end{align}
The function $\phi^{(a,b)}(\vec{z},\vec{\bar{z}},\tau)$ appearing in the
integral \eqref{eq:1} can be written as a homogeneous polynomial in the
$f^{(a)}$ and $\overline{f^{(b)}}$ evaluated at differences of the $z_{i}$ for
any massless amplitude of closed-string states in type-II, heterotic or bosonic
theories \cite{dolan2009, broedel2015, gerken2019e}. For these differences, we
introduce the notation
$f^{(a)}_{ij}=f^{(a)}(z_{ij},\tau)=f^{(a)}(z_{i}{-}z_{j},\tau)$ and similarly
for $\overline{f^{(b)}}$.  We will refer to an integral of the form \eqref{eq:1}
with $\phi$ a polynomial in $f^{(a)}_{ij}$ and $\overline{f^{(b)}_{k\ell}}$ as a
\emph{Koba--Nielsen integral}.

An important class of polynomials in the $f^{(a)}_{ij}$ which appears e.g.\ in
the computation of four-gluon scattering in the heterotic
string~\cite{gerken2019f} is given by the $V_{a}$ functions defined by
\begin{align}
  \Omega(z_{12},\eta,\tau)\Omega(z_{23},\eta,\tau)\dotsm
  \Omega(z_{n-1,n},\eta,\tau)\Omega(z_{n,1},\eta,\tau)
  &=\eta^{-n}\sum_{a=0}^{\infty}\eta^{a}V_{a}(1,2,\dots,n)\,,
  \label{eq:107}
\end{align}
where the labels of $V_{a}$ refer to the order of the punctures in the product
of Kronecker--Eisenstein series. Using the expansion \eqref{eq:102}, the $V_{a}$
can be written in terms of the $f^{(a)}_{ij}$, e.g.
\begin{align}
  \begin{split}
    V_{0}(1,2,\dots,n) &=1\,,\qquad
    V_{1}(1,2,\dots,n) = \sum_{j=1}^{n} f^{(1)}_{j,j+1}\\
    V_2(1,2,\dots,n) &= \sum_{j=1}^n f^{(2)}_{j,j+1} +\sum_{i=1}^n
    \sum_{j=i{+}1}^n f^{(1)}_{i,i+1}f^{(1)}_{j,j+1} \quad\te{etc},
  \end{split}
  \label{eq:109}
\end{align}
where we set $f^{(a)}_{n,n+1}=f^{(a)}_{n,1}$.

\subsection{Modular graph forms}
\label{sec:sub-mgfs}

\emph{Modular graph forms} (\acp{MGF}) are the expansion coefficients in the
Mandelstam expansion of \eqref{eq:1}. In order to define \acp{MGF}, consider a
generalization of the sums in \eqref{eq:3} and \eqref{eq:85},
\begin{align}
  C^{(a,b)}(z,\tau)=
  \sum_{p}' \frac{e^{2\pi i\langle p,z \rangle}}{p^{a}\bar{p}^{b}}\,.
  \label{eq:126}
\end{align}
These functions were previously studied in \cite{zagier1990, dhoker2015a,
  broedel2019}. The Green function and the $f^{(a)}$ and $\overline{f^{(b)}}$
are special cases of the $C^{(a,b)}$ since
\begin{alignat}{2}
  G(z,\tau)&=\frac{\tau_{2}}{\pi}C^{(1,1)}(z,\tau)&&\nonumber\\
  f^{(a)}(z,\tau)&=(-1)^{a-1}C^{(a,0)}(z,\tau)\quad&a&>0\label{eq:127}\\
  \overline{f^{(b)}(z,\tau)}&=-C^{(0,b)}(z,\tau)&b&>0\nonumber\,.
\end{alignat}
Using \eqref{eq:126}, the expansion coefficient of \eqref{eq:1} (for one
monomial in $\phi^{(a,b)}$) at a certain order in $\ap$ has the form
\begin{align}
  \mathcal{C}_{\Gamma}(\tau)=
  \int\dd\mu_{n-1}\prod_{e\in E_{\Gamma}} C^{(a_{e},b_{e})}(z_{e},\tau)\,,
  \label{eq:128}
\end{align}
which is the integral representation of the modular graph form
$\mathcal{C}_{\Gamma}$. The notation in \eqref{eq:128} is suggestive of the
graphical representation of \acp{MGF}~\cite{dhoker2016a}: We can associate an
$n$-vertex graph $\Gamma$ to the integral in \eqref{eq:128} by identifying the
$C^{(a,b)}_{ij}=C^{(a,b)}(z_{ij})$ with an edge form vertex $i$ to vertex $j$,
labeled by $(a,b)$,
\begin{align}
  C^{(a,b)}_{ij}\quad\leftrightarrow\quad
  \begin{tikzpicture}[mgf]
    \node (i) at (0,0) [vertexdot]{};
    \node[vertex,anchor=north] at (i.south) {$i$};
    \node (j) at (2.5,0) [vertexdot]{};
    \node[vertex,anchor=north] at (j.south) {$j$};
    \draw[dedge=0.9] (i) to node[label,pos=0.5]{$(a,b)$} (j);
  \end{tikzpicture}\quad.
  \label{eq:129}
\end{align}
In this notation, $E_{\Gamma}$ in \eqref{eq:128} is the edge set of the
graph. Using the notation
\begin{alignat}{2}
  |A|&=\sum_{e\in E_{\Gamma}}a_{e}\quad&\quad|B|&=\sum_{e\in E_{\Gamma}}b_{e}\,,
\label{eq:133}
\end{alignat}
the \ac{MGF} in \eqref{eq:128} is a non-holomorphic modular form of weight
$(|A|,|B|)$. If the holomorphic and antiholomorphic edge labels are equal,
$a_{e}=b_{e}\ \forall\ e\in E_{\Gamma}$, the \ac{MGF} can be turned into a
modular function by multiplication with $\tau_{2}^{|A|}$. In this case, we call
the \ac{MGF} a \emph{modular graph function}~\cite{dhoker2015a}. Note that also
the weaker condition $|A|=|B|$ is used to define modular graph functions in the
literature.

Since the integrand in \eqref{eq:128} depends on $z$ only through the
exponential factors $e^{2\pi i \langle p,z \rangle}$, we can perform this
integral trivially, leading to conservation of the momenta $p$ at the
vertices. This leads to the sum representation~\cite{dhoker2016a}
\begin{align}
  \mathcal{C}_\Gamma(\tau) =  \sum_{\{p_e\}}'\prod_{e\in E_{\Gamma}}
  \frac{1}{p_e^{a_e} \bar{p}^{b_e}_e} 
  \prod_{i\in V_{\Gamma}} \delta\left(
    \sum_{e'\in E_{\Gamma}}\Gamma_{ie'}p_{e'}\right)
  \label{eq:131}
\end{align}
of \acp{MGF}, where $E_{\Gamma}$ is the set of edges of $\Gamma$, $V_{\Gamma}$
is the set of vertices and
\begin{align}
  \Gamma_{ie}=
  \begin{cases}
    1 & \te{if $e$ is directed into $i$}\\
    -1 & \te{if $e$ is directed out of $i$}\\
    0 & \te{if $e$ is not connected to $i$}
  \end{cases}
  \label{eq:132}
\end{align}
is the incidence matrix of vertex $i$.

A simple example of a modular graph function is given by a two-point graph with
$\ell$ edges with label $(1,1)$ each between the vertices. This \ac{MGF} is
denoted by $D_{\ell}$ and given by~\cite{dhoker2015}
\begin{align}
  D_{\ell}(\tau)=\imtaupi^{\ell}
  \sum_{p_{1},\dots,p_{\ell}}'
  \frac{\delta(p_{1}+\dotsm+p_{\ell})}{|p_{1}|^{2}\dotsm|p_{\ell}|^{2}}
  \label{eq:124}
\end{align}
in the sum representation. Here, the sum was multiplied by a suitable factor of
$\frac{\tau_{2}}{\pi}$ to make $D_{\ell}$ modular invariant, as is customary in
the literature.

If we assign arbitrary labels to the edges, the resulting \ac{MGF} is called
\emph{dihedral} and given by~\cite{dhoker2016a}
\begin{align}
  \cform{a_{1}&\dotsm&a_{R}\\b_{1}&\dotsm&b_{R}}=
  \int\dd\mu_{1}\prod_{i=1}^{R} C^{(a_{i},b_{i})}_{12}=
  \begin{tikzpicture}[mgf]
    \node (1) at (0,0) [vertex] {$1$};
    \node (2) at (3.5,0) [vertex] {$2$};
    \draw[dedge=0.85] (1) to[bend left=50]
    node[label]{$(a_{1},b_{1})$} (2);
    \draw[dedge=0.85] (1) to[bend left=20]
    node[label]{$(a_{2},b_{2})$} (2);
    \node at (1.75,-0.1) {$\vdots$};
    \draw[dedge=0.85] (1) to[bend right=40]
    node[label]{$(a_{R},b_{R})$} (2);
  \end{tikzpicture}\,.
  \label{eq:130}
\end{align}
The $D_{\ell}$ from \eqref{eq:124} are in this notation given by
\begin{align}
  D_{\ell}=\imtaupi^{\ell}\cform{1_{\ell}\\1_{\ell}}\,,
\end{align}
where $1_{\ell}$ denotes the row vector with $\ell$ entries of $1$. Further
special cases of \eqref{eq:130} are the modular graph
functions~\cite{dhoker2015}
\begin{align}
  C_{a,b,c}&=\imtaupi^{a+b+c}\cform{a&b&c\\a&b&c}
  \label{eq:135}\\
  C_{a,b,c,d}&=\imtaupi^{a+b+c+d}
  \cform{a&b&c&d\\a&b&c&d}\,.
  \label{eq:170}
\end{align}

To write one-loop graphs in the notation \eqref{eq:130}, we need a
$\sbmatrix{0\\0}$-column since otherwise the omission of the origin in the sum
sets the \ac{MGF} to zero. Consequently, we have
\begin{subequations}
  \begin{align}
    \cform{a&0\\b&0}&=\sum_{p}' \frac{1}{p^{a}\bar{p}^{b}}\label{eq:136}\\
    \cform{k&0\\0&0}&=\GG_{k}\,,\qquad k>2\label{eq:330}\\
    \cform{s&0\\s&0}&=\piimtau^{s}\EE_{s}\,,\label{eq:331}
    \qquad \Re(s)>1\,.
  \end{align}
\end{subequations}
Here, we have introduced the \emph{holomorphic Eisenstein series} $\GG_{k}$ and
their non-holomorphic counterparts $\EE_{s}$ which are defined by
\begin{align}
  \GG_{k}(\tau)&=\sum'_{(m,n)\in\ZZ^{2}}\frac{1}{(m\tau+n)^k}\,,
  \quad k\geq3\in\NN  \label{eq:45}\\
  \EE_{s}(\tau)&=\left( \frac{\tau_{2}}{\pi} \right)^{s}
  \sum'_{(m,n)\in\ZZ^{2}}\frac{1}{|m\tau+n|^{2s}}\,,
  \quad s\in\CC\,,\quad \Re(s)>1\,.
  \label{eq:51}
\end{align}
We will also use the modular, but non-holomorphic version $\GGhat_{2}$ of the
Eisenstein series $\GG_{2}$, defined by
\begin{align}
  \GGhat_{2}(\tau)=\lim_{s\rightarrow0}
  \sum_{(m,n)\in\ZZ^{2}}' \frac{1}{(m\tau+n)^{2}|m\tau+n|^{s}}\,,
  \label{eq:316}
\end{align}
which can be written as
\begin{align}
  \GGhat_{2}(\tau)=\GG_{2}(\tau)-\frac{\pi}{\tau_{2}}
  \label{eq:50}
\end{align}
with
\begin{align}
  \GG_{2}(\tau)=\sum_{n\neq0}\frac{1}{n^{2}}
  +\sum_{m\neq0}\sum_{n\in\ZZ}\frac{1}{(m\tau+n)^{2}}\,.
  \label{eq:49}
\end{align}
We will assume the regularization \eqref{eq:316} for conditionally convergent
\acp{MGF} throughout and therefore have
\begin{align}
  \cform{2&0\\0&0}=\GGhat_{2}\,.
  \label{eq:332}
\end{align}
More details on the convergence properties of \eqref{eq:128} and \eqref{eq:131}
are given in Section~\ref{sec:divMGF}.

The definition \eqref{eq:131} of \acp{MGF} implies a number of
properties~\cite{dhoker2015}, some of which we want to mention here. Firstly,
the modular behavior of the \acp{MGF} implies that $\mathcal{C}_{\Gamma}=0$ if
$|A|+|B|$ is odd. Secondly, the graphs of all non-zero \acp{MGF} are
one-particle irreducible vacuum bubbles. Furthermore, two-valent vertices can be
dropped by adding the labels of their edges,
\begin{align}
  \begin{tikzpicture}[mgf]
    \node (i) at (0,0) [vertexdot]{};
    \node[vertex,anchor=north] at (i.south) {$i$};
    \node (j) at (2.5,0) [vertexdot]{};
    \node[vertex,anchor=north] at (j.south) {$j$};
    \node (k) at (5,0) [vertexdot]{};
    \node[vertex,anchor=north] at (k.south) {$k$};
    \draw[dedge=0.9] (i) to node[label,pos=0.4]{$(a_{1},b_{1})$} (j);
    \draw[dedge=0.9] (j) to node[label,pos=0.4]{$(a_{2},b_{2})$} (k);
  \end{tikzpicture}
  \,=\,
  \begin{tikzpicture}[mgf]
    \node (i) at (0,0) [vertexdot]{};
    \node[vertex,anchor=north] at (i.south) {$i$};
    \node (k) at (4,0) [vertexdot]{};
    \node[vertex,anchor=north] at (k.south) {$k$};
    \draw[dedge=0.94] (i) to
    node[label,pos=0.45]{$(a_{1}{+}a_{2},b_{1}{+}b_{2})$} (k);
  \end{tikzpicture}
  \quad.
  \label{eq:137}
\end{align}
Finally, if a graph has connectivity one (i.e.\ it can be disconnected by
removing one vertex), the associated \ac{MGF} factorizes,
\begin{align}
  \begin{tikzpicture}[baseline=-3,scale=0.7,line width=0.3mm]
    \node (i) at (2,0) [vertex]{$i$};
    \draw[edge](0,0.75)..controls(1.2,0.75)..(i);
    \draw[edge](0,-0.75)..controls(1.2,-0.75)..(i);
    \node at (1,0.15) {$\vdots$};
    \filldraw[fill=BGColorH,draw=black](0,0) ellipse (0.5 and 1);
    \node at (0,0) {$\Gamma_{1}$};
    \draw[edge](i)..controls(2.8,0.75)..(4,0.75);
    \draw[edge](i)..controls(2.8,-0.75)..(4,-0.75);
    \node at (3,0.15) {$\vdots$};
    \filldraw[fill=BGColorH,draw=black](4,0) ellipse (0.5 and 1);
    \node at (4,0) {$\Gamma_{2}$};
  \end{tikzpicture}
  \quad=\quad
  \begin{tikzpicture}[baseline=-3,scale=0.7,line width=0.3mm]
    \node (i) at (2,0) [vertex]{$i$};
    \draw[edge](0,0.75)..controls(1.2,0.75)..(i);
    \draw[edge](0,-0.75)..controls(1.2,-0.75)..(i);
    \node at (1,0.15){$\vdots$};
    \filldraw[fill=BGColorH,draw=black](0,0) ellipse (0.5 and 1);
    \node at (0,0){$\Gamma_{1}$};
  \end{tikzpicture}
  \,\times\,
  \begin{tikzpicture}[baseline=-3,scale=0.7,line width=0.3mm]
    \node (i) at (2,0) [vertex]{$i$};
    \draw[edge](i)..controls(2.8,0.75)..(4,0.75);
    \draw[edge](i)..controls(2.8,-0.75)..(4,-0.75);
    \node at (3,0.15){$\vdots$};
    \filldraw[fill=BGColorH,draw=black](4,0) ellipse (0.5 and 1);
    \node at (4,0){$\Gamma_{2}$};
  \end{tikzpicture}
  \quad.
  \label{eq:190}
\end{align}

Using the definitions above, we can now discuss the basics of the
\texttt{Mathematica} package \mmapackage in the next section, which can be used
to perform numerous manipulations on \acp{MGF}.


\section{The \protect\mmapackage Mathematica package}
\label{sec:math-pack}
As mentioned in the introduction, we will present a number of simplification
techniques for \acp{MGF} in this paper, which will allow us to derive basis
decompositions for a large number of \acp{MGF}, as discussed in
Section~\ref{sec:conjectured-basis}. To make the resulting decompositions
accessible, it is convenient to have a computer database of them, together with
an implementation for the various techniques to be discussed. This is realized
in the \texttt{Mathematica} package \mmapackage which is included in the arXiv
submission of this paper. It contains $11079$ identities to decompose all two-
and three-point \acp{MGF} of total modular weight $a+b\leq12$ into the basis
given in Section~\ref{sec:conjectured-basis} and functions for basic
manipulations of four-point graphs. The package furthermore contains routines to
automatically expand Koba--Nielsen integrals in terms of \acp{MGF}. In this
section, we will outline the basic usage of the package and, as we discuss the
manipulations for \acp{MGF} in the following sections, we will also describe how
they are implemented in the \mmapackage package. A complete reference of all
defined symbols as well as all functions and their options is given in
Appendix~\ref{cha:mma-reference}.

\subsection{Basics}
The \mmapackage package consists of the package itself in the
\texttt{ModularGraphForms.m} file and two files containing identities between
two- and three-point \acp{MGF} which were generated using the techniques
presented in this paper. To load the package, copy all files into the directory
in which the current notebook is saved and run
\begin{mmaCell}[moredefined={NotebookDirectory}]{Input}
  Get[NotebookDirectory[]\,<>\,"ModularGraphForms.m"]
\end{mmaCell}
\begin{mmaCell}{Print}
  Dihedral identity file found at /home/user/DiIds.txt
\end{mmaCell}
\vspace{-2em}
\begin{mmaCell}{Print}
  Trihedral identity file found at /home/user/TriIds.txt
\end{mmaCell}
\vspace{-2em}
\begin{mmaCell}{Print}
  Loaded 1559 identities for dihedral convergent MGF.
\end{mmaCell}
\vspace{-2em}
\begin{mmaCell}{Print}
  Loaded 9520 identities for trihedral convergent MGF.
\end{mmaCell}
\vspace{-2em}
\begin{mmaCell}[morelst={breakatwhitespace=true}]{Print}
  Successfully loaded the ModularGraphForms package. Have fun!
\end{mmaCell}

The notation used for $\tau$ is \mma{tau}, $\bar{\tau}$ is \mma{tauBar} and
$\tau_{2}$ is \mma{tau[2]}. The zeta values \eqref{eq:28} are written as e.g.\
\mma{zeta[3]} and the holomorphic Eisenstein series \eqref{eq:45} and their
complex conjugates are \mma{g[4]} and \mma{gBar[4]}, respectively. For the
modular version $\GGhat_{2}$ of $\GG_{2}$, we use \mma{gHat[2]} and
\mma{gBarHat[2]}. The non-holomorphic Eisenstein series \eqref{eq:51} and their
higher-depth generalizations defined below in \eqref{eq:172} are denoted for
instance by \mma{e[2,2]}. The normalizations are as described in
Section~\ref{sec:mgfs}.

The modular weight of an expression is determined by the function
\mma{CModWeight}, e.g.\
\begin{mmaCell}{Input}
  CModWeight\mmabigso{}g[4]+\mmaSup{gHat[2]}{2}+\mmaSup{\Big(\mmaFrac{tau[2]}{\mmaDef{\(\pi\)}}\Big)}{4}\,e[2,2]\,gBar[4]\,g[8]\mmabigsc{}
\end{mmaCell}
\begin{mmaCell}{Output}
  \{4,0\} \textnormal{.}
\end{mmaCell}
Complex conjugation is performed by the function
\mma{CComplexConj}, e.g.\
\begin{mmaCell}{Input}
  CComplexConj\mmabigso{}g[4]+\mmaSup{gHat[2]}{2}+\mmaSup{\Big(\mmaFrac{tau[2]}{\mmaDef{\(\pi\)}}\Big)}{4}\,e[2,2]\,gBar[4]\,g[8]\mmabigsc{}
\end{mmaCell}
\begin{mmaCell}{Output}
  \mmaSub{\mmaOver{G}{\(\mmaoverline\)}}{4}+\mmaSubSup{\mmaGGhatb}{2}{2}+\mmaFrac{\mmaSub{E}{2,2}\,\mmaSub{G}{4}\,\mmaSub{\mmaOver{G}{\(\mmaoverline\)}}{8}\,\mmaSubSup{\(\tau\)}{2}{4}}{\mmaSup{\(\pi\)}{4}} \textnormal{.}
\end{mmaCell}
The most important function of the \mmapackage package is the function
\mma{CSimplify}, which performs all known simplifications for \acp{MGF} on the
expression in the argument, e.g.\ the identity \eqref{eq:176} is hard-coded into
the package and can be used as follows,
\begin{mmaCell}{Input}
  CSimplify\mmabigso{}\mmaincform{1 1 1\\1 1 1}\mmabigsc{}
\end{mmaCell}
\begin{mmaCell}{Output}
  \mmaFrac{\mmaSup{\(\pi\)}{3}\,\mmaSub{E}{3}}{\mmaSubSup{\(\tau\)}{2}{3}}+\mmaFrac{\mmaSup{\(\pi\)}{3}\,\mmaSub{\(\zeta\)}{3}}{\mmaSubSup{\(\tau\)}{2}{3}} \textnormal{,}
\end{mmaCell}
where the notation for modular graph forms will be explained in
Section~\ref{sec:graph-topol-not} below. The function \mma{CSimplify} calls the
functions \mma{DiCSimplify}, \mma{TriCSimplify} and \mma{TetCSimplify}, which
perform simplifications on two- three- and four-point graphs, respectively.

The remaining functions in the \mmapackage package will be discussed in the
following sections, along with the manipulations for \acp{MGF} they implement. A
complete reference for all functions and their options is given in
Appendix~\ref{cha:mma-reference}. Within \texttt{Mathematica}, short
explanations for the various objects can be obtained using the \mma{Information}
function, e.g.\ by running \mma{?CModWeight}. A complete list of all available
objects can be printed by running \mma{?ModularGraphForms\`{}*}.

\subsection{Expanding Koba--Nielsen integrals}
\label{sec:exp-KN-ints}
\begin{table}
  \centering
  {\renewcommand{\arraystretch}{1.5}
    \begin{tabular}{ccc}
      \toprule
      function & definition & \texttt{Mathematica} representation\\
      \midrule
      $f^{(a)}_{ij}$ & \eqref{eq:103} & \mma{fz[a,\,i,\,j]}\\
      $\overline{f^{(b)}_{ij}}$ & \eqref{eq:104} & \mma{fBarz[b,\,i,\,j]}\\
      $G_{ij}$ & \eqref{eq:85} & \mma{gz[i,\,j]}\\
      $C^{(a,b)}_{ij}$ & \eqref{eq:126} & \mma{cz[a,\,b,\,i,\,j]}\\
      $V_{a}(k_{1},\dots,k_{r})$ & \eqref{eq:107}
      & \mma{vz[k\textsubscript{1},\dots,k\textsubscript{r}]}\\
      $\overline{V_{b}(k_{1},\dots,k_{r})}$ & \eqref{eq:107}
      & \mma{vBarz[k\textsubscript{1},\dots,k\textsubscript{r}]}\\
      \bottomrule
    \end{tabular}
  }
  \caption{Various $z$-dependent functions defined in
    Section~\ref{sec:mgfs} and their representation in
    \texttt{Mathematica}.}
  \label{tab:zfcts}
\end{table}
As explained in Section~\ref{sec:mgfs}, in string theory, modular graph
forms arise as coefficients in the expansion of Koba--Nielsen integrals. The
\mmapackage package also contains the function \mma{zIntegrate} which performs
this expansion automatically. The syntax is as follows: \mma{zIntegrate} has
three arguments, the first one is the prefactor in front of the Koba--Nielsen
factor, the second one is the number of points in the Koba--Nielsen factor
\eqref{eq:91} and the last one is the order in Mandelstam variables which is
written in terms of \acp{MGF}. E.g.\ the second order in Mandelstams of the
three-point integral
\begin{align}
  \int\dd\mu_{2}\KN_{3}
\end{align}
is computed by
\begin{mmaCell}{Input}
  zIntegrate[1,\,3,\,2]\,//\,Factor
\end{mmaCell}
\begin{mmaCell}{Output}
  \mmaFrac{1}{2}\,\mmaSub{E}{2}\,(\mmaSubSup{s}{1,2}{2}+\mmaSubSup{s}{1,3}{2}+\mmaSubSup{s}{2,3}{2}) \textnormal{.}
\end{mmaCell}
Note that all Mandelstam variables are treated as independent, no momentum
conservation is imposed. A Koba--Nielsen factor which does not contain all Green
functions and Mandelstam variables of \eqref{eq:91} can be represented by
replacing the second argument with a list of point pairs, corresponding to the
Green functions appearing in the Koba--Nielsen factor. E.g.\
$\exp(s_{12}G_{12}+s_{13}G_{13})$ is represented by
\mma{\{\{1,2\},\{1,3\}\}}. For an integral without Koba--Nielsen factor, we can
set the second argument of \mma{zIntegrate} to an arbitrary number and the
third argument to zero.

For the integrand in front of the Koba--Nielsen factor, the functions listed in
Table~\ref{tab:zfcts} are available. To indicate their $z$ dependence, they all
carry a suffix \mma{z}. An arbitrary polynomial in these functions can be given
as the first argument to \mma{zIntegrate}. E.g.\ the first order in Mandelstams
of the integral
\begin{align}
  \int\dd\mu_{3}V_{2}(1,2,3,4)\KN_{4}
\end{align}
is computed by
\begin{mmaCell}{Input}
  zIntegrate[vz[2,\,\{1,\,2,\,3,\,4\}],\,4,\,1]\,//\,Factor
\end{mmaCell}
\begin{mmaCell}{Output}
  \mmam\mmaFrac{\mmaoutcform{3 0\\1 0}(\mmaSub{s}{1,2}\mmam2\,\mmaSub{s}{1,3}+\mmaSub{s}{1,4}+\mmaSub{s}{2,3}\mmam2\,\mmaSub{s}{2,4}+\mmaSub{s}{3,4})\,\mmaSub{\(\tau\)}{2}}{\(\pi\)} \textnormal{.}
\end{mmaCell}

The function \mma{zIntegrate} returns \acp{MGF} in the notation introduced in
Section~\ref{sec:graph-topol-not} below for \acp{MGF} with up to four points,
while exploiting some basic properties of \acp{MGF} as the ones listed in
Section~\ref{sec:sub-mgfs}. If \acp{MGF} with more than four points appear in the
expansion and they cannot be reduced by using these properties, they are printed
as a graph, e.g.\
\begin{mmaCell}{Input}
  zIntegrate[\mmaSup{gz[1,2]}{2}\mmaSup{gz[2,3]}{2}\mmaSup{gz[3,4]}{2}\mmaSup{gz[4,5]}{2}gz[5,1],5,0]
\end{mmaCell}
\begin{mmaCell}[morelst={label=mma:14},moregraphics={moreig={scale=.25}}]{Output}
  \mmaFrac{\mmaGraphics{graphics/5ptGraph}\,\mmaSubSup{\(\tau\)}{2}{9}}{\mmaSup{\(\pi\)}{9}} \textnormal{.}
\end{mmaCell}
Note that if the Koba--Nielsen integral expanded using \mma{zIntegrate} contains
kinematic poles due to $f^{(1)}_{ij}\overline{f^{(1)}_{ij}}$ terms in the
integrand, \mma{zIntegrate} will contain divergent \acp{MGF}, as will be
discussed in detail in Section~\ref{sec:div-MGF-KN-poles}. 

Using the function \mma{zIntegrate} and the decompositions discussed in
Section~\ref{sec:conjectured-basis} below, the two- and three-point generating
functions for Koba--Nielsen integrals were evaluated in terms a few
basis-\acp{MGF} up to total modular weight $12$.


\section{Graph topologies and notation}
\label{sec:graph-topol-not}
The general definition \eqref{eq:131} for modular graph forms depends on a graph
$\Gamma$ with decorated and directed edges, where the decoration has the form
$(a,b)$ with $a,b\in\ZZ$. Since it is inconvenient to specify the entire graph
for every \ac{MGF}, we introduce commonly used notations for graphs with up to
four vertices, the only ones considered in this paper.

\subsection{Two-point modular graph forms}
As introduced in \eqref{eq:130}, dihedral graphs have two vertices and all edges
directed in the same way. They are denoted by~\cite{dhoker2016a}
\begin{align}
  \cform{a_{1}&\dotsm&a_{R}\\b_{1}&\dotsm&b_{R}}=
  \begin{tikzpicture}[mgf]
    \node (1) at (0,0) [vertex] {$1$};
    \node (2) at (3.5,0) [vertex] {$2$};
    \draw[dedge=0.85] (1) to[bend left=50]
    node[label]{$(a_{1},b_{1})$} (2);
    \draw[dedge=0.85] (1) to[bend left=20]
    node[label]{$(a_{2},b_{2})$} (2);
    \node at (1.75,-0.1) {$\vdots$};
    \draw[dedge=0.85] (1) to[bend right=40]
    node[label]{$(a_{R},b_{R})$} (2);
  \end{tikzpicture}
  =\sum_{p_{1},\dots,p_{R}}'\frac{\delta(p_{1}+\dotsm+p_{R})}
  {p_{1}^{a_{1}}\bar{p}_{1}^{b_{1}}\dotsm p_{R}^{a_{R}}\bar{p}_{R}^{b_{R}}}\,.
  \label{eq:177}
\end{align}
This class of functions, as well as many special cases, were studied extensively
in the literature \cite{dhoker2015, dhoker2015a, dhoker2016a, dhoker2016,
  basu2019, dhoker2019, dhoker2017, dhoker2019a, dorigoni2019a, basu2020,
  dhoker2019b, zagier2019}. Since we will frequently encounter a bundle of
parallel edges, we write
\begin{align}
  \cform{A\\B}=\cform{a_{1}&\dotsm&a_{R}\\b_{1}&\dotsm&b_{R}}
  \label{eq:134}
\end{align}
and call $\sbmatrix{A\\B}$ a \emph{block}. In graphs, we draw
\begin{align}
  \begin{tikzpicture}[mgf]
    \node (1) at (0,0) [vertex] {$1$};
    \node (2) at (3,0) [vertex] {$2$};
    \draw[dedge=0.85] (1) to[bend left=15] (2);
    \draw[dedge=0.85] (1) to[bend right=15] (2);
    \draw[dedge=0.85] (1)to node[biglabel]{$\sbmatrix{A\\B}$}(2);
  \end{tikzpicture}
  =
  \begin{tikzpicture}[mgf]
    \node (1) at (0,0) [vertex] {$1$};
    \node (2) at (3.5,0) [vertex] {$2$};
    \draw[dedge=0.85] (1) to[bend left=50]
    node[label]{$(a_{1},b_{1})$} (2);
    \draw[dedge=0.85] (1) to[bend left=20]
    node[label]{$(a_{2},b_{2})$} (2);
    \node at (1.75,-0.1) {$\vdots$};
    \draw[dedge=0.85] (1) to[bend right=40]
    node[label]{$(a_{R},b_{R})$} (2);
  \end{tikzpicture}\,.
  \label{eq:180}
\end{align}

For $|A|=|B|$, (cf.~\eqref{eq:133}) we also introduce the antisymmetric version
\begin{align}
  \aform{A\\B}=\cform{A\\B}-\cform{B\\A}\,,
  \label{eq:433}
\end{align}
which is purely imaginary and was first studied in \cite{dhoker2019a}. Under the
transformation $\tau\rightarrow-\bar{\tau}$, any \ac{MGF} satisfies
$\mathcal{C}_{\Gamma}(-\bar{\tau})=\vmsmash{\overline{\mathcal{C}_{\Gamma}(\tau)}}$
and hence we have that $\aform{A\\B}(-\bar{\tau})=-\aform{A\\B}(\tau)$. Since
$\tau_{2}$ is invariant under this transformation and the Laurent polynomial is
mapped to its negative, the Laurent polynomial of $\aform{A\\B}$ has to vanish,
i.e.\ $\aform{A\\B}$ is a cusp form.

In the \mmapackage package, \acp{MGF} have head \mma{c}, i.e.\ they are formally
given by the function \mma{c} applied to various arguments. Dihedral \acp{MGF}
have one argument which is a $2\times R$ matrix which can, as any other matrix,
be inserted in two-dimensional form or as a nested list,
\begin{mmaCell}[morelst={label=mma:13}]{Input}
  \mmaincform{1 2 3\\1 1 1}+c[\{\{1,2,3\},\{1,1,1\}\}]
\end{mmaCell}
\begin{mmaCell}{Output}
  2\,\mmaoutcform{1 2 3\\1 1 1} \textnormal{.}
\end{mmaCell}
Imaginary cusp forms of the form \eqref{eq:433} have head \mma{a},
\begin{mmaCell}{Input}
  \mmainaform{0 2 3\\3 0 2}
\end{mmaCell}
\begin{mmaCell}{Output}
  \mmaoutaform{0 2 3\\3 0 2} \textnormal{.}
\end{mmaCell}

\subsection{Three-point modular graph forms}

\emph{Trihedral} graphs have three vertices. The notation we use is~\cite{dhoker2016a}
\begin{align}
  \cformtri{A_{1}\\B_{1}}{A_{2}\\B_{2}}{A_{3}\\B_{3}}
  =
  \begin{tikzpicture}[mgf,baseline=(zero.base)]
    \node (1) at (0,0) [vertex] {$1$};
    \node (2) at (60:3.5) [vertex] {$2$};
    \node (3) at (3.5,0) [vertex] {$3$};
    \draw[dedge=0.85] (1) to[bend left=15] (2);
    \draw[dedge=0.85] (1) to[bend right=15] (2);
    \draw[dedge=0.85] (1) to
    node[biglabel](zero){$\sbmatrix{A_{1}\\B_{1}}$}(2);
    \draw[dedge=0.85] (2) to[bend left=15] (3);
    \draw[dedge=0.85] (2) to[bend right=15] (3);
    \draw[dedge=0.85] (2) to
    node[biglabel]{$\sbmatrix{A_{2}\\B_{2}}$}(3);
    \draw[dedge=0.85] (3) to[bend left=15] (1);
    \draw[dedge=0.85] (3) to[bend right=15] (1);
    \draw[dedge=0.85] (3) to
    node[biglabel]{$\sbmatrix{A_{3}\\B_{3}}$}(1);
  \end{tikzpicture}
  \label{eq:181}
\end{align}
and hence
\begin{align}
  \!\!\cformtri{A_{1}\\B_{1}}{A_{2}\\B_{2}}{A_{3}\\B_{3}}\!=\hspace{-0.7ex}
  \sum_{\{p_{i}^{(j)}\}}'\left(\prod_{j=1}^{3}\prod_{i=1}^{R_{j}}
    \frac{1}{(p_{i}^{(j)})^{a_{i}^{(j)}}(\bar{p}_{i}^{(j)})^{b_{i}^{(j)}}}\!
  \right)
  \delta\!\left(\sum_{i=1}^{R_{1}}p_{i}^{(1)}{-}\sum_{i=1}^{R_{2}}p_{i}^{(2)}\!\!
  \right)
  \delta\!\left(\sum_{i=1}^{R_{2}}p_{i}^{(2)}{-}\sum_{i=1}^{R_{3}}p_{i}^{(3)}\!\!
  \right),
  \label{eq:185}
\end{align}
where the block $\sbmatrix{A_{j}\\B_{j}}$ has $R_{j}$ columns. We will use this
notation henceforth. If two vertices are not connected by any
edges, we write\footnote{This graph factorizes according to \eqref{eq:210} into
  $\cform{A_{1}\\B_{1}}\cform{A_{2}\\B_{2}}$.}
\begin{align}
  \cformtri{A_{1}\\B_{1}}{A_{2}\\B_{2}}{\noblock}=
  \begin{tikzpicture}[mgf]
    \node (1) at (0,0) [vertex] {$1$};
    \node (2) at (3.5,0) [vertex] {$2$};
    \node (3) at (7,0) [vertex] {$3$};
    \draw[dedge=0.85] (1) to[bend left=15] (2);
    \draw[dedge=0.85] (1) to[bend right=15] (2);
    \draw[dedge=0.85] (1) to
    node[biglabel]{$\sbmatrix{A_{1}\\B_{1}}$}(2);
    \draw[dedge=0.85] (2) to[bend left=15] (3);
    \draw[dedge=0.85] (2) to[bend right=15] (3);
    \draw[dedge=0.85] (2) to
    node[biglabel]{$\sbmatrix{A_{2}\\B_{2}}$}(3);
  \end{tikzpicture}\,.
  \label{eq:186}
\end{align}

In the \mmapackage package, the function \mma{c} with three
matrix-arguments is used,
\begin{mmaCell}{Input}
  \mmaincformtri{1 1\\1 1}{2 3\\1 1}{4 5\\1 1}
\end{mmaCell}
\begin{mmaCell}{Output}
  \mmaoutcformtri{1 1\\1 1}{2 3\\1 1}{4 5\\1 1} \textnormal{.}
\end{mmaCell}
The edge directions and normalization are as in \eqref{eq:181} and
\eqref{eq:185}, respectively. For empty blocks, we use empty lists,
\begin{mmaCell}{Input}
  \mmaincformtri{\mmanothing}{1 2\\1 1}{3 4\\1 1}
\end{mmaCell}
\begin{mmaCell}{Output}
  \mmaoutcformtri{\{\}}{1 2\\1 1}{3 4\\1 1} \textnormal{.}

\end{mmaCell}

\subsection{Four-point modular graph forms}
\label{sec:4ptMGFs}
Due to their different symmetry properties, it is convenient to distinguish the
following three topologies among four-point graphs.

\emph{Box graphs} have four edges in one cycle
and are denoted by
\begin{align}
  \cformbox{A_{1}\\B_{1}}{A_{2}\\B_{2}}{A_{3}\\B_{3}}{A_{4}\\B_{4}}
  =
  \begin{tikzpicture}[mgf,baseline=(zero.base)]
    \node (1) at (0,0) [vertex] {$1$};
    \node (2) at (0,3) [vertex] {$2$};
    \node (3) at (3,3) [vertex] {$3$};
    \node (4) at (3,0) [vertex] {$4$};
    \draw[dedge=0.85] (1) to[bend left=15] (2);
    \draw[dedge=0.85] (1) to[bend right=15] (2);
    \draw[dedge=0.85] (1) to
    node[biglabel](zero){$\sbmatrix{A_{1}\\B_{1}}$}(2);
    \draw[dedge=0.85] (2) to[bend left=15] (3);
    \draw[dedge=0.85] (2) to[bend right=15] (3);
    \draw[dedge=0.85] (2) to
    node[biglabel]{$\sbmatrix{A_{2}\\B_{2}}$}(3);
    \draw[dedge=0.85] (3) to[bend left=15] (4);
    \draw[dedge=0.85] (3) to[bend right=15] (4);
    \draw[dedge=0.85] (3) to
    node[biglabel]{$\sbmatrix{A_{3}\\B_{3}}$}(4);
    \draw[dedge=0.85] (4) to[bend left=15] (1);
    \draw[dedge=0.85] (4) to[bend right=15] (1);
    \draw[dedge=0.85] (4)
    to node[biglabel]{$\sbmatrix{A_{4}\\B_{4}}$}(1);
  \end{tikzpicture}\,.
  \label{eq:182}
\end{align}
The lattice sum representation similarly to \eqref{eq:185} can be read off
straightforwardly from the graph. In \texttt{Mathematica}, we use \mma{c} with
four arguments,
\begin{mmaCell}{Input}
  \mmaincformbox{1 2\\1 1}{3 4\\1 1}{5 6\\1 1}{7 8\\1 1}
\end{mmaCell}
\begin{mmaCell}{Output}
  \mmaoutcformbox{1 2\\1 1}{3 4\\1 1}{5 6\\1 1}{7 8\\1 1} \textnormal{.}
\end{mmaCell}

\emph{Kite graphs}  have five edges: The cyclic
ones from the box plus one diagonal. We write:
\begin{align}
  \cformkite{A_{1}\\B_{1}}{A_{2}\\B_{2}}{A_{3}\\B_{3}}{A_{4}\\B_{4}}
  {A_{5}\\B_{5}}=
  \begin{tikzpicture}[mgf,rotate=-45]
    \node (1) at (0,0) [vertex] {$1$};
    \node (2) at (0,3.2) [vertex] {$2$};
    \node (3) at (3.2,3.2) [vertex] {$3$};
    \node (4) at (3.2,0) [vertex] {$4$};
    \draw[dedge=0.85] (1) to[bend left=15] (2);
    \draw[dedge=0.85] (1) to[bend right=15] (2);
    \draw[dedge=0.85] (1) to
    node[biglabel]{$\sbmatrix{A_{1}\\B_{1}}$}(2);
    \draw[dedge=0.85] (2) to[bend left=15] (3);
    \draw[dedge=0.85] (2) to[bend right=15] (3);
    \draw[dedge=0.85] (2) to
    node[biglabel]{$\sbmatrix{A_{2}\\B_{2}}$}(3);
    \draw[dedge=0.85] (1) to[bend left=15] (4);
    \draw[dedge=0.85] (1) to[bend right=15] (4);
    \draw[dedge=0.85] (1) to
    node[biglabel]{$\sbmatrix{A_{3}\\B_{3}}$}(4);
    \draw[dedge=0.85] (4) to[bend left=15] (3);
    \draw[dedge=0.85] (4) to[bend right=15] (3);
    \draw[dedge=0.85] (4) to
    node[biglabel]{$\sbmatrix{A_{4}\\B_{4}}$}(3);
    \draw[dedge=0.8] (1) to[bend left=10] (3);
    \draw[dedge=0.8] (1) to[bend right=10] (3);
    \draw[dedge=0.8] (1) to
    node[biglabel]{$\sbmatrix{A_{5}\\B_{5}}$}(3);
  \end{tikzpicture}\,.
  \label{eq:183}
\end{align}
Note that the direction of the four outer edges is different from the box graph.

For kite graphs, \mma{c} has five arguments,
\begin{mmaCell}{Input}
  \mmaincformkite{1 2\\1 1}{1 3\\1 1}{1 4\\1 1}{1 5\\1 1}{1 6\\1 1}
\end{mmaCell}
\begin{mmaCell}{Output}
  \mmaoutcformkite{1 2\\1 1}{1 3\\1 1}{1 4\\1 1}{1 5\\1 1}{1 6\\1 1} \textnormal{.}
\end{mmaCell}

Finally, the full \emph{tetrahedral graph} (also known as \emph{Mercedes graph})
 has six edges connecting all pairs of
points. The Laplace eigenvalue equations of modular graph functions of this
topology were studied in~\cite{kleinschmidt2017}. As will become clear in the
next section, due to its symmetry properties, it is convenient to arrange the
six blocks in three columns as follows:\footnote{In the conventions
  of~\cite{kleinschmidt2017}, the direction of the edges in third block is
  reversed.}
\begin{align}
  \cformtet{A_{1}\\B_{1}}{A_{2}\\B_{2}}{A_{3}\\B_{3}}{A_{4}\\B_{4}}
  {A_{5}\\B_{5}}{A_{6}\\B_{6}}
  =
  \begin{tikzpicture}[mgf,baseline=(4.base)]
    \node (1) at (0,0) [vertex] {$1$};
    \node (2) at (60:5.5) [vertex] {$2$};
    \node (3) at (5.5,0) [vertex] {$3$};
    \draw[opacity=0,name path = ray1](0,0)--(30:5);
    \draw[opacity=0,name path = vert](2)--(2 |- 0,0);
    \node[name intersections={of=ray1 and vert}] (4) at (intersection-1) {$4$};
    \draw[dedge=0.8] (4) to[bend left=10] (2);
    \draw[dedge=0.8] (4) to[bend right=10] (2);
    \draw[dedge=0.8] (4) to
    node[biglabel]{$\sbmatrix{A_{2}\\B_{2}}$}(2);
    \draw[dedge=0.85] (1) to[bend left=10] (4);
    \draw[dedge=0.85] (1) to[bend right=10] (4);
    \draw[dedge=0.85] (1) to
    node[biglabel]{$\sbmatrix{A_{6}\\B_{6}}$}(4);
    \draw[dedge=0.8] (1) to[bend left=7] (2);
    \draw[dedge=0.8] (1) to[bend right=7] (2);
    \draw[dedge=0.8] (1) to
    node[biglabel]{$\sbmatrix{A_{1}\\B_{1}}$}(2);
    \draw[dedge=0.85] (4) to[bend left=10] (3);
    \draw[dedge=0.85] (4) to[bend right=10] (3);
    \draw[dedge=0.85] (4) to
    node[biglabel]{$\sbmatrix{A_{4}\\B_{4}}$}(3);
    \draw[dedge=0.8] (3) to[bend left=7] (2);
    \draw[dedge=0.8] (3) to[bend right=7] (2);
    \draw[dedge=0.8] (3) to
    node[biglabel]{$\sbmatrix{A_{3}\\B_{3}}$}(2);
    \draw[dedge=0.8] (3) to[bend left=7] (1);
    \draw[dedge=0.8] (3) to[bend right=7] (1);
    \draw[dedge=0.8] (3) to
    node[biglabel]{$\sbmatrix{A_{5}\\B_{5}}$}(1);
  \end{tikzpicture}\,.
  \label{eq:184}
\end{align}
Note that in this notation, edge bundles which do not share a common vertex
correspond to blocks written in one column.

Tetrahedral graphs are written in the \mmapackage package as
\mma{c} with six arguments,
\begin{mmaCell}{Input}
  \mmaincformtet{1 2\\1 1}{1 3\\1 1}{1 4\\1 1}{1 5\\1 1}{1 6\\1 1}{1 7\\1 1}
\end{mmaCell}
\begin{mmaCell}{Output}
  \mmaoutcformtet{1 2\\1 1}{1 3\\1 1}{1 4\\1 1}{1 5\\1 1}{1 6\\1 1}{1 7\\1 1} \textnormal{.}
\end{mmaCell}

For all four-point graphs, we will again use the symbol $\noblock$ to denote
blocks without any edges. In \texttt{Mathematica}, we again use empty lists.


\section{Simple relations}
\label{sec:simple-relations}
There are a number of relations between modular graph forms that follow directly
from their definition in terms of graphs and lattice sums. These are easy to
see, yet very powerful and already generate a lot of identities.

\subsection{Symmetries}
\label{sec:MGF-symm}

Given the graph of a modular graph form, the associated $\mathcal{C}$-function
as defined in the previous section is ambiguous and this generates relations
between $\mathcal{C}$-functions with different labels. In the simplest instance,
permutations of the columns of a dihedral graph leave the \ac{MGF}
invariant. The same is true for permutations of columns in any block of the
higher-point graphs.

If a vertex is connected to only two edge bundles, their total momenta have to
agree and hence the two bundles can be swapped without changing the lattice sum
associated to the graph~\cite{dhoker2016a}. For trihedral- and box graphs this
implies invariance under permutations of the blocks~\cite{dhoker2016a},
\begin{align}
  \cformtri{A_{1}\\B_{1}}{A_{2}\\B_{2}}{A_{3}\\B_{3}}
  =\cformtri{A_{2}\\B_{2}}{A_{1}\\B_{1}}{A_{3}\\B_{3}}
  =\cformtri{A_{1}\\B_{1}}{A_{3}\\B_{3}}{A_{2}\\B_{2}}
\end{align}
and similarly for block graphs.

For the same reason, kite graphs are invariant under swapping blocks 1 and 2 as
well as 3 and 4. Furthermore, swapping the vertices 2 and 4 leaves the graph
invariant, so in total the symmetries are
\begin{align}
  \begin{split}
    &\eqspace\!\cformkite{A_{1}\\B_{1}}{A_{2}\\B_{2}}{A_{3}\\B_{3}}{A_{4}\\B_{4}}
    {A_{5}\\B_{5}}
    =\cformkite{A_{2}\\B_{2}}{A_{1}\\B_{1}}{A_{3}\\B_{3}}{A_{4}\\B_{4}}
    {A_{5}\\B_{5}}\\
    &=\cformkite{A_{1}\\B_{1}}{A_{2}\\B_{2}}{A_{4}\\B_{4}}{A_{3}\\B_{3}}
    {A_{5}\\B_{5}}
    =\cformkite{A_{3}\\B_{3}}{A_{4}\\B_{4}}{A_{1}\\B_{1}}{A_{2}\\B_{2}}
    {A_{5}\\B_{5}}\,.
  \end{split}
\label{eq:178}
\end{align}
The double-line notation was chosen to make this intuitive. Note that the
vertices in kite graphs are not all equivalent and this gives rise to the more
complex symmetry properties \eqref{eq:178}.

Tetrahedral graphs have an $S_{4}$ permutation symmetry from relabeling the four
equivalent vertices. These 24 permutations are generated by six permutations:
\begin{itemize}
\item three permutations of columns: Flipping a column comprised of two
  $(A_{i},B_{i})$-blocks in \eqref{eq:184} with any other column produces a sign
  $(-1)^{|1|+|2|+|3|}$ where
  $|1|+|2|+|3|=|A_{1}|+|B_{1}|+|A_{2}|+|B_{2}|+|A_{3}|+|B_{3}|$ is a shorthand
  for the combined modular weight of the top row.\footnote{The sign does not
    depend on if we take the modular weight of the top- or bottom row since the
    total modular weight is even for non-vanishing \acp{MGF}.}
  Explicitly:
  \begin{align}
    \begin{split}
      \cformtet{A_{1}\\B_{1}}{A_{2}\\B_{2}}{A_{3}\\B_{3}}{A_{4}\\B_{4}}
      {A_{5}\\B_{5}}{A_{6}\\B_{6}}&=
      (-1)^{|1|+|2|+|3|}
      \cformtet{A_{2}\\B_{2}}{A_{1}\\B_{1}}{A_{3}\\B_{3}}{A_{5}\\B_{5}}
      {A_{4}\\B_{4}}{A_{6}\\B_{6}}\\
      &=(-1)^{|1|+|2|+|3|}
      \cformtet{A_{3}\\B_{3}}{A_{2}\\B_{2}}{A_{1}\\B_{1}}{A_{6}\\B_{6}}
      {A_{5}\\B_{5}}{A_{4}\\B_{4}}\\
      &=(-1)^{|1|+|2|+|3|}
      \cformtet{A_{1}\\B_{1}}{A_{3}\\B_{3}}{A_{2}\\B_{2}}{A_{4}\\B_{4}}
      {A_{6}\\B_{6}}{A_{5}\\B_{5}}\,.
    \end{split}
    \label{eq:179}
  \end{align}
\item three flips of two top/bottom pairs: Flipping the top/bottom blocks in any
  two columns changes the tetrahedral graph by a sign $(-1)^{|k|+|l|}$, where
  $k$ and $l$ in $|k|+|l|=|A_{k}|+|B_{k}|+|A_{l}|+|B_{l}|$ are given by the
  following prescription: Permute the three columns cyclically until the two
  columns in which top and bottom blocks are swapped are next to each other. The
  blocks in the left one of these has indices $k$ and $l$. Explicitly:
  \begin{align}
    \begin{split}
      \cformtet{A_{1}\\B_{1}}{A_{2}\\B_{2}}{A_{3}\\B_{3}}{A_{4}\\B_{4}}
      {A_{5}\\B_{5}}{A_{6}\\B_{6}}&=
      (-1)^{|1|+|4|}\cformtet{A_{4}\\B_{4}}{A_{5}\\B_{5}}{A_{3}\\B_{3}}
      {A_{1}\\B_{1}}{A_{2}\\B_{2}}{A_{6}\\B_{6}}\\
      &=(-1)^{|2|+|5|}\cformtet{A_{1}\\B_{1}}{A_{5}\\B_{5}}{A_{6}\\B_{6}}
      {A_{4}\\B_{4}}{A_{2}\\B_{2}}{A_{3}\\B_{3}}\\
      &=(-1)^{|3|+|6|}\cformtet{A_{4}\\B_{4}}{A_{2}\\B_{2}}
      {A_{6}\\B_{6}}{A_{1}\\B_{1}}{A_{5}\\B_{5}}{A_{3}\\B_{3}}\,.
    \end{split}
    \label{eq:187}
  \end{align}
\end{itemize}
The arrangement of the blocks in two rows of three columns was chosen to make
these symmetries intuitive. For tetrahedral graphs, although all vertices are
equivalent, the symmetry of the graph is broken by the direction of the edges,
i.e.\ it is not possible to assign the directions in such a way that every
vertex has the same number of ingoing and outgoing edges. Adjusting the edge
direction when relabeling vertices leads to the signs in \eqref{eq:179} and
\eqref{eq:187}. These signs also mean that tetrahedral graphs can vanish by
symmetry although their sum of holomorphic and antiholomorphic labels is
even. E.g., according to \eqref{eq:179},
\begin{align}
  \cformtet{A\\B}{A\\B}{A\\B}{A\\B}{A\\B}{A\\B}=0\,,
  \label{eq:336}
\end{align}
if $|A|+|B|$ odd, although $6(|A|+|B|)$ is even. This form of vanishing by
symmetry does not exist for any of the other discussed graphs since no signs
appear in their symmetry transformations.

In light of the above symmetry properties it is convenient to define a
\emph{canonical representation} \index{modular graph form!canonical
  representation} for the graph topologies discussed so far such that graphs
related by a symmetry transformation are represented by the same arguments of
the $\mathcal{C}$-function. To this end, we define an ordering on the set of
two-row columns and on the set of $2\times R$ matrices. This will allow us to
define an ordering on the \acp{MGF} of a certain topology and the smallest
element in the symmetry orbit of an \ac{MGF} will be the canonical
representation of that graph.

The columns within an $\sbmatrix{A\\B}$-block can be permuted arbitrarily for
all graphs introduced above. The canonical representation of the \acp{MGF}
therefore has the columns in each block in lexicographic order\footnote{In
  lexicographic order, the sequence $a_{1},a_{2},\dots,a_{n}$ is smaller than
  $b_{1},b_{2},\dots,b_{n}$ if $a_{i}<b_{i}$ for the first $i$ for which
  $a_{i}\neq b_{i}$.} w.r.t. the ordering defined by
\begin{itemize}
\item If $a_{1}<a_{2}$ then $\sbmatrix{a_{1}\\b_{1}}<\sbmatrix{a_{2}\\b_{2}}$.
\item If $a_{1}=a_{2}$ then $\sbmatrix{a_{1}\\b_{1}}<\sbmatrix{a_{2}\\b_{2}}$
  if $b_{1}<b_{2}$.
\end{itemize}
Given two blocks $\sbmatrix{A_{1}\\B_{1}}$ and
$\sbmatrix{A_{2}\\B_{2}}$ with canonical column order and $R_{1}$ and
$R_{2}$ columns, respectively, we can define a canonical ordering of the two
blocks by
\begin{itemize}
\item If $R_{1}<R_{2}$ then $\sbmatrix{A_{1}\\B_{1}}<\sbmatrix{A_{2}\\B_{2}}$.
\item If $R_{1}=R_{2}$ then $\sbmatrix{A_{1}\\B_{1}}<\sbmatrix{A_{2}\\B_{2}}$ if
  $A_{1}<A_{2}$ in lexicographic order.
\item If $A_{1}=A_{2}$ then $\sbmatrix{A_{1}\\B_{1}}<\sbmatrix{A_{2}\\B_{2}}$ if
  $B_{1}<B_{2}$ in lexicographic order.
\end{itemize}

Using this ordering, we define
\begin{align}
  \cform{A_{1}\\B_{1}}<\cform{A_{2}\\B_{2}}
  \qquad\te{if}\qquad \sbmatrix{A_{1}\\B_{1}}<\sbmatrix{A_{2}\\B_{2}}
  \label{eq:219}
\end{align}
for dihedral graphs, unless the graph at hand is a one-loop graph. In this case,
we write $\cform{a&0\\b&0}$ instead of $\cform{0&a\\0&b}$, to be consistent with
the previous literature. For graphs with several blocks, we use lexicographic
ordering on the set of blocks, hence
\begin{align}
  \cformtri{A_{1}\\B_{1}}{A_{2}\\B_{2}}{A_{3}\\B_{3}}<
  \cformtri{C_{1}\\D_{1}}{C_{2}\\D_{2}}{C_{3}\\D_{3}}
  \qquad\te{if}\qquad\left(\sbmatrix{A_{1}\\B_{1}},\sbmatrix{A_{2}\\B_{2}},
    \sbmatrix{A_{3}\\B_{3}}\right) < \left(\sbmatrix{C_{1}\\D_{1}},
    \sbmatrix{C_{2}\\D_{2}},\sbmatrix{C_{3}\\D_{3}}\right)
  \label{eq:220}
\end{align}
in lexicographic order and similarly for all four-point graphs with the
numbering of the blocks as in Section~\ref{sec:4ptMGFs}.

For trihedral and box graphs, this just means that the canonical representation
has the blocks (and in each block the columns) in lexicographic ordering. For
kite graphs, the fifth block cannot be moved by the symmetries \eqref{eq:178}
and hence in the canonical representation, the smallest block out of the
remaining four comes first, fixing the second one. The third block is the
smaller one out of the remaining two, fixing the last block. Canonically
represented tetrahedral graphs have the smallest block in the upper left slot,
fixing the lower left block. The smallest block out of the remaining four blocks
sits in the upper middle slot, fixing all remaining entries. The following
examples are all in their canonical representation
\begin{subequations}
  \label{eq:329}
  \begin{gather}
    \cform{3&0\\1&0}\\
    \cform{1&2&3\\7&5&4}\\
    \cformtri{2\\2}{1&1\\1&1}{0&0&1\\2&4&1}\\
    \cformkite{2\\2}{1&2&3\\7&5&4}{1&1\\1&1}{0&0&1\\2&4&1}{1\\1}\\
    \hspace{0.6em}
    \cformtet{1\\1}{2\\2}{2&3\\1&2}{1&2&3\\7&5&4}{0&0&1\\2&4&1}{1&1\\1&1}\,.
  \end{gather}
\end{subequations}

In the \mmapackage package, the function \mma{CSort} brings
\acp{MGF} into their canonical form, using the symmetries discussed above. For
the \acp{MGF} in \eqref{eq:329}, we have e.g.\
\begin{mmaCell}{Input}
  CSort\mmabigso{}\mmabigco{}\mmaincform{0 3\\0 1},\,\mmaincform{2 1 3\\5 7 4},\,\mmaincformtri{1 0 0\\1 4 2}{2\\2}{1 1\\1 1},\,\\
  \mmaincformkite{0 0 1\\4 2 1}{1 1\\1 1}{3 2 1\\4 5 7}{2\\2}{1\\1},\,\mmaincformtet{2 3\\1 2}{1 0 0\\1 4 2}{3 2 1\\4 5 7}{1 1\\1 1}{2\\2}{1\\1}\mmabigcc{}\mmabigsc{}
\end{mmaCell}
\begin{mmaCell}{Output}
  \big\{\mmaoutcform{3 0\\1 0},\,\mmaoutcform{1 2 3\\7 5 4},\,\mmaoutcformtri{2\\2}{1 1\\1 1}{0 0 1\\2 4 1},\,\mmaoutcformkite{2\\2}{1 2 3\\7 5 4}{1 1\\1 1}{0 0 1\\2 4 1}{1\\1},\\
  \quad\mmaoutcformtet{1\\1}{2\\2}{2 3\\1 2}{1 2 3\\7 5 4}{0 0 1\\2 4 1}{1 1\\1 1}\big\} \textnormal{.}
\end{mmaCell}
The output of the function \mma{CSimplify} is always in canonical form. The
property, that tetrahedral graphs can vanish by symmetry, as in the example
\eqref{eq:336}, is implemented in the function \mma{TetCSimplify}. E.g., we have
\begin{mmaCell}{Input}
  TetCSimplify\mmabigso{}\mmaincformtet{1 2\\2 2}{1 2\\2 2}{1 2\\2 2}{1 2\\2 2}{1 2\\2 2}{1 2\\2 2}\mmabigsc{}
\end{mmaCell}
\begin{mmaCell}{Output}
  0 \textnormal{.}
\end{mmaCell}

\subsection{Topological simplifications}
\label{sec:MGF-top-simp}

For certain special cases of the graphs defined in
Section~\ref{sec:graph-topol-not}, the \ac{MGF} simplifies.

For the dihedral case, the fact that one-valent vertices lead to vanishing
\acp{MGF} can be expressed as
\begin{align}
  \cform{a\\b}=0\,.
  \label{eq:188}
\end{align}
It is furthermore convenient to define
\begin{align}
  \cform{\noblock}=1\,.
  \label{eq:193}
\end{align}
The property \eqref{eq:137} that two-valent vertices can be dropped translates
for one-loop dihedral graphs into
\begin{align}
  \cform{a_{1}&a_{2}\\b_{1}&b_{2}}=
  (-1)^{a_{2}+b_{2}}\cform{a_{1}+a_{2}&0\\b_{1}+b_{2}&0}\,.
  \label{eq:139}
\end{align}
For trihedral graphs, \eqref{eq:137} implies
\begin{align}
  \cformtri{a_{1}\\b_{1}}{a_{2}\\b_{2}}{A_{3}\\B_{3}}=
  (-1)^{a_{1}+b_{1}+a_{2}+b_{2}}\cform{a_{1}+a_{2}&A_{3}\\b_{1}+b_{2}&B_{3}}
  \label{eq:189}
\end{align}
and the factorization of one-particle reducible graphs \eqref{eq:190} means that
trihedral graphs with one empty block factorize into dihedral graphs,
\begin{align}
  \cformtri{A_{1}\\B_{1}}{A_{2}\\B_{2}}{\noblock}
  =\cform{A_{1}\\B_{1}}\cform{A_{2}\\B_{2}}\,.
  \label{eq:210}
\end{align}
Via \eqref{eq:193}, this also captures the case of two empty blocks.

Since two- and three-point graphs are special cases of four-point graphs,
topological simplifications of four-point graphs should allow for
simplifications down to dihedral graphs. We will provide a hierarchy of
simplifications from tetrahedral graphs to box graphs which, if applied
repeatedly together with \eqref{eq:188} to \eqref{eq:210}, allow to identify any
lower-point graph which is given as a tetrahedral \ac{MGF}.

Tetrahedral graphs with one empty block are kite graphs,
\begin{align}
  \cformtet{\noblock}{A_{2}\\B_{2}}{A_{3}\\B_{3}}{A_{4}\\B_{4}}
  {A_{5}\\B_{5}}{A_{6}\\B_{6}}
  =(-1)^{|A_{2}|+|B_{2}|+|A_{4}|+|B_{4}|}\cformkite{A_{2}\\B_{2}}
  {A_{3}\\B_{3}}{A_{5}\\B_{5}}{A_{6}\\B_{6}}{A_{4}\\B_{4}}\,.
  \label{eq:192}
\end{align}

A kite graph with one empty block is either a box graph or factorizes,\!
\begin{subequations}
  \label{eq:195}
  \begin{align}
    \cformkite{\noblock}{A_{2}\\B_{2}}{A_{3}\\B_{3}}{A_{4}\\B_{4}}{A_{5}\\B_{5}}
    &=(-1)^{|A_{5}|+|B_{5}|}\cform{A_{2}\\B_{2}}
    \cformtri{A_{3}\\B_{3}}{A_{4}\\B_{4}}{A_{5}\\B_{5}}\\
    \cformkite{A_{1}\\B_{1}}{A_{2}\\B_{2}}{A_{3}\\B_{3}}{A_{4}\\B_{4}}{\noblock}
    &=(-1)^{|A_{3}|+|B_{3}|+|A_{4}|+|B_{4}|}
    \cformbox{A_{1}\\B_{1}}{A_{2}\\B_{2}}{A_{3}\\B_{3}}{A_{4}\\B_{4}}\,.
  \end{align}
\end{subequations}
If the two blocks in the first (or second) pair of blocks have only one column
each, the vertex $2$ (or $4$) becomes two-valent end the kite graph simplifies
into a trihedral graph,
\begin{align}
  \cformkite{a_{1}\\b_{1}}{a_{2}\\b_{2}}{A_{3}\\B_{3}}{A_{4}\\B_{4}}
  {A_{5}\\B_{5}}
  &=(-1)^{|A_{3}|+|B_{3}|+|A_{4}|+|B_{4}|}
  \cformtri{a_{1}+a_{2}&A_{5}\\b_{1}+b_{2}&B_{5}}{A_{3}\\B_{3}}{A_{4}\\B_{4}}\,.
  \label{eq:196}
\end{align}

A box graph with one (or more) empty blocks factorizes into dihedral graphs,
\begin{align}
  \cformbox{\noblock}{A_{2}\\B_{2}}{A_{3}\\B_{3}}{A_{4}\\B_{4}}
  &=\cform{A_{2}\\B_{2}}\cform{A_{3}\\B_{3}}\cform{A_{4}\\B_{4}}
  \label{eq:197}
\end{align}
and a box graph with two blocks of only one column each has a two-valent vertex
and simplifies is a trihedral graph,
\begin{align}
  \cformbox{a_{1}\\b_{1}}{a_{2}\\b_{2}}{A_{3}\\B_{3}}{A_{4}\\B_{4}}
  =\cformtri{a_{1}+a_{2}\\b_{1}+b_{2}}{A_{3}\\B_{3}}{A_{4}\\B_{4}}\,.
  \label{eq:198}
\end{align}

Combined, the relations above show e.g.\ that
\begin{align}
  \cformtet{\noblock}{1\\1}{1\\1}{1\\1}{1\\1}{1\\1}=\cform{1&2&2\\1&2&2}\,.
  \label{eq:191}
\end{align}

In the \texttt{Mathematica} package \mmapackage, the dihedral relations
\eqref{eq:188}--\eqref{eq:139} are implement in the function \mma{DiCSimplify},
\begin{mmaCell}{Input}
  DiCSimplify\mmabigso{}c[\{\}]\mmaincform{0 3\\1 0}+\mmaincform{3\\1}\mmabigsc{}
\end{mmaCell}
\begin{mmaCell}{Output}
  \mmam\!\mmaoutcform{3 0\\1 0} \textnormal{.}
\end{mmaCell}
\mma{DiCSimplify} also rewrites the special cases $\GGhat_{2}$, $\GG_{k}$ and
$\EE_{k}$ of one-loop graphs according to \eqref{eq:330} and \eqref{eq:331}, as
well as \eqref{eq:332}, whereas the one-loop simplification \eqref{eq:139} is
also performed by \mma{CSort}. The function \mma{DiCSimplify} has a Boolean
option \mma{basisExpandG} which, if set to \mma{True}, causes \mma{DiCSimplify}
to expand all holomorphic Eisenstein series in the ring of $\GG_{4}$ and
$\GG_{6}$, e.g.\
\begin{mmaCell}{Input}
  DiCSimplify[g[24],\,basisExpandG\mmainrarrow{}True]
\end{mmaCell}
\begin{mmaCell}{Output}
  \mmaFrac{270\,\mmaSubSup{G}{4}{6}}{66079}+\mmaFrac{5400000\,\mmaSubSup{G}{4}{3}\,\mmaSubSup{G}{6}{2}}{151915621}+\mmaFrac{375\,\mmaSubSup{G}{6}{4}}{73853} \textnormal{.}
\end{mmaCell}
The default value of \mma{basisExpandG} is \mma{False}.

The trihedral simplifications \eqref{eq:189} and \eqref{eq:210} are performed by
\mma{TriCSimplify},
\begin{mmaCell}{Input}
  TriCSimplify\mmabigso{}\mmaincformtri{\mmanothing}{1 2\\1 1}{1 4\\1 1}+\mmaincformtri{1\\1}{2\\2}{1 3\\1 1}\mmabigsc{}
\end{mmaCell}
\begin{mmaCell}[morelst={label=mma:1}]{Output}
  \mmaoutcform{1 2\\1 1}\,\mmaoutcform{1 4\\1 1}+\mmaoutcform{3 1 3\\3 1 1} \textnormal{.}
\end{mmaCell}
Note that the dihedral graphs in \mmaoutref{mma:1} are not simplified or
canonically represented, since \mma{TriCSimplify} only acts
on trihedral graphs. To simplify \mmaoutref{mma:1} further, we can apply
\mma{DiCSimplify},
\renewcommand{\temprefa}{\ref*{mma:1}}
\begin{mmaCell}{Input}
  DiCSimplify\mmabigso{}Out[\temprefa],\,useIds\mmainrarrow{}False\mmabigsc{}
\end{mmaCell}
\begin{mmaCell}{Output}
  \mmaoutcform{1 3 3\\1 1 3} \textnormal{,}
\end{mmaCell}
where the Boolean option \mma{useIds} was set to suppress the expansion using
the result in the basis decompositions to be discussed in
Section~\ref{sec:conjectured-basis}. The hierarchy of topological four-point
simplifications \eqref{eq:192}--\eqref{eq:198} is implemented in the function
\mma{TetCSimplify}. Combining these functions, one can reproduce the example
\eqref{eq:191},
\begin{mmaCell}{Input}
  TetCSimplify\mmabigso{}\mmaincformtet{\mmanothing}{1\\1}{1\\1}{1\\1}{1\\1}{1\\1}\mmabigsc{}
  TriCSimplify[
  CSort[
\end{mmaCell}
\begin{mmaCell}{Output}
  \mmaoutcformtri{2 1\\2 1}{1\\1}{1\\1}
\end{mmaCell}
\begin{mmaCell}{Output}
  \mmaoutcform{2 1 2\\2 1 2}
\end{mmaCell}
\begin{mmaCell}{Output}
  \mmaoutcform{1 2 2\\1 2 2} \textnormal{.}
\end{mmaCell}
The function \mma{CSimplify} acts on \acp{MGF} of all topologies
and calls\\
\mma{DiCSimplify}, \mma{TriCSimplify} and
\mma{TetCSimplify}. It also inherits the option \mma{basisExpandG} from
\mma{DiCSimplify}. We have e.g.\
\begin{mmaCell}{Input}
  CSimplify\mmabigso{}\mmaincformtet{\mmanothing}{1 2\\1 2}{\mmanothing}{1\\2}{2\\1}{2\\2}\mmabigsc{}
\end{mmaCell}
\begin{mmaCell}{Output}
  \mmaFrac{\mmaSup{\(\pi\)}{8}\,\mmaSub{E}{3}\,\mmaSub{E}{5}}{\mmaSubSup{\(\tau\)}{2}{8}} \textnormal{.}
\end{mmaCell}

\subsection{Momentum conservation}
\label{sec:mom-cons}
Momentum conservation~\cite{dhoker2016a} will be the central tool in our
derivation of identities between modular graph forms and can be derived in the
lattice sum representation \eqref{eq:131} as well as the integral representation
\eqref{eq:128} of the \ac{MGF}. As long as all graphs involved are convergent,
as we will assume in this section, both approaches result in the same
expression. If divergent graphs are involved, the integral representation allows
one to use the tools of complex analysis to derive meaningful results, cf.\
Section~\ref{sec:div-mom-cons}.

Starting from the lattice-sum representation~\eqref{eq:131} of an \ac{MGF} with
$|A|+|B|$ odd (hence, a vanishing \ac{MGF}), which we will refer to as the
\emph{seed}, we have for each $j\in V_{\Gamma}$ the \emph{momentum conservation
  identities}
\begin{subequations}
  \label{eq:201}
  \begin{align}
    0&=\sum_{e'\in E_{\Gamma}}\Gamma_{je'}\sum_{\{p_e\}}'\prod_{e\in E_{\Gamma}}
    \frac{p_{e'}}
    {p_e^{a_e} \bar{p}^{b_e}_e} \prod_{i\in V_{\Gamma}}
    \delta\left(\sum_{e''\in E_{\Gamma}}\Gamma_{ie''}p_{e''}\right)
    \label{eq:199}\\
    0&=\sum_{e'\in E_{\Gamma}}\Gamma_{je'}\sum_{\{p_e\}}'\prod_{e\in E_{\Gamma}}
    \frac{\bar{p}_{e'}}
    {p_e^{a_e} \bar{p}^{b_e}_e} \prod_{i\in V_{\Gamma}}
    \delta\left(\sum_{e''\in E_{\Gamma}}\Gamma_{ie''}p_{e''}\right)
    \label{eq:200}
  \end{align}
\end{subequations}
due to the momentum conserving delta functions. We will refer to \eqref{eq:199}
as the \emph{holomorphic-} and to \eqref{eq:200} as the \emph{antiholomorphic
  momentum conservation identity}. By canceling the momenta from the numerators,
\eqref{eq:201} can be expressed entirely as a manipulation of the decorations of
the graph and are therefore identities between \acp{MGF},
\begin{align}
  0=\sum_{e\in E_{\Gamma}}\Gamma_{je}
  \mathcal{C}_{\Gamma_{a_{e}\,\rightarrow\, a_{e}-1}}\,,\qquad
  0=\sum_{e\in E_{\Gamma}}\Gamma_{je}
  \mathcal{C}_{\Gamma_{b_{e}\,\rightarrow\, b_{e}-1}}\,,
  \qquad \forall\, j&\in V_{\Gamma}\,.
  \label{eq:228}
\end{align}
If we had chosen a seed with $|A|+|B|$ even, the resulting \acp{MGF} would have
all vanished trivially. Note that exchanging the sums over $e'$ and the $p_{e}$
in \eqref{eq:201} required all sums to be convergent.

In the integral representation \eqref{eq:128}, the momentum conservation
identities \eqref{eq:201} correspond to integration-by-parts identities
w.r.t. the puncture positions. To see this, note that due to
\eqref{eq:126},\footnote{The $a=1,b=0$ case of $\partial_{\bar{z}}C^{(a,b)}$ is
  compatible with \eqref{eq:364} upon using \eqref{eq:230} below.}
\begin{align}
  \partial_{z}C^{(a,b)}(z)&=-\frac{\pi}{\tau_{2}}C^{(a,b-1)}(z) &
  \partial_{\bar{z}}C^{(a,b)}(z)&=\frac{\pi}{\tau_{2}}C^{(a-1,b)}(z)\,.
  \label{eq:194}
\end{align}
If the integrand in \eqref{eq:128} has no poles, the integral over the total
derivative w.r.t. $z_{j}$ for each $j\in V_{\Gamma}$ vanishes and we have
\begin{subequations}
  \label{eq:208}
  \begin{align}
    0&=\sum_{e'\in E_{\Gamma}}\Gamma_{je'}\int\dd\mu_{n-1}\,
    C^{(a_{e'},b_{e'}-1)}(z_{e'})\prod_{\substack{e\in E_{\Gamma}\\e\neq e'}}
    C^{(a_{e},b_{e})}(z_{e})\\
    0&=\sum_{e'\in E_{\Gamma}}\Gamma_{je'}\int\dd\mu_{n-1}\,
    C^{(a_{e'}-1,b_{e'})}(z_{e'})\prod_{\substack{e\in E_{\Gamma}\\e\neq e'}}
    C^{(a_{e},b_{e})}(z_{e})\,,
  \end{align}
\end{subequations}
agreeing with \eqref{eq:228}.

For dihedral graphs, the identities \eqref{eq:228} for both vertices are
identical and can be written as
\begin{align}
  0=\sum_{i=1}^{R}\cform{A-S_{i}\\B}=\sum_{i=1}^{R}\cform{A\\B-S_{i}}\,,
  \label{eq:202}
\end{align}
the $j$\textsuperscript{th} component of the row vector $S_{i}$ is
$\delta_{ij}$. For trihedral \acp{MGF}, the momentum conservation identities
involve two out of the three blocks and are given by
\begin{align}
  0&=\sum_{i=1}^{R_{1}}\cformtri{A_{1}-S_{i}\\B_{1}}{A_{2}\\B_{2}}{A_{3}\\B_{3}}
  -\sum_{i=1}^{R_{2}}\cformtri{A_{1}\\B_{1}}{A_{2}-S_{i}\\B_{2}}{A_{3}\\B_{3}}
  \label{eq:203}
\end{align}
and similarly for the complex conjugated identities. For box graphs, we have
\begin{align}
  0=\sum_{i=1}^{R_{1}}
  \cformbox{A_{1}-S_{i}\\B_{1}}{A_{2}\\B_{2}}{A_{3}\\B_{3}}{A_{4}\\B_{4}}
  -\sum_{i=1}^{R_{2}}
  \cformbox{A_{1}\\B_{1}}{A_{2}-S_{i}\\B_{2}}{A_{3}\\B_{3}}{A_{4}\\B_{4}}
  \quad\te{and c.c.}
  \label{eq:204}
\end{align}
For kite graphs, we have to distinguish the cases in which the momentum
conservation of vertex $2$ or $4$ is used, yielding
\begin{align}
  0&=\sum_{i=1}^{R_{1}}
  \cformkite{A_{1}-S_{i}\\B_{1}}{A_{2}\\B_{2}}{A_{3}\\B_{3}}
  {A_{4}\\B_{4}}{A_{5}\\B_{5}}
  -\sum_{i=1}^{R_{2}}
  \cformkite{A_{1}\\B_{1}}{A_{2}-S_{i}\\B_{2}}{A_{3}\\B_{3}}
  {A_{4}\\B_{4}}{A_{5}\\B_{5}}\quad\te{and c.c.}
  \label{eq:205}
\end{align}
and the case in which the momentum conservation of vertex $1$ or $3$ is used,
resulting in the identity
\begin{align}
    0&=\sum_{i=1}^{R_{1}}\cformkite{A_{1}-S_{i}\\B_{1}}{A_{2}\\B_{2}}
    {A_{3}\\B_{3}}{A_{4}\\B_{4}}{A_{5}\\B_{5}}
    +\sum_{i=1}^{R_{4}}\cformkite{A_{1}\\B_{1}}{A_{2}\\B_{2}}
    {A_{3}-S_{i}\\B_{3}}{A_{4}\\B_{4}}{A_{5}\\B_{5}}\nonumber\\
    &\qquad+\sum_{i=1}^{R_{5}}\cformkite{A_{1}\\B_{1}}{A_{2}\\B_{2}}
    {A_{3}\\B_{3}}{A_{4}\\B_{4}}{A_{5}-S_{i}\\B_{5}}
    \quad\te{and c.c.}
  \label{eq:206}
\end{align}
The topology of tetrahedral graphs is completely symmetric, hence the momentum
conservation identity for vertex $2$,
\begin{align}
    0&=\sum_{i=1}^{R_{1}}\cformtet{A_{1}-S_{i}\\B_{1}}{A_{2}\\B_{2}}
    {A_{3}\\B_{3}}{A_{4}\\B_{4}}{A_{5}\\B_{5}}{A_{6}\\B_{6}}
    +\sum_{i=1}^{R_{2}}\cformtet{A_{1}\\B_{1}}{A_{2}-S_{i}\\B_{2}}
    {A_{3}\\B_{3}}{A_{4}\\B_{4}}{A_{5}\\B_{5}}{A_{6}\\B_{6}}
   +\sum_{i=1}^{R_{3}}\cformtet{A_{1}\\B_{1}}{A_{2}\\B_{2}}
    {A_{3}-S_{i}\\B_{3}}{A_{4}\\B_{4}}{A_{5}\\B_{5}}{A_{6}\\B_{6}}
  \label{eq:207}
\end{align}
and its complex conjugate are related to those of all other vertices by the
transformations \eqref{eq:179} and \eqref{eq:187}.

In the \mmapackage package, momentum conservation for dihedral and trihedral
graphs is implemented in the functions \mma{DiHolMomConsId} and
\mma{TriHolMomConsId} and their antiholomorphic versions \mma{DiAHolMomConsId}
and \mma{TriAHolMomConsId}. In the dihedral case \eqref{eq:202}, the function
\mma{DiHolMomConsId} takes the seed as its only argument and we have e.g.
\begin{mmaCell}{Input}
  DiHolMomConsId\mmabigso{}\mmaincform{1 1 2\\1 1 1}\mmabigsc{}
\end{mmaCell}
\begin{mmaCell}[morelst={label=mma:2}]{Output}
  \mmaoutcform{0 1 2\\1 1 1}+\mmaoutcform{1 0 2\\1 1 1}+\mmaoutcform{1 1 1\\1 1 1}==0 \textnormal{.}
\end{mmaCell}
For trihedral momentum conservation \eqref{eq:203}, we have to specify which of
the three vertices we use and hence which pair of blocks has its labels
changed. The list of these blocks is passed as a second argument to
\mma{TriHolMomConsId}, e.g.\
\begin{mmaCell}{Input}
  TriHolMomConsId\mmabigso{}\mmaincformtri{1 2\\1 1}{1 3\\1 1}{1 4\\1 1},\,\{2,3\}\mmabigsc{}
\end{mmaCell}
\begin{mmaCell}{Output}
  \mmaoutcformtri{1 2\\1 1}{0 3\\1 1}{1 4\\1 1}+\mmaoutcformtri{1 2\\1 1}{1 2\\1 1}{1 4\\1 1}\mmam\mmaoutcformtri{1 2\\1 1}{1 3\\1 1}{0 4\\1 1}\mmam\mmaoutcformtri{1 2\\1 1}{1 3\\1 1}{1 3\\1 1}==0 \textnormal{.}
\end{mmaCell}

Note that the functions discussed here do not apply \mma{CSort} to the resulting
equation, so that it is more transparent which exponents were lowered. E.g.\
\mmaoutref{mma:2} simplifies to
\renewcommand{\temprefa}{\ref*{mma:2}}
\begin{mmaCell}{Input}
  CSort\mmabigso{}Out[\temprefa]\mmabigsc{}
\end{mmaCell}
\begin{mmaCell}{Output}
  2\,\mmaoutcform{0 1 2\\1 1 1}+\mmaoutcform{1 1 1\\1 1 1}==0 \textnormal{.}
\end{mmaCell}

\subsection{Factorization}
\label{sec:fact}
Consider a modular graph form with a $(0,0)$-edge. In this case, the
graph factorizes~\cite{dhoker2016a}. To see this, consider two vertices $x$ and
$y$ and an edge from $x$ to $y$ with momentum $p$ and decoration
$(0,0)$. Furthermore assume that all other edges connected to $x$ are directed
away from $x$ and have momentum sum $p_{x}$ and all other edges connected to $y$
are directed away from $y$ and have momentum sum $p_{y}$ ,
\begin{align}
  \begin{tikzpicture}[mgf]
    \node (x) at (0,0) [vertex] {$x$};
    \node (y) at (3,0) [vertex,inner sep=0.03cm] {$y$\rule[0.3cm]{0pt}{0pt}};
    \draw[dedge=0.85] (x) to node[label,pos=0.45]{$(0,0)$}
    node[mlabel,anchor=north,pos=0.8]{$p$} (y);
    \node (xlt) at (-1.2,0.5) {};
    \node (xlb) at (-1.2,-0.5) {};
    \node at (-0.7,0.11) {$\vdots$};
    \draw[dedge] (x) to[in=0,out=120] (xlt);
    \draw[dedge] (x) to[in=0,out=240] (xlb);
    \node at (-1.5,0)
    {$p_{x}\,\left\{\rule[0.8cm]{0pt}{0pt}\right.$};
    \node (yrt) at (4.2,0.5) {};
    \node (yrb) at (4.2,-0.5) {};
    \node at (3.7,0.11) {$\vdots$};
    \draw[dedge] (y) to[in=180,out=60] (yrt);
    \draw[dedge] (y) to[in=180,out=300] (yrb);
    \node at (4.5,0)
    {$\left.\rule[0.8cm]{0pt}{0pt}\right\}\,p_{y}$};
  \end{tikzpicture}\,,
\label{eq:209}
\end{align}
where the $(0,0)$-edge is not necessarily the only edge between $x$ and $y$.
In the sum representation, the momentum $p$ only appears in the
momentum-conserving delta functions for the vertices $x$ and $y$. Isolating this
contribution, we get
\begin{align}
    \sum_{p}'\delta(p_{x}{+}p)\delta(p_{y}{-}p)
    &=\sum_{p}\delta(p_{x}{+}p)\delta(p_{y}{-}p)-\delta(p_{x})\delta(p_{y})
    =\delta(p_{x}{+}p_{y})-\delta(p_{x})\delta(p_{y})\,,
  \label{eq:221}
\end{align}
where we added $p=0$ to the sum to evaluate the deltas. When \eqref{eq:221}
appears in the nested lattice sum of an \ac{MGF}, the first term gives rise to
the original \ac{MGF} with the vertices $x$ and $y$ identified, whereas the
second term can be associated to the original \ac{MGF} with the $(0,0)$-edge
removed. Schematically, if the edge $e$ between vertices $x$ and $y$ carries
decoration $(0,0)$, we have
\begin{align}
  \mathcal{C}_{\Gamma_{a_{e}=b_{e}=0}}
  =\mathcal{C}_{\Gamma_{x=y}}-\mathcal{C}_{\Gamma\setminus e}\,.
  \label{eq:229}
\end{align}
If the vertices $x$ and $y$ are connected by more edges than just $e$, these
will factorize as one-loop graphs in the first term of \eqref{eq:229}.

In the integral representation, a $(0,0)$-edge is represented by a factor
$C^{(0,0)}(z)$ in the integrand, which as special case of \eqref{eq:126} can be
simplified to
\begin{align}
    C^{(0,0)}(z)&=\sum_{m,n\in\ZZ}e^{2\pi i(mv-nu)}-1
    =\delta(v)\delta(u)-1=\tau_{2}\delta^{(2)}(z,\bar{z})-1\,,
  \label{eq:230}
\end{align}
where we used \eqref{eq:232}. Note that \eqref{eq:230} is not the $a=0$ case of
\eqref{eq:127}, since $f^{(0)}(z)=1$, but is implied by the $a=1,b=0$ case of
\eqref{eq:194} and \eqref{eq:364}. The interpretation of \eqref{eq:230} is
exactly as in the sum representation: The delta identifies the two vertices
connected by the $(0,0)$ edge and in the second term the $(0,0)$ edge is
removed. In this way, we get again \eqref{eq:229}.

For dihedral \acp{MGF}, \eqref{eq:229} implies\footnote{For one-loop graphs
  $\cform{a&0\\b&0}$, \eqref{eq:222} is trivial upon using \eqref{eq:188}.}
\begin{align}
  \cform{0&A\\0&B}=\prod_{j=1}^{R}\cform{a_j&0\\b_j&0}-\cform{A\\B}
  \label{eq:222}
\end{align}
for higher-point graphs we have
\begin{align}
  \cformtri{0&A_{1}\\0&B_{1}}{A_{2}\\B_{2}}{A_{3}\\B_{3}}
  &=(-1)^{|2|}\cform{A_{2}&A_{3}\\B_{2}&B_{3}}
  \prod_{i=1}^{R_{1}}\cform{a_{1}^{(i)}&0\\b_{1}^{(i)}&0}
  -\cformtri{A_{1}\\B_{1}}{A_{2}\\B_{2}}{A_{3}\\B_{3}}
  \label{eq:223}\displaybreak[0]\\
  \cformbox{0&A_{1}\\0&B_{1}}{A_{2}\\B_{2}}{A_{3}\\B_{3}}{A_{4}\\B_{4}}
  &=\cformtri{A_{2}\\B_{2}}{A_{3}\\B_{3}}{A_{4}\\B_{4}}
  \prod_{i=1}^{R_{1}}\cform{a_{1}^{(i)}&0\\b_{1}^{(i)}&0}
  -\cformbox{A_{1}\\B_{1}}{A_{2}\\B_{2}}{A_{3}\\B_{3}}{A_{4}\\B_{4}}
  \label{eq:224}\displaybreak[0]\\
  \begin{split}
    \cformkite{0&A_{1}\\0&B_{1}}{A_{2}\\B_{2}}
    {A_{3}\\B_{3}}{A_{4}\\B_{4}}{A_{5}\\B_{5}}
    &=(-1)^{|2|+|5|}\cformtri{A_{2}&A_{5}\\B_{2}&B_{5}}
    {A_{3}\\B_{3}}{A_{4}\\B_{4}}
    \prod_{i=1}^{R_{1}}\cform{a_{1}^{(i)}&0\\b_{1}^{(i)}&0}\\
    &\qquad-\cformkite{A_{1}\\B_{1}}{A_{2}\\B_{2}}
    {A_{3}\\B_{3}}{A_{4}\\B_{4}}{A_{5}\\B_{5}}
  \end{split}\label{eq:225}\displaybreak[0]\\[1ex]
  \begin{split}
    \cformkite{A_{1}\\B_{1}}{A_{2}\\B_{2}}
    {A_{3}\\B_{3}}{A_{4}\\B_{4}}{0&A_{5}\\0&B_{5}}
    &=(-1)^{|1|+|3|}\cform{A_{1}&A_{2}\\B_{1}&B_{2}}
    \cform{A_{3}&A_{4}\\B_{3}&B_{4}}
    \prod_{i=1}^{R_{5}}\cform{a_{5}^{(i)}&0\\b_{5}^{(i)}&0}\\
    &\qquad-\cformkite{A_{1}\\B_{1}}{A_{2}\\B_{2}}
    {A_{3}\\B_{3}}{A_{4}\\B_{4}}{A_{5}\\B_{5}}
  \end{split} \label{eq:226}\displaybreak[0]\\[1ex]
  \cformtet{0&A_{1}\\0&B_{1}}{A_{2}\\B_{2}}
  {A_{3}\\B_{3}}{A_{4}\\B_{4}}{A_{5}\\B_{5}}{A_{6}\\B_{6}}
  &=(-1)^{|2|}\cformtri{A_{2}&A_{6}\\B_{2}&B_{6}}{A_{3}&A_{5}\\B_{3}&B_{5}}
  {A_{4}\\B_{4}}
  \prod_{i=1}^{R_{1}}\cform{a_{1}^{(i)}&0\\b_{1}^{(i)}&0}
  -\cformtet{A_{1}\\B_{1}}{A_{2}\\B_{2}}
  {A_{3}\\B_{3}}{A_{4}\\B_{4}}{A_{5}\\B_{5}}{A_{6}\\B_{6}}\,,\label{eq:227}
\end{align}
where we used the abbreviation $|i|=|A_{i}|+|B_{i}|$ as above. Note that the
\acp{RHS} have one vertex less in the first term and one loop order less in the
second term and hence \eqref{eq:222} to \eqref{eq:227} are powerful identities
to simplify \acp{MGF}. Together with the momentum conservation identities from
Section~\ref{sec:mom-cons}, these identities form the backbone of all the
simplifications we will carry out in the following.

In the \mmapackage package, factorization of $(0,0)$-edges for dihedral and
trihedral graph is also done by the functions \mma{DiCSimplify} and
\mma{TriCSimplify}. E.g.\ in the trihedral case \eqref{eq:223}, we have
\begin{mmaCell}{Input}
  TriCSimplify\mmabigso{}\mmaincformtri{0 2 1\\0 1 2}{1 2\\1 1}{1 4\\1 1}\mmabigsc{}
\end{mmaCell}
\begin{mmaCell}{Output}
  \mmam\mmaoutcform{1 0\\2 0}\,\mmaoutcform{2 0\\1 0}\,\mmaoutcform{2 2 1 4\\1 2 1 1}\mmam\mmaoutcformtri{1 2\\2 1}{1 4\\1 1}{2 2\\1 2} \textnormal{.}
\end{mmaCell}
If several $(0,0)$-edges are present, the factorization is repeated until no
more $(0,0)$-edges in the respective topology appear. E.g.\ we have
\begin{mmaCell}{Input}
  TriCSimplify\mmabigso{}\mmaincformtri{0 2 1\\0 1 2}{0 2 2\\0 1 2}{1 4\\1 1}\mmabigsc{}
\end{mmaCell}
\begin{mmaCell}{Output}
  \mmaoutcform{2 0\\1 0}\,\mmaoutcform{2 0\\2 0}\,\mmaoutcform{1 2 1 4\\2 1 1 1}\mmam\mmaoutcform{1 0\\2 0}\,\mmaoutcform{2 0\\1 0}\,\mmaoutcform{0 2 2 1 4\\0 1 2 1 1}+\mmaoutcformtri{1 2\\2 1}{1 4\\1 1}{2 2\\1 2} \textnormal{,}
\end{mmaCell}
where the remaining dihedral factorization can be preformed by applying
\mma{DiCSimplify}.

\subsection{Taking derivatives}
\label{sec:MGFDers}
On top of momentum conservation and factorization, another way to obtain new
identities for \acp{MGF} is by taking derivatives of known identities
w.r.t. $\tau$.

In order to take derivatives of modular functions and -forms, we use the
\emph{Maa\ss{} operators}~\cite{maass1983}
\begin{align}
  \nabla^{(a)}&=2i\tau_{2}\partial_{\tau}+a &
  \nablab^{(b)}&=-2i\tau_{2}\partial_{\bar{\tau}}+b\,.
  \label{eq:73}
\end{align}
When these act on modular forms of weight $(a,b)$ they transform them into
modular forms of shifted modular weight according to
\begin{align}
  \nabla^{(a)}:&\ (a,b)\rightarrow(a+1,b-1) &
  \nablab^{(b)}:&\ (a,b)\rightarrow(a-1,b+1)\,.
  \label{eq:74}
\end{align}
These operators satisfy a product rule when acting on a product $fg$ of modular
forms $f$ and $g$ of holomorphic modular weights $a$ and $a'$, respectively,
\begin{align}
  \nabla^{(a+a')} (fg) = (\nabla^{(a)} f) g +f (\nabla^{(a')}  g)\qquad\te{and
  c.c.}\,.
  \label{eq:75}
\end{align}
For later convenience, we introduce the notation
\begin{align}
  \nabla^{(a)^{\scriptstyle n}}&=\nabla^{(a+n)}\nabla^{(a+n-1)}\dotsm
  \nabla^{(a)}&
  \nablab^{(b)^{\scriptstyle n}}&=\nablab^{(b+n)}\nablab^{(b+n-1)}\dotsm
  \nablab^{(b)}
  \label{eq:321}
\end{align}
for higher derivatives. We will also use the operators
\begin{align}
  \nabladg&=\tau_{2}\nabla^{(0)}=2i\tau_{2}^{2}\partial_{\tau} &
  \nabladgb&=\tau_{2}\nablab^{(0)}=-2i\tau_{2}^{2}\partial_{\bar{\tau}}\,,
  \label{eq:76}
\end{align}
which act on modular forms of vanishing holomorphic and antiholomorphic modular
weight,
\begin{align}
  \nabladg:&\ (0,b)\rightarrow(0,b-2) &
  \nabladgb:&\ (a,0)\rightarrow(a-2,0)\,.
  \label{eq:77}
\end{align}
Using these operators, we will now discuss derivatives of identities between
\acp{MGF}. Note that since the Maa\ss{} operators change the modular weight, one
obtains an identity between \acp{MGF} of different weights.

Consider the action of $\nabla^{(|A|)}$ and $\nablab^{(|B|)}$ and on an \ac{MGF}
of weight $(|A|,|B|)$ in its lattice sum representation \eqref{eq:131}. Using
the product rule \eqref{eq:75} and
\begin{align}
  \nabla^{(a)} \left(\frac{1}{p^{a}}\right)
  = a \,\frac{1}{p^{a+1}\bar{p}^{-1}}\qquad
  \nablab^{(b)} \left(\frac{1}{\bar{p}^{b}}\right)
  =b \,\frac{1}{p^{-1}\bar{p}^{b+1}}\,,
  \label{eq:234}
\end{align}
the derivatives are given by~\cite{dhoker2016a}
\begin{align}
  \nabla^{(|A|)}\mathcal{C}_{\Gamma}
  &=\sum_{e\in E_{\Gamma}}a_{e}
  \mathcal{C}_{\Gamma_{(a_{e},b_{e})\,\rightarrow\,(a_{e}+1,b_{e}-1)}} & 
  \nablab^{(|B|)}\mathcal{C}_{\Gamma}
  &=\sum_{e\in E_{\Gamma}}b_{e}
  \mathcal{C}_{\Gamma_{(a_{e},b_{e})\,\rightarrow\,(a_{e}-1,b_{e}+1)}}\,.
  \label{eq:233}
\end{align}

In the integral representation, $\nabla^{(|A|)}$ and $\nablab^{(|B|)}$ act on
the Jacobi forms $C^{(a,b)}(z,\tau)$ given in \eqref{eq:126}. According to
\eqref{eq:234}, we have
\begin{align}
  \nabla^{(a)}C^{(a,b)}(z,\tau)&=a\, C^{(a+1,b-1)}(z,\tau)&
  \nablab^{(b)}C^{(a,b)}(z,\tau)&=b\, C^{(a-1,b+1)}(z,\tau)
  \label{eq:235}
\end{align}
and using this together with the product rule \eqref{eq:75}, we obtain again
\eqref{eq:233}.

For a dihedral \ac{MGF}, \eqref{eq:233} implies~\cite{dhoker2016a}
\begin{align}
  \nabla^{(|A|)}\cform{A\\B}&=\sum_{i=1}^{R}a_{i}\cform{A+S_{i}\\B-S_{i}}&
  \nabla^{(|B|)}\cform{A\\B}&=\sum_{i=1}^{R}b_{i}\cform{A-S_{i}\\B+S_{i}}\,,
  \label{eq:236}
\end{align}
where the $j$\textsuperscript{th} component of $S_{i}$ is $\delta_{ij}$ as
above. A special case of \eqref{eq:236} is the important relation
\begin{align}
  \nabladg^{n}\EE_{k}
  &=\frac{\tau_{2}^{k+n}}{\pi^{k}}\frac{(k+n-1)!}{(k-1)!}\cform{k+n&0\\k-n&0}\,,
  \label{eq:328}
\end{align}
where $\nabladg$ is defined in \eqref{eq:76}. Since \eqref{eq:233} does not
depend on the topology of the graph, the higher-point versions of \eqref{eq:236}
are completely analogous, so for trihedral graphs, we have
e.g.~\cite{dhoker2016a}
\begin{align}
  \nabla^{(|A|)}\cformtri{A_{1}\\B_{1}}{A_{2}\\B_{2}}{A_{3}\\B_{3}}&=
  \sum_{i=1}^{R_{1}}a_{1}^{(i)}
  \cformtri{A_{1}+S_{i}\\B_{1}-S_{i}}{A_{2}\\B_{2}}{A_{3}\\B_{3}}
  +\sum_{i=1}^{R_{2}}a_{2}^{(i)}
  \cformtri{A_{1}\\B_{1}}{A_{2}+S_{i}\\B_{2}-S_{i}}{A_{3}\\B_{3}}\nonumber\\
  &\qquad+\sum_{i=1}^{R_{3}}a_{3}^{(i)}
  \cformtri{A_{1}\\B_{1}}{A_{2}\\B_{2}}{A_{3}+S_{i}\\B_{3}-S_{i}}
  \quad\te{and c.c.}\,,
  \label{eq:237}
\end{align}
where in the complex conjugation, we swap all $a$ and $b$ labels everywhere and
replace $S_{i}\rightarrow -S_{i}$. Similar identities hold for all four-point
graphs.

When taking the Cauchy--Riemann derivative of a holomorphic Eisenstein series,
one obtains
\begin{align}
  \nabla^{(2k)}\GG_{2k}=2k\cform{2k+1&0\\-1&0}\,,\quad k\geq2\,,
  \label{eq:310}
\end{align}
which cannot be simplified further with the methods presented so far. However,
the $\bar{\tau}$-derivative of the weight $(2k+2,0)$ modular form
\begin{align}
  \frac{\pi}{\tau_{2}}\cform{2k+1&0\\-1&0}+\GG_{2k}\GGhat_{2}\,,
  \quad k\geq2
  \label{eq:311}
\end{align}
vanishes, and hence it can be expanded in the ring of holomorphic Eisenstein
series. To this end, we calculate the $q$ expansion ($q=e^{2\pi i \tau}$)
\begin{align}
  \frac{1}{2k}\nabla^{(2k)}\GG_{2k}=2\zeta_{2k}-\frac{4\zeta_{2k}}{B_{2k}}
  \sum_{n=1}^{\infty}\sigma_{2k-1}(n)(2k-4\pi n\tau_{2})q^{n}\,, \quad k\geq1\,,
  \label{eq:312}
\end{align}
by taking the Cauchy--Riemann derivative of the $q$ expansion of
$\GG_{2k}$. Now, by comparing a finite number of terms, we can expand
\eqref{eq:310} in the ring of $\GG_{4}$ and $\GG_{6}$. Since for low weights
this ring is one-dimensional, we can give a closed formula in these cases,
\begin{align}
  \cform{2k+1&0\\-1&0}
  =\frac{\tau_{2}}{\pi}\left(\frac{2\zeta_{2}\zeta_{2k}}{\zeta_{2k+2}}\GG_{2k+2}
    -\GG_{2k}\GGhat_{2}\right)\,, \quad k=2,3,4\,.
  \label{eq:313}
\end{align}
For the non-holomorphic but modular version
$\GGhat_{2}=\GG_{2}-\frac{\pi}{\tau_{2}}$ of $\GG_{2}$, we obtain
\begin{align}
  \nabla^{(2)}\GGhat_{2}=2\cform{3&0\\-1&0}
  =\frac{\tau_{2}}{\pi}\left(5\GG_{4}-\GGhat_{2}^{2}\right)\,,
  \label{eq:314}
\end{align}
as can be verified by explicitly comparing the $q$ expansions term by term. Note
that \eqref{eq:314} and \eqref{eq:313} for $k=2,3$ are equivalent to the classic
Ramanujan identities
\begin{subequations}
  \label{eq:315}
  \begin{align}
    q\frac{\ddd\GG_2}{\ddd q}&=\frac{ \GG_2^2- 5 \GG_4 }{4\pi^2}\\
    q\frac{\ddd\GG_4}{\ddd q}&=\frac{2 \GG_2 \GG_4 - 7 \GG_6 }{2\pi^2}\\
    q\frac{\ddd\GG_6}{\ddd q}&=\frac{21 \GG_2 \GG_6 - 30 \GG^2_4 }{14\pi^2}\,.
  \end{align}
\end{subequations}
Since the expressions above allow to write the derivative of any holomorphic
Eisenstein series back into a polynomial in holomorphic Eisenstein series, we
can iterate these expressions and simplify arbitrarily high derivatives of
holomorphic Eisenstein series. E.g.\ we have
\begin{subequations}
  \label{eq:317}
  \begin{align}
    \cform{4&0\\-2&0}&=\frac{1}{6}\nabla^{(3)}\nabla^{(2)}\GGhat_{2}
    =\Big(\frac{\tau_{2}}{\pi}\Big)^{2}\left(\frac{35}{3}\GG_{6}
      -5\GG_{4}\GGhat_{2}+\frac{1}{3}\GGhat_{2}^3\right)\\
    \cform{6&0\\-2&0}&=\frac{1}{20}\nabla^{(5)}\nabla^{(4)}\GG_{4}
    =\Big(\frac{\tau_{2}}{\pi}\Big)^{2}\left(\GG_{4} \GGhat_{2}^2
      -7\GG_{6}\GGhat_{2}+5\GG_{4}^2\right)\,.
  \end{align}
\end{subequations}

In the \mmapackage package, the Cauchy--Riemann derivatives \eqref{eq:233} are
implemented in the function \mma{CHolCR} for the holomorphic case and
\mma{CAHolCR} for the antiholomorphic case. For clarity, the result is returned
as it comes out of the action \eqref{eq:233} of $\nabla^{(a)}$ hence, to obtain
the derivative in canonical representation, we have to apply \mma{CSort}, e.g.\ 
\begin{mmaCell}{Input}
  CHolCR\mmabigso{}\mmaincformkite{1 1\\1 1}{1 1\\1 1}{1 1\\1 1}{1 1\\1 1}{1 1\\1 1}\mmabigsc{}//CSort
\end{mmaCell}
\begin{mmaCell}{Output}
  8\,\mmaoutcformkite{1 1\\1 1}{1 1\\1 1}{1 1\\1 1}{1 2\\1 0}{1 1\\1 1}+2\,\mmaoutcformkite{1 1\\1 1}{1 1\\1 1}{1 1\\1 1}{1 1\\1 1}{1 2\\1 0} \textnormal{.}
\end{mmaCell}
The functions \mma{CHolCR} and \mma{CAHolCR} can also be used to calculate
derivatives of holomorphic Eisenstein series,
\begin{mmaCell}{Input}
  CHolCR[g[4]]
  CHolCR[
  CHolCR[gHat[2]]
\end{mmaCell}
\begin{mmaCell}[morelst={label=mma:4}]{Output}
  4\,\mmaoutcform{\verbcent{5} 0\\-1 0}
\end{mmaCell}
\begin{mmaCell}[morelst={label=mma:5}]{Output}
  20\,\mmaoutcform{\verbcent{6} 0\\-2 0}
\end{mmaCell}
\begin{mmaCell}[morelst={label=mma:6}]{Output}
  2\,\mmaoutcform{\verbcent{3} 0\\-1 0} \textnormal{.}
\end{mmaCell}
The simplifications of these expressions by means of the Ramanujan identities
\eqref{eq:313}, \eqref{eq:314} and \eqref{eq:317} and higher-weight
generalizations is performed by the function \mma{DiCSimplify}, if the option
\mma{basisExpandG} is set to \mma{True}, e.g.\
\renewcommand{\temprefa}{\ref*{mma:4}}
\renewcommand{\temprefb}{\ref*{mma:5}}
\renewcommand{\temprefc}{\ref*{mma:6}}
\begin{mmaCell}{Input}
  DiCSimplify[Out[\temprefa],\,basisExpandG\mmainrarrow{}True]
  DiCSimplify[Out[\temprefb],\,basisExpandG\mmainrarrow{}True]
  DiCSimplify[Out[\temprefc],\,basisExpandG\mmainrarrow{}True]
\end{mmaCell}
\begin{mmaCell}{Output}
  \mmaFrac{14\,\mmaSub{G}{6}\,\mmaSub{\(\tau\)}{2}}{\(\pi\)}\mmam\mmaFrac{4\,\mmaSub{G}{4}\,\mmaSub{\mmaGGhat}{2}\,\mmaSub{\(\tau\)}{2}}{\(\pi\)}
\end{mmaCell}
\begin{mmaCell}{Output}
  \mmaFrac{100\,\mmaSubSup{G}{4}{2}\,\mmaSubSup{\(\tau\)}{2}{2}}{\mmaSup{\(\pi\)}{2}}\mmam\mmaFrac{140\,\mmaSub{G}{6}\,\mmaSub{\mmaGGhat}{2}\,\mmaSubSup{\(\tau\)}{2}{2}}{\mmaSup{\(\pi\)}{2}}+\mmaFrac{20\,\mmaSub{G}{4}\,\mmaSubSup{\mmaGGhat}{2}{2}\,\mmaSubSup{\(\tau\)}{2}{2}}{\mmaSup{\(\pi\)}{2}}
\end{mmaCell}
\begin{mmaCell}{Output}
  \mmaFrac{5\,\mmaSub{G}{4}\,\mmaSub{\(\tau\)}{2}}{\(\pi\)}\mmam\mmaFrac{\mmaSubSup{\mmaGGhat}{2}{2}\,\mmaSub{\(\tau\)}{2}}{\(\pi\)} \textnormal{.}
\end{mmaCell}
Using the techniques outlined above, \mma{DiCSimplify} can decompose any
\ac{MGF} of the form $\cform{k&0\\-n&0}$ or $\cform{-n&0\\k&0}$ with
$k,n\in\NN_{0}$ and $k>n$ into the ring of holomorphic Eisenstein series and
powers of $\GGhat_{2}$ and $\frac{\tau_{2}}{\pi}$ (or c.c.).


\section{Holomorphic Subgraph Reduction}
\label{sec:HSR}
Using the relatively straightforward techniques discussed in the previous
section, many identities between \acp{MGF} can be derived. However, an important
class of identities is still missing to decompose all relevant \acp{MGF} into
the basis to be presented in Section~\ref{sec:conjectured-basis}, namely
holomorphic subgraph reduction. In this section, we will review \ac{HSR} as it
was introduced first for dihedral graphs~\cite{dhoker2016a} and the extension of
this technique to higher-point graphs~\cite{gerken2019f}.\footnote{In the
  references, a different convention for \acp{MGF} was used, which differs from
  the one used here by factors of $\tau_{2}$ and $\pi$.}

The basic idea behind \ac{HSR} is the following: If an \ac{MGF} has a closed
subgraph (i.e.\ a subgraph which forms a loop) in which all edges have only
holomorphic momenta (i.e.\ the decorations are all of the form $(a,0)$), then
one can apply the partial-fraction decomposition
\begin{align}
  \frac{1}{p^{a}(q-p)^{b}}
  &=\sum_{k=1}^{a}\binom{a+b-k-1}{a-k}\frac{1}{p^{k}q^{a+b-k}}
  +\sum_{k=1}^{b}\binom{a+b-k-1}{b-k}\frac{1}{q^{a+b-k}(q-p)^{k}}
  \label{eq:238}
\end{align}
to the summand and perform the sum over the loop momentum explicitly. Since for
certain values of $a$, $b$ this sum is only conditionally convergent, it has to
be supplied with a summation prescription, which we will choose to be
\emph{Eisenstein summation}, to be defined below in \eqref{eq:239}. This
procedure however breaks the modular transformation properties at the level of
the individual contributions. As shown in~\cite{dhoker2016a} for two-point
graphs and in~\cite{gerken2019f} for general graphs, the terms with incorrect
modular properties cancel out in the final expression and one obtains a
decomposition of the original \ac{MGF} into terms which all have at least one
loop order less.

As an example, consider the trihedral graph
$\cformtri{1&2\\0&1}{1&2\\0&1}{1&2\\0&1}$, which has a closed three-point
holomorphic subgraph. Using the techniques discussed in this section, it can be
decomposed into
\begin{align}
  \cformtri{1&2\\0&1}{1&2\\0&1}{1&2\\0&1}&=6\cformtri{2\\1}{1&2\\0&1}{2&2\\0&1}
  -3\cform{2&3&4\\1&0&2}+3\cform{1&2&4\\0&1&2}\GGhat_{2}
  +\frac{\pi}{\tau_{2}}\cform{2&2&4\\-1&1&2}
  -2\frac{\pi}{\tau_{2}}\cformtri{2\\1}{1&2\\-1&1}{1&2\\0&1}\,.
  \label{eq:6}
\end{align}

\subsection{Dihedral holomorphic subgraph reduction}
\label{sec:di-HSR}
Dihedral graphs with a holomorphic subgraph have the form
$\cform{a_{+}&a_{-}&A\\0&0&B}$. The edges with labels $(a_{+},0)$ and
$(a_{-},0)$ form the holomorphic subgraph and the sum over the loop momentum
associated to this subgraph can be isolated using the partial-fraction
decomposition \eqref{eq:238}. This sum takes the form
\begin{align}
  Q_{k}(p_{0})=\sum_{p\neq p_{0}}' \frac{1}{p^{k}}\,,\qquad k\geq1
  \label{eq:243}
\end{align}
which is not absolutely convergent for $k=1,2$. Using the Eisenstein summation
prescription
\begin{align}
    \esum_{p\neq r+s\tau}
    f(p)&=\lim_{N\rightarrow\infty}\sum_{\substack{n=-N\\n\neq s}}^{N}\left(
      \lim_{M\rightarrow\infty}\sum_{m=-M}^{M}f(m+n\tau)\right)
    +\lim_{M\rightarrow\infty} \sum_{\substack{m=-M\\m\neq r}}^{M}f(m+s \tau)\,,
  \label{eq:239}
\end{align}
we assign the values
\begin{subequations}
  \label{eq:244}
  \begin{alignat}{2}
    Q_1(p_0) &= -{1 \over p_0} - {\pi \over 2 \tau_2} (p_0- \bar p_0) &&\label{eq:4}\\
    Q_2(p_0)& = -{1 \over {p_0}^2} +\GGhat_{2} + {\pi \over \tau_2}&&\label{eq:5}\\
    Q_k(p_0) &= - {1 \over {p_0}^k} + \GG_k & k&\geq 3
  \end{alignat}
\end{subequations}
to the sums \eqref{eq:243}. Note that the terms $\frac{\pi}{\tau_{2}}p_{0}$ in
\eqref{eq:4} and $\frac{\pi}{\tau_{2}}$ in \eqref{eq:5} do not the have modular
weight we associate to the corresponding \acp{LHS}. However, these terms cancel
out in the final result and we obtain~\cite{dhoker2016a}
\begin{align}
  \cform{a_{+}&a_{-}&A\\0&0&B}&=(-1)^{a_{+}}\GG_{a_{0}}\cform{A\\B}
  -\binom{a_{0}}{a_{-}}\cform{a_{0}&A\\0&B}\nonumber\\
  &\qquad+\sum_{k=4}^{a_{+}}\binom{a_{0}-1-k}{a_{+}-k}
  \GG_{k}\cform{a_{0}-k&A\\0&B}
  +\sum_{k=4}^{a_{-}}\binom{a_{0}-1-k}{a_{-}-k}
  \GG_{k}\cform{a_{0}-k&A\\0&B}
  \label{eq:245}\\
  &\qquad+\binom{a_{0}-2}{a_{+}-1}\left\{\GGhat_{2}\cform{a_{0}-2&A\\0&B}
    +\frac{\pi}{\tau_{2}}\cform{a_{0}-1&A\\-1&B}\right\}\,.\nonumber
\end{align}
For instance, the two-loop graph $\cform{1&2&2\\0&0&1}$ is decomposed into
one-loop graphs by \eqref{eq:245},
\begin{align}
  \cform{1&2&2\\0&0&1}&=3 \cform{5&0\\1&0}-\GGhat_{2}\cform{3&0\\1&0}
  -\frac{\pi}{\tau_{2}}\GG_{4}\,.
  \label{eq:513}
\end{align}

In the \mmapackage \texttt{Mathematica} package, the dihedral \ac{HSR}
\eqref{eq:245} is performed by the function \mma{DiCSimplify}. With the default
options, \mma{DiCSimplify} also applies all known dihedral basis decompositions
to the result and uses momentum conservation to remove negative entries where
possible as will be detailed in Section~\ref{sec:constr-ids}. Both features can
be disabled by setting the Boolean options \mma{momSimplify} and \mma{useIds} to
\mma{False} (they are \mma{True} by default). Hence, in order to get just the
result of the formula \eqref{eq:245}, we can run e.g.\
\begin{mmaCell}{Input}
  DiCSimplify\mmabigso{}\mmaincform{2 2 3 6\\1 2 0 0},\,momSimplify\mmainrarrow{}False,\,useIds\mmainrarrow{}False\mmabigsc{}
\end{mmaCell}
\begin{mmaCell}{Output}
  \mmam84\,\mmaoutcform{2 2 9\\1 2 0}+6\,\mmaoutcform{2 2 5\\1 2 0}\,\mmaSub{G}{4}+\mmaoutcform{2 2 3\\1 2 0}\,\mmaSub{G}{6}+21\,\mmaoutcform{2 2 7\\1 2 0}\,\mmaSub{\mmaGGhat}{2}+\mmaFrac{21\,\(\pi\)\,\mmaoutcform{2 2 \verbcent{8}\\1 2 -1}}{\mmaSub{\(\tau\)}{2}} \textnormal{.}
\end{mmaCell}
The function \mma{DiCSimplify} applies the formula \eqref{eq:245} always to the
two leftmost holomorphic columns. It also performs antiholomorphic subgraph
reduction by applying the complex conjugate of \eqref{eq:245} to graphs with a
closed antiholomorphic subgraph. In order to deactivate dihedral \ac{HSR} in
\mma{DiCSimplify} or \mma{CSimplify}, one can set the Boolean option \mma{diHSR}
to \mma{False} (the default is \mma{True}).

\subsection{Higher-point holomorphic subgraph reduction}
\label{sec:higher-pt-hsr}
In \cite{gerken2019f}, \ac{HSR} for higher-point graphs was worked out. Again,
we can separate the sum over the loop momentum of the holomorphic subgraph using
partial-fraction decomposition. However, as a novelty from three points onward,
on top of sums of the form
\begin{align}
  Q_{k}(p_{1},\dots,p_{n})=\sum_{p\neq p_{1},\dots,p_{n}}' \frac{1}{p^{k}}\,,
  \qquad k\geq1\,,
  \label{eq:265}
\end{align}
which are a straightforward generalization of the sums \eqref{eq:243}, also
shifted lattice sums of the form
\begin{align}
  \sum_{p\neq p_{1},\dots,p_{n}}' \frac{1}{p_{i}-p}
\end{align}
appear. Due to a subtlety in the Eisenstein summation prescription, we cannot
just shift the expressions on the \ac{RHS} of \eqref{eq:4} and add the
additional excluded points. Accounting for this, we have for $k\geq2$
\begin{subequations}
  \label{eq:267}
  \begin{align}
    Q_2(p_1,\dots, p_n) &= \GGhat_2 + {\pi \over \tau_2}
    - \sum_{i=1}^n {1 \over p_i^2}\\
    Q_{k}(p_{1},\dots,p_{n})&=\GG_{k}-\sum_{i=1}^n{1\over p_i^k}\,,
    \quad k\geq3\,.
  \end{align}
\end{subequations}
and for $k=1$, we replace
\begin{align}
  \sum_{p\neq p_{1},\dots,p_n}' \frac{1}{p}
  &\rightarrow Q_{1}(p_{1},\dots,p_n)&
  \sum_{p\neq p_{1},\dots,p_n}' \frac{1}{p_{i}-p} &\rightarrow
  Q_{1}(p_{i},\underbrace{p_{i}-p_{1},\dots,p_{i}-p_{n}}_{ \te{omit
      $p_{i}-p_{i}$}})
  \label{eq:266}
\end{align}
and set
\begin{align}
  Q_1(p_1, \dots, p_n) &= -\sum_{i=1}^n{1 \over p_i}
  -{\pi \over(n+1) \tau_2}\sum_{i=1}^n(p_i -\bar p_i)\,.
  \label{eq:268}
\end{align}

With the expressions \eqref{eq:267} and \eqref{eq:268}, any modular graph form
with an $n$-point holomorphic subgraph can be decomposed. In particular,
in~\cite{gerken2019f}, a closed expression for three-point \ac{HSR} was derived,
cf.\ e.g.\ \eqref{eq:6}. For this case, we illustrate a general feature of
higher-point \ac{HSR}: Since there are several ways in which the
partial-fraction decomposition can be done, different expressions for the
decomposition can be obtained. For instance the graph
$\cformtri{1\\0}{1&1\\0&1}{1&2\\1&0}$ can be decomposed into
\begin{align}
  \cformtri{1\\0}{1&1\\0&1}{1&2\\1&0}
  &=\cform{3&0\\1&0}^{2}+\cform{1&2&3\\1&0&1}-3\cform{1&2&3\\1&1&0}
  +\GGhat_{2}\cform{1&1&2\\0&1&1}
  +\frac{\pi}{\tau_{2}}\cform{1&2&2\\1&-1&1}
  \label{eq:298}
\end{align}
as well as into
\begin{align}
  \cformtri{1\\0}{1&1\\0&1}{1&2\\1&0}&=
  4\cform{6&0\\2&0}{-}\GGhat_{2}\cform{4&0\\2&0}{+}3\cform{1&1&4\\0&1&1}
  {-}\cform{1&2&3\\1&1&0}
  {-}\GGhat_{2}\cform{1&1&2\\0&1&1}
  {-}\frac{\pi}{\tau_{2}}\cform{5&0\\1&0}
  {-}\frac{\pi}{\tau_{2}}\cform{1&1&3\\0&1&0}\,,
  \label{eq:299}
\end{align}
by changing how the partial-fraction decomposition is
executed~\cite{gerken2019f}.

Performing the \ac{HSR} using the expressions for the $Q_{i}$ derived in this
section is laborious and it may be challenging to write the final expression
back into \acp{MGF} in the general case. For this reason, we provide a different
procedure to compute $n$-point \ac{HSR} in Section~\ref{sec:HSR-and-Fay}.

In the \texttt{Mathematica} package \mmapackage, the trihedral two-point
\ac{HSR} formula from \cite{gerken2019f} is implemented in the function
\mma{TriCSimplify}. Again, with the default options, negative entries are
removed via momentum conservation and identities from the database are applied,
so in order to just apply \ac{HSR}, we run
\begin{mmaCell}{Input}
  TriCSimplify\mmabigso{}\mmaincformtri{2 2 1\\0 0 1}{1 1\\1 1}{1 1\\1 1},\,momSimplify\mmainrarrow{}False,\,useIds\mmainrarrow{}False\mmabigsc{}
\end{mmaCell}
\begin{mmaCell}{Output}
  \mmaoutcform{2 2\\0 0}\,\mmaoutcformtri{1\\1}{1 1\\1 1}{1 1\\1 1}\mmam6\,\mmaoutcformtri{1 1\\1 1}{1 1\\1 1}{1 4\\1 0}+2\,\mmaoutcformtri{1 1\\1 1}{1 1\\1 1}{1 2\\1 0}\,\mmaSub{\mmaGGhat}{2}+\\
  \quad{}\mmaFrac{2\,\(\pi\)\,\mmaoutcformtri{1 1\\1 1}{1 1\\1 1}{1 \verbcent{3}\\1 -1}}{\mmaSub{\(\tau\)}{2}} \textnormal{.}
\end{mmaCell}
The three-point \ac{HSR} detailed in~\cite{gerken2019f} is also performed by the
function \mma{TriCSimplify}. For instance,
\begin{mmaCell}{Input}
  DiCSimplify\mmabigso{}TriCSimplify\mmabigso{}\mmaincformtri{1 1\\1 0}{1 2\\1 0}{1\\0}\mmabigsc{},\,momSimplify\mmainrarrow{}False,\,useIds\mmainrarrow{}False\mmabigsc{}
\end{mmaCell}
\begin{mmaCell}{Output}
  \mmaSup{\mmaoutcform{3 0\\1 0}}{2}+\mmaoutcform{1 2 3\\1 0 1}\mmam3\,\mmaoutcform{1 2 3\\1 1 0}+\mmaoutcform{1 1 2\\0 1 1}\,\mmaSub{\mmaGGhat}{2}+\mmaFrac{\(\pi\)\,\mmaoutcform{1 \verbcent{2} 2\\1 -1 1}}{\mmaSub{\(\tau\)}{2}} \textnormal{,}
\end{mmaCell}
reproducing \eqref{eq:298}. \mma{TriCSimplify} performs \ac{HSR} on the first
suitable holomorphic subgraph. It first performs the two-point version, then the
three-point version, also antiholomorphic subgraphs are simplified. With the
Boolean option \mma{triHSR}, trihedral \ac{HSR} can be deactivated (its default
value is \mma{True}) and with the Boolean options \mma{tri2ptHSR} and
\mma{tri3ptHSR}, the two- and three-point versions can be deactivated
individually.

\subsection{Holomorphic subgraph reduction and Fay identities}
\label{sec:HSR-and-Fay}

The discussion of holomorphic subgraph reduction has so far been exclusively in terms of the sum
representation of the \acp{MGF}. In the integral representation, \ac{HSR}
corresponds to certain identities for products of the $f^{(n)}(z,\tau)$
\eqref{eq:103}. These descend from the \emph{Fay identity} of the
Kronecker--Eisenstein series~\cite{fay1973, brown2011}
\begin{align}
  \Omega(z_1, \eta_1, \tau) \Omega(z_2,\eta_2,\tau) &=
  \Omega(z_{1}{-}z_{2},\eta_1,\tau)\Omega(z_2,\eta_1{+}\eta_2,\tau)
  +\Omega(z_{2}{-}z_{1}, \eta_2,\tau)\Omega(z_1,\eta_1{+}\eta_2,\tau)
  \label{eq:571}
\end{align}
by means of the expansion \eqref{eq:102} and are given by~\cite{broedel2015}
\begin{align}
  \label{eq:301}
  \begin{split}
    f^{(a_1)}_{12} f^{(a_2)}_{13} = (-1)^{a_1-1} f^{(a_1+a_2)}_{23}
    &+ \sum_{j=0}^{a_1} \binom{a_2+j-1}{j} f_{32}^{(a_1-j)} f_{13}^{(a_2+j)}\\
    &+ \sum_{j=0}^{a_2} \binom{a_1+j-1}{j}f_{12}^{(a_1+j)} f_{23}^{(a_2-j)}\,,
  \end{split}
\end{align}
where $a_{1},a_{2}\geq0$. According to \eqref{eq:127}, a factor $f^{(a)}_{ij}$
in a Koba--Nielsen integral corresponds to a (holomorphic) $(a,0)$-edge. Hence,
when \eqref{eq:301} is applied in a Koba--Nielsen integrand, it generates an
identity between modular graph forms with holomorphic edges.

\subsubsection{Holomorphic subgraphs with more than two vertices}

Consider an \ac{MGF} with an $n$-point holomorphic subgraph ($n>2$) given by a
Koba--Nielsen integral over
$C_{12}^{(a_{1},0)}C_{13}^{(a_{2},0)}=\pm f_{12}^{(a_{1})}f_{13}^{(a_{2})}$ and
$n-2$ further factors $C_{ij}^{(a_{k},0)}$, as well as additional
non-holomorphic edges. In this case, by focusing on the holomorphic edges, the
\ac{MGF}-identity implied by \eqref{eq:301} can be written graphically as
\begin{align}
  \begin{tikzpicture}[mgf,baseline=(zero.base),scale=0.55]
    \node (1) at (0,0) [vertex] {$1$};
    \node (2) at (2.25,3.032) [vertex] {$2$};
    \node (3) at (4.5,0) [vertex] {$3$};
    \node (zero) at (2.25,1.516){};
    \path (2) --node[pos=0.33,vertexdot](4){} node[pos=0.66,vertexdot](5){} (3);
    \draw[dedge] (2)--(4);
    \draw[dedge] (5)--(3);
    \draw[loosely dotted,thick,dash phase=1pt] (4)--(5);
    \draw[dedge=0.85] (1)--node[label,pos=0.4]{$\scriptstyle(a_{1}{,}0)$}(2);
    \draw[dedge=0.95] (1)--node[label,pos=0.5]{$\scriptstyle(a_{2}{,}0)$}(3);
  \end{tikzpicture}
  &=(-1)^{a_{1}}
  \begin{tikzpicture}[mgf,baseline=(2.base),scale=0.55]
    \node (2) at (0,0) [vertex] {$2$};
    \node (3) at (4.5,0) [vertex] {$3$};
    \path (2) to[bend left=60] node[pos=0.33,vertexdot](4){} node[pos=0.66,vertexdot](5){} (3);
    \draw[dedge] (2) to [bend left=23](4);
    \draw[dedge] (5) to [bend left=23](3);
    \draw[loosely dotted,thick,dash phase=1pt] (4) to [bend left=5](5);
    \draw[dedge=0.97]
    (2)--node[label,pos=0.44]{$\scriptstyle(a_{1}{+}a_{2}{,}0)$}(3);
  \end{tikzpicture}\nonumber\\[-1em]
  &\hspace{-0.5em}-\binom{a_{1}{+}a_{2}{-}1}{a_{1}}
  \begin{tikzpicture}[mgf,baseline=(zero.base),scale=0.55]
    \node (1) at (0,0) [vertex] {$1$};
    \node (2) at (2.25,3.032) [vertex] {$2$};
    \node (3) at (4.5,0) [vertex] {$3$};
    \node (zero) at (30:2.021){};
    \path (2) --node[pos=0.33,vertexdot](4){} node[pos=0.66,vertexdot](5){} (3);
    \draw[dedge] (2)--(4);
    \draw[dedge] (5)--(3);
    \draw[loosely dotted,thick,dash phase=1pt] (4)--(5);
    \draw[dedge=0.97]
    (1)--node[label,pos=0.44]{$\scriptstyle(a_{1}{+}a_{2}{,}0)$}(3);
  \end{tikzpicture}
  -\binom{a_{1}{+}a_{2}{-}1}{a_{2}}
  \begin{tikzpicture}[mgf,baseline=(zero.base),scale=0.55]
    \node (1) at (0,0) [vertex] {$1$};
    \node (2) at (2.25,3.032) [vertex] {$2$};
    \node (3) at (4.5,0) [vertex] {$3$};
    \node (zero) at (30:2.021){};
    \path (2) --node[pos=0.33,vertexdot](4){} node[pos=0.66,vertexdot](5){} (3);
    \draw[dedge] (2)--(4);
    \draw[dedge] (5)--(3);
    \draw[loosely dotted,thick,dash phase=1pt] (4)--(5);
    \draw[dedge=0.95]
    (1)--node[label,pos=0.4]{$\scriptstyle(a_{1}{+}a_{2}{,}0)$}(2);
  \end{tikzpicture}\label{eq:302}\\
  &\hspace{-0.5em}+\sum_{j=0}^{a_{1}-1}\binom{a_{2}{+}j{-}1}{j}
  \begin{tikzpicture}[mgf,baseline=(zero.base),scale=0.55]
    \node (1) at (0,0) [vertex] {$1$};
    \node (2) at (2.25,3.032) [vertex] {$2$};
    \node (3) at (4.5,0) [vertex] {$3$};
    \node (zero) at (30:2.021){};
    \path (2) to[bend left=60] node[pos=0.33,vertexdot](4){} node[pos=0.66,vertexdot](5){} (3);
    \draw[dedge] (2) to [bend left=23](4);
    \draw[dedge] (5) to [bend left=23](3);
    \draw[loosely dotted,thick,dash phase=1pt] (4) to [bend left=5](5);
    \draw[dedge=0.85]
    (3)--node[label,pos=0.4,inner sep=1pt]{$\scriptstyle(a_{1}{-}j{,}0)$}(2);
    \draw[dedge=0.95]
    (1)--node[label,pos=0.45]{$\scriptstyle(a_{2}{+}j{,}0)$}(3);
  \end{tikzpicture}
  +\sum_{j=0}^{a_{2}-1}\binom{a_{1}{+}j{-}1}{j}
  \begin{tikzpicture}[mgf,baseline=(zero.base),scale=0.55]
    \node (1) at (0,0) [vertex] {$1$};
    \node (2) at (2.25,3.032) [vertex] {$2$};
    \node (3) at (4.5,0) [vertex] {$3$};
    \node (zero) at (30:2.021){};
    \path (2) to[bend left=60] node[pos=0.33,vertexdot](4){} node[pos=0.66,vertexdot](5){} (3);
    \draw[dedge] (2) to [bend left=23](4);
    \draw[dedge] (5) to [bend left=23](3);
    \draw[loosely dotted,thick,dash phase=1pt] (4) to [bend left=5](5);
    \draw[dedge=0.85]
    (1)--node[label,pos=0.45]{$\scriptstyle(a_{1}{+}j{,}0)$}(2);
    \draw[dedge=0.95]
    (2)--node[label,pos=0.55,inner sep=1pt]{$\scriptstyle(a_{2}{-}j{,}0)$}(3);
  \end{tikzpicture}\,,\nonumber
\end{align}
where the ellipsis denotes a sequence of holomorphic edges such that the
\ac{LHS} forms a closed $n$-point holomorphic graph. The full \ac{MGF} has
additional non-holomorphic edges which can in general connect any vertices and
are not drawn in \eqref{eq:302}. In going from \eqref{eq:301} to \eqref{eq:302},
we have separated the contributions form $f^{(0)}=1$ since
$f^{(0)}\neq-C^{(0,0)}$, according to \eqref{eq:230}. In the representation
\eqref{eq:302} it is clear that on the \ac{RHS} each term has either one edge
less than the \ac{LHS} (terms two and three) or the closed holomorphic subgraph
has one edge less (terms four and five) or both (the first term). If a term on
the \ac{RHS} has an edge less than the \ac{LHS}, the associated \ac{MGF} has one
loop order less when accounting for the non-holomorphic edges suppressed in
\eqref{eq:302} as well. Hence, the Fay identities \eqref{eq:301} allow to reduce
$n$-point \ac{HSR} to $(n-1)$-point \ac{HSR} plus graphs of lower loop order.

As an example, consider the tetrahedral graph
\begin{align}
  \cformtet{1\\0}{1\\0}{1\\1}{1\\0}{1\\0}{1\\1}=
  \begin{tikzpicture}[mgf,baseline=(4.base),scale=0.7]
    \node (1) at (0,0) [vertex] {$1$};
    \node (2) at (60:5.5) [vertex] {$2$};
    \node (3) at (5.5,0) [vertex] {$3$};
    \node (4) at (30:3.18) [vertex] {$4$};
    \draw[dedge] (1) to node[inner sep=1,label]{$(1,0)$} (2);
    \draw[dedge] (4) to node[inner sep=1,label,pos=0.4]{$(1,0)$} (2);
    \draw[dedge] (3) to node[inner sep=1,label]{$(1,1)$} (2);
    \draw[dedge] (4) to node[inner sep=1,label,pos=0.3]{$(1,0)$} (3);
    \draw[dedge] (3) to node[inner sep=1,label]{$(1,0)$} (1);
    \draw[dedge] (1) to node[inner sep=1,label,pos=0.36]{$(1,1)$} (4);
  \end{tikzpicture}\,,
  \label{eq:487}
\end{align}
which has a four-point holomorphic subgraph and appears in the four-gluon
amplitude in the heterotic string at the order
$\alpha'^{2}$~\cite{gerken2019e}. By applying \eqref{eq:302} to the two
holomorphic edges connected to vertex $4$, we obtain the decomposition (with the
graphs not yet in their canonical representation)
\begin{align}
  \cformtet{1\\0}{1\\0}{1\\1}{1\\0}{1\\0}{1\\1}&=
  -\cformtet{1\\0}{\noblock}{1&2\\1&0}{1\\0}{1\\0}{1\\1}
  -\cformtet{1\\0}{\noblock}{1\\1}{2\\0}{1\\0}{1\\1}
  -\cformtet{1\\0}{2\\0}{1\\1}{\noblock}{1\\0}{1\\1}
  +\cformtet{1\\0}{\noblock}{1&1\\0&1}{1\\0}{1\\0}{1\\1}
  -\cformtet{1\\0}{1\\0}{1&1\\0&1}{\noblock}{1\\0}{1\\1}\,.
  \label{eq:488}
\end{align}
In this expression, every graph has one empty block and can be simplified using
the topological simplifications of Section~\ref{sec:MGF-top-simp} to
\begin{align}
  \cformtet{1\\0}{1\\0}{1\\1}{1\\0}{1\\0}{1\\1}&=
  2\cform{1&2&3\\0&1&1}+2\cformtri{1\\0}{1&1\\0&1}{1&2\\0&1}\,.
\end{align}
In this way, the four-point \ac{HSR} in the original graph was reduced to
three-point \ac{HSR}. The three-point \ac{HSR} can be performed either via
another Fay identity or using the formula from \cite{gerken2019f}. Together with
the basis decompositions to be discussed in Section~\ref{sec:conjectured-basis},
we obtain the final result
\begin{align}
  \cformtet{1\\0}{1\\0}{1\\1}{1\\0}{1\\0}{1\\1}&=2 \cform{6&0\\2&0}-4
  \cform{3&0\\1&0}^{2}+2\GGhat_{2}\cform{4&0\\2&0}
  - 12\frac{\pi}{\tau_{2}}\cform{5&0\\1&0}+
  4\frac{\pi}{\tau_{2}}\GGhat_{2}\cform{3&0\\1&0}+ 4\piimtau^{2}\GG_{4} \,.
  \label{eq:489}
\end{align}

In general, the closed holomorphic subgraph is of course not necessary for the
identity \eqref{eq:301} to hold. Hence, if we remove the edges between vertices
$2$ and $3$ from \eqref{eq:302}, this generates identities between modular graph
forms which have at least two non-parallel holomorphic edges both connected to
the same vertex. For trihedral graphs, we have e.g.
\begin{align}
  \begin{split}
    \cformtri{A_{1}&a_{1}\\B_{1}&0}{A_{2}&a_{2}\\B_{2}&0}{A_{3}\\B_{3}}&=
    (-)^{a_{1}+a_{2}}
    \cformtri{A_{1}\\B_{1}}{A_{2}\\B_{2}}{A_{3}&a_{1}+a_{2}\\B_{3}&0}\\
    &\quad+(-)^{a_{1}+1}\binom{a_{1}{+}a_{2}{-}1}{a_{1}}
    \cformtri{A_{1}\\B_{1}}{A_{2}&a_{1}+a_{2}\\B_{2}&0}{A_{3}\\B_{3}}\\
    &\quad+(-)^{a_{2}+1}\binom{a_{1}{+}a_{2}{-}1}{a_{2}}
    \cformtri{A_{1}&a_{1}+a_{2}\\B_{1}&0}{A_{2}\\B_{2}}{A_{3}\\B_{3}}\\
    &\quad+(-)^{a_{1}}\sum_{j=0}^{a_{1}-1}\binom{a_{2}{+}j{-}1}{j}
    \cformtri{A_{1}\\B_{1}}{A_{2}&a_{2}+j\\B_{2}&0}{A_{3}&a_{1}-j\\B_{3}&0}
    \hspace{-0.7em}\\
    &\quad+(-)^{a_{2}}\sum_{j=0}^{a_{2}-1}\binom{a_{1}{+}j{-}1}{j}
    \cformtri{A_{1}&a_{1}+j\\B_{1}&0}{A_{2}\\B_{2}}{A_{3}&a_{2}-j\\B_{3}&0}\,.
    \hspace{-0.7em}
  \end{split}
  \label{eq:303}
\end{align}
This identity will be a key ingredient in deriving the basis decompositions for
all dihedral and trihedral modular graph forms of total modular weight at most
$12$ is Section~\ref{sec:conjectured-basis}.  If the
$\sbmatrix{A_{3}\\B_{3}}$-block contains a holomorphic edge, \eqref{eq:303} is a
reduction of three-point \ac{HSR} to two-point \ac{HSR} and graphs of lower loop
order. In this case, the Fay identity could be used on any pair of non-parallel
holomorphic edges and this choice corresponds to the different ways to perform
the partial-fraction decomposition in Section~\ref{sec:higher-pt-hsr}, leading
to interesting identities between \acp{MGF} in general. As an example, consider
the graph $\cformtri{1\\0}{1&1\\0&1}{1&2\\1&0}$ which was decomposed using the
traditional \ac{HSR} method in Section~\ref{sec:higher-pt-hsr}. Applying
\eqref{eq:303} to the first two holomorphic columns of this graph leads to
\begin{align}
  \cformtri{1\\0}{1&1\\0&1}{1&2\\1&0}
  &=\cformtri{\varnothing}{1\\1}{1&2&2\\1&0&0}
  -\cformtri{\varnothing}{1&1\\0&1}{1&1&2\\0&1&0}
  +\cformtri{\varnothing}{1&2\\1&0}{1&2\\1&0}
  -\cformtri{1\\0}{1\\1}{1&1&2\\0&1&0}
  +\cformtri{1\\1}{2\\0}{1&2\\1&0}\,,
  \label{eq:337}
\end{align}
which can be shown to be equal to the decomposition \eqref{eq:298} upon using
the topological simplifications from Section~\ref{sec:MGF-top-simp} and the
dihedral \ac{HSR} formula \eqref{eq:245}. On the other hand, we can also apply
\eqref{eq:303} to the second and third holomorphic edges, yielding
\begin{align}
  \cformtri{1\\0}{1&1\\0&1}{1&2\\1&0}
  &=\cformtri{1\\0}{1\\1}{1&3\\1&0}
  -\cformtri{1\\1}{1\\1}{1&3\\0&0}
  +\cformtri{1\\1}{1&1\\0&1}{1&2\\0&0}\,.
  \label{eq:338}
\end{align}
This can be simplified to \eqref{eq:299} by topological identities.

\subsubsection{Holomorphic subgraphs with two vertices}

The restriction of \eqref{eq:302} to holomorphic edges which are not parallel
arises because the Fay identity for Kronecker--Eisenstein series \eqref{eq:571}
involves the three different $z_{1}$, $z_{2}$ and
$z_{1}{-}z_{2}$. As discussed in Appendix~A of \cite{gerken2019d}, by taking the
limit $z_{1}\rightarrow z_{2}$ we obtain the Fay identity
\begin{align}
  f^{(a_{1})}(z)f^{(a_{2})}(z)
  &=(-1)^{a_{2}}\Theta(a_{1}{+}a_{2}{-}4)\GG_{a_{1}+a_{2}}
  +\binom{a_{1}{+}a_{2}}{a_{2}}f^{(a_{1}+a_{2})}(z)\nonumber\\
  &-\sum_{k=4}^{a_{1}}\binom{a_{1}{+}a_{2}{-}1{-}k}{a_{2}-1}
  \GG_{k}f^{(a_{1}+a_{2}-k)}(z)
  -\sum_{k=4}^{a_{2}}\binom{a_{1}{+}a_{2}{-}1{-}k}{a_{1}-1}
  \GG_{k}f^{(a_{1}+a_{2}-k)}(z)\nonumber \\
  &-\binom{a_{1}{+}a_{2}{-}2}{a_{2}-1}\left(\GGhat_{2}f^{(a_{1}+a_{2}-2)}(z)
    +\partial_{z}f^{(a_{1}+a_{2}-1)}(z)\right)\,,\label{eq:308}
\end{align}
where $a_{1},a_{2}>0$ and $\Theta$ is the Heaviside step-function
\begin{align}
  \Theta(x)=
  \begin{cases}
    1 & \te{if }x\geq0\\
    0 & \te{if }x<0
  \end{cases}\,.
  \label{eq:309}
\end{align}
Integrating \eqref{eq:308} against a suitable product of $C^{(a,b)}$ functions
yields two-point \ac{HSR} upon using that
\begin{align}
  \partial_{z}f^{(a_{1}+a_{2}-1)}(z)
  =(-1)^{a_{1}+a_{2}+1}\frac{\pi}{\tau_{2}}C^{(a_{1}+a_{2}-1,-1)}(z)
\end{align}
according to \eqref{eq:194}. E.g.\, when \eqref{eq:308} for $a_{1}+a_{2}\geq3$
is integrated against $\prod_{i=1}^{R}C^{(a_{i},b_{i})}$, we obtain the dihedral
\ac{HSR} identity \eqref{eq:245}.

Together, \eqref{eq:301} and \eqref{eq:308} allow to perform holomorphic subgraph reduction of
holomorphic subgraphs with arbitrarily many vertices in a compact way. Note that
when using Fay identities, we circumvent the need to evaluate conditionally
convergent sums with the Eisenstein summation prescription as shown in
Section~\ref{sec:higher-pt-hsr}. For trihedral three-point \ac{HSR}, it was
checked explicitly in many cases that a combination of \eqref{eq:303} and
two-point \ac{HSR} yields an equivalent expression to the one obtained from the
formula in \cite{gerken2019f}.

In the \mmapackage package, the trihedral Fay identities \eqref{eq:303} are
implemented in the function \mma{TriFay} which returns an equation. The first
argument of this function is the trihedral \ac{MGF} to be decomposed, the second
(optional) argument has the form \mma{\{\{b1,c1\},\{b2,c2\}\}}, where \mma{b1}
and \mma{b2} are the blocks of the (anti)holomorphic edges to be used and
\mma{c1} and \mma{c2} are the columns of those edges. If the second argument is
omitted, the first suitable pair of (anti)holomorphic edges is selected
automatically. As an example, we will consider the decomposition of the graph
$\cformtri{1\\0}{1&1\\0&1}{1&2\\1&0}$ as discussed around \eqref{eq:337} and
\eqref{eq:338}. In order to apply \eqref{eq:303} to the first two holomorphic
columns and then simplify the result to obtain \eqref{eq:298}, we run
\begin{mmaCell}[morelst={label=mma:7}]{Input}
  TriFay\mmabigso{}\mmaincformtri{1\\0}{1 1\\0 1}{1 2\\1 0},\,\{\{1,\,1\},\,\{2,\,1\}\}\mmabigsc{}\\
  DiCSimplify[TriCSimplify[
\end{mmaCell}
\begin{mmaCell}{Output}
  \mmaoutcformtri{1\\0}{1 1\\0 1}{1 2\\1 0}==\mmaoutcformtri{\{\}}{1\\1}{1 2 2\\1 0 0}\!\mmam\!\mmaoutcformtri{\{\}}{1 1\\0 1}{1 1 2\\0 1 0}\!+\!\mmaoutcformtri{\{\}}{1 2\\1 0}{1 2\\1 0}\!\mmam\!\mmaoutcformtri{1\\0}{1\\1}{1 1 2\\0 1 0}
\end{mmaCell}
\begin{mmaCell}{Output}
  \mmaSup{\mmaoutcform{3 0\\1 0}}{2}+\mmaoutcform{1 2 3\\1 0 1}\mmam3\,\mmaoutcform{1 2 3\\1 1 0}+\mmaoutcform{1 1 2\\0 1 1}\,\mmaSub{\mmaGGhat}{2}+\mmaFrac{\(\pi\)\,\mmaoutcform{1 \verbcent{2} 2\\1 -1 1}}{\mmaSub{\(\tau\)}{2}}
\end{mmaCell}
reproducing \eqref{eq:298}. Similarly, \eqref{eq:299} can be obtained by
changing the second argument of \mma{TriFay} in \mmainref{mma:7} to
\mma{\{\{2,\,1\},\,\{3,\,2\}\}} and replacing the option
\mma{momSimplify\mmainrarrow False} of \mma{DiCSimplify} by
\mma{diHSR\mmainrarrow False}.

As mentioned above, trihedral three-point \ac{HSR} is performed by the function
\mma{TriCSimplify}, which implements the formula from \cite{gerken2019f}. If the
Boolean option \mma{tri3ptFayHSR} (which is inherited by \mma{CSimplify}), is
set to \mma{True} (the default is \mma{False}), the three-point \ac{HSR} is
instead performed using the Fay identity \eqref{eq:303} and subsequent two-point
\ac{HSR}. The results of applying the two techniques may look different, if the
basis decompositions from Section~\ref{sec:conjectured-basis} are not applied,
but they are in fact equivalent.


\section{The sieve algorithm}
\label{sec:sieve-algorithm}
With the techniques described in the last two sections, many valuable identities
between modular graph forms can be derived. However, if one is interested in
simplifying a particular \ac{MGF}, e.g.\ one which has appeared as an expansion
coefficient of a Koba--Nielsen integral, it is not always clear which techniques
to combine to obtain the desired decomposition. In this situation, the sieve
algorithm, first introduced in \cite{dhoker2016a}, can be used: It allows for a
systematic decomposition (up to an overall constant) of arbitrary \acp{MGF}, as
long as the basis for the decomposition and all \acp{MGF} of lower
total modular weight $a+b$ are known.

\subsection{Constructing identities}
\label{sec:constr-ids}
As a starting point, assume that we have a combination $F$ of \acp{MGF} of
homogeneous modular weight $(|A|,|B|)$ and we want to check whether or not it
vanishes. The idea behind the sieve algorithm is to repeatedly take derivatives
of $F$ using the Maa{\ss} operator $\nabla^{(|A|)}$ defined in
\eqref{eq:73}. Due to an intricate interplay between momentum conservation
identities and \ac{HSR}, described in detail in \cite{dhoker2016a}, every
derivative can be expressed as a linear combination of products of holomorphic
Eisenstein series, \acp{MGF} with non-negative antiholomorphic labels for each
edge, $\tau_{2}$ with non-positive exponent, \acp{MGF} of the form
$\cform{k&0\\-n&0}$ with $k>n$ and modular invariant factors. After taking $|B|$
derivatives, the antiholomorphic modular weight vanishes according to
\eqref{eq:74} and hence each term in the derivative has to factorize, since any
unfactorized \acp{MGF} would have to have vanishing antiholomorphic labels and
therefore be amenable to \ac{HSR}, leading to a factorized expression. Using the
generalized Ramanujan identities from Section~\ref{sec:MGFDers}, the factors of
the form $\cform{k&0\\-n&0}$ can be decomposed as well. Since each term is
factorized, the total modular weight $a+b$ of every leftover \ac{MGF} is
strictly less than $|A|+|B|$ and if we know all identities between \acp{MGF} of
lower total modular weight, it is manifest if the $|B|$\textsuperscript{th}
derivative of $F$ vanishes or not. If $F$ has $|A|=|B|$, then Lemma~1 in
\cite{dhoker2016a} guarantees that if the derivative vanishes, $F=0$ up to an
overall constant. If $|A|\neq|B|$ and $F$ can be written as the derivative of an
expression with $|A|=|B|$, this primitive vanishes up to a constant, so $F=0$ as
well. We conjecture that the same is true if $F$ cannot be written as the
derivative of an expression with $|A|=|B|$, in line with all cases we tested. In
this way, we can generate identities at progressively higher total modular
weight.

The Cauchy--Riemann derivative of a holomorphic Eisenstein series has the form
$\cform{2k+1&0\\-1&0}$, i.e.\ it is a graph with one edge with negative
antiholomorphic weight. In this case, momentum conservation (and \ac{HSR})
cannot be used to remove the negative entry and in the original version
published in \cite{dhoker2016a}, this fact was used to \emph{sieve} the space of
\acp{MGF} for identities: After taking a derivative and trading negative
antiholomorphic entries for holomorphic Eisenstein series, one subtracts the
same derivative of an \ac{MGF} in such a way that all holomorphic Eisenstein
series cancel. Then, one can take the next derivative of the combined expression
without generating irremovable negative antiholomorphic labels. After having
taken $|B|$ derivatives, the result is purely holomorphic (and still modular),
so we can expand it in the ring of holomorphic Eisenstein series. By subtracting
one final \ac{MGF} such that this derivative vanishes, one has constructed an
identity up to an overall constant. In fact, if a combination of modular graph
forms vanishes, then the holomorphic Eisenstein series have to cancel out in
every derivative. This can however only be verified, if the prefactors of the
holomorphic Eisenstein series are linearly independent. Since they carry lower
total modular weight than the complete expression, this means that we need to
know all identities between graphs of lower total modular weight.

In general, finding \acp{MGF} with the correct Cauchy--Riemann derivatives to
cancel the holomorphic Eisenstein series can be challenging but if we want to
find a decomposition of an \ac{MGF} into a set of basis \acp{MGF}, we can just
take the derivatives of a linear combination of the basis elements and adjust
the coefficients so that the holomorphic Eisenstein series cancel. This is what
is done in the implementation of the sieve algorithm in the \mmapackage package.

Instead of canceling holomorphic Eisenstein series in every derivative as
described above and in \cite{dhoker2016a}, one can also use the generalized
Ramanujan identities discussed in Section~\ref{sec:MGFDers} to perform the
derivatives of the holomorphic Eisenstein series. In this way, the highest
derivative of any \ac{MGF} can be written in terms of holomorphic Eisenstein
series and \acp{MGF} of lower total modular weight for which we assume that the
relations are known, hence identities can be found explicitly.

Consider e.g.\ the weight-$(4,4)$ \ac{MGF} $\cform{1&1&2\\1&2&1}$. The fourth
derivatives of this graph and the weight-$(4,4)$ basis elements $C_{1,1,2}$,
$\EE_{2}^{2}$ and $E_{4}$ are
\begin{subequations}
  \label{eq:323}
  \begin{align}
    \begin{split}
      \nabla^{(4)^{\scriptstyle 4}}\cform{1&1&2\\1&2&1}&=
      120\EE_{2}\GG_{4}\GGhat_{2}^2-840\EE_{2}\GG_{6}\GGhat_{2}
      +600\EE_{2}\GG_{4}^2-360\GG_{4}^2\\
      &\quad+840\GG_{6}\frac{\tau_{2}}{\pi}\cform{3&0\\1&0}
      -240\GG_{4}\GGhat_{2}\frac{\tau_{2}}{\pi}\cform{3&0\\1&0}
    \end{split}\displaybreak[0]\\[1ex]
    \nabla^{(4)^{\scriptstyle 4}}\cform{1&1&2\\1&1&2}&=288\GG_{4}^2
    +48\GG_{4}\GGhat_{2}\frac{\tau_{2}}{\pi}\cform{3&0\\1&0}
    -168\GG_{6}\frac{\tau_{2}}{\pi}\cform{3&0\\1&0}\displaybreak[0]\\[1ex]
    \begin{split}
      \nabla^{(4)^{\scriptstyle 4}}\left(\!\piimtau^{2}\EE_{2}^{2}\right)\!&=
      240\EE_{2}\GG_{4}\GGhat_{2}^2-1680\EE_{2}\GG_{6}\GGhat_{2}
      +1200\EE_{2}\GG_{4}^2+216\GG_{4}^2\\
      &\quad-384\GG_{4}\GGhat_{2}\frac{\tau_{2}}{\pi}\cform{3&0\\1&0}
      +1344\GG_{6}\frac{\tau_{2}}{\pi}\cform{3&0\\1&0}
    \end{split}\displaybreak[0]\\[1ex]
    \nabla^{(4)^{\scriptstyle 4}}\left(\!\piimtau^{4}\EE_{4}\right)&=
    180\GG_{4}^2\,,
  \end{align}
\end{subequations}
where we used the notation \eqref{eq:321} and simplified all derivatives of
holomorphic Eisenstein series using the techniques from
Section~\ref{sec:MGFDers}. Setting a linear combination of these four
expressions to zero and requiring the coefficients of the various terms on the
\ac{RHS} to vanish leaves
\begin{align}
  \nabla^{(4)^{\scriptstyle 4}}\left(\cform{1&1&2\\1&2&1}+\cform{1&1&2\\1&1&2}
    -\frac{1}{2}\piimtau^{2}\EE_{2}^{2}
    +\frac{1}{2}\piimtau^{4}\EE_{4}\right)=0
  \label{eq:324}
\end{align}
as the only solution. If no solution had existed, the four \acp{MGF} in
\eqref{eq:323} would have been proven to be linearly independent. Lemma 1 in
\cite{dhoker2016a} now states that this implies
\begin{align}
  \cform{1&1&2\\1&2&1}+\cform{1&1&2\\1&1&2}
  -\frac{1}{2}\piimtau^{2}\EE_{2}^{2}
  +\frac{1}{2}\piimtau^{4}\EE_{4}
  =\piimtau^{4}\te{const.}
  \label{eq:343}
\end{align}
with some $\tau$-independent constant.\footnote{Due to our normalization
  conventions, the graphs with equal total holomorphic and antiholomorphic edge
  labels are not modular invariant, hence the integration constant is multiplied
  by a suitable power of $\frac{\pi}{\tau_{2}}$.} Using the techniques discussed
in the previous sections, one can also decompose $\cform{1&1&2\\1&2&1}$ directly
and finds that the constant vanishes in this case (as expected since there is no
single-valued \ac{MZV} at the expected transcendental weight $4$).

In the \mmapackage package, the removal of edge labels $-1$ for dihedral and
trihedral graphs is done by the functions \mma{DiCSimplify} and
\mma{TriCSimplify}, if the option \mma{momSimplify} is set to \mma{True} (the
default). The sieve algorithm itself is implemented in the function
\mma{CSieveDecomp}, which uses the traditional method of canceling holomorphic
Eisenstein series in every step. If no further options are given, this function
tries to decompose the graph given in its argument into the basis discussed in
Section~\ref{sec:conjectured-basis}, e.g.\ for the graph $\cform{1&1&2\\1&2&1}$
we considered above, we can run
\begin{mmaCell}{Input}
  CSieveDecomp\mmabigso{}\mmaincform{1 1 2\\1 2 1}\mmabigsc{}
\end{mmaCell}
\begin{mmaCell}{Output}
  \mmam\!\mmaoutcform{1 1 2\\1 1 2}+\mmaFrac{\mmaSup{\(\pi\)}{4}\,\mmaSubSup{E}{2}{2}}{2\,\mmaSubSup{\(\tau\)}{2}{4}}\mmam\mmaFrac{\mmaSup{\(\pi\)}{4}\,\mmaSub{E}{4}}{2\,\mmaSubSup{\(\tau\)}{2}{4}}+\mmaFrac{\mmaSup{\(\pi\)}{4}\,\mmaoutintconst{1 1 2\\1 2 1}}{\mmaSubSup{\(\tau\)}{2}{4}} \textnormal{,}
\end{mmaCell}
reproducing \eqref{eq:343}. The last term in the output is an undetermined
integration constant, labeled by the exponent matrix of the original graph. Such
a constant is added for all graphs with equal holomorphic and antiholomorphic
weight. Setting the Boolean option \mma{verbose} of \mma{CSieveDecomp} to
\mma{True} prints a detailed progress report into the notebook with the
expressions appearing in each derivative and the prefactors of the holomorphic
Eisenstein series which are set to zero. For instance, the output for the third
derivative in the computation above is
\begin{mmaCell}{Print}
  3rd derivative:
\end{mmaCell}
\begin{mmaCell}{Print}
  \mmam\!168\,\mmaoutcform{7 0\\1 0}\mmam108\,bCoeff[1]\,\mmaoutcform{7 0\\1 0}\mmam120\,bCoeff[2]\mmaoutcform{7 0\\1 0}+\\
  \quad12\,\mmaoutcform{3 0\\1 0}\,\mmaSub{G}{4}+12\,bCoeff[1]\,\mmaoutcform{3 0\\1 0}\,\mmaSub{G}{4}
\end{mmaCell}
\begin{mmaCell}{Print}
  (Anti-)holomorphic Eisenstein series:
\end{mmaCell}
\begin{mmaCell}{Print}
  \{\mmaSub{G}{4}\}
\end{mmaCell}
\begin{mmaCell}{Print}
  Coefficients that should be zero:
\end{mmaCell}
\begin{mmaCell}{Print}
  \big\{12\,\mmaoutcform{3 0\\1 0}+12\,bCoeff[1]\,\mmaoutcform{3 0\\1 0}\big\}
\end{mmaCell}
\begin{mmaCell}{Print}
  Find solution for all
\end{mmaCell}
\begin{mmaCell}{Print}
  \big\{\mmaoutcform{3 0\\1 0}\big\}
\end{mmaCell}
\begin{mmaCell}{Print}
  Solutions:
\end{mmaCell}
\begin{mmaCell}{Print}
  \{\{bCoeff[1]\mmaoutrarrow\mmam1\}\} \textnormal{.}
\end{mmaCell}
As one can see, \mma{CSieveDecomp} forms a linear combination of the basis
elements with coefficients \mma{bCoeff} and subtracts it from the \ac{MGF} which
is decomposed. Then, derivatives are taken and in each step the coefficients of
the holomorphic Eisenstein series are set to zero by fixing some of the
\mma{bCoeff}.

The basis used for the decomposition is determined by the option \mma{basis} of
\mma{CSieveDecomp}. If \mma{basis} is an empty list (the default), the basis is
determined by the function \mma{CBasis}, to be discussed in more detail in
Section~\ref{sec:conjectured-basis}. Otherwise, one can also supply a list of
\acp{MGF} of the same weight as the \ac{MGF} to be decomposed. E.g.\ we can
reproduce the momentum conservation identity of the seed $\cform{1&2&2\\1&2&1}$
(up to an overall constant) by running
\begin{mmaCell}{Input}
  CSieveDecomp\mmabigso{}\mmaincform{1 1 2\\1 2 1},\,basis\mmainrarrow\mmabigco{}\mmaincform{0 2 2\\1 1 2},\,\mmaincform{1 1 2\\1 1 2}\mmabigcc{}\mmabigsc{}
\end{mmaCell}
\begin{mmaCell}{Output}
  \mmam\!\mmaoutcform{0 2 2\\1 1 2}\mmam\mmaoutcform{1 1 2\\1 1 2}+\mmaFrac{\mmaSup{\(\pi\)}{4}\,\mmaoutintconst{1 1 2\\1 2 1}}{\mmaSubSup{\(\tau\)}{2}{4}} \textnormal{.}
\end{mmaCell}
If not all coefficients can be fixed (e.g.\ because the \mma{basis} provided is
not linearly independent), \mma{bCoeff} will appear in the output. For further
options and the meaning of various error messages, cf.\
Appendix~\ref{cha:mma-reference}.


\section{Divergent modular graph forms}\label{sec:divMGF}
So far, we have not discussed the convergence properties of the lattice sum
\eqref{eq:131} of \acp{MGF}, but, of course, if the edge labels become too low,
the sum \eqref{eq:131} is not absolutely convergent any more. Interestingly,
conditionally convergent or divergent sums can arise even when one applies the
techniques above only to convergent sums. Sometimes, the divergence cannot be
avoided, e.g.\ when using the sieve algorithm to find decompositions of certain
convergent graphs. When deriving identities, one way to deal with this
phenomenon is to just disregard all identities in which divergent graphs
appear. This is the approach taken in Section~\ref{sec:conjectured-basis} for
(convergent) dihedral and trihedral modular graph forms of weight
$a+b\leq10$. However, in this way, one misses many valuable identities and hence
it is desirable to have at least a partial understanding of how to interpret
divergent \acp{MGF}. In this section, we will describe concrete results which go
in this direction. Below, we will use these divergent techniques to obtain all
dihedral and trihedral (convergent) basis decompositions for $a+b=12$.

\subsection{Divergence conditions}
\label{sec:div-cond}
In this section, we will give simple power-counting arguments to determine if a
particular \ac{MGF} is absolutely convergent or not, building on the behavior of
holomorphic Eisenstein series, for which we know that
\begin{align}
  \GG_{a}=\sum_{p}' \frac{1}{p^{a}}
\end{align}
is absolutely convergent for $a\geq3$, conditionally convergent for $a=2$ and
divergent for $a\leq1$. Accordingly, we will call an \ac{MGF} convergent if all
momenta in the sum \eqref{eq:131} have at least three powers in the denominator
(adding powers of $p$ and $\bar{p}$) and divergent if any momentum appears with
two or less powers in the denominator.\footnote{Note that this simple
  power-counting criterion does not constitute a proof of the convergence or
  divergence of the lattice sum of the \ac{MGF}. In fact, as we will discuss
  below, this argument tends to underestimate the convergence of the sum since
  possible cancellations are not accounted for.}

In order to determine the total powers with which a momentum can appear, one has
to perform some of the sums first by using the momentum-conserving delta
functions (cf.\ e.g.\ \eqref{eq:177}). Of course, there is considerable freedom
in which sums we choose for this, hence different final expressions can result,
with different total powers of the momenta. These expressions correspond to
different rotations of the coordinate axes in the lattice spanned by the
momenta. Since by counting the total exponents, we only test the convergence
properties along the coordinate axes, we pick the representation with the lowest
total power. To illustrate this, consider the dihedral graph
\begin{align}
  \cform{1&1&2\\0&0&2}=\sum_{p_{1},p_{2},p_{3}}'
  \frac{\delta(p_{1}+p_{2}+p_{3})}{p_{1}p_{2}|p_{3}|^{4}}\,.
  \label{eq:472}
\end{align}
We can use the delta function to perform either the $p_{3}$ sum or the $p_{2}$
sum, yielding the expressions
\begin{align}
  \cform{1&1&2\\0&0&2}=\sum_{p_{1},p_{2}}' \frac{1}{p_{1}p_{2}|p_{1}+p_{2}|^{4}}
  =-\sum_{p_{1},p_{3}}'\frac{1}{p_{1}(p_{1}+p_{3})|p_{3}|^{4}}\,.
  \label{eq:347}
\end{align}
In the first of these expressions, $p_{1}$ and $p_{2}$ both come with a power of
$5$ in the denominator, hence according to our criterion above,
$\cform{1&1&2\\0&0&2}$ should be convergent. In the second expression in
\eqref{eq:347} however, $p_{1}$ comes with a power of $2$, hence,
$\cform{1&1&2\\0&0&2}$ should be divergent. The reason that the first expression
seems to be convergent is that the divergence lies in the direction of
$p_{1}+p_{2}=\mathrm{const.}$, whereas by counting the powers of $p_{1}$ and
$p_{2}$, we only probed the directions along those two momenta. Therefore,
$\cform{1&1&2\\0&0&2}$ is divergent.

To summarize, an \ac{MGF} is only convergent if the powers of all momenta are at
least three, in all possible ways to solve the delta functions. We will
translate this in the following into conditions on the labels of the two-,
three- and four-point graphs introduced in Section~\ref{sec:graph-topol-not}.

In dihedral graphs, if we perform the sum over momentum $p$ with the delta
function, we will increase the total powers of all other momenta by the total
power of $p$. Hence, our divergence criterion for dihedral graphs, taking into
account that we can use any of the momenta to solve the delta function, is
\begin{align}
  \cform{A\\B}\quad\te{convergent}\quad\Leftrightarrow\quad
  \min_{\substack{i,j\\i\neq j}}(c_{i}+c_{j})>2\,,
  \label{eq:348}
\end{align}
where $c_{i}=a_{i}+b_{i}$ and $i,j$ run over all edges. The basic criterion
\eqref{eq:348} will have to be satisfied for all edge bundles in higher-point
graphs as well, but the global structure of these graphs adds further criteria.

In general, solving delta functions is equivalent to assigning loop momenta
consistently to the edges of the graph. Hence, by going through the
topologically distinct assignments, we can see to which edges a certain momentum
can propagate and hence what the convergence conditions for this graph should
be. When considering graphs with edge bundles between the vertices (like the
graphs introduced in Section~\ref{sec:graph-topol-not}), we first assign the
total momenta of the bundles consistently. Then, in a bundle of total momentum
$\mfp$, with edges carrying momenta $p_{1},\dots, p_{R}$, we can choose any edge
to solve the momentum conservation constraint, e.g.\ we can drop momentum
$p_{1}$ and assign momentum $\mfp-\sum_{i=2}^{R}p_{i}$ to this edge. For the
convergence conditions, the implications of this are twofold: First, each
momentum can appear in any other edge of the same bundle, implying the condition
\eqref{eq:348} for each bundle. Second, the total momenta of the edge bundles
can appear in any edge, hence we should count the lowest total power for each
edge bundle when determining the convergence condition due to the total
momenta. We will go through this procedure for the three-point and all
four-point graphs in Section~\ref{sec:graph-topol-not} in the following.

For trihedral graphs, there is just one way to assign the bundle momenta, namely
\begin{align}
  \begin{tikzpicture}[mgf,baseline=(zero.base)]
    \node (1) at (0,0) [vertex] {$1$};
    \node (2) at (60:2) [vertex] {$2$};
    \node (3) at (2,0) [vertex] {$3$};
    \draw[dedge=0.9] (1)-- node[inner sep=1.5pt,biglabel](zero){$\mfp$}(2);
    \draw[dedge=0.9] (2)-- node[inner sep=1.5pt,biglabel]{$\mfp$}(3);
    \draw[dedge=0.9] (3)-- node[inner sep=1.5pt,biglabel]{$\mfp$}(1);
  \end{tikzpicture}\,,
  \label{eq:351}
\end{align}
i.e.\ the graph $\cformtri{A_{1}\\B_{1}}{A_{2}\\B_{2}}{A_{3}\\B_{3}}$ is
convergent iff
\begin{align}
  \min_{\substack{i,j\\i\neq j}}\big(c_{i}^{(k)}+c_{j}^{(k)}\big)&>2
  \quad\forall\ k\in\{1,2,3\}&
  \te{an}&\te{d}&\check{c}_{1}+\check{c}_{2}+\check{c}_{3}&>2\,,
  \label{eq:349}
\end{align}
where $c_{i}^{(k)}=a_{i}^{(k)}+b_{i}^{(k)}$ for $i=1,\dots,R_{k}$ in the
notation of \eqref{eq:185} and
$\check{c}_{k}=\min_{i}(a_{i}^{(k)}{+}b_{i}^{(k)})$, where $i$ runs over all
edges in block $k$. As described above, the first condition is due to the
individual momenta in the edge bundles, whereas the second condition is due to
the total bundle momentum $\mfp$. If
$\cformtri{A_{1}\\B_{1}}{A_{2}\\B_{2}}{A_{3}\\B_{3}}$ carries only non-negative
edge labels and does not contain a $(0,0)$-edge (i.e.\ is not factorizable),
then $c_{i}\geq1$ for all edges and the second condition in \eqref{eq:349} is
always satisfied. The same will be true for all other conditions on top of
\eqref{eq:348} for every block in the following.

As a straightforward extension of the trihedral result, the box graph
$\cformbox{A_{1}\\B_{1}}{A_{2}\\B_{2}}{A_{3}\\B_{3}}{A_{4}\\B_{4}}$ is
convergent iff
\begin{align}
  \min_{\substack{i,j\\i\neq j}}\big(c_{i}^{(k)}+c_{j}^{(k)}\big)&>2
  \quad\forall\ k\in\{1,2,3,4\}&
  \te{an}&\te{d}&\check{c}_{1}+\check{c}_{2}+\check{c}_{3}+\check{c}_{4}&>2\,,
  \label{eq:350}
\end{align}
with the same notation as in \eqref{eq:349}.

In kite graphs, there are two topologically distinct ways of assigning the total
momenta of the edge bundles. They are
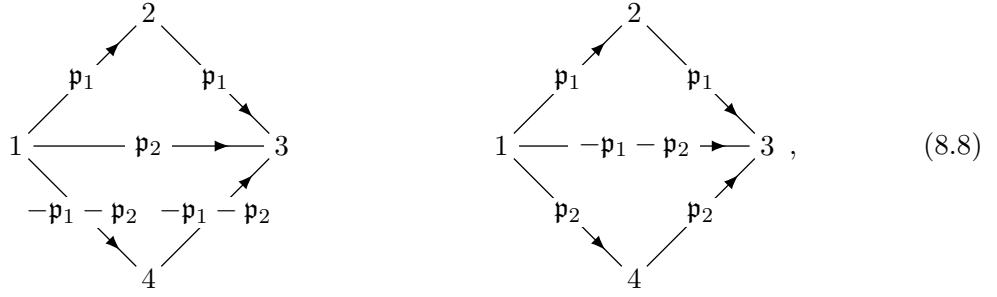
\begin{align}
  &
  \begin{tikzpicture}[mgf,rotate=-45]
    \node (1) at (0,0) [vertex] {$1$};
    \node (2) at (0,2.5) [vertex] {$2$};
    \node (3) at (2.5,2.5) [vertex] {$3$};
    \node (4) at (2.5,0) [vertex] {$4$};
    \draw[dedge=0.85] (1)--node[inner sep=1.5pt,biglabel]{$\mfp_{1}$} (2);
    \draw[dedge=0.85] (2)--node[inner sep=1.5pt,biglabel]{$\mfp_{1}$} (3);
    \draw[dedge=0.85] (1)--node[biglabel]{$\mfp_{2}$} (3);
    \draw[dedge=0.85] (1)
    --node[inner sep=1.5pt,biglabel]{$-\mfp_{1}-\mfp_{2}$} (4);
    \draw[dedge=0.85] (4)
    --node[inner sep=1.5pt,biglabel]{$-\mfp_{1}-\mfp_{2}$} (3);
  \end{tikzpicture}
  &
  \begin{tikzpicture}[mgf,rotate=-45]
    \node (1) at (0,0) [vertex] {$1$};
    \node (2) at (0,2.5) [vertex] {$2$};
    \node (3) at (2.5,2.5) [vertex] {$3$};
    \node (4) at (2.5,0) [vertex] {$4$};
    \draw[dedge=0.85] (1)--node[inner sep=1.5pt,biglabel]{$\mfp_{1}$} (2);
    \draw[dedge=0.85] (2)--node[inner sep=1.5pt,biglabel]{$\mfp_{1}$} (3);
    \draw[dedge=0.9] (1)--node[biglabel]{$-\mfp_{1}-\mfp_{2}$} (3);
    \draw[dedge=0.85] (1)--node[inner sep=1.5pt,biglabel]{$\mfp_{2}$} (4);
    \draw[dedge=0.85] (4)--node[inner sep=1.5pt,biglabel]{$\mfp_{2}$} (3);
  \end{tikzpicture}\,,
  &
  \label{eq:352}
\end{align}
implying that
$\cformkite{A_{1}\\B_{1}}{A_{2}\\B_{2}}{A_{3}\\B_{3}}{A_{4}\\B_{4}}{A_{5}\\B_{5}}$
is convergent iff
\begin{alignat}{2}
    \begin{aligned}
      &&\min_{\substack{i,j\\i\neq j}}\big(c_{i}^{(k)}+c_{j}^{(k)}\big)&>2
      \quad\forall\ k\in\{1,2,3,4,5\}\\
      &\te{and}&\check{c}_{i}+\check{c}_{j}+\check{c}_{5}&>2
      \quad\forall\ (i,j)\in\{(1,2),(3,4)\}\\
      &\te{and}&\check{c}_{1}+\check{c}_{2}+\check{c}_{3}+\check{c}_{4}&>2\,.
    \end{aligned}
  \label{eq:353}
\end{alignat}

For tetrahedral graphs, there are again two topologically distinct ways to
assign the three independent total edge-bundle momenta,
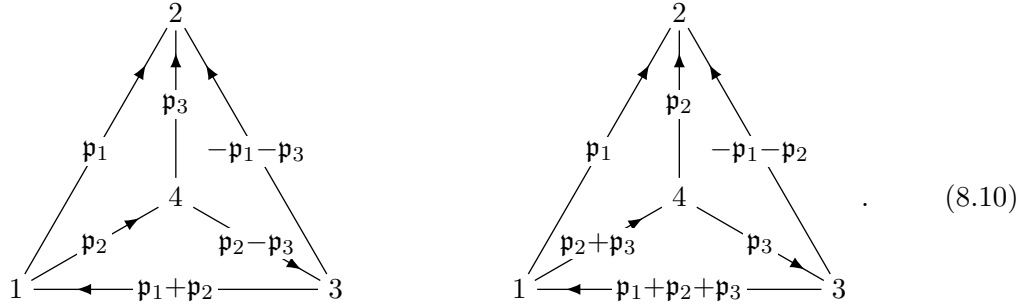
\begin{align}
  &
  \begin{tikzpicture}[mgf,baseline=(4.base)]
    \node (1) at (0,0) [vertex] {$1$};
    \node (2) at (60:4.25) [vertex] {$2$};
    \node (3) at (4.25,0) [vertex] {$3$};
    \draw[opacity=0,name path = ray1](0,0)--(30:5);
    \draw[opacity=0,name path = vert](2)--(2 |- 0,0);
    \node[name intersections={of=ray1 and vert}] (4) at (intersection-1) {$4$};
    \draw[dedge=0.85] (1) --node[inner sep=1.5pt,biglabel]{$\mfp_{1}$} (2);
    \draw[dedge=0.85] (1) --node[inner sep=1.5pt,biglabel]{$\mfp_{2}$} (4);
    \draw[dedge=0.85] (4) --node[inner sep=1.5pt,biglabel]{$\mfp_{3}$} (2);
    \draw[dedge=0.85] (4)
    --node[inner sep=1.5pt,biglabel]{$\mfp_{2}{-}\mfp_{3}$} (3);
    \draw[dedge=0.85] (3)
    --node[inner sep=1.5pt,biglabel]{$-\mfp_{1}{-}\mfp_{3}$} (2);
    \draw[dedge=0.85] (3)
    --node[inner sep=1.5pt,biglabel]{$\mfp_{1}{+}\mfp_{2}$} (1);
  \end{tikzpicture}
  &
  \begin{tikzpicture}[mgf,baseline=(4.base)]
    \node (1) at (0,0) [vertex] {$1$};
    \node (2) at (60:4.25) [vertex] {$2$};
    \node (3) at (4.25,0) [vertex] {$3$};
    \draw[opacity=0,name path = ray1](0,0)--(30:5);
    \draw[opacity=0,name path = vert](2)--(2 |- 0,0);
    \node[name intersections={of=ray1 and vert}] (4) at (intersection-1) {$4$};
    \draw[dedge=0.85] (1) --node[inner sep=1.5pt,biglabel]{$\mfp_{1}$} (2);
    \draw[dedge=0.85] (4) --node[inner sep=1.5pt,biglabel]{$\mfp_{2}$} (2);
    \draw[dedge=0.85] (4) --node[inner sep=1.5pt,biglabel]{$\mfp_{3}$} (3);
    \draw[dedge=0.85] (1)
    --node[inner sep=1.5pt,biglabel]{$\mfp_{2}{+}\mfp_{3}$} (4);
    \draw[dedge=0.85] (3)
    --node[inner sep=1.5pt,biglabel]{$-\mfp_{1}{-}\mfp_{2}$} (2);
    \draw[dedge=0.9] (3)
    --node[inner sep=1.5pt,biglabel]{$\mfp_{1}{+}\mfp_{2}{+}\mfp_{3}$} (1);
  \end{tikzpicture}\,.
  &
\end{align}
This implies that the tetrahedral graph
\begin{align}
  \cformtet{A_{1}\\B_{1}}{A_{2}\\B_{2}}{A_{3}\\B_{3}}{A_{4}\\B_{4}}
  {A_{5}\\B_{5}}{A_{6}\\B_{6}}
\end{align}
is convergent iff
\begin{alignat}{2}
  &&\hspace{-1em}\min_{\substack{i,j\\i\neq j}}
  \big(c_{i}^{(k)}+c_{j}^{(k)}\big)&>2
  \ \forall\ k\in\{1,2,3,4,5,6\}\label{eq:354}\\
  &\te{and}&\check{c}_{i}{+}\check{c}_{j}{+}\check{c}_{k}&>2
  \ \forall\ (i,j,k)\in\{(1,2,6),(1,3,5),(2,3,4),(4,5,6)\}\nonumber\\
  &\te{and}&\check{c}_{i}{+}\check{c}_{j}{+}\check{c}_{k}{+}\check{c}_{\ell}&>2
  \ \forall\ (i,j,k,\ell)\in\{(1,2,4,5),(1,3,4,6),(2,3,5,6)\}\nonumber\,.
\end{alignat}
Here, the penultimate line corresponds to all closed three-point subgraphs, the
last line corresponds to all closed four-point subgraphs.

The convergence conditions discussed so far only depend on the sums of the
holomorphic and antiholomorphic labels of the edges. That this view tends to
underestimate the convergence of the sum can be seen by considering the two
one-loop graphs $\cform{1&0\\0&1}$ and $\cform{1&1\\0&0}$. According to our
condition \eqref{eq:348}, both graphs should be equally divergent. But of
course, while the sum $\cform{1&0\\0&1}$ is divergent, the sum
$\cform{1&1\\0&0}$ is only conditionally convergent and we regularize it by
introducing additional powers of the momentum as in \eqref{eq:316}, yielding
$\GGhat_{2}$. In general, graphs containing a $\sbmatrix{1&1\\0&0}$ subblock can
be simplified using the divergent \ac{HSR} discussed in
Section~\ref{sec:div-HSR}.

In the integral representation, this can be seen as follows:
$f^{(1)}(z,\tau)\sim\frac{1}{z}$ is the only one out of the $f^{(a)}$ which has
a pole. The fact that $\cform{1&1\\0&0}$ is conditionally convergent is
reflected in the fact that the integral of $\frac{1}{z^{2}}$ over a ball around
the origin vanishes, whereas the divergence of $\cform{1&0\\0&1}$ is reflected
in the divergence of the integral of
$|f^{(1)}(z)|^{2}\sim\frac{1}{|z|^{2}}$.

In the \mmapackage package, the function \mma{CCheckConv} checks for convergence
of the argument using the criteria \eqref{eq:348} and \eqref{eq:349} on dihedral
and trihedral graphs. The return value is either \mma{True} for convergent
\acp{MGF} or \mma{False} for divergent \acp{MGF}, e.g.
\begin{mmaCell}{Input}
  CCheckConv\mmabigso{}\mmaincform{0 1 2\\1 0 2}\mmabigsc{}\\
  CCheckConv\mmabigso{}\mmaincformtri{-1 2\\\verbcent{0} 2}{1 1\\0 1}{1 1\\0 1}\mmabigsc{}
\end{mmaCell}
\begin{mmaCell}{Output}
  False
\end{mmaCell}
\begin{mmaCell}{Output}
  False \textnormal{.}
\end{mmaCell}
On top of dihedral and trihedral graphs, \mma{CCheckConv} also checks for
$\EE_{k}$, $\GG_{k}$ and $\overline{\GG}_{k}$ with $k<2$, all other expressions
are treated as convergent. As soon as any divergent object is detected in the
argument, \mma{CCheckConv} returns \mma{False}.

\subsection[Divergent MGFs from Koba--Nielsen integrals]
{Divergent modular graph forms from Koba--Nielsen integrals}
\label{sec:div-MGF-KN-poles}
We study \acp{MGF} in order to expand Koba--Nielsen integrals comprising the
Koba--Nielsen factor \eqref{eq:91} and a polynomial in the functions
$f^{(a)}(z,\tau)$ and $\overline{f^{(b)}(z,\tau)}$ given in \eqref{eq:103} and
\eqref{eq:104}. If this polynomial contains a factor $|f_{ij}^{(1)}|^{2}$ (where
$f_{ij}^{(1)}=f^{(1)}(z_{i}-z_{j})$), the \acp{MGF} in the expansion of the
Koba--Nielsen integral are all divergent since $|f_{ij}^{(1)}|^{2}$ leads to a
$\sbmatrix{1&0\\0&1}$ subblock, which violates the criterion \eqref{eq:348}.

However, the Koba--Nielsen factor regulates this divergence: Since the Jacobi
theta function satisfies $\theta_{1}(z,\tau)\sim z$ for small $z$,
$\exp(s_{ij}G_{ij})\sim |z_{ij}|^{-2s_{ij}}$ for small $z_{ij}$. Using
integration-by-parts identities for the Koba--Nielsen integral, one can in fact
show that a Koba--Nielsen integral with a $|f_{ij}^{(1)}|^{2}$ prefactor
has a pole in the Mandelstams. Hence, the appearance of divergent \acp{MGF} is
merely a signal that one has tried to Taylor-expand around a pole.

As an example, consider the two-point Koba--Nielsen integral
\begin{align}
    \int\dd\mu_{1}\big|f^{(1)}_{12}\big|^{2}\KN_{2}\,,
  \label{eq:355}
\end{align}
whose naive $\alpha'$ expansion
\begin{align}
  \frac{\pi}{\tau_{2}}\EE_{1}-s_{12}\frac{\tau_{2}}{\pi}\cform{0&1&1\\1&0&1}
  -\frac{1}{2}s_{12}^{2}\imtaupi^{2}\cform{0&1&1&1\\1&0&1&1}
  +\mathcal{O}(s_{12}^{3})
  \label{eq:356}
\end{align}
exhibits divergent \acp{MGF} at every order in $s_{12}$. In order to make the
pole in $s_{12}$ manifest, consider the derivative \cite{gerken2019d}
\begin{align}
  \partial_{\bar{z}_{2}}\big(f_{12}^{(1)}\KN_{2}\big)\,.
  \label{eq:357}
\end{align}
We now use \eqref{eq:364} and
\begin{align}
  \partial_{z_j} \KN_{n}&=
  \sum_{i\neq j}  s_{ij} f^{(1)}(z_{ij},\tau)  \KN_{n}\,,
  \label{eq:359}
\end{align}
which follows from \eqref{eq:345}, to evaluate \eqref{eq:357}. With this, we
obtain
\begin{align}
  \partial_{\bar{z}_{2}}\big(f_{12}^{(1)}\KN_{2}\big)=
  \left(\frac{\pi}{\tau_{2}}-\pi\delta^{(2)}(z_{12},\bar{z}_{12})\right)\KN_{2}
  +s_{12}\big|f_{12}^{(1)}\big|^{2}\KN_{2}\,.
  \label{eq:362}
\end{align}
Integrating over $z_{2}$ and solving for \eqref{eq:355} yields (since
$\KN_{2}\rightarrow0$ for $z_{12}\rightarrow0$ the term with the delta function
does not contribute)
\begin{align}
  \int\dd\mu_{1}\big|f^{(1)}_{12}\big|^{2}\KN_{2}
  =-\frac{1}{s_{12}}\frac{\pi}{\tau_{2}}\int\dd\mu_{1}\KN_{2}\,,
  \label{eq:360}
\end{align}
making the pole in $s_{12}$ explicit. The remaining Koba--Nielsen integral in
\eqref{eq:360} has an expansion in convergent \acp{MGF}. Variations of this
technique to expose the kinematic poles in Koba--Nielsen integrals can be found
in countless examples in the literature.

At two points, the integral \eqref{eq:360} is the only Koba--Nielsen integral
with a pole in the Mandelstams and it is associated to the collision of the two
punctures. At three point, several different pole structures can appear,
including nested poles incorporating the three-particle Mandelstam variable
\begin{align}
  s_{123}=s_{12}+s_{13}+s_{23}
  \label{eq:7}
\end{align}
due to the collision of all three punctures. The rewriting of all relevant
three-point integrals making the pole structure manifest and reducing divergent
expansions to convergent ones as above, is summarized in
Appendix~\ref{sec:3pt-IBPs}.

In general, we can use the Fay identity \eqref{eq:301} to rewrite the
$f_{ij}^{(1)}$ contributions to the integrand in terms of
$f_{ij}^{(a)}$ with $a>1$ and derivatives of the Koba--Nielsen factor as
in \eqref{eq:359}. When integrating these expressions by parts, we make one pole
explicit and obtain an expression with poles of lower multiplicity.

\subsection{Divergent modular graph forms from momentum conservation}
\label{sec:div-MGF-mom-cons}
Apart from the expansion of Koba--Nielsen integrals, divergent modular graph
forms can also appear in momentum-conservation identities of convergent
graphs. In the sum representation \eqref{eq:201} of momentum conservation, this
means that the exchange of the sum over edges $e'$ and the sum over momenta
$p_{e}$ is not allowed in this case. Performing it anyway leads to the
decomposition of a convergent series into a sum of divergent series. As an
example, consider the convergent seed $\cform{0&1&2\\1&1&2}$, whose
antiholomorphic momentum-conservation identity is
\begin{align}
  \cform{0&1&2\\1&0&2}+\cform{0&1&2\\1&1&1}
  +\piimtau^{3}(\EE_{1}\EE_{2}-\EE_{3})=0\,,
  \label{eq:369}
\end{align}
after factorization. The graph $\cform{0&1&2\\1&0&2}$ and the Eisenstein series
$\EE_{1}$ are both divergent.

When dealing only with convergent \acp{MGF}, momentum-conservation identities
involving divergent graphs should be discarded. However, as we will discuss
shortly, it is sometimes desirable to have identities between divergent
\acp{MGF} and momentum-conservation identities involving divergent \acp{MGF} can
be used to define those divergent \acp{MGF}. In this framework, we treat the
divergent non-holomorphic Eisenstein series $\EE_{1}$ as a basis element for
divergent \acp{MGF} and find decompositions in the same way as we did for
convergent \acp{MGF}. E.g.\ \eqref{eq:369}, together with the (convergent)
identity
\begin{align}
  \cform{0&1&2\\1&1&1}=-\frac{1}{2}\piimtau^{3}(\EE_{3}-\zeta_{3})\,,
  \label{eq:380}
\end{align}
can be used to decompose the divergent graph $\cform{0&1&2\\1&0&2}$,
\begin{align}
  \cform{0&1&2\\1&0&2}=\piimtau^{3}
  \Big(\frac{3}{2}\EE_{3}-\EE_{1}\EE_{2}+\frac{1}{2}\zeta_{3}\Big)\,.
  \label{eq:377}
\end{align}
Note that this does not extend to momentum-conservation identities of divergent
seeds which have to be treated separately, cf. Section~\ref{sec:div-mom-cons}
below.

In particular, momentum-conservation identities involving divergent graphs can
appear in the sieve algorithm, when removing entries of $-1$ as described in
Section~\ref{sec:constr-ids}. As an example for this phenomenon, consider the
graph $\cform{0&1&2&3\\1&1&2&0}$, whose Cauchy--Riemann derivative is given by
\begin{align}
  \nabla^{(6)}\cform{0&1&2&3\\1&1&2&0}=3\cform{0&1&2&4\\1&1&2&-1}
  +2\cform{0&1&3&3\\1&1&1&0}+\cform{0&2&2&3\\1&0&2&0}\,.
  \label{eq:378}
\end{align}
The $-1$-entry in the first term can be removed by a momentum-conservation
identity which yields, after factorization and divergent \ac{HSR} (to be
discussed below in Section~\ref{sec:div-HSR}),
\begin{align}
  \begin{split}
    \cform{0&1&2&4\\1&1&2&-1}&=
    5\cform{0&2&5\\1&2&0}+\cform{1&2&4\\1&2&0}-\cform{0&1&2&4\\1&1&1&0}
    -\GG_{4}\cform{0&1&2\\1&0&2}-\GGhat_{2}\cform{0&2&3\\1&2&0}\\
    &\qquad+\frac{\pi}{\tau_{2}}\Big(\cform{0&2&4\\1&1&0}-\cform{6&0\\2&0}\Big)
    +\piimtau^{3}(\EE_{2}-\EE_{1}\EE_{2})\,\GG_{4}\,.
  \end{split}
  \label{eq:379}
\end{align}
As explained in Section~\ref{sec:constr-ids}, when constructing identities with
the sieve algorithm, we seek to cancel holomorphic Eisenstein series by adding
suitable \acp{MGF}. In order to do this consistently, we need to know all
relations for the \acp{MGF} in the prefactor of the holomorphic Eisenstein
series. In the example \eqref{eq:379}, however, the prefactor of $\GG_{4}$ is
\begin{align}
  -\cform{0&1&2\\1&0&2}+\piimtau^{3}(\EE_{2}-\EE_{1}\EE_{2})
\end{align}
and hence in particular involves divergent \acp{MGF}. I.e.\ in this case, we
need to know the decomposition \eqref{eq:377} to see explicitly that the
divergence cancels out and to continue with the sieve algorithm.

In general, since (according to \eqref{eq:233}) the action of the
Cauchy--Riemann operator on modular graph forms leaves the sum of holomorphic
and antiholomorphic labels for each edge invariant and the divergence conditions
in Section~\ref{sec:div-cond} are all functions of this sum only, each term in
the derivative of an \ac{MGF} $\mathcal{C}_{\Gamma}$ will have the same
convergence properties as $\mathcal{C}_{\Gamma}$. Momentum conservation however
increases the sum of the labels in one edge and decreases it in another edge in
each term, therefore changing the convergence properties. But since the \ac{MGF}
decomposed in this way is convergent, the divergences have to cancel out upon
plugging in identities for the divergent graphs.

For the remainder of this discussion, we will restrict to dihedral graphs, where
the edge labels are written as columns in one block, but the arguments
generalize straightforwardly to higher-point graphs. In \cite{dhoker2016a},
where the sieve algorithm was introduced, the authors restricted to the case of
strictly positive holomorphic labels and non-negative antiholomorphic labels. In
this case, the column sum for all edges is at least $2$, with at most one
$(1,0)$ edge since we assume that \ac{HSR} is already performed. After taking
the Cauchy--Riemann derivative, momentum conservation is only necessary in the
term in which the $(1,0)$ edge is replaced by a $(2,-1)$ edge. In the momentum
conservation identity, this edge will become a $(2,0)$ edge in each term, hence
the column sum for each edge is again $2$ with at most one edge of sum $1$,
i.e.\ each term is convergent. In this way, the problem of divergent \acp{MGF}
in the sieve algorithm is avoided in \cite{dhoker2016a} and the present
discussion can therefore be regarded as an extension of the previously known
techniques.

\subsection{Divergent holomorphic subgraph reduction}
\label{sec:div-HSR}
On top of momentum conservation and factorization, holomorphic subgraph
reduction is a central technique to derive identities for modular graph
forms. It is therefore desirable to extend \ac{HSR} to divergent graphs. To this
end, we will distinguish the case in which the divergence appears within the
holomorphic subgraph, i.e.\ the sum of the labels of the edges forming the
holomorphic subgraph is at most $2$, from the case in which the divergence
appears outside the holomorphic subgraph, i.e.\ the sum of labels within the
holomorphic subgraph is at least $3$, but the entire \ac{MGF} is still
divergent.

In the case of a divergence outside the holomorphic subgraph, the sum over the
loop momentum, which is performed when doing \ac{HSR}, is convergent. I.e.\ the
divergence acts merely as a spectator and the formulas for two- and three-point
\ac{HSR} discussed in Section~\ref{sec:HSR} are still valid. E.g.\ dihedral
graphs in which the divergence lies outside the holomorphic subgraph are given
by $\cform{0&1&a&A\\1&0&0&B}$ with $a\geq2$ and all column sums in
$\sbmatrix{A\\B}$ at least two. In this case, we can apply the two-point
\ac{HSR} formula \eqref{eq:245} and obtain results consistent with momentum
conservation. For the graph $\cform{0&1&a&A\\1&0&0&B}$ with $a\geq3$ we can see
this explicitly by using the holomorphic momentum-conservation identity of the
convergent seed $\cform{1&1&a&A\\1&0&0&B}$,
\begin{align}
  \cform{0&1&a&A\\1&0&0&B}&=-\cform{1&1&a-1&A\\1&0&0&B}
  -\sum_{i=1}^{R}\cform{1&1&a&A-S_{i}\\1&0&0&B}
  +\cform{1&a&A\\1&0&B}-\frac{\pi}{\tau_{2}}
  \EE_{1}\GG_{a}\prod_{i=1}^{R}\cform{a_{i}&0\\b_{i}&0}\,,
  \label{eq:381}
\end{align}
and applying the \ac{HSR} formula \eqref{eq:245} to the two convergent graphs on
the \ac{RHS}. Similar calculations can be done at three point and the extension
of the \ac{HSR} formulas to divergent graphs in this way was checked empirically
for many cases.

If the holomorphic subgraph itself is divergent, the sum over the loop momentum
which we perform when doing \ac{HSR} is not convergent any more and hence we
cannot use the usual \ac{HSR} formulas in this case. If we restrict to only
non-negative edge labels and assume that the graph under consideration has
already been factorized (i.e.\ it does not contain any $(0,0)$ edges), then
holomorphic subgraphs with more than two edges cannot be divergent. For this
reason, we will restrict to the case of divergent two-point holomorphic
subgraphs. In the sum representation, in which the two-point \ac{HSR} formula
\eqref{eq:245} was derived first, it is unclear how to proceed in the case of
divergent sums. In the integral representation, however, in which the two-point
\ac{HSR} formula was derived from the coincident limit \eqref{eq:308} of the Fay
identity, it is straightforward to generalize \eqref{eq:245} to divergent
holomorphic subgraphs: We can just take the $a_{1}=a_{2}=1$ case of
\eqref{eq:308},
\begin{align}
  \big(f^{(1)}(z)\big)^{2}=2f^{(2)}(z)-\GGhat_{2}-\partial_{z}f^{(1)}(z)
  \label{eq:579}
\end{align}
and integrate it against a product of $C^{(a,b)}(z)$ functions,
as defined in \eqref{eq:126}, yielding
\begin{align}
  \cform{1&1&A\\0&0&B}=-2\cform{2&A\\0&B}-\GGhat_{2}\cform{A\\B}
  +\frac{\pi}{\tau_{2}}\cform{1&A\\-1&B}\,.
  \label{eq:382}
\end{align}
Note that \eqref{eq:245} has an additional term $\GGhat_{2}\cform{0&A\\0&B}$
when naively extended to $a_{+}=a_{-}=1$. Empirically, we found that
\eqref{eq:382} is compatible with momentum conservation in a large number of
cases. Furthermore, \eqref{eq:382} agrees with the special cases
\begin{align}
  \cform{1&1&a\\0&0&b}&=
  -2\cform{a+2&0\\b&0}+\frac{\pi}{\tau_{2}}\cform{a+1&0\\b-1&0}\\
  \cform{1&1&1&1\\0&0&1&1}&=-2\cform{4&0\\2&0}
  -\piimtau^{2}\GGhat_{2}(\EE_{2}+2)+4\frac{\pi}{\tau_{2}}\cform{3&0\\1&0}
\end{align}
which were obtained in \cite{gerken2019e}, where \eqref{eq:579} was derived in a
different way than from the coincident limit of Fay identities.

The divergent two-point \ac{HSR} identity \eqref{eq:382} has a straightforward
generalization to trihedral (and higher-point graphs),
\begin{align}
  \cformtri{1&1&A_{1}\\0&0&B_{1}}{A_{2}\\B_{2}}{A_{3}\\B_{3}}&=
  -2\cformtri{2&A_{1}\\0&B_{1}}{A_{2}\\B_{2}}{A_{3}\\B_{3}}
  -\GGhat_{2}\cformtri{A_{1}\\B_{1}}{A_{2}\\B_{2}}{A_{3}\\B_{3}}
  +\frac{\pi}{\tau_{2}} \cformtri{1&A_{1}\\-1&B_{1}}{A_{2}\\B_{2}}{A_{3}\\B_{3}}\,.
  \label{eq:434}
\end{align}
The only kind of divergent \ac{HSR} which cannot be treated in this way occurs
if the holomorphic subgraph has a higher-point divergence, since this
necessarily means that the holomorphic subgraph involves negative labels.

One might be tempted to also extend the trihedral Fay identity \eqref{eq:303} to
divergent graphs. However, this was found to lead to contradictions, as
illustrated in the following: Consider the divergent trihedral graph
$\cformtri{1\\0}{0&1\\1&0}{0&1\\2&0}$ and simplify it once by performing
three-point \ac{HSR} and once by applying \eqref{eq:303} to the first column and
to the second column of the third block, yielding the decompositions
\begin{subequations}
  \begin{align}
    \cformtri{1\\0}{0&1\\1&0}{0&1\\2&0}&\stackrel{?}{=}
    \GGhat_{2}\cform{1&0\\3&0}+\frac{1}{2}\piimtau^{3}
    (\EE_{1}(\EE_{1}-4\EE_{2}+2)-3\EE_{2}+5\EE_{3}-\zeta_{3})
    \label{eq:383}\\
    \cformtri{1\\0}{0&1\\1&0}{0&1\\2&0}&\stackrel{?}{=}
    \GGhat_{2}\cform{1&0\\3&0}+\frac{\pi}{\tau_{2}}\cform{0&1&1\\1&-1&2}
    +\frac{1}{2}\piimtau^{3}(4\EE_{1}\EE_{2}-5\EE_{3}+\zeta_{3})\,.
    \label{eq:384}
  \end{align}
\end{subequations}
Applying the Fay identity \eqref{eq:303} to any other pair of holomorphic or
antiholomorphic columns also leads to \eqref{eq:383}. Together, \eqref{eq:383}
and \eqref{eq:384} imply
\begin{align}
  \cform{0&1&1\\1&-1&2}
  \stackrel{?}{=}\frac{1}{2}\piimtau^{2}(\EE_{1}^{2}+2\EE_{1}-3\EE_{2})\,.
  \label{eq:385}
\end{align}
Next, consider the divergent trihedral graph
$\cformtri{0\\1}{0&1\\2&0}{1&1\\0&0}$ which can be decomposed via two-point
\ac{HSR} and Fay into
\begin{subequations}
  \begin{align}
    \cformtri{0\\1}{0&1\\2&0}{1&1\\0&0}&\stackrel{?}{=}
    -\piimtau^{3}(\EE_{1}-2 \EE_{2}+\EE_{3}-\zeta_{3})\label{eq:386}\\
    \cformtri{0\\1}{0&1\\2&0}{1&1\\0&0}&\stackrel{?}{=}
    -\frac{\pi}{\tau_{2}}\cform{0&1&1\\1&-1&2}
    +\frac{1}{2}\piimtau^{3}
    (\EE_{1}^2-2\EE_{1}+\EE_{2}-2\EE_{3}+2\zeta_{3})\,,
    \label{eq:387}
  \end{align}
\end{subequations}
respectively, yielding the identity
\begin{align}
  \cform{0&1&1\\1&-1&2}\stackrel{?}{=}
  \frac{1}{2}\piimtau^{2}(\EE_{1}^{2}-3\EE_{2})\,,
  \label{eq:388}
\end{align}
differing form \eqref{eq:385} by a term $\frac{\pi}{\tau_{2}}\EE_{1}$. For this
reason, we will not apply the Fay identity \eqref{eq:303} to divergent graphs.

In the \texttt{Mathematica} package \mmapackage, divergent \ac{HSR} is
implemented in the functions \mma{DiCSimplify} and \mma{TriCSimplify}, along
with the convergent \ac{HSR}. If divergent \ac{HSR} is performed or not, is
controlled by the Boolean option \mma{divHSR}. Dihedral and trihedral \ac{HSR}
can be activated and deactivated individually with the Boolean options
\mma{diDivHSR} and \mma{triDivHSR}. The default values of all these options are
\mma{True}.

\subsection{Taking derivatives of divergent graphs}
It would be desirable to apply the sieve algorithm discussed in
Section~\ref{sec:sieve-algorithm} also to divergent \acp{MGF} to derive
decompositions of divergent \acp{MGF} which are e.g.\ useful to perform the
sieve algorithm on convergent \acp{MGF}. In order to do this, we have to take
derivatives of divergent \acp{MGF}. Unfortunately, this is not straightforward
and, if done naively, contradictions to momentum-conservation identities can
arise. As above, we will restrict in this section to two-point divergences
occurring within one edge bundle since higher-point divergences are only
relevant for graphs with negative entries.

Empirically, we found that taking derivatives of divergent \acp{MGF} using the
formula \eqref{eq:233} is consistent with momentum conservation if the
divergence has the form $\sbmatrix{1&0\\0&1}$, however, a complete understanding
of the structure of derivatives of these divergences is still lacking. If the
divergence has the form $\sbmatrix{1&1\\0&0}$, we can first apply the divergent
\ac{HSR} formula \eqref{eq:382}, leading to a modification of the usual
derivative expression \eqref{eq:233}. E.g.\ consider the graph
$\cform{0&0&A\\1&1&B}$ with all column sums in $\sbmatrix{A\\B}$ at least
2. Using divergent \ac{HSR} \eqref{eq:382}, it can be rewritten to
\begin{align}
  \cform{0&0&A\\1&1&B}=-2\cform{0&A\\2&B}-\GGhatb_{2}\cform{A\\B}
  +\frac{\pi}{\tau_{2}}\cform{-1&A\\1&B}\,.
  \label{eq:389}
\end{align}
Taking the derivative via \eqref{eq:236} and using \eqref{eq:382} to write the
result back into a graph with a holomorphic subgraph yields
\begin{align}
  \nabla^{(|A|)}\cform{0&0&A\\1&1&B}=
  \sum_{i=1}^{R}\cform{0&0&A+S_{i}\\1&1&B-S_{i}}
  -\frac{\pi}{\tau_{2}}\prod_{i=1}^{R}\cform{a_{i}&0\\b_{i}&0}\,,
  \label{eq:390}
\end{align}
with an additional term as compared to a naive application of \eqref{eq:236} on
$\cform{0&0&A\\1&1&B}$.

Aside from \ac{HSR}, this additional term can also be understood as arising from
the derivative of the regularization term implicitly contained in
$\cform{0&0&A\\1&1&B}$. To see this, we first write the regularization term
explicitly,
\begin{align}
  \cform{0&0&A\\1&1&B}=\lim_{s\rightarrow0}\cform{s&0&A\\s+1&1&B}
  \label{eq:391}
\end{align}
and exchange the limit and the differential, resulting in
\begin{align}
  \nabla^{(|A|)}\cform{0&0&A\\1&1&B}
  =&\lim_{s\rightarrow0} s \cform{s+1&0&A\\s&1&B}
  +\sum_{i=1}^{R}\cform{0&0&A+S_{i}\\1&1&B-S_{i}}\,.
  \label{eq:394}
\end{align}
Next, we rewrite the first term using the momentum-conservation identity of the
seed-\ac{MGF} $\cform{s+1&0&A\\s+1&1&B}$, which is convergent for all $s\geq0$,
and factorization, yielding
\begin{align}
  \lim_{s\rightarrow0} s \cform{s+1&0&A\\s&1&B}
  =&{-}\lim_{s\rightarrow0} s\Big(\cform{s+1&0&A\\s+1&0&B}
  -\sum_{i=1}^{R}\cform{s+1&0&A\\s+1&0&B-S_{i}}\Big)\nonumber\\
  =&{-}\lim_{s\rightarrow0} s\bigg(\piimtau^{s{+}1}\mathrm{E}_{s+1}
  \prod_{i=1}^{R}\cform{a_{i}&0\\b_{i}&0}{-}\cform{s+1&A\\s+1&B}
  -\sum_{i=1}^{R}\cform{s+1&0&A\\s+1&0&B-S_{i}}\bigg)\,.
  \label{eq:392}
\end{align}
The last two terms in \eqref{eq:392} are convergent for all $s\geq0$ and hence
drop out after taking the limit. $\EE_{1}$ however is divergent and with the
first Kronecker limit formula
\begin{align}
  \EE_{s+1}=\frac{1}{s}+\mathcal{O}(s^{0})\,,
  \label{eq:393}
\end{align}
we obtain
\begin{align}
  \lim_{s\rightarrow0} s \cform{s+1&0&A\\s&1&B}
  =-\frac{\pi}{\tau_{2}}\prod_{i=1}^{R}\cform{a_{i}&0\\b_{i}&0}\,.
\end{align}
Plugging this into \eqref{eq:394} yields \eqref{eq:390}, the result previously
obtained from divergent \ac{HSR}.\footnote{Note that, to find this agreement, it
  is crucial that we do not simplify \eqref{eq:389} using
  $\cform{2&0\\0&0}=\GGhat_{2}$ before taking the derivative, since this
  contains a regularization term again, whose derivative we would have to take
  into account.}

Similarly to \eqref{eq:390}, we take the
derivative of terms of the form $\cform{1&1&A\\0&0&B}$ by first applying the
formula \eqref{eq:382} and then the usual expression \eqref{eq:236} for the
derivative. The generalization to higher-point graphs with two-point
divergences is straightforward.

Since the techniques outlined in this section to take derivatives of divergent
\acp{MGF} are conjectural and subtle, in the implementation in the \mmapackage
package, a warning is issued whenever the functions \mma{CHolCR} and
\mma{CAHolCR} encounter a divergent graph in their argument. If the Boolean
option \mma{divDer} of these functions is set to \mma{False} (the default is
\mma{True}), \mma{Nothing} is returned if it is divergent. If \mma{divDer} is
set to \mma{True}, divergent derivatives are treated exactly like convergent
ones, only (divergent) \ac{HSR} is performed on the input before the derivative
is taken.

\subsection{Divergent momentum conservation and factorization}
\label{sec:div-mom-cons}
Naively performing momentum conservation of divergent seeds and factorization
leads to inconsistencies, e.g.\ consider the holomorphic momentum-conservation
identity of the seed $\cform{1&1&2\\0&0&3}$ which is naively\footnote{We will
  see below that the first two equalities are not correct for divergent
  graphs. The last equality is too naive because $\cform{1&0\\0&0}$ is
  conditionally convergent and can yield infinity, depending on the summation
  prescription used.}
\begin{align}
  \cform{1&1&1\\0&0&3}
  \stackrel{?}{=}-2\cform{0&1&2\\0&0&3}
  \stackrel{?}{=}-2\cform{1&0\\0&0}\cform{2&0\\0&3}-2\piimtau^{3}\EE_{3}
  \stackrel{?}{=}-2\piimtau^{3}\EE_{3}\,,
  \label{eq:422}
\end{align}
where the first term vanishes due to odd label sums in both \acp{MGF}. The
divergent \ac{HSR} formula \eqref{eq:382} however (and also momentum
conservation of the convergent seed $\cform{1&1&2\\0&3&0}$) leads to
\begin{align}
  \cform{1&1&1\\0&0&3}=\piimtau^{3}(\EE_{2}-2\EE_{3})\,,
  \label{eq:423}
\end{align}
contradicting \eqref{eq:422}. In this section, we will discuss some of the
phenomena that arise in divergent momentum conservation and factorization but
leave a complete understanding to the future.

The additional term in \eqref{eq:423} can be understood in the integral
representation of the \ac{MGF} as follows: Consider the graph
\begin{align}
  \cform{0&1&A\\0&0&B}=\int_{\Sigma}\frac{\dd^{2}z}{\tau_{2}}
  C^{(0,0)}(z)f^{(1)}(z)\prod_{i=1}^{R}C^{(a_{i},b_{i})}(z)\,,
  \label{eq:424}
\end{align}
where $\sbmatrix{A\\B}$ contains no $\sbmatrix{1\\0}$, $\sbmatrix{0\\1}$ or
$\sbmatrix{1\\1}$ columns. We saw in \eqref{eq:230} that
$C^{(0,0)}(z)=\tau_{2}\delta(z,\bar{z})-1$, leading to the usual factorization
rule. In \eqref{eq:424}, the delta function instructs to take the
$z\rightarrow0$ limit of $f^{(1)}(z)\prod_{i=1}^{R}C^{(a_{i},b_{i})}(z)$. But
since $f^{(1)}(z)$ has Laurent expansion
\begin{align}
  f^{(1)}(z)=\frac{1}{z}-z\GGhat_{2}-\bar{z}\frac{\pi}{\tau_{2}}
  +\mathcal{O}(z,\bar{z})^{3}
  \label{eq:425}
\end{align}
and in particular a pole at $0$, we have to expand the product to first
order to obtain
\begin{align}
  \lim_{z\rightarrow0}f^{(1)}(z)\prod_{i=1}^{R}C^{(a_{i},b_{i})}(z)&=
  \Big(\partial_{z}\prod_{i=1}^{R}C^{(a_{i},b_{i})}(z)\Big)_{z=0}
  =-\frac{\pi}{\tau_{2}}\sum_{i=1}^{R}\prod_{j=1}^{R}
  \cform{a_{j}&0\\b_{j}-\delta_{ij}&0}\,,
  \label{eq:426}
\end{align}
using \eqref{eq:194} and the fact that the product vanishes at zero since
$|A|+|B|$ is odd if $\cform{0&1&A\\0&0&B}$ is non-trivial. This yields the
modified factorization rule
\begin{align}
  \cform{0&1&A\\0&0&B}=-\frac{\pi}{\tau_{2}}\sum_{i=1}^{R}\prod_{j=1}^{R}
  \cform{a_{j}&0\\b_{j}-\delta_{ij}&0}-\cform{1&A\\0&B}\,.
  \label{eq:429}
\end{align}
If more $\sbmatrix{1\\0}$ columns are present, higher derivatives of the
remaining graphs have to be taken. If $\sbmatrix{A\\B}$ contains a
$\sbmatrix{1\\1}$ column, corresponding to a Green function in the integral, we
have to iterate this procedure, since the derivative of the Green function is
$f^{(1)}$ (cf.\ \eqref{eq:345}) and hence contains again a pole. In this way we
obtain for the \ac{MGF} $\cform{0&1&1_{n}&A\\0&0&1_{n}&B}$, where $1_{n}$ is
the row vector with $n$ entries of $1$, the factorization rule
\begin{align}
  \cform{0&1&1_{n}&A\\0&0&1_{n}&B}&=
  \piimtau^{n+1}\sum_{i=1}^{R}\prod_{j=1}^{R}
  \cform{a_{j}&0\\b_{j}-\delta_{ij}&0}\sum_{k=0}^{n}
  (-1)^{k+1}\frac{n!}{(n{-}k)!}\mathrm{E}_{1}^{n-k}
  -\cform{1&1_{n}&A\\0&1_{n}&B}\,.
  \label{eq:431}
\end{align}
For trihedral graphs, we have similarly
\begin{align}
  \cformtri{0&1&1_{n}&A_{1}\\0&0&1_{n}&B_{1}}{A_{2}\\B_{2}}{A_{3}\\B_{3}}
   &=(-1)^{|2|}\piimtau^{n+1}\cform{A_{2}&A_{3}\\B_{2}&B_{3}}
  \sum_{i=1}^{R_{1}}\prod_{j=1}^{R_{1}}
  \cform{a_{1}^{(j)}&0\\b_{1}^{(j)}-\delta_{ij}&0}\sum_{k=0}^{n}
  (-1)^{k+1}\frac{n!}{(n{-}k)!}\mathrm{E}_{1}^{n-k}\nonumber\\
  &\qquad-\cformtri{1&1_{n}&A_{1}\\0&1_{n}&B_{1}}{A_{2}\\B_{2}}{A_{3}\\B_{3}}\,.
  \label{eq:432}
\end{align}
In general, the Laurent expansion of $f^{(n)}$ contains a term
$\sim\frac{\bar{z}^{n-1}}{z}$, which vanishes at the origin for $n\geq3$. The
$z\rightarrow0$ limit of $f^{(2)}$ depends on the direction in which the origin
is approached, but $\frac{\bar{z}}{z}$ vanishes when integrated against a delta
function due to the angular part of the integration. Therefore, only the case of
$f^{(1)}$ yields additional terms as discussed above.

When \eqref{eq:429} is used in \eqref{eq:422}, we obtain the correct additional
term, up to a factor of $2$, which arose in the momentum-conservation identity
from the product rule of $\partial_{\bar{z}}$ acting on $f^{(1)}$ (cf.\
\eqref{eq:208}). This spurious factor of $2$ is again due to the pole in
$f^{(1)}$, as can be understood by considering the integral
\begin{align}
  \int_{B_{r}(0)}\!\dd^{2}z\, \partial_{\bar{z}}\bigg(\frac{1}{z^{2}} \bigg)z\,,
  \label{eq:430}
\end{align}
where $B_{r}(0)$ is the ball of radius $r$ around $0$. Evaluating \eqref{eq:422}
using $\partial_{\bar{z}}\left( \frac{1}{z} \right)=\pi\delta^{(2)}(z)$ and the
product rule leads to
\begin{align}
  \int_{B_{r}(0)}\!\dd^{2}z\, \partial_{\bar{z}}\bigg(\frac{1}{z^{2}} \bigg)z\stackrel{?}{=}
  2\int_{B_{r}(0)}\!\dd^{2}z\,\frac{1}{z}\partial_{\bar z}\bigg(\frac{1}{z}\bigg)z
  =2\pi\,,
\end{align}
whereas the factor of $2$ is absent if we apply Stokes' theorem,
\begin{align}
  \int_{B_{r}(0)}\!\dd^{2}z\, \partial_{\bar{z}}\bigg(\frac{1}{z^{2}} \bigg)z
  =\frac{1}{2i}\oint_{\partial B_{r}(0)}\!\dd z\, \frac{1}{z}
  =\pi\res_{z=0}\bigg(\frac{1}{z} \bigg)=\pi\,.
\end{align}

Empirically, momentum-conservation identities of seeds with a divergence of the
form $\sbmatrix{1&0\\0&1}$ seem to be consistent, but we have not investigated
them any further. For trihedral graphs, if the two blocks adjacent to the vertex
used for momentum conservation are convergent and no three-point divergence
appears in the graph, the resulting momentum-conservation identity is valid. If
these conditions are not met, the same care has to be taken as with the dihedral
graphs.

In the \mmapackage package, the modified factorization rules \eqref{eq:431} and
\eqref{eq:432} are implemented in the functions \mma{DiCSimplify} and
\mma{TriCSimplify}, but since they are not tested as thoroughly as the
convergent manipulations, a warning is issued if these special cases are
encountered.  If more than one $\sbmatrix{1\\0}$ or $\sbmatrix{0\\1}$ column
appears next to a $\sbmatrix{0\\0}$ column, the input is returned. The momentum
conservation functions \mma{DiHolMomConsId} and \mma{TriHolMomConsId} and their
complex conjugates issue a warning when the seed is divergent.

As we will see in the next section, the basis decompositions of \acp{MGF}
obtained in this paper rely on manipulations of divergent \acp{MGF} only for the
modular weights $(6,6)$ and $(7,5)$ (and its complex conjugate). The expansion
of the generating function of Koba--Nielsen integrals at two- and three points
involving these sectors was checked to satisfy the Cauchy--Riemann equations
derived in \cite{gerken2019d}. Furthermore, the Laurent polynomials of this
expansion were checked against the closed formula for two-point Laurent
polynomials given in \cite{gerken2020c}.


\section{Basis decompositions}
\label{sec:conjectured-basis}

By combining the techniques discussed in the sections above, we can
systematically generate identities for modular graph forms, starting from a
small number of known relations. In the end, we obtain decompositions of a large
class of complicated \acp{MGF} into a small number of simple graphs. That these
actually a basis for all \acp{MGF} can be proven using techniques from
iterated Eisenstein integrals discussed in \cite{gerken2020c}.

In the \mmapackage \texttt{Mathematica} package, decompositions for all dihedral
and trihedral convergent \acp{MGF} with non-negative edge labels of modular
weight $(a,b)$ with $a+b\leq12$ are given, starting just from the dihedral
identities
\begin{align}
  D_{3}&=\EE_{3}+\zeta_{3}\label{eq:396}\\
  D_{5}&=60C_{1,1,3}+10D_{3}\EE_{2}-48\EE_{5}+16\zeta_{5}\,,\label{eq:397}
\end{align}
where $D_{\ell}$ is defined in \eqref{eq:124} and $C_{a,b,c}$ in
\eqref{eq:135}. These two identities are also the only source of zeta-values in
the basis decompositions.

\subsection{Systematic derivation of identities}
\label{sec:sys-der-MGF-ids}
\begin{table}
  \centering
  \begin{tabular}{ccccc}
    \toprule
    weight & dihedral non-\ac{HSR} & dihedral \ac{HSR}
    & trihedral non-\ac{HSR} & trihedral \ac{HSR}\\
    \midrule
    $(1,1)$ & $0$ & $0$ & $0$ & $0$ \\
    \midrule
    $(2,2)$ & $1$ & $0$ & $0$ & $0$ \\
    $(3,1)$ & $1$ & $0$ & $0$ & $0$ \\
    \midrule
    $(3,3)$ & $7$ & $2$ & $0$ & $0$ \\
    $(4,2)$ & $5$ & $3$ & $0$ & $0$ \\
    $(5,1)$ & $1$ & $4$ & $0$ & $0$ \\
    \midrule
    $(4,4)$ & $27$ & $10$ & $28$ & $20$ \\
    $(5,3)$ & $22$ & $12$ & $17$ & $25$ \\
    $(6,2)$ & $11$ & $16$ & $0$ & $29$ \\
    $(7,1)$ & $1$ & $14$ & $0$ & $12$ \\
    \midrule
    $(5,5)$ & $83$ & $40$ & $326$ & $248$ \\
    $(6,4)$ & $73$ & $44$ & $247$ & $291$ \\
    $(7,3)$ & $47$ & $50$ & $91$ & $322$ \\
    $(8,2)$ & $19$ & $50$ & $0$ & $243$ \\
    $(9,1)$ & $1$ & $35$ & $0$ & $94$ \\
    \midrule
    $(6,6)$ & $228$ & $138$ & $2236$ & $2044$ \\
    $(7,5)$ & $206$ & $142$ & $1844$ & $2191$ \\
    $(8,4)$ & $150$ & $154$ & $990$ & $2359$ \\
    $(9,3)$ & $83$ & $149$ & $276$ & $2008$ \\
    $(10,2)$ & $29$ & $124$ & $0$ & $1207$ \\
    $(11,1)$ & $1$ & $74$ & $0$ & $439$ \\
    \midrule
    total & $996$ & $1061$ & $6055$ & $11532$ \\
    \bottomrule
  \end{tabular}
  \caption{Number of convergent dihedral and trihedral \acp{MGF} with
    non-negative edge labels, excluding products. For graphs containing closed
    holomorphic subgraphs, no basis decompositions need to be found
    independently, they are implied by \ac{HSR} and the basis decompositions
    of the non-\ac{HSR} graphs.}
  \label{tab:MGFs}
\end{table}

In order to apply the techniques discussed above systematically, we consider
subspaces with total modular weight $a+b=\mathrm{const.}$ of the space of all
\acp{MGF} and derive all identities in one subspace before continuing to the
next higher total weight.

Within each subspace, we start by considering weight $a=b$ which corresponds to
\acp{MGF} which are modular invariant after multiplication by
$\tau_{2}^{a}$. The identities in this space are generated by combining momentum
conservation with Fay identities:
\begin{itemize}
\item We write down all convergent dihedral and trihedral \acp{MGF} of weight
  $(a{+}1,a)$ and $(a,a{+}1)$ without closed holomorphic subgraphs and use them
  as seeds to generate holomorphic and antiholomorphic momentum-conservation
  identities, respectively. Closed holomorphic subgraphs in the seeds would
  necessarily lead to negative labels in the identity which could not be removed
  by momentum conservation.
\item We write down all convergent trihedral \acp{MGF} of weight $(a,a)$,
  including those which contain closed holomorphic subgraphs and apply the Fay
  identity \eqref{eq:303} in all possible ways.
\end{itemize}
Afterwards, we remove all relations which contain divergent \acp{MGF} after
topological simplifications and factorizations. Then, we simplify the remaining
identities using \ac{HSR}, the (generalized) Ramanujan identities discussed in
Section~\ref{sec:MGFDers} and identities known from lower total modular weight
and expand holomorphic Eisenstein series in the ring spanned by $\GG_{4}$ and
$\GG_{6}$. The resulting large system of linear equations, together with the
identities \eqref{eq:396} and \eqref{eq:397} can then be solved for all
convergent dihedral and trihedral \acp{MGF} which do not appear in the basis.

After the $a=b$ sector, we continue with the weight-$(a{+}k,a{-}k)$ sectors with
$k=1,\dots,a-1$ as follows: In addition to the momentum-conservation and Fay
identities for these sectors, we also take the Cauchy--Riemann derivative of all
basis decompositions in the $(a{+}k{-}1,a{-}k{+}1)$ sector (excluding \acp{MGF}
containing closed holomorphic subgraphs), which were found before. Again, we
remove all relations containing divergent \acp{MGF}. Finally, we take the
complex conjugate of all identities obtained, to also cover the $k<0$ sectors.

In this way, basis decompositions for all convergent dihedral and trihedral
\acp{MGF} can be found with total modular weight $a+b\leq10$. The number of
these \acp{MGF} is listed in Table~\ref{tab:MGFs}. Note that we did not need to
use the sieve algorithm in this process, hence we do not have undetermined
integration constants in the basis decompositions.

Although the strategy outlined above is successful in the $a+b\leq10$ sectors,
at weight $(6,6)$, it is not sufficient to decompose all trihedral \acp{MGF}. To
obtain the decompositions of these graphs as well, we keep the
momentum-conservation identities containing divergent graphs and simplify them
using the divergent \ac{HSR} outlined in Section~\ref{sec:div-HSR} if possible
(both divergent holomorphic subgraphs and divergences outside of the holomorphic
subgraph appear). In this way, we can decompose all graphs in the $(6,6)$ and
$(7,5)$ sectors. For the remaining sectors in Table~\ref{tab:MGFs}, the
convergent identities are sufficient again.

In this way, basis decompositions for $1646$ dihedral and $9520$ trihedral
convergent \acp{MGF} with non-negative edge labels and without closed
holomorphic subgraphs were found and implemented in the functions
\mma{DiCSimplify} and \mma{TriCSimplify} of the \mmapackage package. Since
\mma{CSimplify} calls \mma{DiCSimplify} and \mma{TriCSimplify}, we have e.g.
\begin{mmaCell}{Input}
  CSimplify\mmabigso{}\mmaincform{1 1 1 1\\1 1 1 1}\mmabigsc{}\\
  CSimplify\mmabigso{}\mmaincformtri{1\\1}{1 1\\1 1}{1 1\\1 1}\mmabigsc{}
\end{mmaCell}
\begin{mmaCell}{Output}
  24\,\mmaoutcform{1 1 2\\1 1 2}+\mmaFrac{3\,\mmaSup{\(\pi\)}{4}\,\mmaSubSup{E}{2}{2}}{\mmaSubSup{\(\tau\)}{2}{4}}\mmam\mmaFrac{18\,\mmaSup{\(\pi\)}{4}\,\mmaSub{E}{4}}{\mmaSubSup{\(\tau\)}{2}{4}}
\end{mmaCell}
\begin{mmaCell}{Output}
  2\,\mmaoutcform{1 1 3\\1 1 3}\mmam\mmaFrac{2\,\mmaSup{\(\pi\)}{5}\,\mmaSub{E}{5}}{5\,\mmaSubSup{\(\tau\)}{2}{5}}+\mmaFrac{3\,\mmaSup{\(\pi\)}{5}\,\mmaSub{\(\zeta\)}{5}}{10\,\mmaSubSup{\(\tau\)}{2}{5}} \textnormal{.}
\end{mmaCell}

All the basis decompositions contained in the \mmapackage package were checked
to be compatible with the Cauchy--Riemann equation of the generating series of
Koba--Nielsen integrals discussed in \cite{gerken2019d} at two- and three
points. The decompositions of \acp{MGF} with $a+b\leq10$ were used in
\cite{gerken2020c} to find representations of \acp{MGF} in terms of iterated
Eisenstein integrals via this generating series.

\subsection{Bases for modular graph forms}
\label{sec:MGF-bases}
\begin{table}
  \centering
  {
    \renewcommand{\arraystretch}{1.1}
    \setlength{\tabcolsep}{2pt}
    \begin{tabular}{ccc}
      \toprule
      weight & \ \# basis elements & basis elements\\
      \midrule
      $(2,2)$ & $1$ & $\piimtauil^{2}\EE_{2}$\\
      $(3,1)$ & $1$ & $\cform{3&0\\1&0}$\\
      \midrule
      $(3,3)$ & $2$ & $\piimtauil^{3}\EE_{3},\,\piimtauil^{3}\zeta_{3}$\\
      $(4,2)$ & $1$ & $\cform{4&0\\2&0}$ \\
      $(5,1)$ & $1$ & $\cform{5&0\\1&0}$ \\
      \midrule
      $(4,4)$ & $4$ & $\,\piimtauil^{4}\EE_{4},\,\cform{1&1&2\\1&1&2},
      \,\piimtauil^{4}\EE_{2}^{2},\,\cform{1&0\\3&0}\cform{3&0\\1&0}$\\
      $(5,3)$ & $3$ & $\cform{5&0\\3&0},\,\cform{1&1&3\\1&1&1},
      \,\piimtauil^{2}\EE_{2}\cform{3&0\\1&0}$ \\
      $(6,2)$ & $2$ & $\cform{6&0\\2&0},\,\cform{3&0\\1&0}^{2}$ \\
      $(7,1)$ & $1$ & $\cform{7&0\\1&0}$\\
      \midrule
      \multirow{2}{*}{$(5,5)$} & \multirow{2}{*}{$9$}
      & $\piimtauil^{5}\EE_{5},\,\cform{1&1&3\\1&1&3},
      \,\aform{0&2&3\\3&0&2},\,\aform{0&1&2&2\\1&1&0&3},
      \,\piimtauil^{5}\zeta_{5},\,\piimtauil^{5}\EE_{2}\EE_{3},$\\
      & & $\piimtauil^{5}\EE_{2}\zeta_{3},
      \,\cform{1&0\\3&0}\cform{4&0\\2&0},
      \,\cform{3&0\\1&0}\cform{2&0\\4&0}$\\[0.5em]
      \multirow{2}{*}{$(6,4)$} & \multirow{2}{*}{$8$} & $\cform{6&0\\4&0},
      \,\cform{1&1&4\\1&1&2},\,\cform{1&2&3\\1&0&3},
      \,\cform{1&1&1&3\\0&1&1&2},$\\
      & & $\piimtauil^{3}\EE_{3}\cform{3&0\\1&0},
      \,\piimtauil^{3}\EE_{2}\cform{4&0\\2&0},\,
      \cform{3&0\\1&0}\zeta_{3},\,\cform{1&0\\3&0}\cform{5&0\\1&0}$\\[0.5em]
      $(7,3)$  & $5$ & $\cform{7&0\\3&0},\,\cform{1&1&5\\1&1&1},
      \,\cform{0&2&5\\1&0&2},\,\cform{3&0\\1&0}\cform{4&0\\2&0},
      \,\piimtauil^{2}\EE_{2}\cform{5&0\\1&0}$\\
      $(8,2)$ & $3$ & $\cform{8&0\\2&0},\,
      \cform{0&3&5\\1&0&1},\,\cform{3&0\\1&0}\cform{5&0\\1&0}$\\
      $(9,1)$ & $1$ & $\cform{9&0\\1&0}$\\
      \midrule
      & & $\piimtauil^{6}\EE_{6},\,\cform{1&1&4\\1&1&4},
      \,\cform{1&2&3\\1&2&3},\,\cform{2&2&2\\2&2&2},
      \,\cform{1&1&2&2\\1&1&2&2},\,\aform{0&2&4\\5&0&1},$\\
      & &  $\aform{0&2&2&2\\3&0&1&2},\,\aform{0&1&2&3\\2&1&3&0},
      \,\piimtauil^{6}\zeta_{3}^{2},\,\piimtauil^{6}\EE_{3}^{2},
      \,\piimtauil^{6}\EE_{3}\zeta_{3},\,\piimtauil^{6}\EE_{2}\EE_{4},$\\
      $(6,6)$ & $21$ & $\piimtauil^{2}\EE_{2}\cform{1&1&2\\1&1&2},
      \,\piimtauil^{6}\EE_{2}^{3},
      \,\cform{4&0\\2&0}\cform{2&0\\4&0},\,\cform{5&0\\1&0}\cform{1&0\\5&0},$\\
      & & $\cform{3&0\\1&0}\cform{3&0\\5&0},\,\cform{1&0\\3&0}\cform{5&0\\3&0},
      \,\cform{3&0\\1&0}\cform{1&1&1\\1&1&3},$\\
      & & $\cform{1&0\\3&0}\cform{1&1&3\\1&1&1},
      \,\piimtauil^{2}\EE_{2}\cform{3&0\\1&0}\cform{1&0\\3&0}$\\[0.5em]
      & & $\cform{7&0\\5&0},\,\cform{0&1&6\\1&4&0},\,\cform{0&1&6\\2&3&0},
      \,\cform{0&2&5\\2&3&0},\cform{0&3&4\\4&0&1},
      \,\cform{1&1&2&3\\1&1&2&1},$\\
      & & $\cform{1&2&2&2\\1&0&2&2},\,\cform{0&1&2&4\\2&1&2&0},
      \,\piimtauil^{3}\EE_{3}\cform{4&0\\2&0},
      \,\piimtauil^{3}\cform{4&0\\2&0}\zeta_{3},$\\
      $(7,5)$ & $18$ & $\piimtauil^{4}\EE_{4}\cform{3&0\\1&0},
      \,\piimtauil^{2}\EE_{2}\cform{5&0\\3&0},
      \,\cform{3&0\\1&0}\cform{1&1&2\\1&1&2},$\\
      & & $\piimtauil^{2}\EE_{2}\cform{1&1&3\\1&1&1},
      \,\cform{3&0\\1&0}^{2}\cform{1&0\\3&0},
      \,\cform{5&0\\1&0}\cform{2&0\\4&0},$\\
      & & $\cform{1&0\\3&0}\cform{6&0\\2&0},
      \,\piimtauil^{4}\EE_{2}^{2}\cform{3&0\\1&0}$\\[0.5em]
      & & $\cform{8&0\\4&0},\,\cform{0&2&6\\2&2&0},
      \,\cform{0&3&5\\2&2&0},\,\cform{0&4&4\\3&0&1},
      \,\cform{1&2&2&3\\1&1&2&0},\,\cform{1&2&2&3\\1&1&2&0}$\\
      $(8,4)$&$14$& $\cform{4&0\\2&0}^{2},\piimtauil^{3}\EE_{3}\cform{5&0\\1&0},
      \,\piimtauil^{3}\cform{5&0\\1&0}\zeta_{3},
      \,\cform{3&0\\1&0}\cform{5&0\\3&0},$\\
      & & $\piimtauil^{2}\EE_{2}\cform{6&0\\2&0},
      \,\cform{3&0\\1&0}\cform{1&1&3\\1&1&1},
      \,\piimtauil^{2}\EE_{2}\cform{3&0\\1&1}^{2},
      \,\cform{3&0\\1&0}\cform{7&0\\1&0}$\\[0.5em]
      \multirow{2}{*}{$(9,3)$} & \multirow{2}{*}{$8$}
      & $\cform{9&0\\3&0},\,\cform{0&3&6\\1&2&0},
      \,\cform{0&3&6\\2&1&0},\,\cform{0&4&5\\2&1&0},
      \,\cform{4&0\\2&0}\cform{5&0\\1&0},$\\
      & & $\,\cform{3&0\\1&0}\cform{6&0\\2&0},
      \,\piimtauil^{2}\EE_{2}\cform{7&0\\1&0},
      \,\cform{3&0\\1&0}^{2}$\\[0.5em]
      $(10,2)$  & $4$ & $\cform{10&0\\2&0},\,\cform{0&4&6\\1&1&0},
      \,\cform{5&0\\1&0}^{2},\,\cform{3&0\\1&0}\cform{7&0\\1&0}$\\
      $(11,1)$  & $1$ & $\cform{11&0\\1&0}$\\
      \bottomrule
    \end{tabular}
  }
  \caption{Basis elements used in the \mmapackage package for (convergent)
    modular graph forms of weight $a+b\leq12$, excluding holomorphic Eisenstein
    series. The counting includes zeta values.}
  \label{tab:Cbasis}
\end{table}

\begin{table}
  \centering
  {
    \renewcommand{\arraystretch}{1.1}
    \setlength{\tabcolsep}{2pt}
    \begin{tabular}{ccc}
      \toprule
      weight & \ \# basis elements & basis elements\\
      \midrule
      $(2,2)$ & $1$ & $\EE_{2}$ \\[0em]
      $(3,1)$ & $1$ & $\nabladg \EE_{2}$ \\
      \midrule
      $(3,3)$ & $2$ & $\EE_{3},\,\zeta_{3}$ \\[0em]
      $(4,2)$ & $1$ & $\nabladg \EE_{3}$  \\[0em]
      $(5,1)$ & $1$ & $\nabladg^{2} \EE_{3}$  \\
      \midrule
      $(4,4)$ & $4$ &  $\EE_{4},\,\EE_{2,2},\,\EE_{2}^{2},
      \,\tau_{2}^{-2}\nabladg \EE_{2}\nabladgb \EE_{2}$ \\[0em]
      $(5,3)$ & $3$ & $\nabladg \EE_{4},\,\nabladg \EE_{2,2},
      \,\EE_{2}\nabladg \EE_{2}$  \\[0em]
      $(6,2)$ & $2$ & $\nabladg^{2} \EE_{4},\,(\nabladg \EE_{2})^{2}$  \\[0em]
      $(7,1)$ & $1$ & $\nabladg^{3}\EE_{4}$ \\
      \midrule
      \multirow{2}{*}{$(5,5)$} & \multirow{2}{*}{$9$} & $\EE_{5},\,\EE_{2,3},
      \,\BB_{2,3},\,\BB'_{2,3},\,\zeta_{5},$ \\
      & & $\,\EE_{2}\EE_{3},\,\EE_{2}\zeta_{3},
      \,\tau_{2}^{-2}\nabladgb\EE_{2}\nabladg \EE_{3},
      \,\tau_{2}^{-2}\nabladg\EE_{2}\nabladgb \EE_{3}$ \\[0.7em]
      \multirow{2}{*}{$(6,4)$} & \multirow{2}{*}{$8$} & $\nabladg\EE_{5},
      \,\nabladg \EE_{2,3},\,\nabladg\BB_{2,3},\,\nabladg\BB'_{2,3},$ \\
      & & $\,\nabladg\EE_{2}\EE_{3},\,\EE_{2}\nabladg\EE_{3},
      \,\nabladg\EE_{2}\zeta_{3},
      \,\tau_{2}^{-2}\nabladgb\EE_{2}\nabladg^{2}\EE_{3}$ \\[0.7em]
      $(7,3)$  & $5$ & $\nabladg^{2}\EE_{5},\,\nabladg^{2}\EE_{2,3},
      \,\nabladg^{2}\BB'_{2,3},
      \,\nabladg\EE_{2}\nabladg\EE_{3},\,\EE_{2}\nabladg^{2}\EE_{3}$\\[0em]
      $(8,2)$ & $3$ & $\nabladg^{3}\EE_{5},\,\nabladg^{3}\BB'_{2,3},
      \,\nabladg\EE_{2}\nabladg^{2}\EE_{3}$\\[0em]
      $(9,1)$ & $1$ & $\nabladg^{4}\EE_{5}$\\
      \midrule
      \multirow{4}{*}{$(6,6)$} & \multirow{4}{*}{$21$} & $\EE_{6},\,\EE_{2,4},
      \,\EE_{3,3},\,\EE'_{3,3},\,\EE_{2,2,2},\,\BB_{2,4},
      \,\BB'_{2,4},\,\BB_{2,2,2},\,\zeta_{3}^{2},$ \\
      & &  $\EE_{3}^{2},\,\EE_{3}\zeta_{3},\,\EE_{2}\EE_{4},
      \,\EE_{2}\EE_{2,2},\,\EE_{2}^{3},
      \,\tau_{2}^{-2}\nabladg\EE_{3}\nabladgb\EE_{3},
      \,\tau_{2}^{-4}\nabladg^{2}\EE_{3}
      \nabladgb\hspace{-0.3ex}\rule[1.7ex]{0pt}{0pt}^{2}\EE_{3}$ \\
      & & $\,\tau_{2}^{-2}\nabladg\EE_{2}\nabladgb\EE_{4},
      \,\tau_{2}^{-2}\nabladgb\EE_{2}\nabladg\EE_{4},
      \tau_{2}^{-2}\nabladg\EE_{2}\nabladgb\EE_{2,2},
      \,\tau_{2}^{-2}\nabladgb\EE_{2}\nabladg\EE_{2,2},$\\
      & &  $\,\tau_{2}^{-2}\EE_{2}\nabladg\EE_{2}\nabladgb\EE_{2}$\\[0.7em]
      & &  $\nabladg\EE_{6},\,\nabladg\EE_{2,4},\,\nabladg\EE_{3,3},
      \,\nabladg\EE'_{3,3},\,\nabladg\EE_{2,2,2},\,\nabladg\BB_{2,4},
      \,\nabladg\BB'_{2,4},\,\nabladg\BB_{2,2,2},$\\
      $(7,5)$ & $18$ & $\EE_{3}\nabladg\EE_{3},\,\nabladg\EE_{3}\zeta_{3},
      \,\nabladg\EE_{2}\EE_{4},\,\EE_{2}\nabladg\EE_{4},
      \,\nabladg\EE_{2}\EE_{2,2},\,\EE_{2}\nabladg\EE_{2,2},
      \,\EE_{2}^{2}\nabladg\EE_{2},$\\
      & & $\tau_{2}^{-2}\nabladg^{2}\EE_{3}\nabladgb\EE_{3},
      \,\tau_{2}^{-2}\nabladgb\EE_{2}\nabladg^{2}\EE_{4},
      \,\tau_{2}^{-2}(\nabladg\EE_{2})^{2}\nabladgb\EE_{2}$\\[0.7em]
      & & $\nabladg^{2}\EE_{6},\,\nabladg^{2}\EE_{2,4},
      \,\nabladg^{2}\EE'_{3,3},\,\nabladg^{2}\BB_{2,4},
      \,\nabladg^{2}\BB'_{2,4},\,\nabladg^{2}\BB_{2,2,2}$\\
      $(8,4)$ & $14$ & $(\nabladg\EE_{3})^{2},\,\EE_{3}\nabladg^{2}\EE_{3},
      \,\nabladg^{2}\EE_{3}\zeta_{3},\,\nabladg\EE_{2}\nabladg\EE_{4},
      \,\EE_{2}\nabladg^{2}\EE_{4},\,\nabladg\EE_{2}\nabladg\EE_{2,2},$\\
      & & $\EE_{2}(\nabladg\EE_{2})^{2},
      \,\tau_{2}^{-2}\nabladgb\EE_{2}\nabladg^{3}\EE_{4}$\\[0.7em]
      \multirow{2}{*}{$(9,3)$} & \multirow{2}{*}{$8$} & $\nabladg^{3}\EE_{6},
      \,\nabladg^{3}\EE'_{3,3},\,\nabladg^{3}\BB_{2,4},
      \,\nabladg^{3}\BB'_{2,4},\,\nabladg\EE_{3}\nabladg^{2}\EE_{3},
      \,\nabladg\EE_{2}\nabladg^{2}\EE_{4},$\\
      & & $\EE_{2}\nabladg^{3}\EE_{4},\,(\nabladg\EE_{2})^{3}$\\[0.7em]
      $(10,2)$  & $4$ & $\nabladg^{4}\EE_{6},\,\nabladg^{4}\BB'_{2,4},
      \,(\nabladg^{2}\EE_{3})^{2},\,\nabladg\EE_{2}\nabladg^{3}\EE_{4}$\\
      $(11,1)$  & $1$ & $\nabladg^{5}\EE_{6}$\\
      \bottomrule
    \end{tabular}
  }
  \caption{Basis of (convergent) modular graph forms of weight $a+b\leq12$,
    excluding holomorphic Eisenstein series. The prefactors of $\tau_{2}$ were
    chosen such that the modular weight in the sector $(a{+}k,a{-}k)$ is
    $(0,-2k)$ for $0\leq k<a$. The counting includes zeta values.}
  \label{tab:Ebasis}
\end{table}

Using the procedure outlined in Section~\ref{sec:sys-der-MGF-ids}, we obtain
decompositions for many modular graph forms, which leave as independent
\acp{MGF} only the ones listed in Table~\ref{tab:Cbasis}. That these form indeed
a basis of all \acp{MGF} (not just two- and three-point graphs) at the
corresponding weights can be proven using iterated Eisenstein integrals and
generating functions of Koba--Nielsen integrals \cite{gerken2020c}. The basis
elements in the sector $(a,b)$ with $a<b$ are given by complex
conjugation. Furthermore, basis elements containing a holomorphic Eisenstein
series are not listed in Table~\ref{tab:Cbasis}, since they can be constructed
from the bases at lower weights, e.g.\ the $(6,4)$ sector contains the
additional basis elements $\GG_{4}\cform{2&0\\4&0}$ and
$\GG_{6}\overline{\GG}_{4}$. In the following, we will refer to basis elements
given as products as \emph{reducible} and to the remaining ones as
\emph{irreducible}. On top of various modular graph forms, we have included in
Table~\ref{tab:Cbasis} also the constants $\zeta_{3}$, $\zeta_{5}$ and
$\zeta_{3}^{2}$ in the relevant sectors.

Note that starting from total modular weight $10$, the sector with equal
holomorphic and antiholomorphic weight contains cusp forms. Specifically,
in the basis of the $(5,5)$ sector, the three cusp forms
\begin{subequations}
  \label{eq:398}
  \begin{gather}
    \cform{1&0\\3&0}\cform{4&0\\2&0}-\cform{3&0\\1&0}\cform{2&0\\4&0}
    \label{eq:400}\\
    \aform{0&2&3\\3&0&2}\label{eq:401}\\
    \aform{0&1&2&2\\1&1&0&3}\label{eq:402}
  \end{gather}
\end{subequations}
appear. Similarly, the $(6,6)$ basis contains the cusp forms
\begin{subequations}
  \label{eq:399}
  \begin{gather}
    \cform{3&0\\1&0}\cform{3&0\\5&0}-\cform{1&0\\3&0}\cform{5&0\\3&0}\\
    \cform{3&0\\1&0}\cform{1&1&1\\1&1&3}-\cform{1&0\\3&0}\cform{1&1&3\\1&1&1}\\
    \aform{0&2&4\\5&0&1}\\
    \aform{0&2&2&2\\3&0&1&2}\\
    \,\,\aform{0&1&2&3\\2&1&3&0}\,.
  \end{gather}
\end{subequations}
The remaining basis elements in these sectors are real.\footnote{Note that if we
  form antisymmetric combinations $\aform{A\\B}$ in the $(a,a)$ sectors with
  $a\leq4$, these vanish since all basis elements are real.} The cusp forms
\eqref{eq:400} and \eqref{eq:401} were discussed in \cite{dhoker2019a}, whereas
\eqref{eq:402} has higher loop order than the graphs studied in the
reference. In the weight $(6,6)$ sector, the dimension of the space of two-loop
imaginary cusp forms was found to be $2$ in \cite{dhoker2019a}, in agreement
with \eqref{eq:399}.

The basis of \acp{MGF} has an intricate structure which is closely related to
the counting of iterated Eisenstein integrals, but this structure is not
manifest in the basis given in Table~\ref{tab:Cbasis}. To make the relation to
iterated Eisenstein integrals more transparent, we will use a second basis,
summarized in Table~\ref{tab:Ebasis}. The basis has been multiplied by
$\tau_{2}^{a+k}/\pi^{a}$ in the $(a+k,a-k)$ sector of Table~\ref{tab:Ebasis} as
compared to Table~\ref{tab:Cbasis} for ease of notation. This means in
particular that the basis elements given for the $a=b$ sectors are rendered
modular invariant.

The structure of the basis in Table~\ref{tab:Ebasis} is the following: In the
modular invariant sectors, we split the irreducible basis elements into real and
complex \acp{MGF}. The real ones are denoted by $\EE$, the complex ones by
$\BB$, where the subscript refers to the holomorphic Eisenstein series appearing
in the Cauchy--Riemann equations of the respective basis element. If several
basis elements belong to the same sector w.r.t. these holomorphic Eisenstein
series, we use a prime to distinguish them.

The non-holomorphic Eisenstein series $\EE_{k}$ defined in \eqref{eq:51} belong
to the real basis elements. The remaining real basis elements of higher depth
were defined in~\cite{broedel2018} to streamline their Cauchy--Riemann equations
as detailed below and are given in terms of the \acp{MGF} defined previously by
{\allowdisplaybreaks
\begin{subequations}
  \label{eq:172}
  \begin{align}
    \EE_{2,2}&=C_{1,1,2}-\frac{9}{10}\EE_{4}\\
    \EE_{2,3}&=C_{1,1,3}-\frac{43}{35}\EE_{5}\\
    \EE_{3,3}&=3C_{1,2,3}+C_{2,2,2}-\frac{15}{14}\EE_{6}\\
    \EE_{3,3}'&=C_{1,2,3}+\frac{17}{60}C_{2,2,2}-\frac{59}{140}\EE_{6}\\
    \EE_{2,4}&=\vphantom{\frac{1}{1}}9C_{1,1,4}+3C_{1,2,3}+C_{2,2,2}-13\EE_{6}\\
    \EE_{2,2,2}&=-C_{1,1,2,2}+\frac{232}{45}C_{2,2,2}+\frac{292}{15}C_{1,2,3}
    +\frac{2}{5}C_{1,1,4}+2\EE_{3}^{2}+\EE_{2}\EE_{4}-\frac{466}{45}\EE_{6}\,,
  \end{align}
\end{subequations}
}
where $C_{a,b,c}$ and $C_{a,b,c,d}$ were defined in \eqref{eq:135} and
\eqref{eq:170}, respectively. A subscript $k$ means in this notation that the
holomorphic Eisenstein series $\GG_{2k}$ appears in the Cauchy--Riemann
equations, i.e.\ in the lowest Cauchy--Riemann derivative in which a holomorphic
Eisenstein series appears. This determines the sector of iterated Eisenstein
integrals that appear in the expansion of the basis element, cf.\ the discussion
in Section~5 of \cite{gerken2020c}. For instance, the basis element $\EE_{2,4}$
belongs to the $\GG_{4}\GG_{8}$ sector. The Cauchy--Riemann equations which make
this manifest for the real irreducible basis elements are
{\allowdisplaybreaks
\begin{subequations}
  \label{eq:403}
  \begin{align}
    \nabladg^{k}\EE_{k}
    &=\frac{\tau_{2}^{2k}}{\pi^{k}}\frac{(2k-1)!}{(k-1)!}\GG_{2k}\\
    \nabladg^{3}\EE_{2,2}&=
    -6\frac{\tau_{2}^{4}}{\pi^{2}}\GG_{4}\nabladg\EE_{2}\\
    \nabladg^{3}\EE_{2,3}&=-2\nabladg\EE_{2}\nabladg^{2}\EE_{3}
    -4\frac{\tau_{2}^{4}}{\pi^{2}}\GG_{4}\nabladg\EE_{3}\\
    \nabladg^{5}\EE_{3,3}&=
    180\frac{\tau_{2}^{6}}{\pi^{3}}\GG_{6}\nabladg^{2}\EE_{3}\\
    \nabladg^{4}\EE_{3,3}'&=
    -12\frac{\tau_{2}^{6}}{\pi^{3}}\GG_{6}\nabladg\EE_{3}\\
    \nabladg^{3}\EE_{2,4}&=
    -\frac{27}{2}\nabladg\big(\EE_{2}\nabladg^{2}\EE_{4}\big)
    -\frac{27}{4}\nabladg^{3}\BB_{2,4}-\frac{21}{40}\nabladg^{3}\BB'_{2,4}
    -27\frac{\tau_{2}^{4}}{\pi^{2}}\GG_{4}\nabladg\EE_{4} \\
    \nabladg^{3}\EE_{2,2,2}&=(\nabladg\EE_{2})^{3}
    -12\frac{\tau_{2}^{4}}{\pi^{2}}\GG_{4}\nabladg\EE_{2,2}\,,
  \end{align}
\end{subequations}}
where we use the Cauchy--Riemann operator defined in \eqref{eq:76} and the
complex basis elements $\BB_{2,4}$ and $\BB'_{2,4}$ are defined in
\eqref{eq:404}. The right-hand sides in \eqref{eq:403} all lie manifestly in the
same sector of holomorphic Eisenstein series as indicated by the subscripts on
the left-hand side. In \cite{broedel2018}, the real irreducible basis elements
$\EE$ were written in terms of iterated Eisenstein integrals. From this, we can
read off their Laurent polynomials \cite{dhoker2015, dhoker2016}, namely
{\allowdisplaybreaks
\begin{subequations}
  \label{eq:408}
  \begin{align}
    \EE_{k}\big|_{q^{0}\bar{q}^{0}}&=(-1)^{k-1}\frac{B_{2k}}{(2k)!}(4y)^{k}
    +4\binom{2k-3}{k-1}\zeta_{2k-1}(4y)^{1-k}\\
    \EE_{2,2}\big|_{q^{0}\bar{q}^{0}}&=-\frac{y^4}{20250}+\frac{y\zeta_{3}}{45}
    +\frac{5 \zeta_{5}}{12 y}-\frac{\zeta_{3}^2}{4 y^2}\\
    \EE_{2,3}\big|_{q^{0}\bar{q}^{0}}&=-\frac{4 y^5}{297675}
    +\frac{2y^2\zeta_{3}}{945}-\frac{\zeta_{5}}{180}+\frac{7 \zeta_{7}}{16 y^2}
    -\frac{\zeta_{3} \zeta_{5}}{2 y^3}\\
    \EE_{3,3}\big|_{q^{0}\bar{q}^{0}}&=\frac{2 y^6}{6251175}
    +\frac{y \zeta_{5}}{210}+\frac{\zeta_{7}}{16 y}-\frac{7 \zeta_{9}}{64 y^3}
    +\frac{9 \zeta_{5}^2}{64 y^4}\\
    \EE'_{3,3}\big|_{q^{0}\bar{q}^{0}}&=-\frac{y^6}{18753525}
    +\frac{y\zeta_{5}}{630}
    +\frac{3\zeta_{7}}{160y}-\frac{7\zeta_{9}}{480 y^3}\\
    \EE_{2,4}\big|_{q^{0}\bar{q}^{0}}&=-\frac{y^6}{70875}
    +\frac{y^3 \zeta_{3}}{525}+\frac{3 \zeta_{7}}{40 y}
    +\frac{25 \zeta_{9}}{8 y^3}-\frac{135 \zeta_{3} \zeta_{7}}{32 y^4}\\
    \EE_{2,2,2}\big|_{q^{0}\bar{q}^{0}}&=\frac{4 y^6}{9568125}
    {-}\frac{2 y^3 \zeta_{3}}{10125}{+}\frac{y \zeta_{5}}{54}
    {+}\frac{\zeta_{3}^2}{90}{+}\frac{661 \zeta_{7}}{1800 y}
    {-}\frac{5 \zeta_{3} \zeta_{5}}{12 y^2}
    {+}\frac{\zeta_{3}^3}{6 y^3}\,,\!\!
  \end{align}
\end{subequations}
}
where $y=\pi\tau_{2}$, and the Laurent polynomial of $\EE_{k}$ can be read off
from its well-known $q$-expansion, given e.g.\ in
\cite{fleig2018}.

The cusp forms listed in \eqref{eq:398} and \eqref{eq:399} were all of the form
$\mathcal{C}_{\Gamma}-\overline{\mathcal{C}_{\Gamma}}$ and hence purely
imaginary. Using the Laurent polynomials \eqref{eq:408}, and the basis elements
given in Table~\ref{tab:Ebasis}, it is easy to show that there are no real cusp
forms in the space of \acp{MGF} at weight $(a,a)$ with $a\leq5$ and that there
are five real cusp forms at weight $(6,6)$. A basis in this space of real cusp
forms is given by
\begin{subequations}
  \label{eq:8}
  \begin{align}
    \realCusp_{1}&=\frac{8}{15}\EE_{3,3}-4\EE'_{3,3}-\frac{1}{3} \EE_{2}^3
    +\EE_{4} \EE_{2}+\frac{349}{875}\EE_{3}^2+\frac{2}{45}\zeta_3^2
    +\frac{1}{3}\tau_2^{-2}\EE_{2}\nabladgb\EE_{2}\nabladg\EE_{2}\nonumber\\
    &\qquad -\frac{233}{1750}\tau_{2}^{-2}\nabladgb\EE_{3}\nabladg\EE_{3}
    +\frac{1}{10500}\tau_{2}^{-4}\nabladgb^2\EE_{3} \nabladg^2\EE_{3}
    -\frac{1}{6}\tau_{2}^{-2}\left(\nabladg\EE_{2}\nabladgb\EE_{4}
      {+}\nabladgb\EE_{2}\nabladg\EE_{4}\right)\displaybreak[0]\\[0.5em]
    \realCusp_{2}&=\EE_{2,4}+\frac{8748}{175}\EE_{3,3}-\frac{5622}{35}\EE'_{3,3}
    -\frac{269}{50}\EE_{3}^2
    -\frac{3739}{2100}\tau_{2}^{-2}\nabladgb\EE_{3}\nabladg\EE_{3}
    +\frac{1}{840}\tau_{2}^{-4}\nabladgb^2\EE_{3}\nabladg^2\EE_{3}\nonumber\\
    &\qquad+\frac{9}{8}\tau_{2}^{-2}\left(\nabladg\EE_{2}\nabladgb\EE_{4}
      {+}\nabladgb\EE_{2}\nabladg\EE_{4}\right)\displaybreak[0]\\[0.5em]
    \realCusp_{3}&=\EE_{2,2,2}+\frac{5288}{1125}\EE_{3,3}
    -\frac{2644}{75}\EE'_{3,3}+\EE_{2}\EE_{2,2}-\frac{1}{6}\EE_{2}^3
    +\frac{401}{17500}\EE_{3}^2
    +\frac{1}{4}\tau_{2}^{-2}\EE_{2} \nabladgb\EE_{2} \nabladg\EE_{2}\nonumber\\
    &\qquad-\frac{11801}{39375}\tau_{2}^{-2}\nabladgb\EE_{3} \nabladg\EE_{3}
    +\frac{127}{630000}\tau_{2}^{-4}\nabladgb^2\EE_{3}\nabladg^2\EE_{3}
    \displaybreak[0]\\[0.5em]
    \realCusp_{4}&=-2 \EE_{2} \EE_{2,2}-\EE_{2}^3+\frac{8757}{1250}\EE_{3}^2
    +\frac{1}{5}\zeta_3^2
    +\frac{3}{2}\tau_{2}^{-2}\EE_{2}\nabladgb\EE_{2}\nabladg\EE_{2}\nonumber\\
    &\qquad +\tau_{2}^{-2}\left(\nabladg\EE_{2}\nabladgb\EE_{2,2}
      {+}\nabladgb\EE_{2} \nabladg\EE_{2,2}\right)
    {-}\frac{3283}{1875}\tau_{2}^{-2}\nabladgb\EE_{3}\nabladg\EE_{3}
    {-}\frac{7}{15000}\tau_{2}^{-4}\nabladgb^2\EE_{3}\nabladg^2\EE_{3}
    \displaybreak[0]\\[0.5em]
    \realCusp_{5}&=-\frac{9}{5} \EE_{2} \EE_{2,2}-\frac{311}{350} \EE_{2}^3
    +\frac{26187}{12500}\EE_{3}^2+\EE_{3}\zeta_3  +\frac{311}{2625}\zeta_3^2
    +\frac{307}{700}\tau_{2}^{-2}\EE_{2} \nabladgb\EE_{2} \nabladg\EE_{2}\nonumber\\
    &\qquad-\frac{1638}{3125}\tau_{2}^{-2}\nabladgb\EE_{3}\nabladg\EE_{3}
    +\frac{21}{50000}\tau_{2}^{-4}\nabladgb^2\EE_{3}\nabladg^2\EE_{3}\,.
  \end{align}
\end{subequations}

The complex irreducible basis elements follow the same notation regarding the
sectors of holomorphic Eisenstein series. They are defined in terms of lattice
sums by
\begin{subequations}
  \label{eq:404}
  \begin{align}
    \BB_{2,3}&=
    \imtaupi^{5}\Big(\aform{0&1&2&2\\1&1&0&3}+\cform{3&0\\1&0}\cform{2&0\\4&0}
    -\cform{1&0\\3&0}\cform{4&0\\2&0}\Big)\displaybreak[0]\\[1ex]
    \BB'_{2,3}&=\imtaupi^{5}\Big(\frac{1}{2}\aform{0&2&3\\3&0&2}
    {+}\aform{0&1&2&2\\1&1&0&3}{+}\cform{3&0\\1&0}\cform{2&0\\4&0}
    {-}\cform{1&0\\3&0}\cform{4&0\\2&0}\Big)\nonumber\\
    &\qquad+\frac{129}{20}\EE_{5}-\frac{1}{2}\EE_{2}\zeta_{3}
    -\frac{21}{4}C_{1,1,3}\displaybreak[0]\\[1ex]
    \begin{split}
      \BB_{2,4}&=\imtaupi^{6}\Big(\aform{0&2&4\\5&0&1}
      +2\big(\cform{3&0\\1&0}\cform{3&0\\5&0}
      -\cform{1&0\\3&0}\cform{5&0\\3&0}\big)\Big)\\
      &\qquad+C_{1,1,4}+\frac{1}{3}C_{1,2,3}+\frac{1}{9}C_{2,2,2}
      -\EE_{2}\EE_{4}-\frac{13}{9}\EE_{6}
    \end{split}\displaybreak[0]\\[1ex]
    \BB'_{2,4}&=\imtaupi^{6}\aform{0&2&2&2\\3&0&1&2}-30C_{1,1,4}-10C_{1,2,3}
    -\frac{10}{3}C_{2,2,2}
    -3\EE_{3}\zeta_{3}+\frac{130}{3}\EE_{6}
    \displaybreak[0]\\[1ex]
      \BB_{2,2,2}&=\imtaupi^{6}\Big(4\aform{0&1&2&3\\2&1&3&0}
      {+}\frac{121}{50}\aform{0&2&2&2\\3&0&1&2}
      {-}\frac{113}{5}\aform{0&2&4\\5&0&1}
      {+}\frac{266}{5}\big(\cform{1&0\\3&0}\cform{5&0\\3&0}
      {-}\cform{3&0\\1&0}\cform{3&0\\5&0}\big)\nonumber\\
      &\qquad\quad{+}4\big(\cform{3&0\\1&0}\cform{1&1&1\\1&1&3}
      {-}\cform{1&0\\3&0}\cform{1&1&3\\1&1&1}\big)\Big)
      +6C_{1,1,2}\EE_{2}-\frac{27}{5}\EE_{2}\EE_{4}
      -\frac{63}{50}\EE_{3}\zeta_{3}\,,
  \end{align}
\end{subequations}
where the real modular graph functions $C_{a,b,c}$ are defined in
\eqref{eq:135}. The complex basis elements $\BB_{2,3}$ and $\BB_{2,3}'$ of the
$a+b=10$ sector were first mentioned in \cite{gerken2020c}. Only the first of
the basis elements in \eqref{eq:404} is purely imaginary, the others contain
imaginary and real contributions. The complex conjugates of the basis \acp{MGF}
in \eqref{eq:404} are
{\allowdisplaybreaks
\begin{subequations}
  \label{eq:405}
  \begin{align}
    \overline{\BB_{2,3}}&=-\BB_{2,3}\\
    \overline{\BB'_{2,3}}&=-\BB'_{2,3}-\EE_{2}\zeta_{3}-\frac{21}{2}\EE_{2,3}\\
    \overline{\BB_{2,4}}
    &=-\BB_{2,4}-2 \EE_{2}\EE_{4}+\frac{2}{9}\EE_{2,4}\\
    \overline{\BB'_{2,4}}&=-\BB'_{2,4}-6\EE_{3}\zeta_{3}-\frac{20}{3}\EE_{2,4}
    \label{eq:438}\\
    \overline{\BB_{2,2,2}}
    &=-\BB_{2,2,2}-\frac{63}{25}\EE_{3}\zeta_{3}+12\EE_{2}\EE_{2,2}\,.\label{eq:439}
  \end{align}
\end{subequations}}
The definition of the basis elements $\EE$ and $\BB$ was guided by the maxim to
delay the appearance of holomorphic Eisenstein series in the Cauchy--Riemann
equations to higher derivatives and to separate the different sectors of
holomorphic Eisenstein series at the same time. Although this does not fix the
basis elements uniquely, the remaining freedom allows one only to isolate one
purely imaginary basis element,  $\BB_{2,3}$. Similarly to
\eqref{eq:404}, the first Cauchy--Riemann derivatives of the complex basis
elements in which holomorphic Eisenstein series appear, are
{\allowdisplaybreaks
\begin{subequations}
  \label{eq:407}
  \begin{align}
    \nabladg^{2}\BB_{2,3}&=\frac{2}{7}\nabladg^{2}\BB'_{2,3}
    +\frac{3}{2}\big(\nabladg\EE_{2}\nabladg\EE_{3}-\EE_{2}\nabladg^{2}\EE_{3}
    +\nabladg^{2}\EE_{2,3}\big)
    +\frac{\tau_{2}^{4}}{\pi^{2}}\GG_{4}\big(9\EE_{3}+3\zeta_{3}\big)\\
    \nabladg^{4}\BB'_{2,3}&=
    1260\frac{\tau_{2}^{6}}{\pi^{3}}\GG_{6}\nabladg\EE_{2}\\
    \nabladg^{4}\BB_{2,4}&=-\frac{7}{90}\nabladg^{4}\BB'_{2,4}
    -1680\frac{\tau_{2}^{8}}{\pi^{4}}\GG_{8}\EE_{2}\label{eq:410}\\
    \nabladg^{5}\BB'_{2,4}&=
    151200\frac{\tau_{2}^{8}}{\pi^{4}}\GG_{8}\nabladg\EE_{2}\\
    \nabladg^{3}\BB_{2,2,2}&=-9(\nabladg\EE_{2})^{3}
    -\frac{\tau_{2}^{4}}{\pi^{2}}\GG_{4}\big(72\EE_{2}\nabladg\EE_{2}
    +36\nabladg\EE_{2,2}\big)\,.
  \end{align}
\end{subequations}}
Since the complex basis elements are given in \eqref{eq:404} in terms of real
basis elements, for which the Laurent polynomials are listed in \eqref{eq:408},
and cusp forms with vanishing Laurent polynomials, we can assemble the Laurent
polynomials of the $\BB$ as well. They are given by
{\allowdisplaybreaks
\begin{subequations}
  \label{eq:409}
  \begin{align}
    \BB_{2,3}\big|_{q^{0}\bar{q}^{0}}&=0\\
    \BB'_{2,3}\big|_{q^{0}\bar{q}^{0}}&=\frac{y^5}{14175}
    -\frac{y^2 \zeta_{3}}{45}+\frac{7 \zeta_{5}}{240}-\frac{\zeta_{3}^2}{2 y}
    -\frac{147 \zeta_{7}}{64 y^2}+\frac{21 \zeta_{3} \zeta_{5}}{8 y^3}\\
    \BB_{2,4}\big|_{q^{0}\bar{q}^{0}}&=-\frac{4 y^6}{637875}
    -\frac{\zeta_{7}}{180 y}+\frac{25 \zeta_{9}}{72 y^3}
    -\frac{35 \zeta_{3} \zeta_{7}}{32 y^4}\\
    \BB'_{2,4}\big|_{q^{0}\bar{q}^{0}}&=\frac{2 y^6}{42525}
    {-}\frac{4y^3 \zeta_{3}}{315}{-}\frac{\zeta_{7}}{4 y}
    {-}\frac{9 \zeta_{3} \zeta_{5}}{4 y^2}{-}\frac{125 \zeta_{9}}{12 y^3}
    {+}\frac{225 \zeta_{3} \zeta_{7}}{16 y^4}\\
    \BB_{2,2,2}\big|_{q^{0}\bar{q}^{0}}&=-\frac{y^6}{151875}
    +\frac{y \zeta_{5}}{18}+\frac{\zeta_{3}^2}{10}
    +\frac{311 \zeta_{3} \zeta_{5}}{200 y^2}-\frac{3 \zeta_{3}^3}{2 y^3}\,.
  \end{align}
\end{subequations}}

The basis elements $\EE$ and $\BB$ span the irreducible sectors of the modular
invariant subspaces of \acp{MGF}. For the subspaces with modular weight $(a,b)$
with $a>b$, we take the Cauchy--Riemann derivatives of the $\EE$ and $\BB$ as
irreducible basis elements. Since the space of \acp{MGF} of weight $(a+k,a-k)$
shrinks with growing $k$, there are relations between the Cauchy--Riemann
derivatives of the $\EE$ and $\BB$, leading to dropouts in this pattern. In
general, these dropouts are manifest in the Cauchy--Riemann equations
\eqref{eq:403} and \eqref{eq:407}, however some of the real basis elements
satisfy relations at derivatives lower than the one in which the first
holomorphic Eisenstein series appear as stated in \eqref{eq:403}. These
additional relations are
\begin{subequations}
  \label{eq:406}
  \begin{align}
    \nabladg^{2}\EE_{2,2}&=-\frac{1}{2}(\nabladg\EE_{2})^{2}\\
    \nabladg^{2}\EE_{3,3}&=\frac{3}{4}(\nabladg\EE_{3})^{2}
    +\frac{15}{2}\nabladg^{2}\EE'_{3,3}\\
    \nabladg^{2}\EE_{2,2,2}&=-2\nabladg\EE_{2}\nabladg\EE_{2,2}\,.
  \end{align}
\end{subequations}
For the complex basis elements, there are no relations at lower derivatives than
in \eqref{eq:407}.

On top of the irreducible basis elements $\EE$ and $\BB$, there are reducible
basis elements which are products of irreducible basis elements of lower
weights. We also take derivatives of these reducible basis elements to generate
the bases of weight $(a,b)$ with $a>b$. Again, this is constrained by the
relations \eqref{eq:403}, \eqref{eq:407} and \eqref{eq:406}. As for the
irreducible basis elements, the Cauchy--Riemann derivatives of the reducible
basis elements also contain terms with holomorphic Eisenstein series, which are
not written in the basis. Furthermore, the derivative of terms of the form
$\nabladgb^{n}\EE_{k}$ is (up to prefactors) $\nabladgb^{n-1}\EE_{k}$. The
derivative of the only depth-two instance $\nabladgb\EE_{2,2}$ gives rise to
$2\EE_{2,2}-\EE_{2}^{2}$, as follows from the Laplace equation
$(\Delta-2)\EE_{2,2}=-\EE_{2}^{2}$~\cite{dhoker2015}.

Since the action of the derivative operators $\nabladg$ and $\nabladgb$ on $y$
is straightforwardly given by
\begin{align}
  \nabladg y=\nabladgb y = \frac{y^{2}}{\pi}\,,
  \label{eq:411}
\end{align}
using the decompositions into the basis of Table~\ref{tab:Ebasis} and the known
Laurent polynomials \eqref{eq:408} and \eqref{eq:409}, we can easily assemble
the Laurent polynomials of all dihedral and trihedral \acp{MGF} of total weight
$a+b\leq 12$. These computations are made straightforward in the \mmapackage
package as outlined in the following.

Computations in the \mmapackage package are performed in the basis listed in
Table~\ref{tab:Cbasis}. Using the function \mma{CConvertToNablaE}, an expression
can be converted into the basis given in Table~\ref{tab:Ebasis}. The real basis
elements are represented by e.g.\ \mma{e[2,2]}, and \mma{ep[3,3]} for the primed
version. The complex basis elements are given by e.g.\ \mma{b[2,3]} and
\mma{bp[2,3]}. The Cauchy--Riemann derivatives are denoted by the functions
\mma{nablaE}, \mma{nablaEp}, \mma{nablaB} and \mma{nablaBp}. Their complex
conjugates are \mma{nablaBarE}, \mma{nablaBarEp}, \mma{nablaBarBBar} and
\mma{nablaBarBpBar}. The first arguments of these functions is always the order
of the derivative, the second is a list with the subscripts of the basis
element, e.g.\
$\nabladgb\hspace{-0.3ex}\rule[1.7ex]{0pt}{0pt}^{2}\overline{\BB}_{2,4}$ is
denoted by \mma{nablaBarBBar[2,\{2,4\}]}. These basis elements are translated
back into the basis given in Table~\ref{tab:Cbasis} by the function
\mma{CConvertFromNablaE}. Note that only the derivatives appearing in
Table~\ref{tab:Ebasis} can be converted in this way. As an example, the
decomposition of the graph $\cform{1&2&4\\2&2&1}$ can be performed by
\begin{mmaCell}{Input}
  CConvertToNablaE\mmabigso{}CSimplify\mmabigso{}\mmaincform{1 2 4\\2 2 1}\mmabigsc{}\mmabigsc{}
\end{mmaCell}
\begin{mmaCell}[morelst={label=mma:8}]{Output}
  \mmaFrac{3\,\mmaSup{\(\pi\)}{6}\,\(\nabla\)\mmaSub{E}{6}}{28\,\mmaSubSup{\(\tau\)}{2}{7}}\mmam\mmaFrac{5\,\mmaSup{\(\pi\)}{6}\,\(\nabla\)\mmaSub{E}{3,3}}{9\,\mmaSubSup{\(\tau\)}{2}{7}}+\mmaFrac{5\,\mmaSup{\(\pi\)}{6}\,\(\nabla\)\mmaSubSup{E}{3,3}{\(\prime\)}}{3\,\mmaSubSup{\(\tau\)}{2}{7}} \textnormal{.}
\end{mmaCell}

The derivative operator $\nabladg$ is not implemented directly, but since it is
given by $\nabladg=\tau_{2}\nabla^{(0)}$ (cf.\ \eqref{eq:76}), it can be
obtained by acting with \mma{tau[2]CHolCR} on an \ac{MGF} with vanishing modular
weight. E.g.\ the Cauchy--Riemann equation \eqref{eq:410} is reproduced by
\begin{mmaCell}{Input}
  CConvertToNablaE\mmabigso{}Nest\mmabigso{}CSimplify\mmabigso{}tau[2]CHolCR[#]\mmabigsc{}&,b[2,4],4\mmabigsc{}\mmabigsc{}
\end{mmaCell}
\begin{mmaCell}{Output}
  \mmam\mmaFrac{7}{90}\,\mmaSup{\(\nabla\)}{4}\mmaSubSup{B}{2,4}{\(\prime\)}\mmam\mmaFrac{1680\,\mmaSub{E}{2}\,\mmaSub{G}{8}\,\mmaSubSup{\(\tau\)}{2}{8}}{\mmaSup{\(\pi\)}{4}} \textnormal{.}
\end{mmaCell}

The Laurent polynomials \eqref{eq:408} and \eqref{eq:409} are implemented in the
function \mma{CLaurentPoly}, which replaces each of the basis elements by its
Laurent polynomial and performs the necessary Cauchy--Riemann derivatives. E.g.\
the Laurent polynomial of the graph $\cform{1&2&4\\2&2&1}$ decomposed in
\mmaoutref{mma:8} can be obtained via \renewcommand{\temprefa}{\ref*{mma:8}}
\begin{mmaCell}{Input}
  CLaurentPoly[Out[\temprefa]]
\end{mmaCell}
\begin{mmaCell}{Output}
  \mmam\mmaFrac{19\,\mmaSup{\(\pi\)}{12}}{91216125}+\mmaFrac{5\,\mmaSup{\(\pi\)}{12}\,\mmaSubSup{\(\zeta\)}{5}{2}}{16\,\mmaSup{y}{10}}+\mmaFrac{\mmaSup{\(\pi\)}{12}\,\mmaSub{\(\zeta\)}{7}}{288\,\mmaSup{y}{7}}\mmam\mmaFrac{7\,\mmaSup{\(\pi\)}{12}\,\mmaSub{\(\zeta\)}{9}}{64\,\mmaSup{y}{9}}\mmam\mmaFrac{135\,\mmaSup{\(\pi\)}{12}\,\mmaSub{\(\zeta\)}{11}}{512\,\mmaSup{y}{11}} \textnormal{.}
\end{mmaCell}

The basis elements at a certain weight are accessible via the function
\mma{CBasis}. If the option \mma{basis} is set to the string \mmastr{"C"} (the
default value), the basis from Table~\ref{tab:Cbasis} is returned, if it is set
to the string \mmastr{"nablaE"}, the basis from Table~\ref{tab:Ebasis} is
returned, e.g.
\begin{mmaCell}{Input}
  CBasis[3,\,5]\\
  CBasis[3,\,5,\,basis\mmainrarrow"nablaE"]
\end{mmaCell}
\begin{mmaCell}{Output}
  \Big\{\mmaoutcform{1 1 1\\1 1 3},\,\mmaoutcform{3 0\\5 0},\,\mmaFrac{\mmaSup{\(\pi\)}{2}\,\mmaoutcform{1 0\\3 0}\,\mmaSub{E}{2}}{\mmaSubSup{\(\tau\)}{2}{2}}\Big\}
\end{mmaCell}
\begin{mmaCell}{Output}
  \big\{\mmaOver{\(\nabla\)}{\(\mmaoverline\)}\mmaSub{E}{2,2},\mmaOver{\(\nabla\)}{\(\mmaoverline\)}\mmaSub{E}{4},\mmaSub{E}{2}\,\mmaOver{\(\nabla\)}{\(\mmaoverline\)}\mmaSub{E}{2}\big\} \textnormal{.}
\end{mmaCell}

Together with the function \mma{zIntegrate} described in
Section~\ref{sec:exp-KN-ints}, the basis decompositions available in the
\mmapackage package are sufficient to expand all two- and three-point
Koba--Nielsen integrals to the orders which give rise to \acp{MGF} of total
modular weight at most $12$. This was crucial for checking and solving the
differential equation of the generating function of Koba--Nielsen integrals in
\cite{gerken2019d, gerken2020c}. The arXiv submission of this paper includes the
expansion of the two- and three point versions of the generating function
$Y^{\tau}_{\vec{\eta}}$ defined in \cite{gerken2020c} up to order $12$. For the
three-point version, it also contains the Laurent polynomial of the generating
series. At two-point, it was checked that the
Laurent polynomials obtained using the basis decompositions agree with the
closed formula given in \cite{gerken2020c} from genus-zero integrals.


\section{Conclusion and outlook}
\label{sec:conclusion}

In this paper, we systematically studied relations between modular graph forms,
a class of non-holomorphic modular forms used in the computation of the
low-energy expansion of closed-string genus-one amplitudes in type-II, heterotic
or bosonic theories.

We studied \acp{MGF} with two, three and four vertices and introduced in
particular a concise notation for four-point graphs and studied their symmetry
properties systematically. For these graphs, we reviewed how topological
simplifications, momentum-conservation at the vertices, factorization of
$\sbmatrix{0\\0}$-edges and Cauchy--Riemann derivatives lead to relations
between \acp{MGF} and discussed how these can also be understood in the integral
representation of \acp{MGF}. This point of view led us to a new formulation of
holomorphic subgraph reduction which can be understood as integrated Fay
identities of Kronecker--Eisenstein series. This formulation yields an efficient
iterative procedure for higher-point \ac{HSR}, circumventing difficulties in
earlier approaches.

Since divergent \acp{MGF} appear naturally in the expansion of Koba--Nielsen
integrals and in momentum-conservation identities, we initiated a systematic
study of these divergent sums, starting with an analysis of the superficial
degree of divergence for \acp{MGF} with up to four points. We discussed
holomorphic subgraph reduction in divergent \acp{MGF} and Cauchy--Riemann
derivatives as well as momentum conservation and factorization of divergent
graphs.

By constructing all momentum conservation- and Fay identities at the
corresponding weight and applying the techniques described above, we could find
basis decompositions for all (convergent) two- and three-point \acp{MGF} of
total modular weight $a+b\leq12$. The only additional input in this process were
the two well-known identities for $D_{3}$ and $D_{5}$, which are also the source
of the zeta values in the basis decompositions.

We then discussed a particular basis for \acp{MGF} systematically built out of
real and complex basis elements and their derivatives. Since the Laurent
polynomials of these basis elements are known from the literature, we can
compute the Laurent polynomials of all decomposed \acp{MGF}. This allowed us to
identify five linearly independent real cusp forms at weight $(6,6)$ and to show
that no real cusp forms exist in the space of \acp{MGF} at lower weights.

The basis decompositions, as well as implementations of the manipulations
discussed above, are made available in the ancillary files of the arXiv
submission of this paper in the form of the \texttt{Mathematica} package
\mmapackage together with two text files containing the decompositions. Using
this package, we decomposed the generating function for Koba--Nielsen integrals
introduced in \cite{gerken2020c} at two- and three points up to order $12$. The
resulting expansion is also included in the arXiv submissions.

Interestingly, the basis of \acp{MGF} obtained in this work only contains
dihedral graphs. From an argument involving iterated Eisenstein integrals given
in \cite{gerken2020c}, we know that the basis is nevertheless complete and hence
also all higher-point graphs beyond the trihedral ones considered here can be
decomposed into only dihedral graphs at weight $a+b\leq12$. It would be
interesting to see at which weight more complicated topologies have to be
included. A first step in this direction would be an extension of the
\mmapackage package to a complete treatment of four-point graphs, which would
not only allow one to explicitly find decompositions of four-point graphs, but
presumably also to perform the basis decompositions at higher weights.


\section*{Acknowledgments}
I would like to thank Axel Kleinschmidt and Oliver Schlotterer for numerous
enlightening discussions during all stages of this project and for carefully
reading the manuscript. I would also like to thank them for ongoing
collaborations on related projects that initiated this work. I am supported by
the International Max Planck Research School for Mathematical and Physical
Aspects of Gravitation, Cosmology and Quantum Field Theory.

\appendix

\section{Complete reference for the Modular Graph Forms package}
\label{cha:mma-reference}

In this appendix, we give a complete reference of all symbols defined in the
\mmapackage package, all functions and their options and detailed instructions
how to load the package. In Section~\ref{sec:mma-example}, we show how the
integrals appearing in the four-gluon amplitude of the heterotic string
discussed in \cite{gerken2019d} can be computed using the \mmapackage
package.

Within \texttt{Mathematica}, short descriptions of the various symbols,
functions and options can be displayed using the \mma{Information} function,
e.g.\ by running \mma{?g}. A list of all the symbols defined in the package is
printed by running \mma{?ModularGraphForms\`{}*}. The options and default values
for a function are accessible via the \mma{Options} function, e.g.
\begin{mmaCell}{Input}
  Options[CBasis]
\end{mmaCell}
\begin{mmaCell}{Output}
  \{basis\mmaoutrarrow{}C\} \textnormal{.}
\end{mmaCell}

\subsection{Files and loading the package}
The \texttt{Mathematica} package \mmapackage includes the three files
\texttt{ModularGraphForms.m}, \texttt{DiIds.txt} and \texttt{TriIds.txt}. The
first one provides the package itself, whereas the two text files contain the
basis decompositions described in Section~\ref{sec:conjectured-basis} for
dihedral and trihedral graphs, respectively. The package loads the latter files
automatically and expects them in the same directory, in which also the
\texttt{ModularGraphForms.m} file is saved. However, the text files can also be
imported into \texttt{Mathematica} using the \mma{Get} function and can be used
independently of the \mmapackage package.

To load the package, call the \mma{Get} function on the
\texttt{ModularGraphForms.m} file. Either the full path can be provided,
\begin{mmaCell}{Input}
  Get["/home/user/ModularGraphForms.m"]
\end{mmaCell}
or, if the files are placed in one of the directories in \texttt{Mathematica}'s
search path, it is sufficient to run
\begin{mmaCell}{Input}
  Get["ModularGraphForms.m"] \textnormal{.}
\end{mmaCell}
A list of the directories in \texttt{Mathematica}'s search path is available in
the global variable \mma{\$Path} and includes the current directory, which by
default is the directory in which the current Notebook is saved.

\subsection{Symbols}
The \mmapackage package defines a number of symbols used for the various objects
in this paper. For most of these symbols, a 2d-notation is implemented which
makes the output easier to read. E.g.\ $\tau_{2}$ is represented by
\mma{tau[2]}, but printed as
\begin{mmaCell}{Input}
  tau[2]
\end{mmaCell}
\begin{mmaCell}{Output}
  \mmaSub{\(\tau\)}{2} \textnormal{.}
\end{mmaCell}
These 2d-outputs can be copied to input cells and used for further
computations. The input form of the 2d-output can be accessed by the function
\mma{InputForm}, e.g.
\begin{mmaCell}{Input}
  InputForm[\mmaSub{\(\tau\)}{2}]
\end{mmaCell}
\begin{mmaCell}{Output}
  tau[2] \textnormal{.}
\end{mmaCell}

Using the \mma{\$Assumptions} variable, the \mmapackage package sets the global
assumption that $\tau_{2}>0$. This is helpful e.g.\ when simplifying equations.

\subsubsection{General symbols}
Five general symbols used by the \mmapackage package are
\begin{longtable}{ScSc}
  \toprule
  \texttt{Mathematica} symbol & description\\
  \midrule\endfirsthead
  \midrule
  \texttt{Mathematica} symbol & description\\
  \midrule\endhead
  \mma{tau} &  modular parameter $\tau$\\
  \mma{tauBar} & $\bar\tau$\\
  \mma{tau[2]} & $\tau_{2}=\Im\tau$\\
  \mma{y}& $y=\pi\tau_{2}$\\
  \mma{zeta[k]} & $\zeta_{k}$ as defined in \eqref{eq:28}\\
  \mma{bCoeff}
  & \makecell[t]{coefficient in the sieve algorithm,\\cf. \mma{CSieveDecomp}}\\
  \bottomrule
\end{longtable}

\subsubsection{Modular graph forms}
The conventions for two-, three- and four-point modular graph forms were
introduced in detail in Section~\ref{sec:graph-topol-not}. The symbols used to
represent \acp{MGF}, (non-)holomorphic Eisenstein series and real and complex
basis elements are
\begin{longtable}{ScSc}
  \toprule
  \texttt{Mathematica} symbol & description\\
  \midrule\endfirsthead
  \midrule
  \texttt{Mathematica} symbol & description\\
  \midrule\endhead
  \mma{c\mmabigso{}\dots\mmabigsc{}}
  & \ac{MGF}, cf.\ Section~\ref{sec:graph-topol-not}\\
  \mma{a\mmabigso{}\dots\mmabigsc{}} & $\aform{A\\B}$ as defined in \eqref{eq:433}\\
  \mma{intConst\mmabigso{}\dots\mmabigsc{}}&integration constant, cf. \mma{CSieveDecomp}\\
  \mma{intConstBar\mmabigso{}\dots\mmabigsc{}} & complex conjugate of \mma{intConst}\\
  \mma{g[k]} & $\GG_{k}$ as defined in \eqref{eq:45}\\
  \mma{gBar[k]} & $\overline{\GG}_{k}$\\
  \mma{gHat[2]} & $\GGhat_{2}$ as defined in \eqref{eq:316}\\
  \mma{gBarHat[2]} & $\GGhatb_{2}$\\
  \mma{e[k\textsubscript{1},\dots,k\textsubscript{r}]}
  & \makecell[t]{$\EE_{k}$ as defined in \eqref{eq:51}\\
    and $\EE_{k_{1},\dots,k_{r}}$ as defined in \eqref{eq:172}}\\
  \mma{ep[k\textsubscript{1},\dots,k\textsubscript{r}]}
  & $\EE'_{k_{1},\dots,k_{r}}$ as defined in \eqref{eq:172} \\
  \mma{b[k\textsubscript{1},\dots,k\textsubscript{r}]}
  & $\BB_{k_{1},\dots,k_{r}}$ as defined in \eqref{eq:404}\\
  \mma{bp[k\textsubscript{1},\dots,k\textsubscript{r}]}
  & $\BB'_{k_{1},\dots,k_{r}}$ as defined in \eqref{eq:404}\\
  \bottomrule
\end{longtable}
Note that \acp{MGF} are represented by the symbol \mma{c}, but are printed
with a capital \mmaInlineCell{Output}{C}. When copying this output into an input
cell, the capital \mma{C} should not be changed into a lowercase
\mma{c}. Furthermore, the basis elements listed here are meaningful only for the
indices defined in \eqref{eq:172} and \eqref{eq:404}.

The \texttt{Mathematica} symbols used to represent Cauchy--Riemann derivatives
of real and complex basis elements of \acp{MGF} are
\begin{longtable}{ScSc}
  \toprule
  \texttt{Mathematica} symbol & description\\
  \midrule\endfirsthead
  \midrule
  \texttt{Mathematica} symbol & description\\
  \midrule\endhead
  \mma{nablaE[n,\{k\textsubscript{1},\dots,k\textsubscript{r}\}]} 
  & $\nabladg^{n}\EE_{k_{1},\dots,k_{r}}$\\
  \mma{nablaBarE[n,\{k\textsubscript{1},\dots,k\textsubscript{r}\}]} 
  & $\nabladgb^{n}\EE_{k_{1},\dots,k_{r}}$\\
  \mma{nablaEp[n,\{k\textsubscript{1},\dots,k\textsubscript{r}\}]} 
  & $\nabladg^{n}\EE'_{k_{1},\dots,k_{r}}$\\
  \mma{nablaBarEp[n,\{k\textsubscript{1},\dots,k\textsubscript{r}\}]} 
  & $\nabladgb^{n}\EE'_{k_{1},\dots,k_{r}}$\\
  \mma{nablaB[n,\{k\textsubscript{1},\dots,k\textsubscript{r}\}]} 
  & $\nabladg^{n}\BB_{k_{1},\dots,k_{r}}$\\
  \mma{nablaBarBBar[n,\{k\textsubscript{1},\dots,k\textsubscript{r}\}]} 
  & $\nabladgb^{n}\overline{\BB_{k_{1},\dots,k_{r}}}$\\
  \mma{nablaBp[n,\{k\textsubscript{1},\dots,k\textsubscript{r}\}]} 
  & $\nabladg^{n}\BB'_{k_{1},\dots,k_{r}}$\\
  \mma{nablaBarBpBar[n,\{k\textsubscript{1},\dots,k\textsubscript{r}\}]} 
  & $\nabladgb^{n}\overline{\BB'_{k_{1},\dots,k_{r}}}$\\
  \bottomrule
\end{longtable}
The derivative operator $\nabladg$ and its complex conjugate are defined in
\eqref{eq:76}. The zeroth derivative returns the argument, e.g.\
\begin{mmaCell}{Input}
  nablaE[0,\,\{5\}]
\end{mmaCell}
\begin{mmaCell}{Output}
  \mmaSub{E}{5} \textnormal{.}
\end{mmaCell}

\subsubsection{Iterated Eisenstein integrals}
For compatibility with the data provided in the ancillary file of
\cite{gerken2020c}, the \mmapackage package defines the following symbols for
iterated Eisenstein integrals, although no manipulations of these objects can be
performed within this package.
\begin{longtable}{ScSc}
  \toprule
  \texttt{Mathematica} symbol & description\\
  \midrule\endfirsthead
  \midrule
  \texttt{Mathematica} symbol & description\\
  \midrule\endhead
  \mma{
    esv$\mmabigso{}\smatrix{\te{j\textsubscript{1} \dots{} j\textsubscript{$\ell$}}\\
      \te{k\textsubscript{1} \dots{} k\textsubscript{$\ell$}}}\mmabigsc{}$}
  & $\esvtau{j_{1}&\dotsm&j_{\ell}\\k_{1}&\dotsm&k_{\ell}}$\\
  \mma{esvS$\mmabigso{}
    \smatrix{\te{j\textsubscript{1} \dots{} j\textsubscript{$\ell$}}\\
      \te{k\textsubscript{1} \dots{} k\textsubscript{$\ell$}}}\mmabigsc{}$}
  & $\esvStau{j_{1}&\dotsm&j_{\ell}\\k_{1}&\dotsm&k_{\ell}}$\\
  \mma{esvBar$\mmabigso{}
    \smatrix{\te{j\textsubscript{1} \dots{} j\textsubscript{$\ell$}}\\
      \te{k\textsubscript{1} \dots{} k\textsubscript{$\ell$}}}\mmabigsc{}$}
  & $\overline{\esvtau{j_{1}&\dotsm&j_{\ell}\\k_{1}&\dotsm&k_{\ell}}}$\\
  \mma{betasv$\mmabigso{}
    \smatrix{\te{j\textsubscript{1} \dots{} j\textsubscript{$\ell$}}\\
      \te{k\textsubscript{1} \dots{} k\textsubscript{$\ell$}}}\mmabigsc{}$}
  & $\betasvtau{j_{1}&\dotsm&j_{\ell}\\k_{1}&\dotsm&k_{\ell}}$\\
  \mma{betasvS$\mmabigso{}
    \smatrix{\te{j\textsubscript{1} \dots{} j\textsubscript{$\ell$}}\\
      \te{k\textsubscript{1} \dots{} k\textsubscript{$\ell$}}}\mmabigsc{}$}
  & $\betasvStau{j_{1}&\dotsm&j_{\ell}\\k_{1}&\dotsm&k_{\ell}}$\\
  \mma{betasvBar$\mmabigso{}
    \smatrix{\te{j\textsubscript{1} \dots{} j\textsubscript{$\ell$}}\\
      \te{k\textsubscript{1} \dots{} k\textsubscript{$\ell$}}}\mmabigsc{}$}
  & $\overline{\betasvtau{j_{1}&\dotsm&j_{\ell}\\k_{1}&\dotsm&k_{\ell}}}$\\
  \bottomrule
\end{longtable}
\noindent As for \acp{MGF}, the matrices can be inserted in
\texttt{Mathematica} either as nested lists or as 2d input,
cf.~\mmainref{mma:13}. For the definitions of the iterated Eisenstein integrals,
see~\cite{gerken2020c}.

\subsubsection{Koba--Nielsen integrals}
\label{sec:mma-KN-ints}
For the evaluation and representation of Koba--Nielsen integrals and their
generating series, the following symbols are defined.
\begin{longtable}{ScSc}
  \toprule
  \texttt{Mathematica} symbol & description\\
  \midrule\endfirsthead
  \midrule
  \texttt{Mathematica} symbol & description\\
  \midrule\endhead
  \mma{eta[k\textsubscript{1},\dots,k\textsubscript{r}]}
  & $\eta_{k_{1},\dots,k_{r}}$ expansion variable as in \cite{gerken2020c}\\
  \mma{etaBar[k\textsubscript{1},\dots,k\textsubscript{r}]}
  & $\bar{\eta}_{k_{1},\dots,k_{r}}$\\
  \mma{s[k\textsubscript{1},\dots,k\textsubscript{r}]} &
  $s_{k_{1},\dots,k_{r}}$ as defined in \eqref{eq:15} and \eqref{eq:7}\\
  \mma{fz[a,\,i,\,j]} & $f^{(a)}_{ij}$ as defined in \eqref{eq:103}  \\
  \mma{fBarz[b,\,i,\,j]}
  & $\overline{f^{(b)}_{ij}}$ as defined in \eqref{eq:104}  \\
  \mma{gz[i,\,j]} & $G_{ij}$ as defined in \eqref{eq:85}  \\
  \mma{cz[a,\,b,\,i,\,j]} & $C^{(a,b)}_{ij}$ as defined in \eqref{eq:126}  \\
  \mma{vz[a,\{k\textsubscript{1},\dots,k\textsubscript{r}\}]}
  & $V_{a}(k_{1},\dots,k_{r})$ as defined in \eqref{eq:107}  \\
  \mma{vBarz[b,\{k\textsubscript{1},\dots,k\textsubscript{r}\}]}
  & $\overline{V_{b}(k_{1},\dots,k_{r})}$ \\
  \bottomrule
\end{longtable}
Symbols which represent functions which can appear in the integrand of a
Koba--Nielsen integral have the suffix \mma{z}.

\subsection{Functions}

The functions in the \mmapackage package are organized into three main
categories: Dihedral functions only manipulate dihedral \acp{MGF} and carry the
prefix \mma{Di}. Trihedral functions only manipulate trihedral \acp{MGF} and
carry the prefix \mma{Tri}. General functions act on \acp{MGF} of all supported
graph topologies or perform other tasks which are not specific to any graph
topology. They carry a prefix \mma{C}. On top of these, there is limited support
for four-point manipulations in the form of the function \mma{TetCSimplify} and
a function to expand Koba--Nielsen integrals in \acp{MGF}.

\subsubsection{General functions}

\subsubsection*{\mmaHeading{CBasis}}

The function \mma{CBasis} returns a list of basis elements for \acp{MGF}.

\begin{description}
\item[Arguments] \mma{CBasis} accepts two arguments, corresponding to the
  holomorphic and antiholomorphic modular weight of the basis.
  
\item[Return value] \mma{CBasis} returns the basis of \acp{MGF} at the modular
  weight passed as the arguments as listed in Tables~\ref{tab:Cbasis} and
  \ref{tab:Ebasis}, excluding the zeta values $\zeta_{3}$, $\zeta_{5}$ and
  $\zeta_{3}^{2}$. Note that at weight $(a{+}k,a{-}k)$, the basis elements in
  Table~\ref{tab:Cbasis} have weight $(a{+}k,a{-}k)$, whereas in
  Table~\ref{tab:Ebasis}, they have weight $(0,-2k)$.

\item[Options] If the option \mma{basis} is set to the string \mmastr{"C"} (the
  default) the basis from Table~\ref{tab:Cbasis} is returned, if the option
  \mma{basis} is set to the string \mmastr{"nablaE"}, the basis from
  Table~\ref{tab:Ebasis} is returned. No other values for \mma{basis} are
  admissible.

\item[Warnings]\ 
  \begin{itemize}
  \item If the sum of the holomorphic- and antiholomorphic modular weights
    passed in the arguments is odd, the warning
    \mmawarning{CBasis}{incorrModWeight} is issued and \mma{CBasis} returns an
    empty list.
  \item If the sum of the holomorphic- and antiholomorphic modular weights
    passed in the arguments is less than four, the warning
    \mmawarning{CBasis}{tooLowWeight} is issued and \mma{CBasis} returns an
    empty list.
  \item If the basis for the modular weight passed to \mma{CBasis} is not
    implemented, the warning \mmawarning{CBasis}{noBasis} is issued and
    \mma{CBasis} returns an empty list.
  \end{itemize}
\item[Examples]\
\begin{mmaCell}{Input}
  CBasis[3,\,7]
\end{mmaCell}
\begin{mmaCell}{Output}
  \big\{\mmaoutcform{1 1 1\\1 1 5},\,\mmaoutcform{3 0\\7 0},\,\mmaoutcform{1 0\\3 0}\,\mmaoutcform{2 0\\4 0},\,\mmaFrac{\mmaSup{\(\pi\)}{2}\,\mmaoutcform{1 0\\5 0}\,\mmaSub{E}{2}}{\mmaSubSup{\(\tau\)}{2}{2}},\,\mmaoutcform{0 1 2\\2 0 5}\big\}
\end{mmaCell}
\begin{mmaCell}{Input}
  CBasis\mmabigso{}3,\,7,\,basis\mmainrarrow"nablaE"\mmabigsc{}
\end{mmaCell}
\begin{mmaCell}{Output}
  \{\mmaSup{\mmaOver{\(\nabla\)}{\(\mmaoverline\)}}{2}\mmaSub{E}{2,3},\mmaSup{\mmaOver{\(\nabla\)}{\(\mmaoverline\)}}{2}\mmaSub{E}{5},\mmaOver{\(\nabla\)}{\(\mmaoverline\)}\mmaSub{E}{2} \mmaOver{\(\nabla\)}{\(\mmaoverline\)}\mmaSub{E}{3},\mmaSub{E}{2} \mmaSup{\mmaOver{\(\nabla\)}{\(\mmaoverline\)}}{2}\mmaSub{E}{3},\mmaSup{\mmaOver{\(\nabla\)}{\(\mmaoverline\)}}{2}\mmaSubSup{\mmaOver{B}{\(\mmaoverline\)}}{2,3}{\(\prime\)}\}
\end{mmaCell}  
\end{description}

\subsubsection*{\mmaHeading{CCheckConv}}

The function \mma{CCheckConv} tests if \acp{MGF} are convergent or divergent.

\begin{description}
\item[Argument] \mma{CCheckConv} accepts one argument which is an arbitrary
  expression, possibly containing \acp{MGF} of any topology and Eisenstein
  series.

\item[Return value] \mma{CCheckConv} returns \mma{True} or \mma{False}. If the
  argument contains an \ac{MGF} which is divergent according to the conditions
  discussed in Section~\ref{sec:div-cond} or a $\EE_{k}$, $\GG_{k}$ or
  $\overline{\GG}_{k}$ with $k<2$, the function returns \mma{False}, otherwise
  it returns \mma{True}.

\item[Examples]\ 
\begin{mmaCell}{Input}
  CCheckConv\mmabigso{}e[1]\,\mmaincform{2 0\\3 0}\mmabigsc{}
\end{mmaCell}
\begin{mmaCell}{Output}
  False
\end{mmaCell}
Since $\EE_{1}$ is divergent, the return value is \mma{False}, even though
$\cform{2&0\\3&0}=0$.

\begin{mmaCell}{Input}
  CCheckConv\mmabigso{}\mmaincformkite{\verbcent{1} 2\\-2 2}{1 2\\1 2}{\verbcent{1} 2\\-2 2}{1 2\\1 2}{2 2\\2 2}\mmabigsc{}
\end{mmaCell}
\begin{mmaCell}{Output}
  False
\end{mmaCell}
Since the last condition in \eqref{eq:353} is violated, the return value is
\mma{False}.

\begin{mmaCell}{Input}
  CCheckConv\mmabigso{}\mmaincformkite{\verbcent{1} 2\\-2 2}{1 2\\1 2}{\verbcent{1} 2\\-2 2}{2 3\\2 1}{2 2\\2 2}\mmabigsc{}
\end{mmaCell}
\begin{mmaCell}{Output}
  True
\end{mmaCell}
Since here $\check{c}_{4}$ as defined below \eqref{eq:349} is increased, the
last condition in \eqref{eq:353} is also satisfied and the return value is
\mma{True}.
\end{description}

\subsubsection*{\mmaHeading{CComplexConj}}

The function \mma{CComplexConj} computes the complex conjugate of an expression.

\begin{description}
\item[Argument] \mma{CComplexConj} accepts one arbitrary argument.
  
\item[Return value] \mma{CComplexConj} returns its argument with all \acp{MGF}
  complex conjugated and written in their canonical representation. This includes
  Eisenstein series, complex basis elements (according to \eqref{eq:405})
  Cauchy--Riemann derivatives of basis elements and integration constants, unless
  the \ac{MGF} in the argument is real.

\item[Example]\ 
\begin{mmaCell}{Input}
  CComplexConj\mmabigso{}\mmabigco{}g[4],\,b[2,\,4],\,\mmainintconst{1 2 1\\1 1 4},\,nablaB[1,\,\{2,\,4\}]\mmabigcc{}\mmabigsc{}
\end{mmaCell}
\begin{mmaCell}{Output}
  \big\{\mmaSub{\mmaOver{G}{\(\mmaoverline\)}}{4},\,\mmam\mmaSub{B}{2,4}\mmam2\,\mmaSub{E}{2}\,\mmaSub{E}{4}+\mmaFrac{2\,\mmaSub{E}{2,4}}{9},\,$\overline{\te{\mmainintconst{1 2 1\\1 1 4}}}$,\,$\overline{\nabla}$\mmaSub{\mmaOver{B}{\(\mmaoverline\)}}{2,4}\big\}
\end{mmaCell}
\end{description}

\subsubsection*{\mmaHeading{CConvertToNablaE} and \mmaHeading{CConvertFromNablaE}}

The functions \mma{CConvertToNablaE} and \mma{CConvertFromNablaE} convert an
expression between the bases given in Tables~\ref{tab:Cbasis} and
\ref{tab:Ebasis}.

\begin{description}
\item[Argument] Both \mma{CConvertToNablaE} and \mma{CConvertFromNablaE} accept
  one arbitrary argument.
  
\item[Return value] \mma{CConvertToNablaE} replaces all of the basis elements in
  Table~\ref{tab:Cbasis} in its argument with their expansions in the basis of
  Table~\ref{tab:Ebasis}. \mma{CConvertFromNablaE} replaces all of the basis
  elements in Table~\ref{tab:Ebasis} in its argument with their expansions in
  the basis of Table~\ref{tab:Cbasis}. On top of the elements listed explicitly
  in these tables, $\nabladg^{n}\EE_{k}$ and $\cform{k+n&0\\k-n&0}$ are
  rewritten according to \eqref{eq:328} for any $n$ and $k$. The results are not
  manipulated any further and \acp{MGF} in the argument which are not in the
  basis to be converted are left untouched.

\item[Examples]\
\begin{mmaCell}{Input}
  CConvertToNablaE\mmabigso{}\mmaincform{1 1 4\\1 1 2}\mmabigsc{}
\end{mmaCell}
\begin{mmaCell}[morelst={label=mma:11}]{Output}
  \mmaFrac{\mmaSup{\(\pi\)}{5}\,\(\nabla\)\mmaSub{B}{2,3}}{18\,\mmaSubSup{\(\tau\)}{2}{6}}\mmam\mmaFrac{\mmaSup{\(\pi\)}{5}\,\(\nabla\)\mmaSubSup{B}{2,3}{\(\prime\)}}{18\,\mmaSubSup{\(\tau\)}{2}{6}}\mmam\mmaFrac{\mmaSup{\(\pi\)}{5}\,\mmaSub{E}{3}\,\(\nabla\)\mmaSub{E}{2}}{12\,\mmaSubSup{\(\tau\)}{2}{6}}+\mmaFrac{\mmaSup{\(\pi\)}{5}\,\mmaSub{E}{2}\,\(\nabla\)\mmaSub{E}{3}}{12\,\mmaSubSup{\(\tau\)}{2}{6}}+\mmaFrac{41\,\mmaSup{\(\pi\)}{5}\,\(\nabla\)\mmaSub{E}{5}}{140\,\mmaSubSup{\(\tau\)}{2}{6}}\\
  \quad+\mmaFrac{\mmaSup{\(\pi\)}{5}\,\(\nabla\)\mmaSub{E}{2,3}}{24\,\mmaSubSup{\(\tau\)}{2}{6}}\mmam\mmaFrac{\mmaSup{\(\pi\)}{5}\,\(\nabla\)\mmaSub{E}{2}\,\mmaSub{\(\zeta\)}{3}}{36\,\mmaSubSup{\(\tau\)}{2}{6}}
\end{mmaCell}
\begin{mmaCell}{Input}
  CConvertToNablaE\mmabigso{}\mmaincform{1 2 3\\1 1 2}\mmabigsc{}
\end{mmaCell}
\begin{mmaCell}{Output}
  \mmaoutcform{1 2 3\\1 1 2}
\end{mmaCell}
\renewcommand{\temprefa}{\ref*{mma:11}}
\begin{mmaCell}{Input}
  CConvertFromNablaE[Out[\temprefa]]
\end{mmaCell}
\begin{mmaCell}{Output}
  \mmaoutcform{1 1 4\\1 1 2}
\end{mmaCell}

\end{description}

\subsubsection*{\mmaHeading{CHolCR} and \mmaHeading{CAHolCR}}

The functions \mma{CHolCR} and \mma{CAHolCR} compute the holomorphic- and
antiholomorphic Cauchy--Riemann derivative, respectively.

\begin{description}
\item[Argument] Both \mma{CHolCR} and \mma{CAHolCR} accept one argument which
  should be a functional expression (e.g.\ a polynomial) involving \acp{MGF} and
  Eisenstein series.
  
\item[Return value] \mma{CHolCR} returns the holomorphic Cauchy--Riemann
  derivative of its argument, using the derivative operator defined in
  \eqref{eq:73}, by applying \eqref{eq:233}. The result is always given in terms
  of lattice sums, even if the argument involves Cauchy--Riemann derivatives of
  basis elements. The generalized Ramanujan identities from
  Section~\ref{sec:MGFDers} are not applied. If the argument contains a
  divergent graph with a closed holomorphic subgraph, \ac{HSR} is applied before
  the derivative is taken, while $\cform{2&0\\0&0}$ is not replaced by
  $\GGhat_{2}$. The output is not manipulated any further. \mma{CAHolCR}
  returns the antiholomorphic Cauchy--Riemann derivative.

\item[Options] The Boolean option \mma{divDer} specifies if derivatives of
  divergent graphs are taken or not. If it is set to \mma{False} (the default is
  \mma{True}) and a divergent \ac{MGF} appears in the argument, \mma{CHolCR} and
  \mma{CAHolCR} return \mma{Nothing}.

\item[Warnings]\
  \begin{itemize}
  \item If the argument of \mma{CHolCR} contains a divergent \ac{MGF}, the
    warning \mmawarning{CHolCR}{derOfDiv} is issued (and c.c.).
  \item The argument is passed to \mma{CModWeight} (see below), to check if it
    has homogeneous modular weight. If it does not, the warning
    \mmawarning{CModWeight}{WeightNotHom} is issued and \mma{Nothing} is
    returned.
  \end{itemize}

\item[Examples]\
\begin{mmaCell}{Input}
  CHolCR\mmabigso{}\mmabigco{}nablaE[1,\,\{3\}],\,gBarHat[2],\,\mmaincform{1 1 1\\1 1 1}\mmabigcc{}\mmabigsc{}
\end{mmaCell}
\begin{mmaCell}{Output}
  \big\{\mmaFrac{12\,\mmaoutcform{5 0\\1 0}\,\mmaSubSup{\(\tau\)}{2}{4}}{\mmaSup{\(\pi\)}{3}},\,\mmaFrac{\(\pi\)}{\mmaSup{\(\tau\)}{2}},\,\mmaoutcform{1 1 2\\1 1 0}+\mmaoutcform{1 2 1\\1 0 1}+\mmaoutcform{2 1 1\\0 1 1}\big\}
\end{mmaCell}
\begin{mmaCell}{Input}
  CHolCR\mmabigso{}\mmaincform{0 0 3\\1 1 3}\mmabigsc{}
\end{mmaCell}
\begin{mmaCell}{Message}
  CHolCR : Warning: You are generating the holomorphic Cauchy-Riemann derivative of the divergent expression \mmaoutcform{0 0 3\\1 1 3}. This may be problematic.
\end{mmaCell}
\begin{mmaCell}{Output}
  \mmam6\,\mmaoutcform{4 0\\4 0}+\mmaFrac{2\,\(\pi\)\,\mmaoutcform{3 0\\3 0}}{\mmaSub{\(\tau\)}{2}}
\end{mmaCell}
  
\end{description}

\subsubsection*{\mmaHeading{CLaurentPoly}}

The function \mma{CLaurentPoly} replaces basis elements by their Laurent
polynomials.

\begin{description}
\item[Argument] \mma{CLaurentPoly} accepts one arbitrary argument.
  
\item[Return value] \mma{CLaurentPoly} returns its argument with the real basis
  elements \eqref{eq:172}, the complex basis elements \eqref{eq:404}, their
  Cauchy--Riemann derivatives and complex conjugates, as well as all
  non-holomorphic- and holomorphic Eisenstein series (including $\GGhat_{2}$)
  replaced by their Laurent polynomials. The Laurent polynomials of the real and
  complex basis elements are given in \eqref{eq:408} and \eqref{eq:409},
  respectively, their derivatives are obtained using \eqref{eq:411}.

\item[Options] The Boolean option \mma{usey} specifies if the output is given in
  terms of $\tau_{2}$ (\mma{False}) or $y=\pi\tau_{2}$ (\mma{True}, the
  default).

\item[Examples]\ 
\begin{mmaCell}{Input}
  CLaurentPoly[nablaBarBBar[2,\,\{2,\,4\}]]
\end{mmaCell}
\begin{mmaCell}{Output}
  \mmam\mmaFrac{8\,\mmaSup{y}{8}}{30375\,\mmaSup{\(\pi\)}{2}}\mmam\mmaFrac{105\,\mmaSub{\(\zeta\)}{3}\,\mmaSub{\(\zeta\)}{7}}{8\,\mmaSup{\(\pi\)}{2}\,\mmaSup{y}{2}}+\mmaFrac{25\,\mmaSub{\(\zeta\)}{9}}{12\,\mmaSup{\(\pi\)}{2}\,y}
\end{mmaCell}
\begin{mmaCell}{Input}
  CLaurentPoly[\{g[6],\,gHat[2],\,e[7]\},\,usey\mmainrarrow{}False]
\end{mmaCell}
\begin{mmaCell}{Output}
  \big\{\mmaFrac{2\,\mmaSup{\(\pi\)}{6}}{945},\,\mmaFrac{\mmaSup{\(\pi\)}{2}}{3}\mmam\mmaFrac{\(\pi\)}{\mmaSub{\(\tau\)}{2}},\,\mmaFrac{4\,\mmaSup{\(\pi\)}{7}\,\mmaSubSup{\(\tau\)}{2}{7}}{18243225}+\mmaFrac{231\,\mmaSub{\(\zeta\)}{13}}{512\,\mmaSup{\(\pi\)}{6}\,\mmaSubSup{\(\tau\)}{2}{6}}\big\}
\end{mmaCell}
\end{description}

\subsubsection*{\mmaHeading{CListHSRs}}

The function \mma{CListHSRs} lists \acp{MGF} with closed holomorphic subgraphs
in an expression.

\begin{description}
\item[Arguments] \mma{CListHSRs} accepts one arbitrary argument.
  
\item[Return value] \mma{CListHSRs} returns a list with all dihedral and
  trihedral graphs with closed holomorphic subgraphs appearing somewhere in its
  argument. If the argument does not contain any dihedral or trihedral graphs,
  \mma{CListHSRs} returns the empty list.

\item[Examples]\ 
\begin{mmaCell}{Input}
  CListHSRs\mmabigso{}\mmaincformtri{1\\0}{1 2\\0 1}{1 2\\2 0}+\mmaincform{7 0\\3 0}\mmabigsc{}
\end{mmaCell}
\begin{mmaCell}{Output}
  \big\{\mmaoutcformtri{1\\0}{1 2\\0 1}{1 2\\2 0}\big\}
\end{mmaCell}
  
\end{description}

\subsubsection*{\mmaHeading{CModWeight}}

The function \mma{CModWeight} determines the modular weight of an expression.

\begin{description}
\item[Argument] \mma{CModWeight} accepts one argument which can be either a
  modular form (possibly of trivial modular weight), a product of modular forms
  or a sum of products of modular forms.
  
\item[Return value] \mma{CModWeight} returns a list with two elements,
  corresponding to the holomorphic and antiholomorphic modular weight,
  respectively.

\item[Warnings]\
  \begin{itemize}
  \item If a sum is passed to \mma{CModWeight} and the modular weights of the
    summands do not agree, \mma{CModWeight} returns \mma{Null} and the warning
    \mmawarning{CModWeight}{WeightNotHom}, containing a list of the modular
    weights appearing in the sum, is issued.
  \item If symbols appear in the argument of \mma{CModWeight}, for which no
    modular weight is implemented, \mma{CModWeight} returns the modular weight
    which the expression would have if all symbols of unknown weight were
    modular invariant. A list of the terms whose weight could not be determined
    is printed as part of the warning \mmawarning{CModWeight}{UnknownExp}.
  \end{itemize}

\item[Examples]\
\begin{mmaCell}{Input}
  CModWeight[\mmaSup{tau[2]}{-2}\,nablaBarE[1,\,\{2\}]\,nablaE[2,\,\{4\}]+nablaBp[1,\,\{2,\,4\}]]
\end{mmaCell}
\begin{mmaCell}{Output}
  \{0,\,\mmam2\}
\end{mmaCell}
\begin{mmaCell}{Input}
  CModWeight\mmabigso{}e[2]+\mmaoutcform{2 0\\2 0}\mmabigsc{}
\end{mmaCell}
\begin{mmaCell}{Message}
  CModWeight: The modular weight of the argument is not homogeneous, the weights \{2,2\},\,\{0,0\} appear.
\end{mmaCell}
\begin{mmaCell}{Input}
  CModWeight[g[2] g[4]]
\end{mmaCell}
\begin{mmaCell}{Message}
  CModWeight: Expression(s) \{\mmaSub{G}{2}\} found whose modular weight could not be determined. The returned weight assumes them to be modular invariant.
\end{mmaCell}
\begin{mmaCell}{Input}
  \{4,\,0\}
\end{mmaCell}
  
\end{description}

\subsubsection*{\mmaHeading{CSieveDecomp}}

The function \mma{CSieveDecomp} decomposes an \ac{MGF} using the sieve
algorithm.

\begin{description}
\item[Arguments] \mma{CSieveDecomp} accepts one argument which can be either a
  dihedral or a trihedral \ac{MGF} without closed holomorphic subgraph.
  
\item[Return value] \mma{CSieveDecomp} performs the sieve algorithm on its
  argument as discussed in Section~\ref{sec:sieve-algorithm} and returns the
  decomposition obtained. If the holomorphic modular weight is larger than the
  antiholomorphic one, \mma{CSieveDecomp} takes holomorphic derivatives,
  otherwise antiholomorphic ones. If both modular weights of the argument are
  equal, an integration constant \mma{intConst} labeled by the exponent matrix
  of the argument and dressed with an appropriate factor of
  $\frac{\pi}{\tau_{2}}$ is added to the final decomposition. If the basis into
  which the argument is decomposed is not linearly independent, the output
  contains free parameters with head \mma{bCoeff}.

\item[Options]\
  {
  \renewcommand{\arraystretch}{1.3}  
  \begin{longtable}[r]{cccc}
    \toprule option & \makecell{possible\\values} & \makecell{default\\value}
    & \makecell[c]{description}\\
    \midrule \mma{verbose} & \mma{True}, \mma{False} & \mma{False}
    & \makecell[t]{activates verbose output}\\
    \mma{divDer} & \mma{True}, \mma{False} & \mma{False}
    & \makecell[t]{activates decomposition of\\divergent graphs}\\
    \mma{basis} & \makecell[t]{list of \acp{MGF}} & \mma{\{\}}
    & \makecell[t]{basis elements for\\decomposition}\\
    \mma{addIds} & \makecell[t]{list of replacement\\rules for \acp{MGF}} &
    \mma{\{\}}
    & \makecell[t]{additional replacement rules\\ applied to each derivative}\\
    \mma{CSimplifyOpts} & \makecell[t]{option assignments\\of \mma{CSimplify}} &
    see below
    & \makecell[t]{options passed to \mma{CSimplify} when\\ simplifying the derivatives}\\
    \bottomrule
  \end{longtable}
  }
  \vspace{-1em}
  The default value of \mma{CSimplifyOpts} is
  \mma{\{basisExpandG\mmainrarrow{}True\}}. If the option \mma{basis} is set to
  the empty list, the appropriate basis is determined automatically using
  \mma{CBasis}. Since this basis does not contain powers of $\EE_{1}$, it is not
  sufficient for the decomposition of divergent graphs. The basis elements have
  to be \acp{MGF} without closed holomorphic subgraphs of the same modular
  weight as the argument. Divergent basis elements are only admissible if
  \mma{divDer} is set to \mma{True}.
  
\item[Warnings]\
  \begin{itemize}
  \item If the argument of \mma{CSieveDecomp} is divergent, the warning
    \mmawarning{CSieveDecomp}{divArg} is issued. The decomposition proceeds only
    if \mma{divDer} is set to \mma{True}.
  \item If one of the basis elements is divergent, but the argument is not, the
    warning \mmawarning{CSieveDecomp}{divBasis} is issued.
  \item If a holomorphic Eisenstein series could not be canceled in one of the
    derivatives, the warning \mmawarning{CSieveDecomp}{noSol} is issued. This
    happens e.g.\ if the basis is not large enough.
  \item If in one of the derivatives, an undecomposed graph appears in the
    coefficient of a holomorphic Eisenstein series, the warning
    \mmawarning{CSieveDecomp}{holEisenCoeffNoBasis} is issued and the algorithm
    interrupted. \acp{MGF} are considered decomposed if they appear in the basis
    given by \mma{CBasis}. For modular weight $a+b\leq12$, these undecomposed
    graphs will be divergent.
  \end{itemize}

\item[Examples]\
\begin{mmaCell}{Input}
  CSieveDecomp\mmabigso{}\mmaincformtri{1\\1}{1 1\\1 1}{1 1\\1 1}\mmabigsc{}
\end{mmaCell}
\begin{mmaCell}{Output}
  2\,\mmaoutcform{1 1 3\\1 1 3}\mmam\mmaFrac{2\,\mmaSup{\(\pi\)}{5}\,\mmaSub{E}{5}}{5\,\mmaSubSup{\(\tau\)}{2}{5}}+\mmaFrac{\mmaSup{\(\pi\)}{5}\mmaoutintconsttri{1\\1}{1 1\\1 1}{1 1\\1 1}}{\mmaSubSup{\(\tau\)}{2}{5}}
\end{mmaCell}
\begin{mmaCell}{Input}
  CSieveDecomp\mmabigso{}\mmaincform{1 1 1\\1 1 1},\,basis\mmainrarrow\mmabigco{}\mmaincform{0 1 2\\1 1 1},\,\mmaFrac{\mmaSubSup{\(\tau\)}{2}{3}}{\mmaSup{tau[2]}{3}}\,e[3]\mmabigcc{}\mmabigsc{}
\end{mmaCell}
\begin{mmaCell}{Output}
  \mmam\mmaoutcform{0 1 2\\1 1 1}+2\,bCoeff[2]\,\mmaoutcform{0 1 2\\1 1 1}+\mmaFrac{\mmaSup{\(\pi\)}{3}\,bCoeff[2]\,\mmaSub{E}{3}}{\mmaSubSup{\(\tau\)}{2}{3}}+\mmaFrac{\mmaSup{\(\tau\)}{3}\,\mmaoutintconst{1 1 1\\1 1 1}}{\mmaSubSup{\(\tau\)}{2}{3}}
\end{mmaCell}
\begin{mmaCell}{Input}
  CSieveDecomp\mmabigso{}\mmaincform{0 1 1 1\\3 0 1 1}\mmabigsc{}
\end{mmaCell}
\begin{mmaCell}{Message}
  CSieveDecomp : The 1st derivative contains the undecomposed graph(s) \big\{\mmaoutcform{0 1 1\\1 0 1}\big\} as a coefficient of a holomorphic Eisenstein series.
\end{mmaCell}
\begin{mmaCell}{Output}
  \mmaoutcform{0 1 1 1\\3 0 1 1}
\end{mmaCell}
\begin{mmaCell}{Input}
  CSieveDecomp\mmabigso{}\mmaincform{0 1 1 1\\3 0 1 1},\,addIds\mmainrarrow\mmabigco{}\mmaincform{0 1 1\\1 0 1}\mmainrarrow\mmam\mmaFrac{\mmaSup{\(\pi\)}{2}\,\mmaSubSup{E}{1}{2}}{2\,\mmaSup{tau[2]}{2}}+\mmaFrac{\mmaSup{\(\pi\)}{2}\,\mmaSub{E}{2}}{2\,\mmaSup{tau[2]}{2}}\mmabigcc{}\mmabigsc{}
\end{mmaCell}
\begin{mmaCell}{Output}
  2\,\mmaoutcform{3 0\\5 0}\mmam2\,\mmaoutcform{1 1 1\\1 1 3}\mmam\mmaFrac{\mmaSup{\(\pi\)}{2}\,\mmaoutcform{1 0\\3 0}\,\mmaSub{E}{2}}{\mmaSubSup{\(\tau\)}{2}{2}}
\end{mmaCell}

\end{description}

\subsubsection*{\mmaHeading{CSimplify}}

The function \mma{CSimplify} performs all known simplifications for \acp{MGF}.

\begin{description}
\item[Argument] \mma{CSimplify} accepts one arbitrary argument.
  
\item[Return value] \mma{CSimplify} applies, in this order, the specialized
  functions \mma{TetCSimplify}, \mma{TriCSimplify} and \mma{DiCSimplify} to its
  argument until it no longer changes and returns the result.

\item[Options] \mma{CSimplify} accepts all the options of both \mma{TriCSimplify} and
  \mma{DiCSimplify} and passes them to these functions when they are called.
\item[Examples]\
\begin{mmaCell}{Input}
  CSimplify\mmabigso{}\mmaincformtri{1\\0}{1 1\\0 1}{1 2\\1 0}\mmabigsc{}
\end{mmaCell}
\begin{mmaCell}{Output}
  \mmaFrac{3}{2}\,\mmaSup{\mmaoutcform{3 0\\1 0}}{2}\mmam\mmaFrac{1}{2}\,\mmaoutcform{6 0\\2 0}\mmam\mmaFrac{1}{2}\,\mmaoutcform{4 0\\2 0}\,\mmaSub{\mmaGGhat}{2}\mmam\mmaFrac{\mmaSup{\(\pi\)}{2}\,\mmaSub{G}{4}}{\mmaSubSup{\(\tau\)}{2}{2}}+\mmaFrac{3\,\(\pi\)\,\mmaoutcform{5 0\\1 0}}{\mmaSub{\(\tau\)}{2}}\mmam\mmaFrac{\(\pi\)\,\mmaoutcform{3 0\\1 0}\,\mmaSub{\mmaGGhat}{2}}{\mmaSub{\(\tau\)}{2}}
\end{mmaCell}
\begin{mmaCell}{Input}
  CSimplify\mmabigso{}\mmaincformtri{1\\0}{1 1\\0 1}{1 2\\1 0},\,tri3ptFayHSR\mmainrarrow{}True\mmabigsc{}
\end{mmaCell}
\begin{mmaCell}{Output}
  \mmaFrac{3}{2}\,\mmaSup{\mmaoutcform{3 0\\1 0}}{2}\mmam\mmaFrac{1}{2}\,\mmaoutcform{6 0\\2 0}\mmam\mmaFrac{1}{2}\,\mmaoutcform{4 0\\2 0}\,\mmaSub{\mmaGGhat}{2}\mmam\mmaFrac{\mmaSup{\(\pi\)}{2}\,\mmaSub{G}{4}}{\mmaSubSup{\(\tau\)}{2}{2}}+\mmaFrac{3\,\(\pi\)\,\mmaoutcform{5 0\\1 0}}{\mmaSub{\(\tau\)}{2}}\mmam\mmaFrac{\(\pi\)\,\mmaoutcform{3 0\\1 0}\,\mmaSub{\mmaGGhat}{2}}{\mmaSub{\(\tau\)}{2}}
\end{mmaCell}
\begin{mmaCell}{Input}
  CSimplify\mmabigso{}\mmaincformtet{\mmanothing}{1\\1}{1\\1}{1\\1}{\mmanothing}{1\\1}\mmabigsc{}
\end{mmaCell}
\begin{mmaCell}{Output}
  0
\end{mmaCell}
  
\end{description}

\subsubsection*{\mmaHeading{CSort}}

The function \mma{CSort} sorts \acp{MGF} into their canonical representation.

\begin{description}
\item[Argument] \mma{CSort} accepts one arbitrary argument.
  
\item[Return value] \mma{CSort} returns its argument with all \acp{MGF} written
  in their canonical representation as discussed in Section~\ref{sec:MGF-symm}.

\item[Example]\ 
\begin{mmaCell}{Input}
  CSort\mmabigso{}\mmabigco{}\mmaincformtet{2 2\\1 1}{1 1\\1 1}{1 1\\1 1}{1 2\\1 1}{2 2\\1 1}{1 1\\1 2}\mmabigcc{},\,\mmaincformtri{1 1\\1 0}{1\\0}{1 2\\1 0}\mmabigsc{}
\end{mmaCell}
\begin{mmaCell}{Output}
  \big\{\mmaoutcformtet{1 1\\1 1}{1 1\\1 1}{2 2\\1 1}{1 2\\1 1}{2 2\\1 1}{1 1\\1 2},\,\mmaoutcformtri{1\\0}{1 1\\0 1}{1 2\\1 0}\big\}
\end{mmaCell}
\end{description}

\subsubsection{Dihedral functions}

\subsubsection*{\mmaHeading{DiHolMomConsId} and \mmaHeading{DiAHolMomConsId}}

The functions \mma{DiHolMomConsId} and \mma{DiAHolMomConsId} generate holomorphic
and antiholomorphic dihedral momentum-conservation identities, respectively.

\begin{description}
\item[Argument] Both \mma{DiHolMomConsId} and \mma{DiAHolMomConsId} accept a
  dihedral \ac{MGF} as their only argument.

\item[Return value] \mma{DiHolMomConsId} returns the holomorphic momentum
  conservation identity \eqref{eq:202} of the seeds given in the argument as an
  equation with \ac{RHS} $0$. \mma{DiAHolMomConsId} returns the antiholomorphic
  momentum-conservation identity. No further manipulation as e.g.\ sorting into
  the canonical representation are performed on the output.

\item[Warnings] If the argument of \mma{DiHolMomConsId} is divergent according
  to \mma{CCheckConv}, the warning \mmawarning{DiHolMomConsId}{divDiHolMomCons}
  (and c.c.) is issued.

\item[Examples]\ 
\begin{mmaCell}{Input}
  DiHolMomConsId\mmabigso{}\mmaincform{1 1 2\\1 1 1}\mmabigsc{}\\
  DiAHolMomConsId\mmabigso{}\mmaincform{1 1 2\\1 1 1}\mmabigsc{}
\end{mmaCell}
\begin{mmaCell}{Output}
  \mmaoutcform{0 1 2\\1 1 1}+\mmaoutcform{1 0 2\\1 1 1}+\mmaoutcform{1 1 1\\1 1 1}\,==\,0
\end{mmaCell}
\begin{mmaCell}{Output}
  \mmaoutcform{1 1 2\\0 1 1}+\mmaoutcform{1 1 2\\1 0 1}+\mmaoutcform{1 1 2\\1 1 0}\,==\,0
\end{mmaCell}

\begin{mmaCell}{Input}
  DiHolMomConsId\mmabigso{}\mmaincform{0 1 2\\1 0 2}\mmabigsc{}
\end{mmaCell}
\begin{mmaCell}{Message}
  DiHolMomConsId : You are generating the holomorphic momentum-conservation identity of the divergent seed \mmaoutcform{0 1 2\\1 0 2}. Divergent seeds can lead to inconsistent identities.
\end{mmaCell}
\begin{mmaCell}{Output}
  \mmaoutcform{-1 1 2\\\verbcent{1} 0 2}+\mmaoutcform{0 0 2\\1 0 2}+\mmaoutcform{0 1 1\\1 0 2}\,==\,0
\end{mmaCell}
\end{description}

\subsubsection*{\mmaHeading{DiCSimplify}}

The function \mma{DiCSimplify} performs all known dihedral simplifications.

\begin{description}
\item[Argument] \mma{DiCSimplify} accepts one arbitrary argument.

\item[Return value] \mma{DiCSimplify} returns the expression given as the
  argument with all dihedral \acp{MGF} (including one-loop graphs such as
  Eisenstein series) rewritten in a simplified form, if possible. This is done
  by performing the following manipulations on all dihedral graphs, until the
  result does not change any more.
  \begin{enumerate}
  \item \label{item:6} Apply \ac{HSR} \eqref{eq:245} and its divergent analog
    \eqref{eq:382}.
  \item \label{item:1} Set $\cform{\noblock}=1$, cf.\ \eqref{eq:193}.
  \item Factorize on $\sbmatrix{0\\0}$ columns according to \eqref{eq:222} and
    \eqref{eq:431}.
  \item Set $\cform{a\\b}=0$, cf.\ \eqref{eq:188}.
  \item \label{item:4} Remove entries of $-1$ by using momentum conservation as
    described in Section~\ref{sec:sieve-algorithm}.
  \item Sort dihedral \acp{MGF} into their canonical representation as described
    in Section~\ref{sec:MGF-symm}.
  \item Set graphs with odd $|A|+|B|$ to zero.
  \item Rewrite $\cform{k&0\\0&0}=\GG_{k}$ and c.c., cf.\ \eqref{eq:330}.
  \item \label{item:5} Rewrite $\cform{2&0\\0&0}=\GGhat_{2}$ and c.c., cf.\
    \eqref{eq:332}.
  \item Set $\GG_{k}$ with $k$ odd to zero and c.c.
  \item Rewrite $\cform{k&0\\k&0}=\piimtauil^{k}\EE_{k}$, cf.\ \eqref{eq:331}.
  \item \label{item:2} Apply generalized Ramanujan identities discussed in
    Section~\ref{sec:MGFDers} and expand holomorphic Eisenstein series in the
    ring of $\GG_{4}$ and $\GG_{6}$.
  \item \label{item:3} Apply basis decompositions discussed in
    Section~\ref{sec:conjectured-basis}, in the basis listed in
    Table~\ref{tab:Cbasis}.
  \end{enumerate}
  Within this process, the steps \ref{item:1} to \ref{item:2} are repeated
  until the result no longer changes, before step \ref{item:3} is executed.
\item[Options]\
  {
  \begin{longtable}[c]{cccc}
    \toprule option & \makecell{possible\\values} & \makecell{default\\value}
    & \makecell[c]{description}\\
    \midrule
    \mma{basisExpandG} & \mma{True}, \mma{False} & \mma{False}
    & \makecell[c]{activates step \ref{item:2}}\\
    \mma{momSimplify} & \mma{True}, \mma{False} & \mma{True}
    & \makecell[c]{deactivates step \ref{item:4}}\\
    \mma{repGHat2} & \mma{True}, \mma{False} & \mma{True}
    & \makecell[c]{deactivates step \ref{item:5}}\\
    \mma{useIds} & \mma{True}, \mma{False} & \mma{True}
    & \makecell[c]{deactivates step \ref{item:3}}\\
    \mma{diHSR} & \mma{True}, \mma{False} & \mma{True}
    & \makecell[c]{deactivates step \ref{item:6}}\\
    \mma{divHSR} & \mma{True}, \mma{False} & \mma{True}
    & deactivates step \ref{item:6} for divergent graphs\\
    \mma{diDivHSR} & \mma{True}, \mma{False} & \mma{True}
    & deactivates step \ref{item:6} for divergent graphs\\
    \bottomrule
  \end{longtable}
  }\vspace{-1em}
  Both options \mma{divHSR} and \mma{diDivHSR} have to be \mma{True} for
  divergent graphs to be included in step \ref{item:6}.

\item[Warnings]\
  \begin{itemize}
  \item If a graph in the argument contains a $\sbmatrix{0\\0}$ column next to a
    $\sbmatrix{1\\0}$ or $\sbmatrix{1\\0}$ column, the warning
    \mmawarning{DiCSimplify}{dangerousFact} is issued and the modified
    factorization rule \eqref{eq:431} applied.
  \item If a divergent graph with a holomorphic subgraph is encountered but
    \ac{HSR} cannot be performed because either one of the options \mma{divHSR}
    or \mma{diDivHSR} is set to \mma{False}, the warning
    \mmawarning{DiCSimplify}{divHSRNotPossible} is issued.
  \end{itemize}
\item[Examples]\
\begin{mmaCell}{Input}
  DiCSimplify\mmabigso{}\mmaincform{1 2 2 2\\0 0 1 2}\mmabigsc{}
\end{mmaCell}
\begin{mmaCell}{Output}
  3\,\mmaoutcform{3 0\\1 0}\,\mmaoutcform{4 0\\2 0}\mmam15\,\mmaoutcform{7 0\\3 0}\mmam9\mmaoutcform{0 2 5\\1 0 2}+\mmaFrac{21}{2}\,\mmaoutcform{1 1 5\\1 1 1}\mmam\mmaoutcform{5 0\\3 0}\,\mmaSub{\mmaGGhat}{2}+\\
  \mmaFrac{1}{2}\,\mmaoutcform{1 1 3\\1 1 1}\,\mmaSub{\mmaGGhat}{2}\mmam\mmaFrac{2\,\mmaSup{\(\pi\)}{2}\,\mmaoutcform{5 0\\1 0}}{\mmaSubSup{\(\tau\)}{2}{2}}+\mmaFrac{6\,\(\pi\)\,\mmaoutcform{6 0\\2 0}}{\mmaSub{\(\tau\)}}\mmam\mmaFrac{2\,\(\pi\)\,\mmaoutcform{4 0\\2 0}\,\mmaSub{\mmaGGhat}{2}}{\mmaSub{\(\tau\)}{2}}
\end{mmaCell}
\begin{mmaCell}{Input}
  DiCSimplify\mmabigso{}\mmaincform{1 2 2 2\\0 0 1 2},\,momSimplify\mmainrarrow{}False,\,useIds\mmainrarrow{}False\mmabigsc{}
\end{mmaCell}
\begin{mmaCell}{Output}
  \mmam3\,\mmaoutcform{2 2 3\\1 2 0}+\mmaoutcform{1 2 2\\0 1 2}\mmaSub{\mmaGGhat}{2}+\mmaFrac{\(\pi\)\,\mmaoutcform{\verbcent{2} 2 2\\-1 1 2}}{\mmaSub{\(\tau\)}{2}}
\end{mmaCell}
\begin{mmaCell}{Input}
  DiCSimplify\mmabigso{}\mmaincform{0 0 1 1 1 2 4\\0 1 1 1 1 3 4}\mmabigsc{}\,//\,Simplify
\end{mmaCell}
\begin{mmaCell}{Message}
  DiCSimplify : The graph \mmaoutcform{0 0 1 1 1 2 4\\0 1 1 1 1 3 4} is factorized and contains a (1,0) or (0,1) column. This may be problematic.
\end{mmaCell}
\begin{mmaCell}{Output}
  \mmam\mmaFrac{\mmaSup{\(\pi\)}{8}\,\mmaoutcform{1 0\\3 0}\,(-6+6\,\mmaSub{E}{1}\mmam3\,\mmaSubSup{E}{1}{2}+\mmaSubSup{E}{1}{3})\,\mmaSub{E}{4}+\mmaoutcform{0 1 1 1 2 4\\1 1 1 1 3 4}\,\mmaSubSup{\(\tau\)}{2}{8}}{\mmaSubSup{\(\tau\)}{2}{8}}
\end{mmaCell}
  
\end{description}

\subsubsection{Trihedral functions}

\subsubsection*{\mmaHeading{TriHolMomConsId} and \mmaHeading{TriAHolMomConsId}}

The functions \mma{TriHolMomConsId} and \mma{TriAHolMomConsId} generate
trihedral holomorphic and antiholomorphic momentum-conservation identities,
respectively.

\begin{description}
\item[Arguments] Both \mma{TriHolMomConsId} and \mma{TriAHolMomConsId} accept
  two arguments: The first is a trihedral \ac{MGF}, the second is one of the
  lists \mma{\{1,2\}}, \mma{\{2,3\}} or \mma{\{1,3\}}, where the order of the
  elements in the list does not matter.

\item[Return value] \mma{TriHolMomConsId} returns the holomorphic trihedral
  momentum-conservation identity \eqref{eq:203} as an equation with \ac{RHS}
  zero. Due to the permutation symmetry of the three blocks in a trihedral
  \ac{MGF}, any two blocks can be involved in the momentum-conservation identity
  (i.e.\ have their holomorphic weight reduced) and the second argument of
  \mma{TriHolMomConsId} specifies which two blocks should be used to generate
  the identity. \mma{TriAHolMomConsId} generates the antiholomorphic momentum
  conservation identity. No further manipulation is performed on the output.

\item[Warnings] If either one of the blocks in the second argument is divergent
  as a dihedral \ac{MGF} or if the trihedral \ac{MGF} in the first argument has
  a three-point divergence (cf.\ \eqref{eq:349}), the warning
  \mmawarning{TriHolMomConsId}{divTriHolMomCons} (and c.c.) is issued.
  
\item[Examples]\ 
\begin{mmaCell}{Input}
  TriHolMomConsId\mmabigso{}\mmaincformtri{1\\1}{1 1\\1 1}{1 1\\1 1},\,\{1,\,2\}\mmabigsc{}
\end{mmaCell}
\begin{mmaCell}{Output}
  \mmaoutcformtri{0\\1}{1 1\\1 1}{1 1\\1 1}\mmam\mmaoutcformtri{1\\1}{0 1\\1 1}{1 1\\1 1}\mmam\mmaoutcformtri{1\\1}{1 0\\1 1}{1 1\\1 1}\,==\,0
\end{mmaCell}
\begin{mmaCell}{Input}
  TriAHolMomConsId\mmabigso{}\mmaincformtri{1\\1}{1 1\\1 1}{1 1\\1 1},\,\{3,\,2\}\mmabigsc{}
\end{mmaCell}
\begin{mmaCell}{Output}
  \mmam\mmaoutcformtri{1\\1}{1 1\\0 1}{1 1\\1 1}\mmam\mmaoutcformtri{1\\1}{1 1\\1 0}{1 1\\1 1}+\mmaoutcformtri{1\\1}{1 1\\1 1}{1 1\\0 1}+\mmaoutcformtri{1\\1}{1 1\\1 1}{1 1\\1 0}\,==\,0
\end{mmaCell}
\begin{mmaCell}{Input}
  TriHolMomConsId\mmabigso{}\mmaincformtri{1\\1}{0 1\\1 0}{1 1\\1 1},\,\{1,\,3\}\mmabigsc{}
\end{mmaCell}
\begin{mmaCell}{Output}
  \mmaincformtri{0\\1}{0 1\\1 0}{1 1\\1 1}\mmam\mmaincformtri{1\\1}{0 1\\1 0}{0 1\\1 1}\mmam\mmaincformtri{1\\1}{0 1\\1 0}{1 0\\1 1}\,==\,0
\end{mmaCell}
\end{description}

\subsubsection*{\mmaHeading{TriFay}}

The function \mma{TriFay} generates trihedral Fay identities.

\begin{description}
\item[Arguments] \mma{TriFay} accepts up to two arguments. The first (mandatory)
  argument is a trihedral \ac{MGF}, the second (optional) argument is a list of
  the form
  \mma{\{\{b\textsubscript{1},c\textsubscript{1}\},\{b\textsubscript{2},c\textsubscript{2}\}\}},
  where \mma{c\textsubscript{i}} is a column number in the
  \mma{b\textsubscript{i}}\textsuperscript{th} block and the list selects two
  columns, both of the form $\sbmatrix{a\\0}$ with $a\geq1$ or $\sbmatrix{0\\b}$
  with $b\geq1$ in the trihedral graph. If the second argument is omitted,
  \mma{TriFay} selects the first suitable pair of columns automatically,
  starting from the left and trying holomorphic column pairs first.

\item[Return value] \mma{TriFay} returns an equation in which the \ac{LHS} is
  the graph specified in the first argument and the \ac{RHS} is given by
  \eqref{eq:303} (or its complex conjugate), with the columns
  $\sbmatrix{a_{1}\\0}$ and $\sbmatrix{a_{2}\\0}$ selected by the second
  argument or determined automatically. No further manipulations are performed
  on the output.

\item[Warnings] If no second argument is passed to \mma{TriFay} and no suitable
  pair of columns could be found, the warning \mmawarning{TriFay}{noFayCols} is
  issued.

\item[Examples]\
\begin{mmaCell}{Input}
  TriFay\mmabigso{}\mmaincformtri{1\\0}{1 2\\0 2}{1 2\\1 0}\mmabigsc{}
\end{mmaCell}
\begin{mmaCell}{Output}
  \mmaoutcformtri{1\\0}{1 2\\0 2}{1 2\\1 0}\,==\,\mmaoutcformtri{\{\}}{2\\2}{1 2 2\\1 0 0}\mmam\mmaoutcformtri{\{\}}{1 2\\0 2}{1 1 2\\0 1 0}+\\
  \quad\mmaoutcformtri{\{\}}{1 2\\1 0}{2 2\\0 2}\mmam\mmaoutcformtri{1\\0}{2\\2}{1 1 2\\0 1 0}+\mmaoutcformtri{2\\0}{2\\2}{1 2\\1 0}
\end{mmaCell}
\begin{mmaCell}{Input}
  TriFay\mmabigso{}\mmaincformtri{0\\1}{0 1\\2 1}{0 2\\1 2},\,\{\{1,\,1\},\,\{3,\,1\}\}\mmabigsc{}
\end{mmaCell}
\begin{mmaCell}{Output}
  \mmaincformtri{0\\1}{0 1\\2 1}{0 2\\1 2}\,==\,\mmaoutcformtri{\{\}}{2\\2}{0 0 1\\2 2 1}+\mmaoutcformtri{\{\}}{0 1\\2 1}{0 2\\2 2}\mmam\\
  \quad\mmaoutcformtri{\{\}}{0 2\\1 2}{0 0 1\\1 2 1}\mmam\mmaoutcformtri{0\\1}{2\\2}{0 0 1\\1 2 1}+\mmaoutcformtri{0\\2}{2\\2}{0 1\\2 1}
\end{mmaCell}
  
\end{description}

\subsubsection*{\mmaHeading{TriCSimplify}}

The function \mma{TriCSimplify} applies all known trihedral simplifications.

\begin{description}
\item[Argument] \mma{TriCSimplify} accepts one arbitrary argument.

\item[Return value] \mma{TriCSimplify} returns the expression given as the
  argument with all trihedral \acp{MGF} rewritten in a simplified form, if
  possible. This is done by performing the following manipulations on all
  trihedral graphs, until the result does not change any more.
  \begin{enumerate}
  \item \label{item:7}Apply two-point \ac{HSR} using the trihedral
    generalization of \eqref{eq:245} as described in \cite{gerken2019f} and its
    divergent analog \eqref{eq:434}.
  \item \label{item:8}Apply three-point \ac{HSR} using the closed formula in
    \cite{gerken2019f}.
  \item \label{item:9} Set graphs with odd total modular weight $a+b$ to zero.
  \item Apply the topological simplification \eqref{eq:210}.
  \item Apply the topological simplification \eqref{eq:189}.
  \item Factorize on $\sbmatrix{0\\0}$ columns according to \eqref{eq:223} and
    \eqref{eq:432}.
  \item \label{item:10} Remove entries of $-1$ by using momentum conservation as
    described in Section~\ref{sec:sieve-algorithm}.
  \item \label{item:11} Sort trihedral \acp{MGF} into their canonical
    representation as described in Section~\ref{sec:MGF-symm}.
  \item \label{item:12} Apply basis decompositions discussed in
    Section~\ref{sec:conjectured-basis}, in the basis listed in
    Table~\ref{tab:Cbasis}.
  \end{enumerate}
  Within this process, the steps \ref{item:9} to \ref{item:11} are repeated
  until the result no longer changes, before step \ref{item:12} is executed.
  
\item[Options]\ 
  {
  \begin{longtable}[r]{cccc}
    \toprule option & \makecell{possible\\values} & \makecell{default\\value}
    & \makecell[c]{description}\\
    \midrule
    \mma{momSimplify} & \mma{True}, \mma{False} & \mma{True}
    & \makecell[c]{deactivates step \ref{item:10}}\\
    \mma{useIds} & \mma{True}, \mma{False} & \mma{True}
    & \makecell[c]{deactivates step \ref{item:12}}\\
    \mma{triHSR} & \mma{True}, \mma{False} & \mma{True}
    & \makecell[c]{deactivates steps \ref{item:7} and \ref{item:8}}\\
    \mma{tri2ptHSR} & \mma{True}, \mma{False} & \mma{True}
    & \makecell[c]{deactivates step \ref{item:7}}\\
    \mma{tri3ptHSR} & \mma{True}, \mma{False} & \mma{True}
    & \makecell[c]{deactivates step \ref{item:8}}\\
    \mma{tri3ptFayHSR} & \mma{True}, \mma{False} & \mma{False}
    & \makecell[t]{activates three-point \ac{HSR} via the Fay\\identity
      \eqref{eq:303} instead of the forumla in \cite{gerken2019f}}\\
    \mma{divHSR} & \mma{True}, \mma{False} & \mma{True}
    & \hspace{-0.5ex}deactivates steps \ref{item:7} and \ref{item:8} for divergent graphs\\
    \mma{triDivHSR} & \mma{True}, \mma{False} & \mma{True}
    & \hspace{-0.5ex}deactivates steps \ref{item:7} and \ref{item:8} for divergent graphs\\
    \bottomrule
  \end{longtable}
  }\vspace{-1em}
  Both options \mma{divHSR} and \mma{triDivHSR} have to be \mma{True} for
  divergent graphs to be included in steps \ref{item:7} and
  \ref{item:8}. Furthermore, if \mma{tri3ptFayHSR} is set to \mma{True}, setting
  \mma{tri2ptHSR} to \mma{False} also deactivates three-point \ac{HSR} since
  \eqref{eq:303} reduces three-point \ac{HSR} to two-point \ac{HSR}.
  
\item[Warnings]\ 
  \begin{itemize}
  \item If a graph in the argument contains a $\sbmatrix{0\\0}$ column next to a
    $\sbmatrix{1\\0}$ or $\sbmatrix{1\\0}$ column, the warning
    \mmawarning{TriCSimplify}{dangerousFact} is issued and the modified
    factorization rule \eqref{eq:432} applied.
  \item If a divergent graph with a holomorphic subgraph is encountered but
    \ac{HSR} cannot be performed because either one of the options \mma{divHSR}
    or \mma{triDivHSR} is set to \mma{False}, the warning
    \mmawarning{TriCSimplify}{divHSRNotPossible} is issued.
  \item If three-point \ac{HSR} is performed on a divergent graph using Fay
    identities by setting the option \mma{tri3ptFayHSR} to \mma{True}, the
    warning \mmawarning{TriCSimplify}{div3ptFay} is issued.
  \item If three-point \ac{HSR} is performed via the formula in
    \cite{gerken2019f} and there is no ordering of the blocks which prevents a
    divergent expression in the result (cf.\ discussion in Section~4.2.4 of the
    reference), the warning \mmawarning{TriCSimplify}{noConvHSROrder} is
    issued. If one of the options \mma{divHSR} or \mma{triDivHSR} is set to
    \mma{False}, the warning \mmawarning{TriCSimplify}{divHSRNotPossible} is
    issued and the \ac{HSR} is not performed.
  \end{itemize}

\item[Examples]\
\begin{mmaCell}{Input}
  TriCSimplify\mmabigso{}\mmaincformtri{1\\0}{1 1\\0 1}{1 2\\1 0}\mmabigsc{}
\end{mmaCell}
\begin{mmaCell}[morelst={label=mma:9}]{Output}
  \mmam6\,\mmaoutcform{2 4\\2 0}+2\,\mmaoutcform{3 1 2\\1 1 0}\mmam6\,\mmaoutcform{4 1 1\\1 0 1}+\mmaSup{\mmaoutcform{1\\1}}{2}\,\mmaSub{G}{4}+\\
  \quad2\,\mmaoutcform{2 1 1\\1 0 1}\,\mmaSub{\mmaGGhat}{2}+\mmaFrac{2\,\(\pi\)\,\mmaoutcform{2 \verbcent{3}\\2 -1}}{\mmaSub{\(\tau\)}{2}}+\mmaFrac{2\,\(\pi\)\,\mmaoutcform{3 1 1\\0 0 1}}{\mmaSub{\(\tau\)}{2}}
\end{mmaCell}
\begin{mmaCell}{Input}
  TriCSimplify\mmabigso{}\mmaincformtri{1\\0}{1 1\\0 1}{1 2\\1 0},\,tri3ptFayHSR\mmainrarrow{}True\mmabigsc{}
\end{mmaCell}
\begin{mmaCell}[morelst={label=mma:10}]{Output}
  \mmaoutcform{1\\1}\mmaoutcform{1 1\\0 1}\mmam\mmaoutcform{2 1 3\\1 1 0}\mmam\mmaoutcform{2 1 3\\2 0 0}+3\,\mmaoutcform{4 1 1\\1 0 1}\mmam\mmaoutcform{2 1 1\\1 0 1}\,\mmaSub{\mmaGGhat}{2}\mmam\mmaFrac{\(\pi\)\,\mmaoutcform{3 1 1\\0 0 1}}{\mmaSub{\(\tau\)}{2}}
\end{mmaCell}
\renewcommand{\temprefa}{\ref*{mma:9}}
\renewcommand{\temprefb}{\ref*{mma:10}}
\begin{mmaCell}{Input}
  DiCSimplify[Out[\temprefa]-Out[\temprefb]]
\end{mmaCell}
\begin{mmaCell}{Output}
  0
\end{mmaCell}
\begin{mmaCell}{Input}
  TriCSimplify\mmabigso{}\mmaincformtri{1\\0}{1 2\\0 0}{1 2\\1 0},\,tri2ptHSR\mmainrarrow{}False,\,tri3ptFayHSR\mmainrarrow{}True\mmabigsc{}
\end{mmaCell}
\begin{mmaCell}{Output}
  \mmaoutcformtri{1\\0}{1 2\\0 0}{1 2\\1 0}
\end{mmaCell}

\end{description}

\subsubsection{Four-point simplification}

\subsubsection*{\mmaHeading{TetCSimplify}}

The function \mma{TetCSimplify} applies topological simplifications on
four-point graphs.

\begin{description}
\item[Argument] \mma{TetCSimplify} accepts one arbitrary argument.

\item[Return value] \mma{TetCSimplify} returns the expression given as the
  argument with all four-point \acp{MGF} rewritten in a simplified form, if
  possible. This is done by performing the following manipulations on all
  four-point graphs (not only tetrahedral ones), until the result does not
  change any more.
  \begin{enumerate}
  \item Set graphs with odd total modular weight $a+b$ to zero.
  \item Apply the topological simplification \eqref{eq:192}.
  \item Apply the topological simplifications \eqref{eq:195} and \eqref{eq:196}.
  \item Apply the topological simplifications \eqref{eq:197} and \eqref{eq:198}.
  \item Set four-point \acp{MGF} to zero which vanish by symmetry, cf.\
    \eqref{eq:336}.
  \item Sort four-point \acp{MGF} into their canonical
    representation as described in Section~\ref{sec:MGF-symm}.
  \end{enumerate}

\item[Examples]\ 
\begin{mmaCell}{Input}
  TetCSimplify\mmabigso{}\mmaincformtet{1 1\\1 1}{1 1\\1 1}{1 2\\1 1}{2 2\\1 1}{2 2\\1 1}{1 1\\1 2}\mmabigsc{}
\end{mmaCell}
\begin{mmaCell}{Output}
  0
\end{mmaCell}
\begin{mmaCell}{Input}
  TetCSimplify\mmabigso{}\mmaincformtet{\mmanothing}{1\\1}{1\\1}{1\\1}{\mmanothing}{1\\1}\mmabigsc{}
\end{mmaCell}
\begin{mmaCell}{Output}
  \mmaoutcform{1\\1}\mmaoutcformtri{1\\1}{1\\1}{1\\1}
\end{mmaCell}
    
\end{description}

\subsubsection{Koba--Nielsen integration}

\subsubsection*{\mmaHeading{zIntegrate}}

The function \mma{zIntegrate} expands Koba--Nielsen integrals in terms of
\acp{MGF}.

\begin{description}
\item[Arguments] \mma{zIntegrate} represents a Koba--Nielsen integral and
  accepts three arguments. The first argument should be a polynomial in the
  objects with suffix \mma{z} introduced in Section~\ref{sec:mma-KN-ints},
  specifying the prefactor of the Koba--Nielsen factor. The second argument
  should be a natural number specifying the number of punctures in the
  Koba--Nielsen factor or a list of pairs of natural numbers
  \mma{\{\{i,\,j\},\,\{k,\,l\},\,\dots\}}, specifying the Green functions (and
  associated Mandelstam variables) appearing in the Koba--Nielsen factor. The
  third argument should be a natural number specifying the order to which the
  Koba--Nielsen integral is to be expanded.

  \renewcommand{\temprefa}{\ref*{mma:14}}
\item[Return value] \mma{zIntegrate} returns the order specified by the last
  argument of the Koba--Nielsen integral defined by the first two arguments. The
  resulting \acp{MGF} are simplified using the general properties listed below
  \eqref{eq:332} and all the techniques implemented in \mma{CSimplify}, apart
  from \ac{HSR} and the application of the basis decompositions from
  Section~\ref{sec:conjectured-basis}. If the resulting \acp{MGF} require graphs
  with more than four vertices, for which no notation was defined, a graphical
  representation of those graphs is printed, cf.~\mmaoutref{mma:14}. No
  constraints are placed on the Mandelstam variables.

\item[Examples]\ 
\begin{mmaCell}{Input}
  zIntegrate[vz[2,\,\{1,\,2\}]+vz[2,\,\{3,\,4\}],\,4,\,1]\,//\,Simplify
  CSimplify[
\end{mmaCell}
\begin{mmaCell}{Output}
  \mmam\mmaFrac{\big(2\,\mmaoutcform{3 0\\1 0}+\mmaoutcform{1 1 1\\0 0 1}\big)\,(\mmaSub{s}{1,2}+\mmaSub{s}{3,4})\mmaSub{\(\tau\)}{2}}{\(\pi\)}
\end{mmaCell}
\begin{mmaCell}{Output}
  \mmam\mmaSub{\mmaGGhat}{2}\,\mmaSub{s}{1,2}\mmam\mmaSub{\mmaGGhat}{2}\,\mmaSub{s}{3,4}
\end{mmaCell}
\begin{mmaCell}{Input}
  zIntegrate[vz[2,\,\{1,\,2\}]\,vz[2,\,\{3,\,4\}],\,4,\,1]\,//\,Simplify
  CSimplify[
\end{mmaCell}
\begin{mmaCell}{Output}
  \mmam\mmaFrac{\big(2\,\mmaoutcform{3 0\\1 0}+\mmaoutcform{1 1 1\\0 0 1}\big)\,\mmaSub{\mmaGGhat}{2}\,(\mmaSub{s}{1,2}+\mmaSub{s}{3,4})\mmaSub{\(\tau\)}{2}}{\(\pi\)}
\end{mmaCell}
\begin{mmaCell}{Output}
  \mmam\mmaSubSup{\mmaGGhat}{2}{2}\,\mmaSub{s}{1,2}\mmam\mmaSubSup{\mmaGGhat}{2}{2}\,\mmaSub{s}{3,4}
\end{mmaCell}
\begin{mmaCell}{Input}
  zIntegrate[fz[1,\,1,\,2]\,fBarz[1,\,1,\,3],\,3,\,2]\,//\,Simplify
  CSimplify[
\end{mmaCell}
\begin{mmaCell}{Output}
  \mmaFrac{\mmaSub{s}{2,3}\big(\mmam2\,\mmaoutcform{1 1 1\\0 1 2}\,\mmaSub{s}{1,2}\mmam2\,\mmaoutcform{0 1 2\\1 1 1}\,\mmaSub{s}{1,3}+\mmaoutcform{1 1 1\\1 1 1}\,\mmaSub{s}{2,3}\big)\,\mmaSubSup{\(\tau\)}{2}{2}}{2\,\mmaSup{\(\pi\)}{2}}
\end{mmaCell}
\begin{mmaCell}{Output}
  \mmaFrac{\(\pi\) \mmaSub{s}{2,3} (\mmaSub{s}{1,2}+\mmaSub{s}{1,3}+\mmaSub{s}{2,3}) (\mmaSub{E}{3}+\mmaSub{\(\zeta\)}{3})}{2 \mmaSub{\(\tau\)}{2}}
\end{mmaCell}

\end{description}

\subsection{Example: four-gluon scattering in the heterotic string}
\label{sec:mma-example}

In this section, we use the functions introduced above to reproduce the
expansions for the integrals $\II_{1234}^{(2,0)}$ and $\II_{1234}^{(4,0)}$
defined by
\begin{align}
  \II^{(4,0)}_{1234}(s_{ij},\tau) &= \int \dd \mu_3 \, V_4(1,2,3,4)\,\KN_{4}\\
  \II^{(2,0)}_{1234}(s_{ij},\tau) &= \int \dd \mu_3 \, V_2(1,2,3,4)\,\KN_{4} \,,
\end{align}
which appear in the planar sector of four-gluon scattering in the heterotic
string, cf.\ Section~2.4 of \cite{gerken2019e}.

All of the steps in the calculation are automatized, with one exception: The
four-point \ac{HSR}-identity \eqref{eq:488} has to be added by hand. To this
end, we first define the replacement rule
\begin{mmaCell}{Input}
  tetrule=\mmaincformtet{1\\0}{1\\0}{1\\1}{1\\0}{1\\0}{1\\1}\mmainrarrow\mmam\mmaincformtet{1\\0}{\mmanothing}{1 2\\1 0}{1\\0}{1\\0}{1\\1}\mmam\mmaincformtet{1\\0}{\mmanothing}{1\\1}{2\\0}{1\\0}{1\\1}\mmam\\
  \quad\mmaincformtet{1\\0}{2\\0}{1\\1}{\mmanothing}{1\\0}{1\\1}+\mmaincformtet{1\\0}{\mmanothing}{1 1\\1 0}{1\\0}{1\\0}{1\\1}\mmam\mmaincformtet{1\\0}{1\\0}{1 1\\1 0}{\mmanothing}{1\\0}{1\\1};
\end{mmaCell}
In order to bring the output into a nice form, we furthermore define the helper
function
\begin{mmaCell}[morelst={label=mma:15},morepattern={poly_, poly, x__, x, \#, i_, j_, i, j},moredefined={MonomialList,NegativeDegreeLexicographic,DeleteDuplicates,SortBy},morefunctionlocal={ap,mandOrd,result}]{Code}
  prettify[poly_]:=Block[{ap,mandOrd,result},
     mandOrd=MonomialList[poly,{s[1,2],s[2,3]}]/.List[x__]:>Plus[x];
     result=DeleteCases[DeleteDuplicates[Flatten[CoefficientList[mandOrd,
        {s[1,2],s[2,3]}]]],0];
     result=Collect[mandOrd,result];
     result=(SortBy[({Exponent[#/.s[i_,j_]:>ap s[i,j],ap],#}&)
        /@(List@@result),First][[All,2]])/.List[x__]:>HoldForm[Plus[x]];
     Return[result]];
\end{mmaCell}
The integral $\II_{1234}^{(4,0)}$ can now be expanded to second order by running
\begin{mmaCell}[functionlocal=i]{Input}
  Sum\mmabigso{}zIntegrate\mmabigso{}vz[4,\,\{1,\,2,\,3,\,4\}],\,4,\,i\mmabigsc{},\,\{i,\,0,\,2\}\mmabigsc{};
\end{mmaCell}
To this we apply the four-point \ac{HSR}-rule from above, decompose all
resulting \acp{MGF} into the basis from Table~\ref{tab:Cbasis} and change the
basis to Table~\ref{tab:Ebasis},
\begin{mmaCell}{Input}
\end{mmaCell}
Since \mma{zIntegrate} does not apply momentum conservation to the Mandelstam
variables, we do this explicitly,
\begin{mmaCell}{Input}
   \quad{}s[1,\,3]\mmainrarrow{}\,-\,s[1,\,2]\,-\,s[2,\,3]\};
\end{mmaCell}
Finally, we apply the function \mma{prettify} defined in \mmainref{mma:15} to
rearrange the output
\begin{mmaCell}{Input}
  prettify[
\end{mmaCell}
\begin{mmaCell}{Output}
  \mmaSub{G}{4}\!+\!(\mmaSub{s}{1,2}\!+\!\mmaSub{s}{2,3})\,\Big(\!\mmam\!6\,\mmaSub{G}{4}\!\mmam\!\mmaFrac{3\,\(\pi\)\,\mmaSub{\mmaGGhat}{2}\,\(\nabla\)\mmaSub{E}{2}}{\mmaSubSup{\(\tau\)}{2}{2}}\Big)\!+\!\big(\mmaSubSup{s}{1,2}{2}\!+\!\mmaSubSup{s}{2,3}{2}\big)\,\Big(2\,\mmaSub{E}{2}\,\mmaSub{G}{4}\!+\!\mmaFrac{\mmaSup{\(\pi\)}{2}\,\mmaSup{\(\nabla\)}{2}\mmaSub{E}{3}}{6\,\mmaSubSup{\(\tau\)}{2}{4}}\!+\!\mmaFrac{2\,\(\pi\)\,\mmaSub{\mmaGGhat}{2}\,\(\nabla\)\mmaSub{E}{3}}{3\,\mmaSubSup{\(\tau\)}{2}{2}}\Big)\!+\!
  \quad\mmaSub{s}{1,2}\,\mmaSub{s}{2,3}\,\Big(2\,\mmaSub{E}{2}\,\mmaSub{G}{4}+\mmaFrac{2\,\mmaSup{\(\pi\)}{2}\,\mmaSup{\(\nabla\)}{2}\mmaSub{E}{3}}{3\,\mmaSubSup{\(\tau\)}{2}{4}}+\mmaFrac{8\,\(\pi\)\,\mmaSub{\mmaGGhat}{2}\,\(\nabla\)\mmaSub{E}{3}}{3\,\mmaSubSup{\(\tau\)}{2}{2}}\Big) \textnormal{,}
\end{mmaCell}
which agrees with the result found in \cite{gerken2019e}.  The Laurent
polynomial of the first orders of $\II_{1234}^{(4,0)}$ can now easily be
obtained by
\begin{mmaCell}{Input}
  prettify[CLaurentPoly[ReleaseHold[
\end{mmaCell}
\begin{mmaCell}{Output}
  \mmaFrac{\mmaSup{\(\pi\)}{4}}{45}\!\!+\!\!(\mmaSub{s}{1,2}\!\!+\!\!\mmaSub{s}{2,3})\Big(\!\!\mmam\!\!\mmaFrac{2\,\mmaSup{\(\pi\)}{4}\,y}{45}\!\!\mmam\!\!\mmaFrac{3\,\mmaSup{\(\pi\)}{4}\,\mmaSub{\(\zeta\)}{3}}{\mmaSup{y}{3}}\!\!+\!\!\mmaFrac{\mmaSup{\(\pi\)}{4}\,\mmaSub{\(\zeta\)}{3}}{\mmaSup{y}{2}}\Big)\!\!+\!\!\mmaSub{s}{1,2}\mmaSub{s}{2,3}\Big(\mmaFrac{94\,\mmaSup{\(\pi\)}{4}\,\mmaSup{y}{2}}{14175}\!\!+\!\!\mmaFrac{2\,\mmaSup{\(\pi\)}{4}\,\mmaSub{\(\zeta\)}{3}}{45\,y}\!\!+\!\!\mmaFrac{5\,\mmaSup{\(\pi\)}{4}\,\mmaSub{\(\zeta\)}{5}}{\mmaSup{y}{4}}\!\!\mmam\!\!\mmaFrac{\mmaSup{\(\pi\)}{4}\,\mmaSub{4\(\zeta\)}{5}}{3\,\mmaSup{y}{3}}\Big)\!\!+\!\!
  \quad\big(\mmaSubSup{s}{1,2}{2}+\mmaSubSup{s}{2,3}{2}\big)\Big(\mmaFrac{34\,\mmaSup{\(\pi\)}{4}\,\mmaSup{y}{2}}{14175}+\mmaFrac{2\,\mmaSup{\(\pi\)}{4}\,\mmaSub{\(\zeta\)}{3}}{45\,y}+\mmaFrac{5\,\mmaSup{\(\pi\)}{4}\,\mmaSub{\(\zeta\)}{5}}{\mmaSup{y}{4}}\mmam\mmaFrac{\mmaSup{\(\pi\)}{4}\,\mmaSub{\(\zeta\)}{5}}{3\,\mmaSup{y}{3}}\Big) \textnormal{.}
\end{mmaCell}

Similarly, we can expand $\II_{1234}^{(2,0)}$ to third order by running
\begin{mmaCell}[functionlocal=i,morelst={label=mma:12}]{Input}
  Sum\mmabigso{}zIntegrate\mmabigso{}vz[2,\,\{1,\,2,\,3,\,4\}],\,4,\,i\mmabigsc{},\,\{i,\,0,\,3\}\mmabigsc{};\\
   \quad{}s[1,\,3]\mmainrarrow{}\,-\,s[1,\,2]\,-\,s[2,\,3]\};\\
  prettify[
\end{mmaCell}
\begin{mmaCell}{Output}
  \mmam\mmaFrac{3\,\(\pi\,\nabla\)\mmaSub{E}{2}\,(\mmaSub{s}{1,2}+\mmaSub{s}{2,3})}{\mmaSubSup{\(\tau\)}{2}{2}}+\mmaFrac{8\,\(\pi\,\nabla\)\mmaSub{E}{3}\,\mmaSub{s}{1,2}\,\mmaSub{s}{2,3}}{3\,\mmaSubSup{\(\tau\)}{2}{2}}+\mmaFrac{2\,\(\pi\,\nabla\)\mmaSub{E}{3}\,(\mmaSubSup{s}{1,2}{2}+\mmaSubSup{s}{2,3}{2})}{3\,\mmaSubSup{\(\tau\)}{2}{2}}+
  \quad\big(\mmaSubSup{s}{1,2}{2}\mmaSub{s}{2,3}\!+\!\mmaSub{s}{1,2}\mmaSubSup{s}{2,3}{2}\big)\Big(\!\!\mmam\!\mmaFrac{12\,\(\pi\)\,\mmaSub{E}{2}\,\(\nabla\)\mmaSub{E}{2}}{\mmaSubSup{\(\tau\)}{2}{2}}\!\mmam\!\mmaFrac{8\,\(\pi\,\nabla\)\mmaSub{E}{4}}{5\,\mmaSubSup{\(\tau\)}{2}{2}}\!\mmam\!\mmaFrac{24\,\(\pi\,\nabla\)\mmaSub{E}{2,2}}{\mmaSubSup{\(\tau\)}{2}{2}}\Big)\!+
  \quad\big(\mmaSubSup{s}{1,2}{3}\!+\!\mmaSubSup{s}{2,3}{3}\big)\,\Big(\!\!\mmam\!\mmaFrac{6\,\(\pi\)\,\mmaSub{E}{2}\,\(\nabla\)\mmaSub{E}{2}}{\mmaSubSup{\(\tau\)}{2}{2}}\!\mmam\!\mmaFrac{4\,\(\pi\,\nabla\)\mmaSub{E}{4}}{5\,\mmaSubSup{\(\tau\)}{2}{2}}\!\mmam\!\mmaFrac{12\,\(\pi\,\nabla\)\mmaSub{E}{2,2}}{\mmaSubSup{\(\tau\)}{2}{2}}\Big) \textnormal{,}
\end{mmaCell}
in agreement with the result in \cite{gerken2019e}. The next higher order in
$\ap$ of $\II_{1234}^{(2,0)}$ contains two tetrahedral graphs. One of them
vanishes by symmetry, the other one can be reduced to trihedral graphs by means
of the Fay identity \eqref{eq:302},
\begin{align}
  \cformtet{1\\0}{1\\1}{1\\1}{1\\0}{1\\1}{1\\1}&=0\\
  \begin{split}
    \cformtet{1\\0}{1\\0}{1\\1}{1\\1}{1\\1}{1\\1}&=
    -\cformtet{\varnothing}{\varnothing}{1\\1}{1\\1}{1\\1}{1&2\\1&0}
    -\cformtet{\varnothing}{2\\0}{1\\1}{1\\1}{1\\1}{1\\1}
    -\cformtet{2\\0}{\varnothing}{1\\1}{1\\1}{1\\1}{1\\1}\\
    &\qquad+\cformtet{\varnothing}{1\\0}{1\\1}{1\\1}{1\\1}{1&1\\0&1}
    -\cformtet{1\\0}{\varnothing}{1\\1}{1\\1}{1\\1}{1&1\\0&1}\,.
  \end{split}\label{eq:9}
\end{align}
If the Fay identity \eqref{eq:9} is added by hand, similarly to how
\eqref{eq:488} was added above, the expansion of $\II_{1234}^{(2,0)}$ can be
extended to the order $\ap^{4}$.


\section{Kinematic poles in three-point Koba--Nielsen integrals}
\label{sec:3pt-IBPs}

As explained in Section~\ref{sec:div-MGF-KN-poles}, a factor
$|f_{ij}^{(1)}|^{2}$ in a Koba--Nielsen integral leads to a naive expansion of this
integral terms of divergent \acp{MGF}. This signals a pole in one or more of the
Mandelstam variables which can be made explicit by means of integration-by-parts
manipulations. In this appendix we discuss the resulting expressions for all
three-point Koba--Nielsen integrals containing $|f_{ij}^{(1)}|^{2}$ factors
using the notation
\begin{align}
  W^{\tau}_{(a_{2},a_{3}|b_{2},b_{3})}(\sigma|\rho)&=
  \int\!\frac{\ddd^{2}z_{2}}{\tau_{2}}\frac{\ddd^{2}z_{3}}{\tau_{2}}
  \,\rho[f_{12}^{(a_{2})}f_{23}^{(a_{3})}]\sigma[\overline{f_{12}^{(b_{2})}}
  \,\overline{f_{23}^{(b_{3})}}]\KN_{3}\label{eq:561}
\end{align}
introduced in \cite{gerken2019d}. Here, the permutations
$\rho,\sigma \in \mathcal{S}_2$ act on the subscripts $i,j \in \{2,3\}$ of
$f^{(n)}$ and $\overline{f^{(n)}}$ but not on those of $a_i$ and $b_j$.

If only one $|f_{ij}^{(1)}|^{2}$ is present in the integrand and the
other $f^{(a)}$, $\overline{f^{(b)}}$ do not depend on $z_{i}$ or $z_{j}$, we
can use the puncture only occurring in $|f_{ij}^{(1)}|^{2}$ to integrate
by parts, obtaining one more term compared to \eqref{eq:360},
\begin{align}
  W^{\tau}_{(1{,}a|1{,}b)}(2{,}3|2{,}3)&=
  (-)^{a+1}\frac{s_{13}}{s_{12}}W^{\tau}_{(1{,}a|1{,}b)}(2{,}3|3{,}2)
  -\frac{1}{s_{12}}\frac{\pi}{\tau_{2}}
  W^{\tau}_{(0{,}a|0{,}b)}(2{,}3|2{,}3)\,,
  \label{eq:363}
\end{align}
with $a\neq1$ or $b\neq1$. Three more cases can be obtained from \eqref{eq:363}
by relabeling of the Mandelstam variables,
\begin{align}
  W^{\tau}_{(a{,}1|b{,}1)}(2{,}3|2{,}3)
  &=W^{\tau}_{(1{,}a|1{,}b)}(2{,}3|2{,}3)\Big|_{s_{12}\leftrightarrow s_{23}}\\
  W^{\tau}_{(1{,}a|1{,}b)}(3{,}2|3{,}2)
  &=W^{\tau}_{(1{,}a|1{,}b)}(2{,}3|2{,}3)\Big|_{s_{12}\leftrightarrow s_{13}}\\
  W^{\tau}_{(a{,}1|b{,}1)}(3{,}2|3{,}2)
  &=W^{\tau}_{(1{,}a|1{,}b)}(2{,}3|2{,}3)\Big|_{
    \substack{s_{12}\rightarrow s_{23}\\s_{23}\rightarrow s_{13}\\
      s_{13}\rightarrow s_{12}}}\,,
\end{align}
where again $a\neq1$ or $b\neq1$.

If both punctures $i$ and $j$ of $|f_{ij}^{(1)}|^{2}$ also appear in
other $f^{(a)}$, $\overline{f^{(b)}}$ factors, one obtains an additional term
from the action of $\partial_{\bar{z}}$ on the corresponding $f^{(a)}$ according
to \eqref{eq:364}. In this way, we obtain
\begin{align}
  W^{\tau}_{(a{,}1|b{,}1)}(3{,}2|2{,}3)&=
  \bigg\{\frac{s_{23}}{s_{13}}W^{\tau}_{(1{,}a|b{,}1)}(2{,}3|3{,}2)\label{eq:366}\\
  &\qquad+\frac{(-)^{b}}{s_{13}}\frac{\pi}{\tau_{2}}
  \Big[W^{\tau}_{(0{,}a|b{,}0)}(2{,}3|2{,}3)
  +(-)^{a-1}W^{\tau}_{(1{,}a-1|b{,}0)}(2{,}3|3{,}2)\Big]\bigg\}_{
    \substack{s_{12}\rightarrow s_{13}\\s_{13}\rightarrow s_{23}\\
      s_{23} \rightarrow s_{12}}}\nonumber\displaybreak[0]\\[0.7em]
  W^{\tau}_{(a{,}1|b{,}1)}(3{,}2|2{,}3)&=
  \bigg\{\frac{s_{23}}{s_{12}}W^{\tau}_{(a{,}1|1{,}b)}(2{,}3|3{,}2)\label{eq:367}\\
  &\qquad+\frac{1}{s_{12}}\frac{\pi}{\tau_{2}}
  \Big[ W^{\tau}_{(a{,}0|0{,}b)}(2{,}3|3{,}2)
  -W^{\tau}_{(a{,}0|1{,}b-1)}(2{,}3|3{,}2)\Big]\bigg\}_{
    \substack{s_{12}\rightarrow s_{23}\\s_{13}\rightarrow s_{12}
      \\s_{23} \rightarrow s_{13}}}\,,\nonumber
\end{align}
where $a\neq1$ in \eqref{eq:366} and $b\neq1$ in \eqref{eq:367} and we set
$\overline{f^{(-1)}}=0$. With the help of the Mandelstam relabelings, we avoid
the need of a Fay identity to write the \ac{RHS} in terms of the integrals
\eqref{eq:561}. One further case can be obtained by Mandelstam relabelings of
\eqref{eq:366} and \eqref{eq:367},
\begin{align}
  W^{\tau}_{(a{,}1|b{,}1)}(2{,}3|3{,}2)&=
  W^{\tau}_{(a{,}1|b{,}1)}(3{,}2|2{,}3)\Big|_{s_{12}\leftrightarrow s_{13}}\,,
  \label{eq:368}
\end{align}
where $a\neq1$ or $b\neq1$.

If two $|f_{ij}^{(1)}|^{2}$ factors are present in the integrand, we
obtain (on top of the poles for each $|f_{ij}^{(1)}|^{2}$) a three-point
kinematic pole $\sim \frac{1}{s_{123}}$, where the three-point Mandelstam
variable is defined in \eqref{eq:7}. Hence, the
integral
\begin{align}
  W^\tau_{(1,1|1,1)}(2{,}3|2{,}3) &=
  \int\dd\mu_2\,\big|f_{12}^{(1)}\big|^{2}\,\big|f_{23}^{(1)}\big|^{2}\,\KN_{3}
  \label{eq:370}
\end{align}
has pole structure
$\frac{1}{s_{123}}\big(\frac{1}{s_{12}}+\frac{1}{s_{23}}\big)$. Similarly, the
integral
\begin{align}
  W^\tau_{(1,1|1,1)}(3{,}2|2{,}3) &=-\int\dd\mu_2\,
  f_{12}^{(1)}\,\overline{f_{13}^{(1)}}\big|f_{23}^{(1)}\big|^{2}\,\KN_{3}
  \label{eq:371}
\end{align}
has pole structure $\frac{1}{s_{123}s_{23}}$. As discussed in Appendix~D of
\cite{gerken2019d}, these poles can be made explicit by the integration-by-parts
manipulation
\begin{align}
  W^{\tau}_{(1{,}1|1{,}1)}(2{{,}}3|2{{,}}3)
  &={-}\frac{s_{13}}{s_{123}}\left[ W^{\tau}_{(2{,}0|1{,}1)}(2{,}3|2{,}3)
    {+}W^{\tau}_{(0{,}2|1{,}1)}(2{,}3|2{,}3)
    {+}W^{\tau}_{(2{,}0|1{,}1)}(2{,}3|3{,}2) \right]\!\\
  &\qquad{-}\frac{1}{s_{123}}\frac{\pi}{\tau_{2}}\left[
    W^{\tau}_{(1{,}0|1{,}0)}(2{,}3|2{,}3)
    {+}W^{\tau}_{(0{,}1|0{,}1)}(2{,}3|2{,}3)\right]\nonumber\displaybreak[0]\\[1em]
  W^{\tau}_{(1{,}1|1{,}1)}(3{,}2|2{,}3)
  &={-}\frac{s_{13}}{s_{123}}\left[ W^{\tau}_{(2{,}0|1{,}1)}(3{,}2|2{,}3)
    {+}W^{\tau}_{(0{,}2|1{,}1)}(3{,}2|2{,}3)
    {+}W^{\tau}_{(2{,}0|1{,}1)}(3{,}2|3{,}2) \right]\!\\
  &\qquad{+}\frac{1}{s_{123}}\frac{\pi}{\tau_{2}}\left[
    W^{\tau}_{(1{,}0|1{,}0)}(3{,}2|2{,}3)
    {+}W^{\tau}_{(1{,}0|0{,}1)}(2{,}3|2{,}3)
    {+}W^{\tau}_{(0{,}1|0{,}1)}(2{,}3|2{,}3) \right]\,,\nonumber
\end{align}
where the formulas above can be used to manifest the two-particle poles on the
\ac{RHS}. The permutations of \eqref{eq:370}
and \eqref{eq:371} can again be obtained by relabeling the Mandelstam variables,
\begin{align}
  W^{\tau}_{(1{,}1|1{,}1)}(2{,}3|3{,}2)&=
  W^{\tau}_{(1{,}1|1{,}1)}(3{,}2|2{,}3)\Big|_{s_{12}\leftrightarrow s_{13}}\\
  W^{\tau}_{(1{,}1|1{,}1)}(3{,}2|3{,}2)&=
  W^{\tau}_{(1{,}1|1{,}1)}(2{,}3|2{,}3)\Big|_{s_{12}\leftrightarrow s_{13}}\,.
\end{align}


\printbibliography

\end{document}